\documentclass[11pt]{article}
\usepackage{verbatim,amsmath,amssymb}
\usepackage{epsfig,float,color}
\usepackage{geometry}
\usepackage{setspace}
\usepackage{natbib}
\usepackage{amsthm,mathrsfs,amsfonts,dsfont}
\usepackage{hyperref}
\usepackage[yyyymmdd,hhmmss]{datetime}
\usepackage[capitalise]{cleveref}
\usepackage{caption}
\usepackage{subcaption}
\usepackage{tablefootnote}
\usepackage{footmisc}[1] %footnote
\usepackage{placeins}
\usepackage{graphicx}
\usepackage{cmap}
\definecolor{darkblue}{rgb}{0,0,1}
\definecolor{cgn}{rgb}{0,0,0}
\definecolor{cgm}{rgb}{0,0,0}
\definecolor{cgn2}{rgb}{0,0,0}
\definecolor{cgm2}{rgb}{0,0,0}
\hypersetup{pdftex=true, colorlinks=true, breaklinks=true, linkcolor=darkblue, menucolor=darkblue, pagecolor=darkblue, citecolor=darkblue, urlcolor=darkblue}

\newcommand{\bitm}{\begin{itemize}}
\newcommand{\eitm}{\end{itemize}}
\newcommand{\bnumr}{\begin{enumerate}}
\newcommand{\enumr}{\end{enumerate}}
\newcommand{\bolds}[1]{\boldsymbol{#1}}

\newcommand {\aab}{a^{\alpha\beta}}
\newcommand {\auab}{a_{\alpha\beta}}
\newcommand {\agd}{a^{\gamma\delta}}
\newcommand {\abd}{a^{\beta\delta}}
\newcommand {\abg}{a^{\beta\gamma}}
\newcommand {\aad}{a^{\alpha\delta}}
\newcommand {\aag}{a^{\alpha\gamma}}

\newcommand {\augd}{a_{\gamma\delta}}
\newcommand {\augb}{a_{\gamma\beta}}

\newcommand {\adb}{a^{\delta\beta}}

\newcommand {\Aab}{A^{\alpha\beta}}
\newcommand {\Auab}{A_{\alpha\beta}}
\newcommand {\Aag}{A^{\alpha\gamma}}

\newcommand {\Aad}{A^{\alpha\delta}}

\newcommand {\Abg}{A^{\beta\gamma}}

\newcommand {\Agd}{A^{\gamma\delta}}

\newcommand {\Abd}{A^{\beta\delta}}

\newcommand {\Mab}{M^{\alpha\beta}}

\newcommand {\bab}{b^{\alpha\beta}}

\newcommand {\buab}{b_{\alpha\beta}}

\newcommand {\bugd}{b_{\gamma\delta}}

\newcommand {\tauab}{\tau^{\alpha\beta}}

\newcommand{\mrT}{\mathrm{T}}

\newcommand {\eqb}[1]{\begin{equation}\begin{array}{#1}}
\newcommand {\eqe}{\end{array}\end{equation}}

\newcommand {\esb}[1]{\begin{equation*}\begin{array}{#1}}
\newcommand {\ese}{\end{array}\end{equation*}}
\newcommand {\ds}{\displaystyle}

\newcommand {\pa}[2]{\frac{\partial{#1}}{\partial{#2}}}

\newcommand {\paqq}[3]{\frac{\partial^2{#1}}{\partial{#2}\,\partial{#3}}}
\newcommand {\back}{\! \! \!}
\newcommand {\is}{\back &=& \back}
\newcommand {\dis}{\back &:=& \back}

\newcommand {\plus}{\back &+& \back}
\newcommand {\mi}{\back &-& \back}

\newcommand {\tr}{\mathrm{tr}\,}

\newcommand {\dif}{\mathrm{d}}

\newcommand {\II}{{I\kern-.3em I}}
\newcommand {\III}{{I\kern-.3em I\kern-.3em I}}

\newcommand {\mra}{\mathrm{a}}

\newcommand {\mrs}{\mathrm{s}}

\newcommand {\mf}{\mathbf{f}}

\newcommand {\mk}{\mathbf{k}}

\newcommand {\mx}{\mathbf{x}}

\newcommand {\ba}{\boldsymbol{a}}
\newcommand {\bb}{\boldsymbol{b}}
\newcommand {\bc}{\boldsymbol{c}}
\newcommand {\bd}{\boldsymbol{d}}
\newcommand {\be}{\boldsymbol{e}}

\newcommand {\bi}{\boldsymbol{i}}

\newcommand {\bn}{\boldsymbol{n}}

\newcommand {\br}{\boldsymbol{r}}

\newcommand {\bt}{\boldsymbol{t}}
\newcommand {\bu}{\boldsymbol{u}}
\newcommand {\bv}{\boldsymbol{v}}

\newcommand {\bx}{\boldsymbol{x}}
\newcommand {\by}{\boldsymbol{y}}

\newcommand {\mC}{\mathbf{C}}

\newcommand {\mN}{\mathbf{N}}

\newcommand {\mP}{\mathbf{P}}

\newcommand {\mX}{\mathbf{X}}

\newcommand {\bA}{\boldsymbol{A}}

\newcommand {\bC}{\boldsymbol{C}}

\newcommand {\bE}{\boldsymbol{E}}
\newcommand {\bF}{\boldsymbol{F}}

\newcommand {\bI}{\boldsymbol{I}}

\newcommand {\bM}{\boldsymbol{M}}
\newcommand {\bN}{\boldsymbol{N}}

\newcommand {\bP}{\boldsymbol{P}}

\newcommand {\bR}{\boldsymbol{R}}
\newcommand {\bS}{\boldsymbol{S}}
\newcommand {\bT}{\boldsymbol{T}}
\newcommand {\bU}{\boldsymbol{U}}

\newcommand {\bX}{\boldsymbol{X}}
\newcommand {\bY}{\boldsymbol{Y}}
\newcommand {\bZ}{\boldsymbol{Z}}

\newcommand {\vphi}{\varphi}

\newcommand {\bsig}{\mbox{\boldmath$\sigma$}}

\newcommand {\IR}{{\rm\kern.24em
   \vrule width.02em height1.53ex depth-.05ex
   \kern-.3em R}}
\newcommand {\ic}{{\rm\kern.20em
   \vrule width.02em height1.0ex depth-.05ex
   \kern-.22em c}}
\newcommand {\ia}{{\rm\kern.20em
   \vrule width.02em height1.05ex depth-.0ex
   \kern-.25em a}}
\newcommand {\IC}{{\rm\kern.24em
   \vrule width.02em height1.4ex depth-.05ex
   \kern-.26em C}}
\newcommand {\ID}{{\rm\kern.34em
   \vrule width.02em height1.5ex depth-.05ex
   \kern-.36em D}}
\newcommand {\IS}{{\rm\kern.24em
   \vrule width.02em height1.6ex depth.05ex
   \kern-.26em S}}
\newcommand {\IT}{{\rm\kern.50em
   \vrule width.02em height1.55ex depth-.05ex
   \kern-.52em T}}

\newcommand {\IE}{{\rm\kern.24em
   \vrule width.02em height1.55ex depth-.05ex
   \kern-.33em E}}
\newcommand {\IEa}{{\rm\kern.24em
   \vrule width.02em height1.55ex depth-.05ex
   \kern-.33em E}^{1}_{ijkl}}
\newcommand {\IEb}{{\rm\kern.24em
   \vrule width.02em height1.55ex depth-.05ex
   \kern-.33em E}^{2}_{ijkl}}

\newcommand {\sA}{\mathcal{A}}

\newcommand {\sJ}{\mathcal{J}}

\newcommand {\Ass}[2]{\kern 0.9ex \vrule width0.45em height0.2ex depth0ex \kern -2.1ex \bigwedge_{#1}^{#2}}
\newcommand {\ASS}[2]{\kern 1.45ex \vrule width0.5em height0.2ex depth0ex \kern -2.65ex \bigwedge_{#1}^{#2}}

\newcommand {\aabgd}{{a}^{\alpha\beta\gamma\delta}}
\newcommand {\Aabgd}{{A}^{\alpha\beta\gamma\delta}}

\newcommand {\cabgd}{{c}^{\alpha\beta\gamma\delta}}
\newcommand {\dabgd}{{d}^{\alpha\beta\gamma\delta}}
\newcommand {\eabgd}{{e}^{\alpha\beta\gamma\delta}}
\newcommand {\fabgd}{{f}^{\alpha\beta\gamma\delta}}

\graphicspath{ {images/} }
\begin{document}

\begin{center}
%Print Date : \today, \currenttime\\
\Large{\bf{A new shell formulation for graphene structures based on \textcolor{cgn}{existing} ab-initio data}}\\

\end{center}

\begin{center}

\large{Reza Ghaffari\footnote{email: ghaffari@aices.rwth-aachen.de}, Thang X. Duong, Roger A. Sauer\footnote{corresponding author, email: sauer@aices.rwth-aachen.de}}\\
\vspace{4mm}

\small{\textit{
Aachen Institute for Advanced Study in Computational Engineering Science (AICES), \\
RWTH Aachen University, Templergraben 55, 52056 Aachen, Germany}} \\[1.1mm]
%
%$^\dag$Department of Chemical and Biomolecular Engineering, University of California at Berkeley, \\
%110 Gilman Hall, Berkeley, CA 94720-1460, USA \\[1.1mm]
%
%$^\ddag$Department of Mechanical Engineering, University of California at Berkeley, %\\
%6141 Etcheverry Hall, Berkeley, CA 94720-1740, USA

\vspace{4mm}

Published\footnote{This pdf is the personal version of an article whose final publication is available at \href{https://doi.org/10.1016/j.ijsolstr.2017.11.008}{http:/\!/sciencedirect.com}} in \textit{International Journal of Solids and Structures}, \href{https://doi.org/10.1016/j.ijsolstr.2017.11.008}{DOI: 10.1016/j.ijsolstr.2017.11.008}\\
Submitted on 28.~December 2016, Revised on 1.~November 2017, Accepted on 10.~November 2017
\end{center}

\vspace{3mm}

%\doublespacing

\rule{\linewidth}{.15mm}
{\bf Abstract}\\
\textcolor{cgn}{An existing hyperelastic membrane model for graphene calibrated from ab-initio data \citep{Kumar2014_01} is adapted to curvilinear coordinates and extended to a rotation-free shell formulation based on isogeometric finite elements. Therefore, the membrane model is extended by a hyperelastic bending model that reflects the ab-inito data of \citet{Kudin2001_01}. The proposed formulation can be implemented straight-forwardly into an existing finite element package, \textcolor{cgn2}{since it does not require the description of molecular interactions}. It thus circumvents the use of interatomic potentials that tend to be less accurate than ab-initio data.} The proposed shell formulation is verified and analyzed by a set of simple test cases. The results are in agreement to analytical solutions and satisfy the FE patch test. \textcolor{cgn}{The} performance of the shell formulation for graphene structures is illustrated \textcolor{cgn}{by several} numerical examples. The considered examples are indentation and peeling of graphene and torsion\textcolor{cgn}{, bending and axial stretch of} carbon nanotubes. Adhesive substrates are modeled by the Lennard-Jones potential and a coarse grained contact model. \textcolor{cgn}{In principle, the} proposed formulation can be extended to other 2D materials.


\vspace{-4mm}
\rule{\linewidth}{.15mm}
\section{Introduction}\label{s:Introduction}
Due to their extraordinary mechanical, thermal and electrical properties, carbon based structures (\textcolor{cgn}{like} graphene, carbon nanotube\textcolor{cgn}{s} (CNT), fullerene) are of high interest in many industrial applications \citep{Balandin2011_01,Marinho2012_01}. Graphene can be considered as the basic structure to build other carbon based structures. So, the first step to analyze and obtain material properties of these structures is the investigation of graphene.\\
Several methods are used to model and simulate graphene structures. Among them, first principle simulations are the most exact numerical methods. They simulate interaction of electrons, but they are limited to small time and length scales. The tight-binding (TB) method and density functional theory (DFT) can solve larger problems. However, they are still restricted to a few hundred atoms. The high computational cost of the mentioned methods motivates the usage of empirical or semi-empirical potentials in the framework of molecular dynamics (MD) simulations. These potentials concentrate on atomic interactions and ignore electron interactions \citep{Ansari2012_01}. When the size of a problem reaches a few hundreds of nano-meters, MD simulations become impractical and other methods should be used. Continuum based methods are another option, however classical continuum mechanics does not capture \textcolor{cgn}{size and boundary effects} at the nano scale \textcolor{cgn}{\citep{Cauchy1851}. These issues can be resolved by the surface Cauchy-Born rule \citep{Park2006_01}, boundary Cauchy-Born rule \citep{Abdolhosseini2011_01} and modified boundary Cauchy-Born rule \citep{Ghaffari2015_01}}.\\
The time and size restriction of quantum mechanics (QM) and MD methods and the inaccuracy of classical continuum \textcolor{cgn}{mechanical theories} motivate the use of multiscale methods. The most well-known multiscale method to simulate graphene and CNTs is the exponential Cauchy-Born (ECB) rule developed by \cite{Arroyo2002_01, Arroyo2004_01, Arroyo2004_02, Arroyo2005_01}. These works employ the first generation Brenner potential (FGBP) \citep{Brenner1990_01} and second generation Brenner potential (SGBP) \citep{brenner2002_01}. These potentials and the multiscale methods based on them, underestimate the elastic modulus and bending stiffness compared to QM simulations. \textcolor{cgn}{Hence}, the usage of \textcolor{cgn}{QM becomes necessary} \citep{Arroyo2004_01}. The elastic properties obtained by different methods are compared by \citet{Cao2014_01}. QM data can be either used to enrich the atomistic potential \citep{Lindsay2010_01}, or calibrate the strain energy density \citep{Xu2012_01}.\\ \textcolor{cgn}{Classical continuum theories have been extended to lattice structures by \citet{Ericksen1979}, \citet{pitteri1985_01} and \citet{Fadda2000_01}.} \textcolor{cgn}{Based on this, \citet{Sfyris2014_01} use Taylor expansion of the strain energy in order to describe graphene. \citet{Sfyris2014_02} obtain a set of invariants based on the right Cauchy-Green strain tensor, curvature tensor and shift vector and propose a strain energy functional based on a linear combination of those invariants.} \textcolor{cgn}{\citet{Delfani2013_01} and \citet{Delfani2015_01,Delfani2016_01} use Taylor expansion for the strain energy and apply the symmetry operators to the elasticity tensors in order to reduce the number of independent variables. These works consider a finite thickness for graphene and use through the thickness integration of the strain energy in order to obtain the bending stiffness. However, the thickness of graphene is a controversial quantity. On the other hand, the bending modulus can be obtained directly from the change of the bond and dihedral angles \citep{Lu2009_01} without introducing a thickness. \citet{Kumar2014_01} proposed a membrane surface strain energy per unit area that does not account for bending resistance. This is included here by using the Canham bending strain energy \citep{CANHAM1970_01} in conjunction with the nonlinear shell formulation of \citet{Duong2016_01}, which does not require the notion of a thickness.}\\
Experimental tests can be considered as an alternative method to capture the properties of graphene. Such tests are very important to calibrate and validate numerical simulations. The most common experimental tests are uniaxial stretching, pure shear, simple shear, torsion and indentation. The uniaxial and biaxial tests require grips to hold a specimen and it is difficult to build nanoscale grips. The experimental indentation test setup needs no grips and it can be built by laminating a graphene sheet on a substrate, e.g.~$\textnormal{SiO}_2$, via exfoliation. Force-displacement curves can then be obtained by using an atomic-force microscope (AFM). It should be noted that initial stress within the graphene and adhesion to the substrate play an important role in experimental results \citep{Lee2008_01}.\\
The adhesion energy between a substrate and graphene can be obtained from blister tests \citep{Koenig2011_01,Boddeti2013_01}. The atomic structure of graphene on a substrate can be investigated experimentally \citep{Masa2007_01} and theoretically  \textcolor{cgn}{\citep{Zubaer2009_01, Edson2010_01, Lu2010_01,He2013_01, Gao2014_01,Sfyris2017_01}}. The corrugation of substrates is investigated by \citet{Li2010_01} and \citet{Aitken2010_01}. Adhesion effects on micro-mechanical structures are investigated by \citet{Maboudian1997_01}. Recent development on the adhesion of graphene membranes is reviewed by \citet{Bunch2012_01}. A continuum model is developed for multi-buckling of graphene on a substrate by \citet{Gao2016_01}. The effects of substrate morphology on graphene is investigated by a continuum model and verified by an atomistic simulation \citep{boddeti2016_01}. The effects of substrate \textcolor{cgn2}{rippling} on the bending stiffness of a graphene are considered by \citet{Jomehzadeh2015_01}.\\ Results of the indentation test can be used to obtain graphene properties. However, connecting in-plane material properties with the force-displacement curve of the indentation test is a challenging issue. The connection between the indentation results and in-plane material properties is investigated by \citet{Han2015_01}. Moreover, this \textcolor{cgn2}{test} does not yield any direct information about the constitutive model, like the functional form of the strain energy density. It has been verified that graphene behaves isotropically in infinitesimal strain \cite{Cao2014_01}. But under large deformations the material response depends on the relative direction of loading and the chirality. Therefore, graphene can not be considered isotropic in general \citep{Larsson2011_01,Cao2014_01}. The anisotropic response of graphene can be simulated with the structural tensor obtained from the symmetry groups of the lattice structure \citep{Kumar2014_01}. \textcolor{cgn}{In this approach, a set of virtual DFT experiments are conducted and the coefficients of the strain energy are calibrated.} Hence, there is no need to study the deformation of the lattice as it is done in the Cauchy-Born rule.\\
In the research group of the authors, a novel curvilinear finite element (FE) formulation has been developed and applied to \textcolor{cgn2}{computations of} liquid and solid membranes \citep{Sauer2014_01}. It has been extended to rotation-free shells \citep{Sauer2015_02,Duong2016_01} and employed in the modeling of anisotropic materials \citep{Roohbakhshan2016_01}. The new formulation is based on isogeometric FE \citep{cottrell2009} that provides higher accuracy than Lagrange-based FE.
It is extended here to model and simulate graphene based structures.\vspace{2.5mm}\\
The highlights and novelties of the \textcolor{cgn}{proposed new shell model} are
\begin{itemize}
\item \textcolor{cgn}{\textcolor{cgn2}{It combines the existing ab-initio data of \citet{Kudin2001_01} and \citet{Kumar2014_01}} with} the rotation-free isogeometric shell model of \citet{Duong2016_01}.
\item \textcolor{cgn}{It can be implemented straightforwardly within the finite element method, which is more efficient than molecular dynamics for large systems.}
\item \textcolor{cgn}{It} is closer to experimental and quantum results than shell models
       based on interatomic potentials, since those \textcolor{cgm}{often} underestimate the elastic modulus of graphene by about one third.
\item \textcolor{cgn}{\textcolor{cgn2}{It is directly based on a surface strain energy defined in closed form
that avoids the evaluation of molecular interactions as they appear, e.g. in the Cauchy-Born rule.}}
\item \textcolor{cgn}{It} accounts for contact and adhesion, allowing to study indentation and peeling of graphene.
\item \textcolor{cgn}{It} is fully nonlinear and can be used to accurately analyze the buckling behavior of graphene structures.
\item \textcolor{cgn}{It} can be extended to other 2D material models\textcolor{cgm2}{, like those of} \citet{Sfyris2015_01}.
\end{itemize}
The remainder of this paper is organized as follows: In Sec.~\ref{s:kinematics}, the kinematics of deforming surfaces are presented and based on that, in Secs.~\ref{ss:membrane_energy} and \ref{ss:bending_part_graphene}, the membrane and bending constitutive laws are developed. Details of the considered FE formulation are provided in Sec.~\ref{s:Finite_element_formulation}. In Sec.~\ref{s:Verification}, the proposed constitutive law and FE formulation are verified by several elementary benchmark tests. Further, numerical examples are presented in Sec.~\ref{s:Numerical_example} to demonstrate the capability of the proposed model for the simulation of graphene based specimens. The paper is concluded in Sec.~\ref{s:conclusion}.
\section{Kinematics}\label{s:kinematics}
In this section, the kinematics of deforming surfaces are described. This description is then used in \textcolor{cgn2}{the} constitutive modeling and FE formulation.
\subsection{Kinematics of deforming surfaces}
The surface description is essential for membrane and shell formulations. \cite{Sauer2015_02} proposed a \textcolor{cgn2}{rotation-free curvilinear framework for membranes and shells}. The parametric description of the surface in the reference and the current configuration can be written as
\eqb{l}
\bX = \ds \bX(\xi^{1},\xi^{2})~,
\eqe
\eqb{l}
\bx = \ds \bx(\xi^{1},\xi^{2})~,
\eqe
and the corresponding tangent vectors are
\eqb{l}
\bA_{\alpha} = \ds \pa{\bX}{\xi^{\alpha}}~,
\eqe
\eqb{l}
\ba_{\alpha} = \ds \pa{\bx}{\xi^{\alpha}}~.
\eqe
Next, the co-variant components of the metric tensors are defined from the inner product as
\eqb{l}
\Auab = \bA_{\alpha}\cdot\bA_{\beta}~,
\eqe
\eqb{l}
\auab = \ba_{\alpha}\cdot\ba_{\beta}~,
\eqe
and the contra-variant components of the metric tensors are defined as the inverse of the co-variant metric tensors, i.e.
\eqb{l}
[\Aab] = [\Auab]^{-1}~,
\eqe
\eqb{l}
[\aab] = [\auab]^{-1}~.
\eqe
The dual tangent vectors can then be defined as
\eqb{lll}
\bA^{\alpha} \dis \Aab\,\bA_{\beta}~,
\eqe
\eqb{lll}
\ba^{\alpha} \dis \aab\,\ba_{\beta}~.
\eqe
The normal unit vector of the surface in its reference and current configuration can then be written as
\eqb{l}
\bN =\ds \frac{\bA_{1}\times\bA_{2}}{\|\bA_{1}\times\bA_{2}\|}~,
\eqe
\eqb{l}
\bn =\ds \frac{\ba_{1}\times\ba_{2}}{\|\ba_{1}\times\ba_{2}\|}~.
\eqe
The 3D identity tensor $\bolds{1}$ can be written in terms of the surface identity tensors in reference \textcolor{cgn2}{configuration,} $\bI$ \textcolor{cgn2}{, and} current configuration\textcolor{cgn2}{, $\bi$ ,} as
\eqb{l}
\bolds{1} = \bI+\bN\otimes\bN = \bi+\bn\otimes\bn~,
\eqe
where $\bI$ and $\bi$ can be written as
\eqb{l}
\bI = \bA_{\alpha}\otimes\bA^{\alpha} = \Auab\,\bA^{\alpha}\otimes\bA^{\beta} = \Aab\,\bA_{\alpha}\otimes\bA_{\beta}~,
\eqe
\eqb{l}
\bi =\ba_{\alpha}\otimes\ba^{\alpha}  = \auab\,\ba^{\alpha}\otimes\ba^{\beta} = \aab\,\ba_{\alpha}\otimes\ba_{\beta}~.
\eqe
Next, the curvature tensor can be written as
\eqb{l}
\bb = \buab\,\ba^{\alpha}\otimes\ba^{\beta}~,
\eqe
where $\buab$ are the co-variant components of the curvature tensor defined as
\eqb{l}
\buab := \bn\cdot\ba_{\alpha,\beta}=\bn\cdot\ba_{\alpha;\beta}~,
\eqe
and $\ba_{\alpha,\beta}$ and $\ba_{\alpha;\beta}$ are the parametric and co-variant derivatives of the tangent vectors. They are connected by
\eqb{l}
\ba_{\alpha;\beta} = \ba_{\alpha,\beta} -\Gamma^{\gamma}_{\alpha\beta}\,\ba_{\gamma}~,
\eqe
where $\Gamma^{\gamma}_{\alpha\beta}$ is the Christoffel symbol of the second kind, which is defined as
\eqb{lll}
\Gamma^{\gamma}_{\alpha\beta}\ \dis \ba_{\alpha,\beta}\cdot\ba^{\gamma}~.
\eqe
The contra-variant components of $\bb$ are defined as

\eqb{lll}
\bab \dis \aag\,b_{\gamma\delta}\,\adb~.
\eqe

\textcolor{cgn}{The frame indifference of} the material model is an important requirement in continuum mechanics.  \textcolor{cgn}{It} is satisfied if models are written in term of tensor invariants. The mean and Gaussian curvatures are a good set of invariants to characterize the bending energy. They are defined as
\eqb{l}
H := \ds  \frac{1}{2}\bb:\bi = \frac{1}{2}b^{\alpha}_{\alpha}~,
\eqe
\eqb{l}
\kappa := \ds \det(\bb) =\frac{\det[\buab]}{\det[\augd]}~.
\eqe
The definition of the determinant of surface \textcolor{cgn}{tensors} can be found in \cite{Javili2014_01}. Alternatively, the invariants can be defined based on the principal curvatures $\kappa_{\alpha}$ as
\eqb{lll}
H \dis \ds \frac{1}{2}(\kappa_1+\kappa_2)~,
\eqe
\eqb{lll}
\kappa \dis \ds \kappa_1\,\kappa_2~.
\eqe
The principal curvatures are the eigenvalues of matrix $[\buab\,\abg]$.\\
\subsection{Kinematics of deformation}
The first step to develop a material model is the introduction of stress and strain measures. The logarithmic strain $\bE^{(0)}$ is sensitive to material nonlinearities even for small strains. Furthermore, it can be additively decomposed into volumetric and shear parts. These features make it an excellent candidate for the development of material models \citep{Neff2013_01,  Kumar2014_01, Neff2015_01, Neff2015_02, Ghiba2015_01, montella2016_01, Neff2016_01}. The Lagrangian logarithmic strain is defined as the logarithm of the right stretch tensor $\bU$. The standard method to calculate non-integer powers and the logarithm of a tensor is based on the spectral decomposition. The surface deformation gradient can be written as
\eqb{l}
\bF = \bR\,\bU~,
\label{e:decomposition_of_gradient_deformation}
\eqe
where $\bR$  is the rotation tensor. Using the spectral decomposition, $\bU$ can be written based on its eigenvalues $\lambda_i$ and eigenvectors $\textcolor{cgn}{\bY_{\!i}}$ , which are the principal stretches and their directions, respectively. So, the spectral decomposition of $\bU$ is
\eqb{l}
\bU  = \ds \sum\limits_{i=1,2} {\lambda_i\,\textcolor{cgn}{\bY_{\!i}}\otimes\textcolor{cgn}{\bY_{\!i}}}~.
\label{e:stretch_polar_decomposition}
\eqe
Using the same decomposition, the right Cauchy-Green tensor $\bC$ can be written based on its eigenvalues $\Lambda_i$ and eigenvectors $\textcolor{cgn}{\bY_{\!i}}$. It should be noted that $\bU$ and $\bC$ have the same principal directions, i.e.~eigenvectors, and $\Lambda_i = \lambda_i^2$. So, $\bC$ can be written as
\eqb{l}
\bC = \ds \sum\limits_{i=1,2}{\Lambda_i\,\textcolor{cgn}{\bY_{\!i}}\otimes\textcolor{cgn}{\bY_{\!i}}}~.
\eqe
The eigenprojection tensors $\textcolor{cgn}{\bY_{\!i}}\otimes\textcolor{cgn}{\bY_{\!i}}$ can be analytically obtained from Sylvester's formula \citep{itskov2015_01}
\eqb{lll}
\mP_1:= \textcolor{cgn}{\bY_{\!1}}\otimes\textcolor{cgn}{\bY_{\!1}} = \ds\frac{\bC - \Lambda_2\,\bI}{\Lambda_1 - \Lambda_2} =: P^1_{\alpha\beta}\,\bA^\alpha\otimes\bA^\beta~,\\[4mm]
\mP_2:= \textcolor{cgn}{\bY_{\!2}}\otimes\textcolor{cgn}{\bY_{\!2}} = \ds\frac{\bC - \Lambda_1\,\bI}{\Lambda_2 - \Lambda_1} =: P^2_{\alpha\beta}\,\bA^\alpha\otimes\bA^\beta~,
\label{e:P12}
\eqe
where $P^i_{\alpha\beta}$ are defined as
\eqb{lll}
P^1_{\alpha\beta} \dis \ds\frac{1}{\Lambda_1 - \Lambda_2}\,(\auab - \Lambda_2\,\Auab)~,\\[4mm]
P^2_{\alpha\beta} \dis \ds\frac{1}{\Lambda_2 - \Lambda_1}\,(\auab - \Lambda_1\,\Auab)~.
\eqe
Using the spectral decomposition of $\bU$, the logarithmic strain can be written as
\eqb{lll}
\bE^{(0)} \is \ln(\bU) = \ds \frac{1}{2}\ln(J)\,\bI+\ln\lambda\,(\textcolor{cgn}{\bY_{\!1}}\otimes\textcolor{cgn}{\bY_{\!1}} -\textcolor{cgn}{\bY_{\!2}}\otimes\textcolor{cgn}{\bY_{\!2}} )~,
\label{e:logaritmic_strain}
\eqe
where the two terms on the right hand side are the aeral \textcolor{cgn}{(surface dilatation)} and deviatoric parts of the logarithmic strain, and $\lambda$ and $J$ are defined as
\eqb{lll}
\lambda \dis \ds \sqrt{\frac{\lambda_1}{\lambda_2}}~,
\eqe
\eqb{lll}
J\dis\lambda_1\,\lambda_2~.
\eqe
They are tensor invariants, just like $H$ and $\kappa$. \\
Based on the presented kinematics, a hyperelastic strain energy function will be introduced for curvilinear coordinates in the next section.

\section{Material model}\label{s:material_model}
In this section, the membrane constitutive law of \citet{Kumar2014_01} is extended to the curvilinear
shell theory of \citet{Sauer2015_02} and its computational counterpart  \citep{Duong2016_01}. \\
For hyperelastic materials, it is assumed that all energy is stored as elastic energy and the dissipation is zero. The strain energy density \textcolor{cgn}{is defined directly on the surface (as energy per reference area). It should not be confused with strain energy per volume, which is not needed here. For Kirchhoff-Love kinematics, the strain energy can be written as a function of the metric and curvature tensor.} \textcolor{cgn2}{Here, the strain energy is defined analytically in closed form such that the evaluation of molecular interactions is avoided.} In the framework of the shell model of \citet{Sauer2015_02}, the derivative of the strain energy density \textcolor{cgn}{with respect to $\auab$ and $\buab$} gives the stress and the bending moment as
\eqb{l}
\tauab = \ds 2\pa{W}{\auab}~,
\eqe
\eqb{l}
\Mab_{0} = \ds \pa{W}{\buab}~.
\eqe
In addition, the elasticity tensors are defined as
\eqb{lll}
\cabgd \dis \ds 4\paqq{W}{\auab}{\augd}~,
\eqe
\eqb{lll}
\dabgd \dis \ds 2\paqq{W}{\auab}{\bugd}~,
\eqe
\eqb{lll}
\eabgd \dis \ds 2\paqq{W}{\buab}{\augd}~,
\eqe
\eqb{lll}
\fabgd \dis \ds \paqq{W}{\buab}{\bugd}~,
\eqe
where $W$ can be written as summation of membrane and bending parts,
\eqb{l}
W = \ds W_{\text{m}}+W_{\text{b}}~.
\eqe
Given $\tauab$ and $M_0^{\alpha\beta}$, the Cauchy stress tensor can be written as
\eqb{lll}
\bsig \is N^{\alpha\beta}\,\ba_{\alpha}\otimes \ba_{\beta} + S^{\alpha}\,\ba_{\alpha}\otimes\bn~,
\label{e:shell_stress}
\eqe
where
\eqb{lll}
N^{\alpha\beta} \is \ds \sigma^{\alpha\beta}+b_{\gamma}^{\alpha}\,M^{\gamma\beta}~,
\eqe
\eqb{lll}
S^{\alpha} \is \ds -M_{;\beta}^{\beta\alpha}~,
\eqe
and $\sigma^{\alpha\beta} = \tauab/J$ and $M^{\alpha\beta} = M_{0}^{\alpha\beta}/J~.$
The mixed in-plane components of $\bsig$ are defined as
\eqb{lll}
N^{\alpha}_{\beta} \dis N^{\alpha\gamma}\,a_{\gamma\beta}~.
\label{e:mix_N}
\eqe
\subsection{Membrane energy}\label{ss:membrane_energy}
\begin{figure}
        \centering
     \includegraphics[height=55mm]{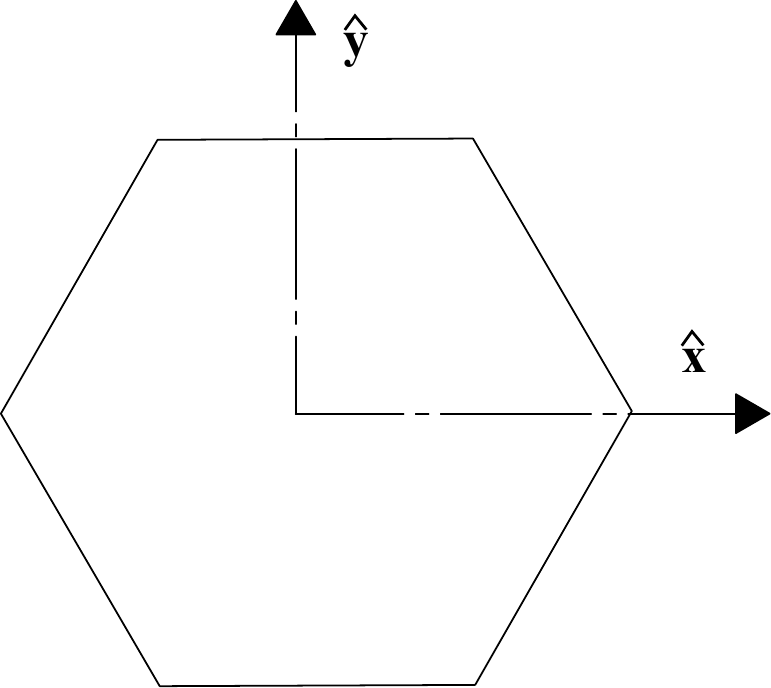}
    \caption{Orthonormal vectors characterizing the \textcolor{cgm2}{undeformed} graphene lattice.}
    \label{f:lattice}
\end{figure}

\begin{figure}[h]
    \begin{subfigure}[t]{0.32\linewidth}
        \centering
     \includegraphics[height=39mm]{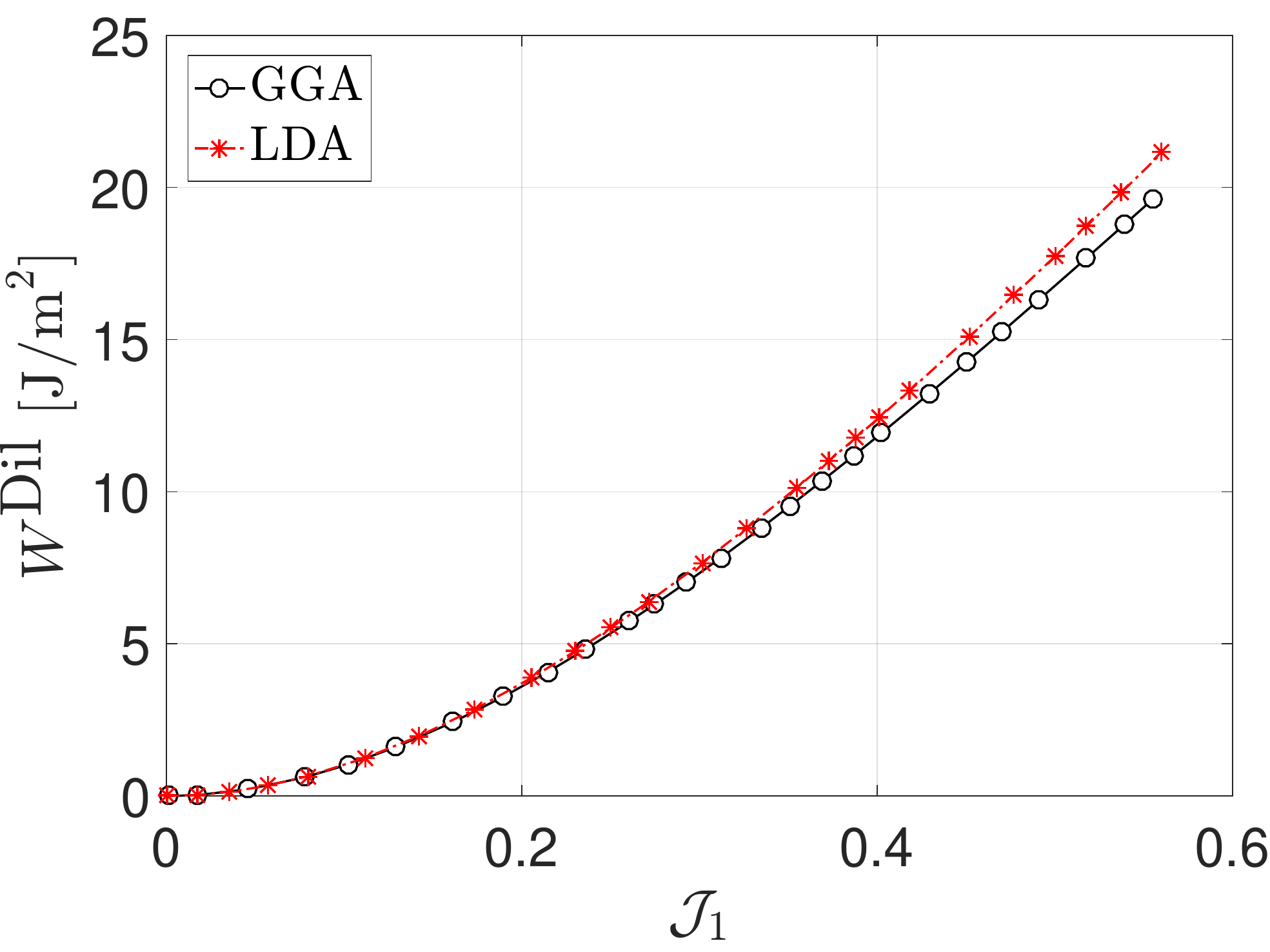}
        \subcaption{}
        \label{f:dilatation_strain_engery_comparsion}
    \end{subfigure}
    \begin{subfigure}[t]{0.32\linewidth}
        \centering
 \includegraphics[height=39mm]{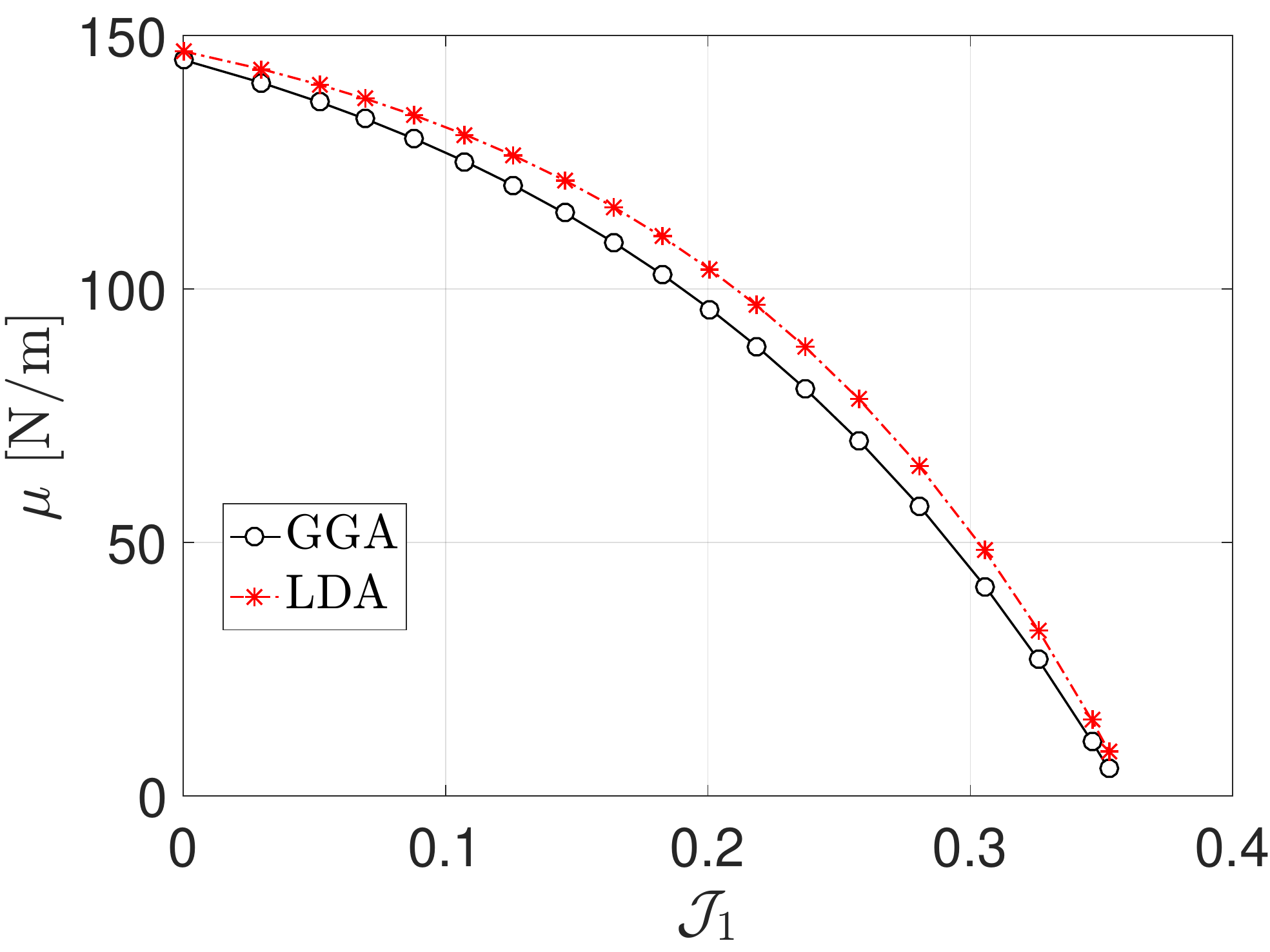}
        \subcaption{}
        \label{f:mu_comparsion}
    \end{subfigure}
    \begin{subfigure}[t]{0.32\linewidth}
        \centering
    \includegraphics[height=39mm]{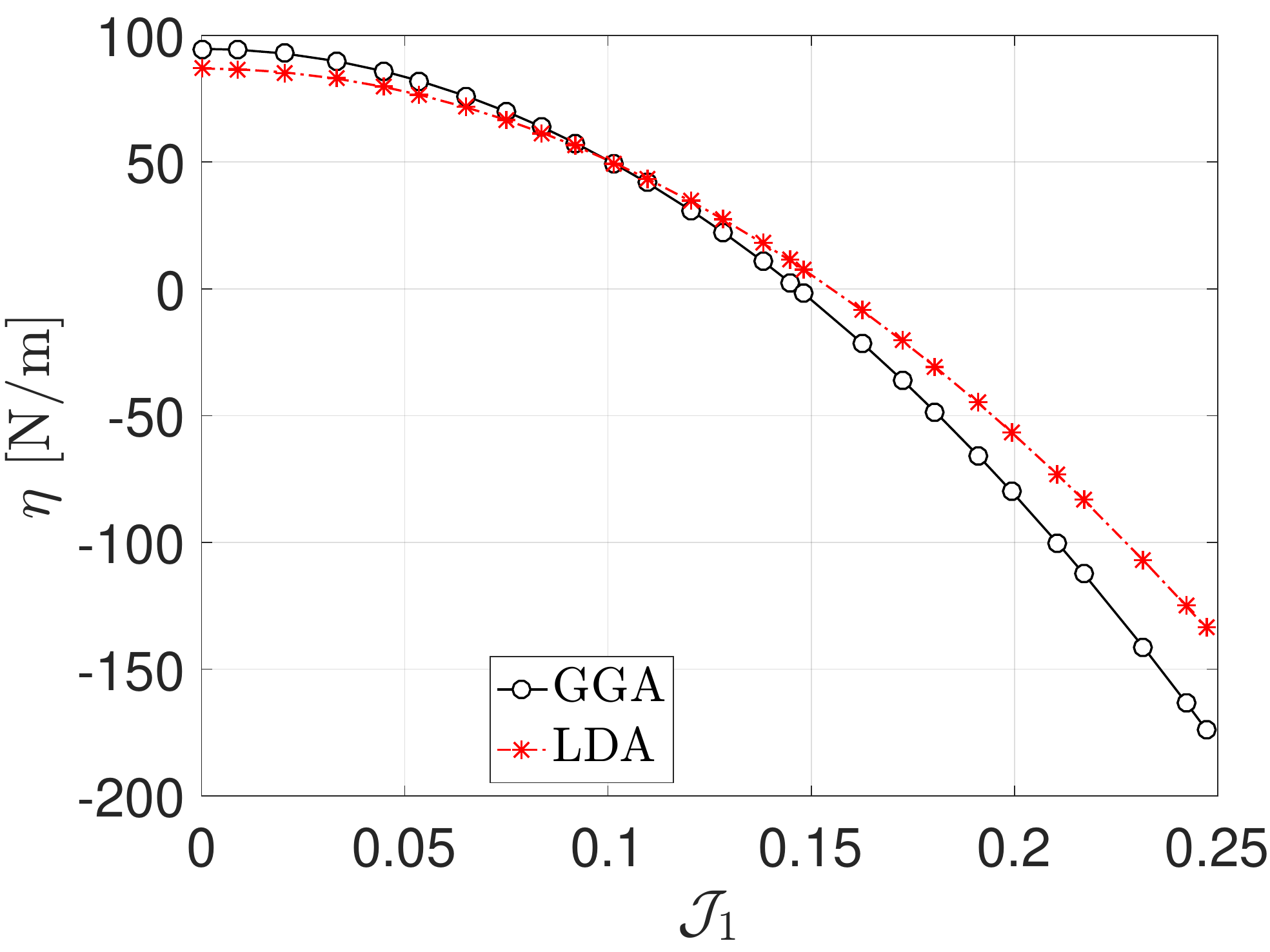}
        \subcaption{}
        \label{f:eta_comparsion}
    \end{subfigure}
    \caption{Membrane behavior: Variation of (\subref{f:dilatation_strain_engery_comparsion}) dilatation Energy $W_{\text{m}}^{\mathrm{dil}}$,
             (\subref{f:mu_comparsion}) shear modulus $\mu$ and
             (\subref{f:eta_comparsion}) anisotropic part $\eta$ with $\sJ_1$ for the model of \citet{Kumar2014_01}.}
    \label{f:material_behaviour}
\end{figure}

Graphene is a thin 2D structure with a thickness of only one atom. This motivates the development of material models based on membrane and shell theories.
A suitable strain energy function should be selected to capture the anisotropic behavior of graphene. Based on the isotropization theorem, an anisotropic tensorial functional can be written as an isotropic tensorial functional by utilizing structural tensors \citep{Zheng1994_01}. So, the strain energy density can be isotropized by using the invariants obtained based on isotropization. Based on the symmetry group of graphene and using the logarithmic strain, the following three invariants can be introduced \citep{Kumar2014_01}
\eqb{lll}
\sJ_1 \dis \epsilon_\mra = \ln J~;\quad J:=\lambda_1\,\lambda_2~, \\[3mm]
\sJ_2 \dis \ds\ \frac{1}{4}\,\gamma_i^2 := \frac{1}{2}\,\textcolor{cgm}{\bE^{(0)}_{\text{dev}}}:\textcolor{cgm}{\textcolor{cgm}{\bE^{(0)}_{\text{dev}}}} =(\ln\lambda)^2;~\lambda_1 \ge \lambda_2~,\\[3mm]
\sJ_3 \dis \ds\frac{1}{8}\,\gamma_\theta^3 :=  \ds\frac{1}{8}\left[ \left(\textcolor{cgm}{\hat{\bM}}:\textcolor{cgm}{\bE^{(0)}_{\text{dev}}}\right)^3 - 3\,\left(\textcolor{cgm}{\hat{\bM}}:\textcolor{cgm}{\bE^{(0)}_{\text{dev}}}\right)\left(\textcolor{cgm}{\hat{\bN}}:\textcolor{cgm}{\bE^{(0)}_{\text{dev}}}\right)^2\right]=(\ln\lambda)^3\cos(6\theta)~,
%= (\ln\lambda)^3\,\cos(6\,\theta)~ ~,
\label{e:defsJ}
\eqe
where $\textcolor{cgm}{\bE^{(0)}_{\text{dev}}}$ is the deviatoric part of the logarithmic strain (Appendix~\ref{s:preliminary}). \textcolor{cgn}{\textcolor{cgn2}{A} similar approach is used by \citet{Sfyris2014_02} to obtain another set of invariants.} Here the first and second invariants model the isotropic features of the material, and the third invariant captures anisotropic behavior of the material. $\textcolor{cgm}{\hat{\bM}}$ and $\textcolor{cgm}{\hat{\bN}}$ are defined as
\eqb{lll}
\textcolor{cgm}{\hat{\bM}} := \ds \hat{\bx} \otimes \hat{\bx} - \hat{\by} \otimes \hat{\by}~,
\eqe
\eqb{lll}
\textcolor{cgm}{\hat{\bN}} := \ds \hat{\bx} \otimes \hat{\by} + \hat{\by} \otimes \hat{\bx}~,
\eqe
where $\hat{\bx}$ and $\hat{\by}$ are two orthonormal vectors. The former is aligned along the armchair direction (Fig.~\ref{f:lattice}). $\theta$ is the maximum stretch angle relative to the armchair direction and defined as
\eqb{lll}
\theta \dis \arccos{(\bY_1\cdot\hat{\bx})}~.
\eqe
Then, the membrane strain energy density, per unit area of the initial configuration, can be decomposed into pure dilatation $W_{\text{m}}^{\mathrm{dil}}$ and deviatoric $W_{\text{m}}^{\mathrm{dev}}$ parts as
\eqb{l}
W_{\text{m}}(\sJ_1,\sJ_2,\sJ_3) = W_{\text{m}}^{\mathrm{dil}}(\sJ_1) + W_{\text{m}}^{\mathrm{dev}}(\sJ_2,\,\sJ_3;\sJ_1)~.
%\label{e:total_strain_energy}
\eqe
The dilatation and deviatoric parts of the strain energy density are
\eqb{lll}
 W_{\text{m}}^{\mathrm{dil}} \dis \varepsilon\big[1 - (1+\hat{\alpha}\,\epsilon_\mra)\,\exp(-\hat{\alpha}\,\epsilon_\mra)\big]~,\\[2mm]
 W_{\text{m}}^{\mathrm{dev}} \dis 2\,\mu(\epsilon_\mra)\,\sJ_2 + \eta(\epsilon_\mra)\,\sJ_3~,
 %\label{e:decomposed_strain_energy}
\eqe
where $\mu$ and $\eta$ are defined as
\eqb{lll}
\mu(\epsilon_\mra) \dis \mu_0 - \mu_1\,e^{\hat{\beta}\,\epsilon_\mra}~,\\[2mm]
\eta(\epsilon_\mra) \dis \eta_0 - \eta_1\,\epsilon_\mra^2~.
%\label{e:mu_and_eta_relation}
\eqe
The membrane model is thus characterized by the seven material constants
$\varepsilon$, $\hat{\alpha}$, $\mu_0$, $\mu_1$, $\hat{\beta}$, $\eta_0$, and $\eta_1$ . They are given in Tab.~\ref{t:Graphene_cons}. In that table, LDA and GGA are abbreviations for local density approximation and generalized gradient approximation. These are two approximations used in DFT simulations in order to compute the material constants.  Using the relations given in the previous section, the contra-variant component of the Kirchhoff stress tensor for the cases of distinct and repeated eigenvalues are
\eqb{lll}
\tauab_{\text{m}}  \is \big[ \varepsilon\,\hat{\alpha}^2\,\epsilon_\mra\,e^{-\hat{\alpha}\,\epsilon_\mra} - 2\,(\eta_1\,\epsilon_\mra\,\sJ_3 + \mu_1\,\hat{\beta}\,\sJ_2\,e^{\hat{\beta}\,\epsilon_\mra})\big]\,\aab + 2\,\mu\,\ln\lambda\,\chi^{\alpha\,\beta} + \ds\frac{1}{4}\,\eta\,\mu^{\alpha\beta},~\text{for} ~\lambda_1\ne\lambda_2~,
\label{e:tau_distinct_curvilinear}
\eqe
\textcolor{cgn2}{and}
\eqb{lll}
\tauab_{\text{m}}  \is \big[ \varepsilon\,\hat{\alpha}^2\,\epsilon_\mra\,e^{-\hat{\alpha}\,\epsilon_\mra} \big]\,\aab,~\text{for}~\lambda_1=\lambda_2~,
\label{e:tau_repeated_curvilinear}
\eqe
where $\mu^{\alpha\beta}$ and $\chi^{\alpha\,\beta}$ are defined as
\eqb{lll}
\mu^{\alpha\beta} \dis
\ds \sum\limits_{i,j=1,2}{f_{ij}\, P_i^{\alpha\gamma}\,T^{(0)}_{\gamma\delta}\,P_j^{\delta\beta}}~,
\eqe
\eqb{lll}
\chi^{\alpha\beta} \dis \ds \frac{1}{\lambda_1^2}\,P_1^{\alpha\beta}\, -\frac{1}{\lambda_2^2}\,P_2^{\alpha\beta}~,
\eqe
and $f_{ij}$ is defined as
\eqb{lll}
f_{ij} \dis \left\lbrace \begin{array}{l}
\ds\frac{1}{2\,\lambda_i^2}~,\quad $if $ i=j~, \\[4mm]
\ds\frac{\ln\lambda_i - \ln\lambda_j}{\lambda_i^2 - \lambda_j^2}~,\quad $if $ i\neq j~.
\end{array} \right.
\eqe
$T^{(0)}_{\gamma\delta}$ and the elasticity tensors related to Eqs.~(\ref{e:tau_distinct_curvilinear})
and (\ref{e:tau_repeated_curvilinear}) are given in Appendix \ref{s:Derivation_membrane_ constitutive_law_based_on_curvilinear_coordinate}. \\ The variation of  $W_{\text{m}}^{\mathrm{dil}}$, $\mu$ and $\eta$ with $\sJ_1$  are presented in Fig.~\ref{f:material_behaviour}. The elastic modulus of the current model is compared with \textcolor{cgm2}{reference data} in Tab.~\ref{t:elastic_modulus_comparsion}. The elastic modulus of the current model is closer to \textcolor{cgn2}{existing} experiments and ab\textcolor{cgm2}{-}initio results \textcolor{cgm}{than the exponential Cauchy-Born rule}. \\
\begin{table}
  \centering
    \begin{tabular}{c c c c c c c c }
      \hline
        & $\hat{\alpha}$ & $\varepsilon~[\textnormal{N/m}]$  &  $\mu_0~[\textnormal{N/m}]$ & $\mu_1~ [\textnormal{N/m}]$ & $\hat{\beta}$ & $\eta_0~[\textnormal{N/m}]$ & $\eta_1~[\textnormal{N/m}]$ \\
     % \hline\hline
      GGA  & 1.53 & 93.84  & 172.18 & 27.03 & 5.16 & 94.65 & 4393.26 \\
      LDA  & 1.38 & 116.43 & 164.17 & 17.31 & 6.22 & 86.9\footref{fn:Sandeep_mat_constant} & 3611.5\footref{fn:Sandeep_mat_constant} \\
      \hline
    \end{tabular}
  \caption{Membrane behavior: Material constants of graphene \citep{Kumar2014_01}.}\label{t:Graphene_cons}
\end{table}

%\footnotetext{This is the labeled footnote\label{note1}}
\footnotetext{\label{fn:Sandeep_mat_constant} These material parameters were provided by S. Kumar and D. M. Parks on a personal request by the authors as the material parameters in \cite{Kumar2014_01} contain errors. }

\begin{table}
\begin{tabular}{ l c }
  \hline
    & Elastic modulus E [$\textcolor{cgm}{\mathrm{N/m}}$] \\
  \textcolor{cgm}{Exp. Cauchy-Born} with FGBP \citep{Arroyo2004_01} & 236 \\
  \textcolor{cgm}{Exp. Cauchy-Born} with SGBP \citep{Arroyo2004_01} & 243 \\
  \textcolor{cgm}{MM3 \citep{Gupta2010_01}} & \textcolor{cgm}{340}\\
  Current model with GGA \citep{Kumar2014_01} & 349 \\
  Current model with LDA \citep{Kumar2014_01} & 354 \\
  Ab initio \citep{Kudin2001_01} & 345 \\
  Experimental \citep{Lee2008_01} & 340\\
  \hline
\end{tabular}
    \caption{Membrane behavior: Comparison of the elastic modulus \textcolor{cgm}{for various methods. FGBP = first generation Brenner potential; SGBP = second generation Brenner potential; LDA = local
    density approximation; GGA = generalized gradient approximation.}}
    \label{t:elastic_modulus_comparsion}
\end{table}

\subsection{Bending energy}\label{ss:bending_part_graphene}
In this section, the membrane model of \cite{Kumar2014_01} is extended to a shell model by including a bending strain energy. In the current work, the uniaxial bending model of \cite{Lu2009_01} is extended to tensorial form using the tensorial invariants introduced in Sec.~\ref{s:kinematics}. \cite{Lu2009_01} show that the bending behavior of graphene is linear up to the curvature 1 $\mathrm{nm}^{-1}$ and the material response is isotropic below this limit. They used FGBP and SGBP to determine the material parameter of their bending model. In the former, only the bond angle is included and in the latter the dihedral angle is included in the potential. \textcolor{cgm2}{\cite{Lu2009_01} report} the bending \textcolor{cgm2}{moduli} 0.133 and 0.225 nN$\cdot$nm obtained from FGBP and SGBP \citep{Brenner1990_01, brenner2002_01}. The latter is much closer to \textcolor{cgn2}{existing} ab\textcolor{cgm2}{-}initio energy calcluations, which determine the bending modulus at 0.238 nN$\cdot$nm~\citep{Kudin2001_01}. These bending parameters are summarized in Tab.~\ref{t:graphene_bending_material_cons}. In all the following simulations the bending modulus is taken as 0.238 nN$\cdot$nm unless otherwise mentioned. The bending stiffness of graphene is not due to its thickness. The main source of the bending stiffness of graphene is the changing of bond angles and dihedral angles. The model of \citet{CANHAM1970_01} is a good candidate for modeling the bending of graphene. The Canham bending strain energy density per unit current area can be written as
\eqb{lll}
w_{\mathrm{b}}= \ds \frac{c}{2}\left(\kappa_1^2+\kappa_2^2\right)~,
\label{e:bend_energy_cur}
\eqe
where $c$ is a material constant. \textcolor{cgn}{Since $\kappa_1$ and $\kappa_2$ depend on the in-plane stretch, model (\ref{e:bend_energy_cur}) introduces coupling between membrane
and bending deformation.} The bending strain energy per unit of current area can be transformed to the reference area as
\eqb{lll}
\ds W_{\mathrm{b}}= \ds J\,\frac{c}{2}\left(\kappa_1^2+\kappa_2^2\right)~.
\eqe
The final form of the Kirchhoff stress and bending moment tensors due to bending are
\eqb{lll}
\ds \tauab_{\mathrm{b}} \is \ds J\,\left[c\,(2H^2+\kappa)\aab-4c\,H\,\bab\right]~,
\eqe
\textcolor{cgn2}{and}
\eqb{lll}
\Mab_0 = c\,J\,\bab~.
\label{e:bending_moment}
\eqe
The elasticity tensors for bending can be found in \citet{Sauer2015_02}. \textcolor{cgn}{As seen from (\ref{e:bending_moment}), the Canham model leads to a simple linear relationship between curvature and bending moment. The coefficient of proportionality, $c$, corresponds to the bending stiffness. An advantage of the Canham model is that it does not require the notion of a thickness. An alternative approach is to use 3D elasticity and apply thickness integration as is considered in the formulation of \citet{Delfani2016_01}.}
\begin{table}
  \centering
     \begin{tabular}{c c c c}
     & FGBP & SGBP & QM \\
     \hline
     \hline
    c~[nN$\cdot$nm] & 0.133 & 0.225 & 0.238 \\
  \end{tabular}
  \caption{Bending behavior: Bending stiffness according to various atomistic \textcolor{cgm}{and quantum} models \textcolor{cgn2}{\citep{Lu2009_01,Kudin2001_01}}.}\label{t:graphene_bending_material_cons}
\end{table}

\section{Finite element formulation}\label{s:Finite_element_formulation}
The presented graphene material model is implemented within the isogeometric rotation-free shell formulation  of \citet{Duong2016_01}, which is briefly summarized here. This section starts with the equilibrium equation and the corresponding  discretization approximations. Then, the weak form is presented. Finally, the stiffness matrices are obtained based on standard linearization.
\subsection{Equilibrium equation}
The equilibrium equation for a Kirchhoff-Love shell can be written as
\eqb{l}
\bT_{;\alpha}^{\alpha} + \bolds{f}=\rho\,\dot{\bv}~~~\forall~\bx~\in~\mathcal{B}~,
\eqe
where \textcolor{cgn2}{``;'' is the co-variant derivative,} $\bT^{\alpha} = \bsig^{\mathrm{T}}\cdot\ba^{\alpha}$ , and $\bolds{f}$ and $\dot{\bv}$ are body force and acceleration vectors. $\bolds{f}$ and the boundary conditions can be defined as
\eqb{llll}
\bolds{f} &:=& f^{\alpha}\,\ba_{\alpha}+p\,\bn~~&\forall~\bx~\in~~\mathcal{B}~,\\[1mm]
\bu &:=& \bar{\bu}~~&\forall~\bx~\in~\partial\mathcal{B}_u~,\\[1mm]
\bn &:=& \bar{\bn}~~&\forall~\bx~\in~\partial\mathcal{B}_n~,\\[1mm]
\bt &:=& \bar{\bt}~~&\forall~\bx~\in~\partial\mathcal{B}_t~,\\[1mm]
m_{\tau} &:=& \bar{m}_{\tau}~~&\forall~\bx~\in~\partial\mathcal{B}_{\tau}~,
\eqe
where $\bar{\bu}$ and $\bar{\bn}$ are prescribed position and normal unit vector at the Dirichlet boundary, and $\bar{\bt}$ and $\bar{m}_{\tau}$ are tractions and the bending moments at the Neumann boundary.

\subsection{Discretization}
The continua in current, $\mathcal{B}$ , and reference, $\mathcal{B}_{0}$ , configuration can be discretized with NURBS shape functions and control points as
\eqb{l}
\bX= \mN\,\mX_{n_e}~;~\bx = \mN\,\mx_{n_e}~,
\label{e:dis_surface}
\eqe
where $\mN$ and $n_e$ are the NURBS shape function matrix and the number of control points per element. Using the same shape functions, the following relations can be obtained
\eqb{lll}
\delta\bx \is \mN\,\delta\mx_{n_e}~,\\
\ba_\alpha  \is  \mN_{,\alpha}\,\mx_{n_e}~,\\
\delta\ba_\alpha  \is  \mN_{,\alpha}\,\delta\mx_{n_e}~,\\
\ba_{\alpha,\beta}  \is  \mN_{,\alpha\beta}\,\mx_{n_e}~,\\
\ba_{\alpha;\beta}  \is  \tilde{\mN}_{;\alpha\beta}\,\mx_{n_e}~,\\
\label{e:dis_surface_object}
\eqe
where ``,'' \textcolor{cgn2}{is} the parametric derivative and $\mN$~, $\mN_{,\alpha}$~, $\mN_{,\alpha\beta}$~, and $\tilde{\mN}_{;\alpha\beta}$ are defined as
\eqb{lll}
\mN \dis [N_1\,\boldsymbol{1}, N_2\,\boldsymbol{1}, \ldots, N_{n_e}\,\boldsymbol{1}]~,\\[1mm]
\mN_{,\alpha} \dis [N_{1,\alpha}\,\boldsymbol{1}, N_{2,\alpha}\,\boldsymbol{1}, \ldots, N_{n_e,\alpha}\,\boldsymbol{1}]~, \\[1mm]
\mN_{,\alpha\beta} \dis [N_{1,\alpha\beta}\,\boldsymbol{1}, N_{2,\alpha\beta}\,\boldsymbol{1}, \ldots, N_{n_e,\alpha\beta}\,\boldsymbol{1}]~,\\ [1mm]
\tilde{\mN}_{;\alpha\beta} \dis \mN_{,\alpha\beta}-\Gamma_{\alpha\beta}^{\gamma}\,\mN_{,\gamma}~,
\eqe
%\subsection{Distretization of Weak form}
where $N_{i}$ are NURBS shape functions. Using approximations (Eqs.~(\ref{e:dis_surface}) and (\ref{e:dis_surface_object})), the weak form can be written as
\eqb{l}
\ds \sum\limits_{e=1}^{n_{\text{el}}} { (G_{\text{in}}^{e}+G_{\text{int}}^{e}+G_{\text{c}}^{e}-G_{\text{ext}}^{e})}=0~, \forall~\delta \mx_{e} \in \mathcal{B}~,
\eqe
where $n_{\text{el}}$ is the number of elements. $G_{\mathrm{c}}^{e}$ is related to contact and is discussed in Appendix \ref{s:contact_stiffness}. $G_{\text{in}}^{e}$, $G_{\text{int}}^{e}$, and $G_{\text{ext}}^{e}$ are inertial, internal and external parts and are defined as
\eqb{llll}
G_{\text{in}}^{e} := \delta\mx_{e}^{\text{T}}\,\mf_{\text{in}}^{e} ~;~~~\text{with}~~\mf_{\text{in}}^{e}= \ds \int\limits_{\mathcal{B}_{0}^{e} }{\mN^{\text{T}}\,\rho\,\dot{\bv}~\dif A}~,
\eqe
\eqb{lll}
G_{\text{int}}^{e} := \delta\mx_{e}^{\text{T}}\,(\mf_{\text{int}\tau}^{e}+\mf_{\text{int}M}^{e})~;~~\text{with}~~\mf_{\text{int}\tau}^{e} =\ds \int\limits_{\mathcal{B}_{0}^{e} }{\tau^{\alpha\beta}\,\mN_{,\alpha}^{\text{T}}\,\ba_{\beta}~\dif A}~;~ \mf_{\text{int}M}^{e} =\ds \int\limits_{\mathcal{B}_{0}^{e} }{M_0^{\alpha\beta}\,\tilde{\mN}_{;\alpha\beta}^{\text{T}}\,\bn~\dif A}~,
\eqe
\eqb{lll}
G_{\text{ext}}^{e} \dis \delta\mx_{e}^{\text{T}}\,\left(\mf_{\text{ext}0}^{e}+ \mf_{\text{ext}p}^{e}+ \mf_{\text{ext}t}^{e}+ \mf_{\text{ext}m}^{e}\right)~.
\label{e:weak_form_external}
\eqe
The terms in Eq. (\ref{e:weak_form_external}) are
\eqb{lll}
\mf_{\text{ext}0}^{e} \is \ds \int\limits_{\mathcal{B}_{0}^{e} }{\mN^{\text{T}}\,\bolds{f}_{0}~\dif A}~,\\[3mm]
\mf_{\text{ext}p}^{e} \is \ds \int\limits_{\mathcal{B}^{e} }{\mN^{\text{T}}\,p\,\bn~\dif a}~,\\[3mm]
\mf_{\text{ext}t}^{e} \is \ds \int\limits_{\partial_t \mathcal{B}^{e} }{\mN^{\text{T}}\,\bt~\dif s}~,\\[3mm]
\mf_{\text{ext}m}^{e} \is \ds -\int\limits_{\partial_m \mathcal{B}^{e} }{\mN_{,\alpha}^{\text{T}}\,\nu^{\alpha}\,m_{\tau}\,\bn~\dif s}~,
\eqe
where along $\partial_m \mathcal{B}$ bending Neumann BCs are applied. $\nu^{\alpha}$ are the contra-variant components of the normal vector on the boundary $\partial \mathcal{B}$.
\subsection{Stiffness matrix}
The stiffness matrix appearing in the linearization of the discretized weak form, can be written as
\eqb{lll}
\mk^{e} \is \mk_{\tau\tau}^{e}+\mk_{\tau M}^{e}+\mk_{M\tau}^{e}+\mk_{MM}^{e} +\mk_{\tau}^{e} +\mk_{M}^{e}+\mk_{\text{extp}}^{e}+\mk_{\text{extm}}^{\text{e}}+\mk_c^{e}~,
\eqe
where the material stiffness matrices are defined by
\eqb{lll}
\mk_{\tau\tau}^{e} \dis \ds \int\limits_{\mathcal{B}_{0}^{e} }{\cabgd\,\mN_{,\alpha}^{\text{T}}\,(\ba_{\beta}\otimes\ba_{\gamma})\,\mN_{,\delta}~\dif A}~,\\[8mm]
\mk_{\tau M}^{e} \dis \ds \int\limits_{\mathcal{B}_{0}^{e} }{\dabgd\,\mN_{,\alpha}^{\text{T}}\,(\ba_{\beta}\otimes\bn)\,\tilde{\mN}_{;\gamma\delta}~\dif A}~,\\[8mm]
\mk_{M\tau}^{e} \dis \ds \int\limits_{\mathcal{B}_{0}^{e} }{\eabgd\,\tilde{\mN}_{;\alpha\beta}^{\text{T}}\,(\bn\otimes\ba_{\gamma})\,\mN_{,\delta}~\dif A}~,\\[8mm]
\mk_{MM}^{e} \dis \ds \int\limits_{\mathcal{B}_{0}^{e} }{\fabgd\,\tilde{\mN}_{;\alpha\beta}^{\text{T}}\,(\bn\otimes\bn)\,\tilde{\mN}_{;\gamma\delta}~\dif A}~,
\eqe
where the elasticity tensors are given in Appendix \ref{s:Derivation_membrane_ constitutive_law_based_on_curvilinear_coordinate} and \cite{Sauer2015_02}. Furthermore, the geometrical stiffness matrices are defined as
\eqb{lll}
\mk_{\tau}^{e} \dis \ds \int\limits_{\mathcal{B}_{0}^{e} }{\mN_{,\alpha}^\text{T} \,\tau^{\alpha\beta}\,\mN_{,\beta}~\dif A}~,
\eqe
\eqb{lll}
\mk_{M}^{e} \dis \ds \mk_{M1}^{e}+\mk_{M2}^{e}+(\mk_{M2}^{e})^{\text{T}}~,
\eqe
with
\eqb{lll}
\mk_{M1}^{e} \dis \ds -\int\limits_{\mathcal{B}_{0}^{e} }{\buab\,\Mab_0\,\agd \,\mN_{,\gamma}^{\text{T}}\,(\bn\otimes\bn)\,\mN_{,\delta}~\dif A}~,\\[8mm]
\mk_{M2}^{e} \dis \ds -\int\limits_{\mathcal{B}_{0}^{e} }{\Mab_0\, \mN_{,\gamma}^{\text{T}}\,(\bn\otimes\,a^{\gamma})\,\tilde{\mN}_{;\alpha\,\beta}~\dif A}~.
\eqe
The external tangent matrices $\mk_{\text{extp}}^{e}$ and $\mk_{\text{extm}}^{e}$ can be found in \citet{Sauer2014_01} and \citet{Duong2016_01}, and the contact stiffness matrix $\mk_c^{e}~$ is given in  Appendix \ref{s:contact_stiffness}.

%\section{Numerical Example}\label{s:Numerical_example}
\section{Elementary model behavior}\label{s:Verification}
In this section, the elementary behavior of the model under some simple deformation states is investigated. Pure dilatation, uniaxial stretch, pure shear and pure bending are considered as test cases, and the FE implementation is verified by analytical solutions.
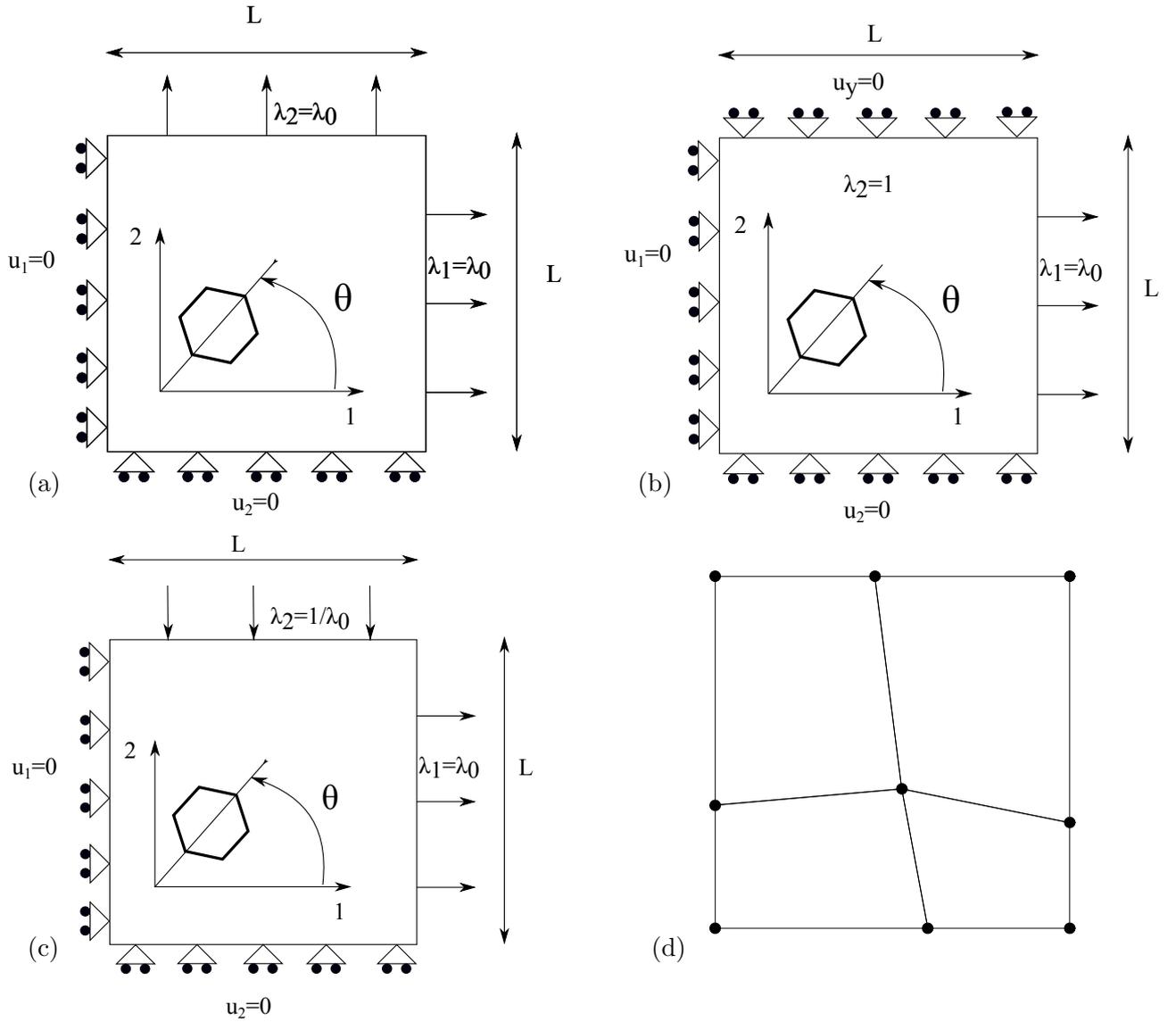
\begin{figure}
\begin{center} \unitlength1cm
\begin{picture}(18,16)
\put(0.0,7.5){\includegraphics[height=85mm]{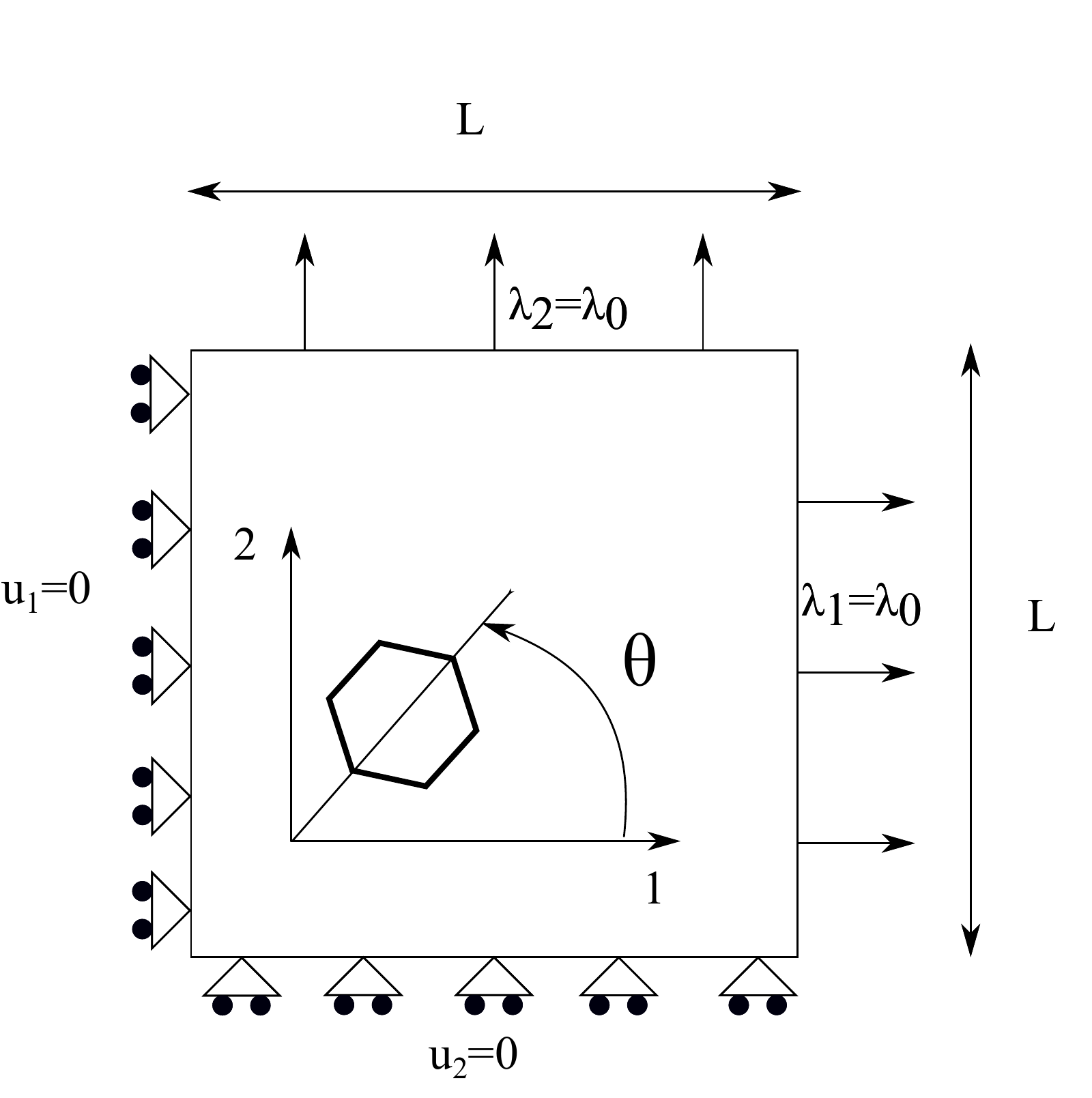}}
\put(0.3,8){(a)}
\put(9,7.5){\includegraphics[height=85mm]{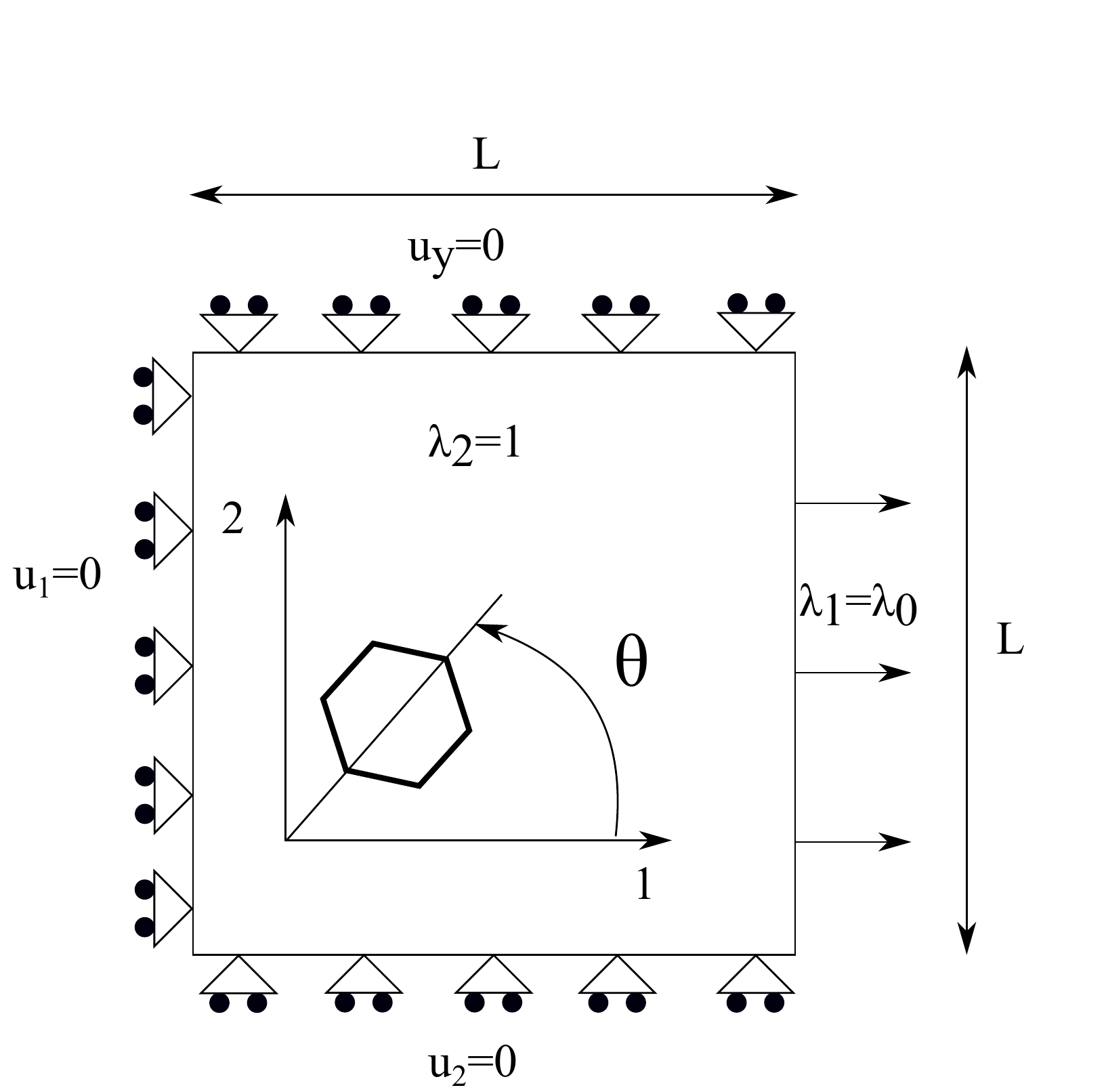}}
\put(9.3,8){(b)}
\put(0.0,0.0){\includegraphics[height=85mm]{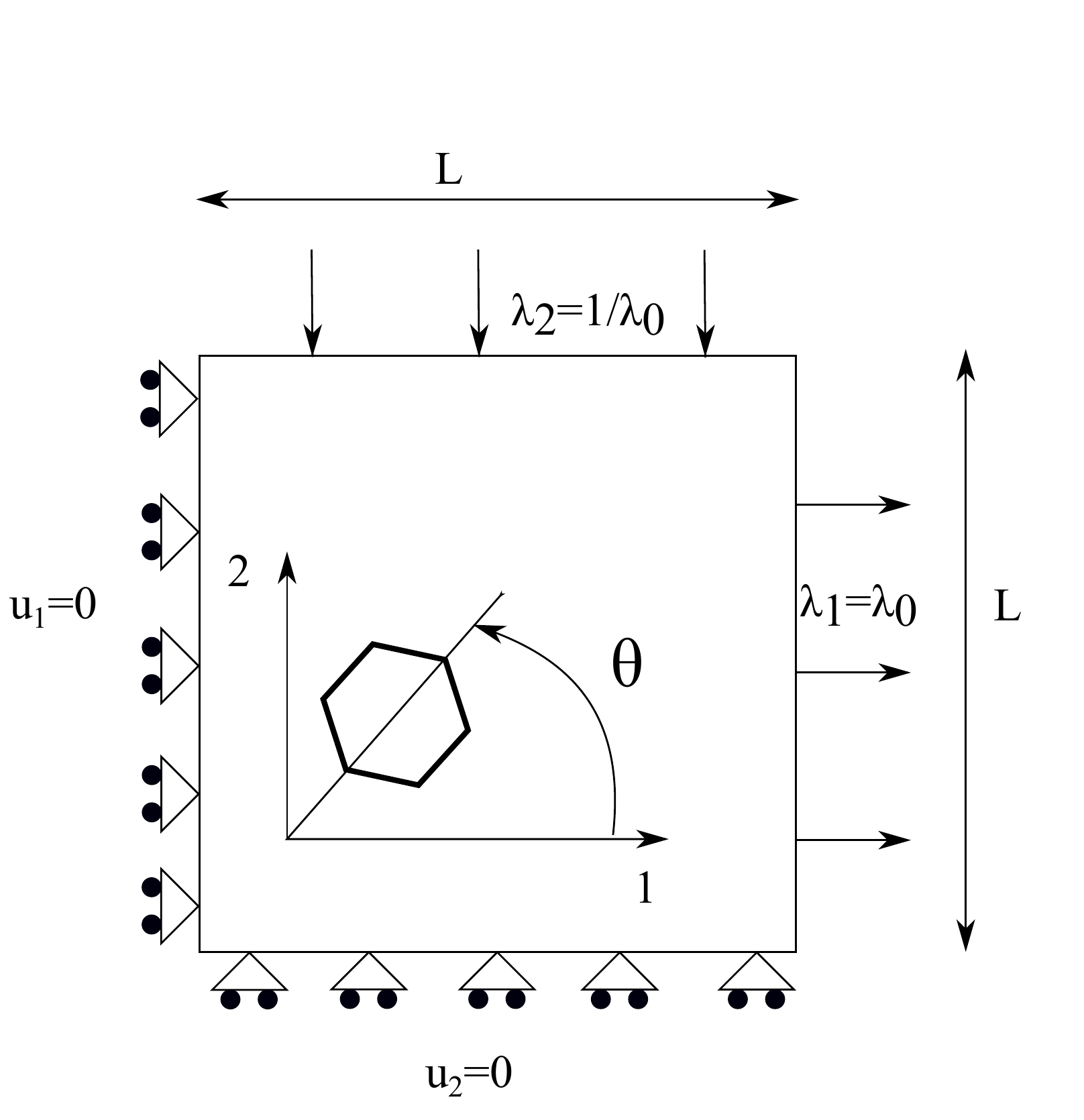}}
\put(0.3,1.1){(c)}
\put(10,1.0){\includegraphics[height=60mm]{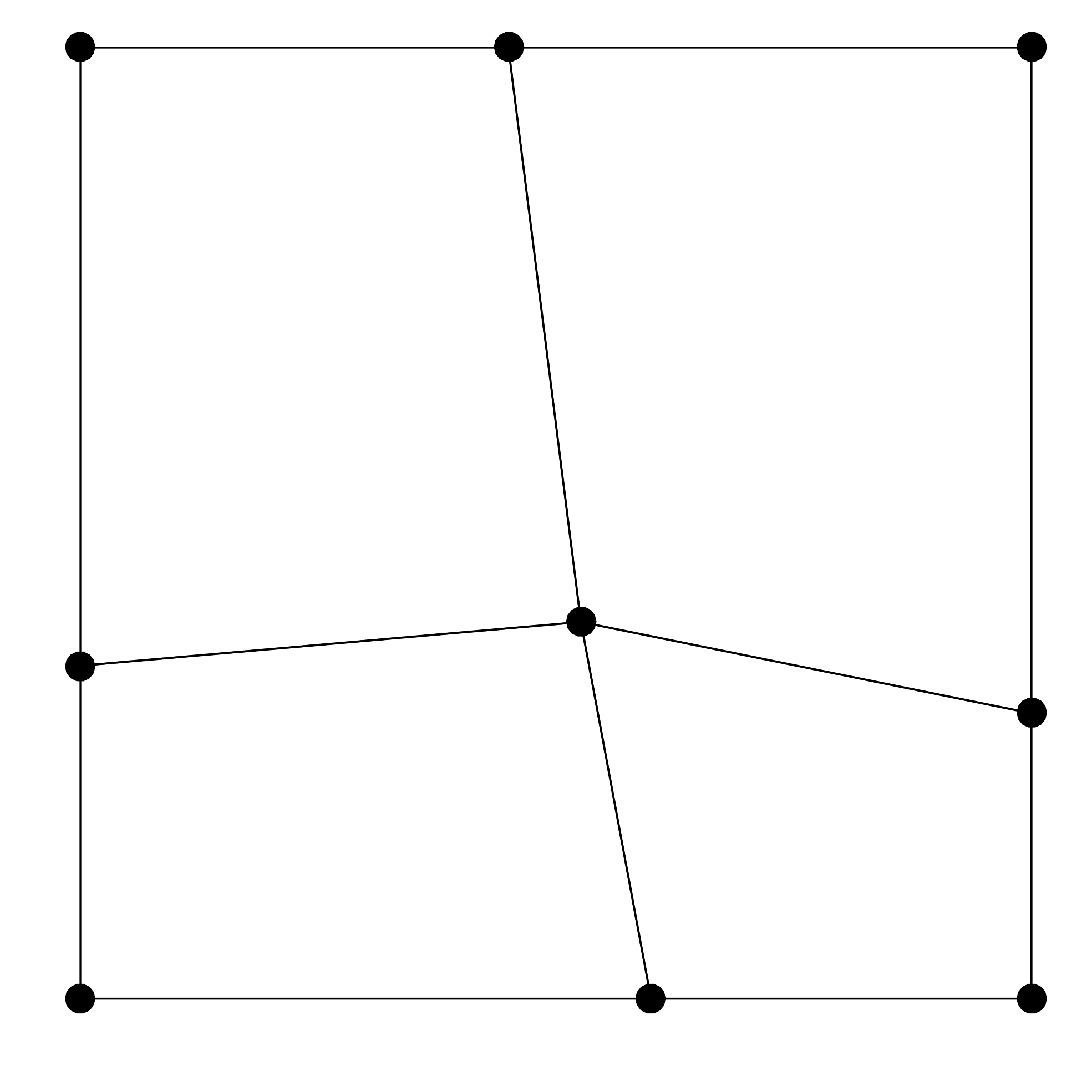}}
\put(9.5,1.1){(d)}
\end{picture}
\caption{Elementary membrane deformations: (a) Pure dilatation; (b) uniaxial stretch; (c) pure shear; (d) considered FE mesh; linear Lagrangian elements are used and the nodes are indicated by $\bullet$. Angle $\theta$ denotes the orientation of the graphene lattice \textcolor{cgn}{with respect to}~the direction of larger stretch $\lambda_1$ , \textcolor{cgn}{and $\lambda_0>1$ is gradually increased.}}
\label{f:Boundary_conditions}
\end{center}
\end{figure}

\subsection{Pure dilatation}
\begin{figure}
    \begin{subfigure}[t]{1\textwidth}
        \centering
    \includegraphics[height=55mm]{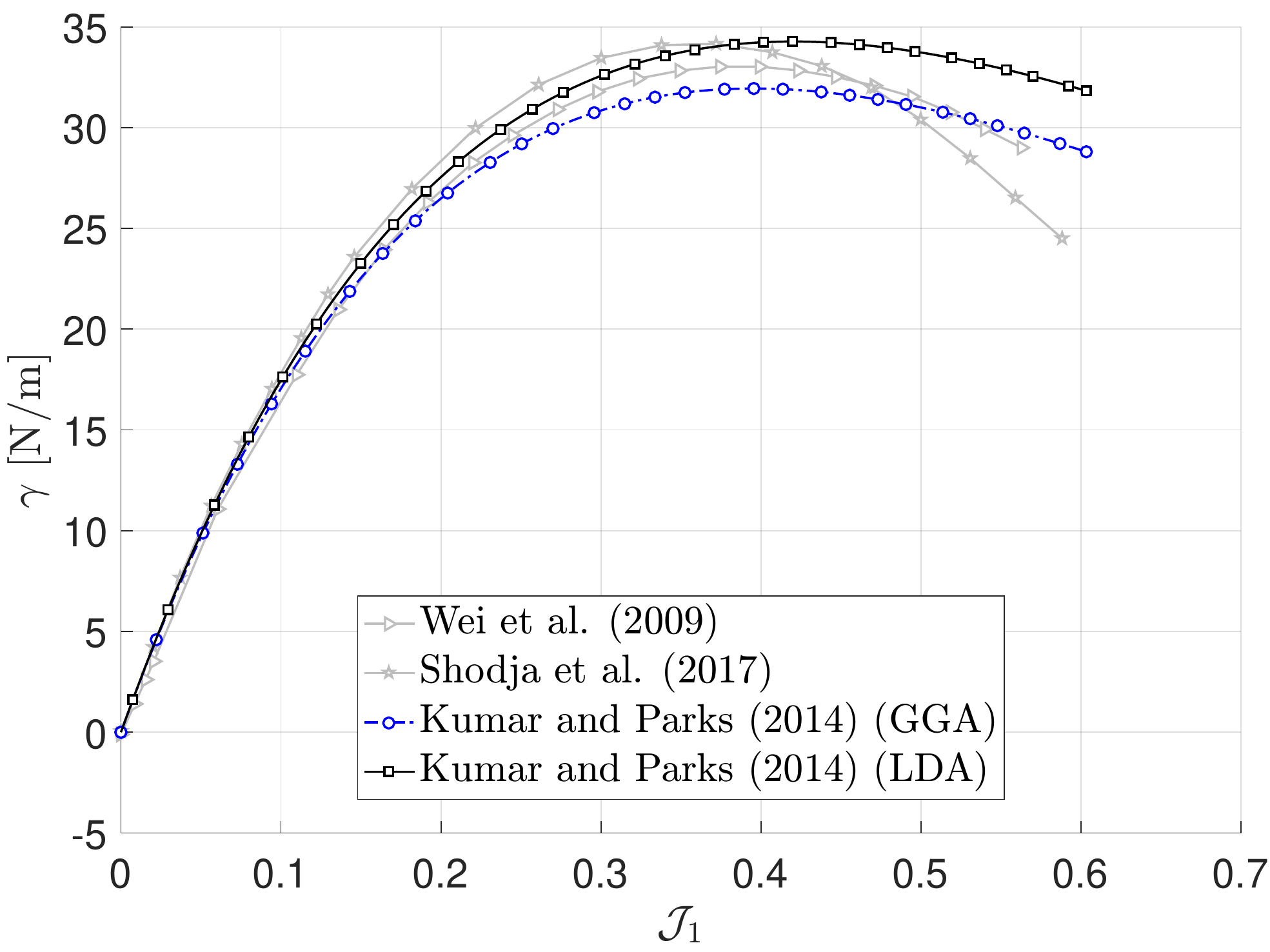}
        %\subcaption{}
        \label{f:Analytical_solution_pure_dilatation}
    \end{subfigure}
    \caption{Pure dilatation: Surface tension. \textcolor{cgn}{The current continuum model is based on the quantum mechanical results of \citet{Kumar2014_01}. Other quantum mechanical results are available in \citet{Wei2009_01} and \citet{Shodja2017_01}.}}
    \label{f:pure_dilatation_example}
\end{figure}

The numerical model consists of a square sheet stretched uniformly in \textcolor{cgm2}{two perpendicular} directions. The boundary conditions and the lattice orientation are shown in Fig.~\ref{f:Boundary_conditions}a. The surface tension\footnote{The surface tension is equal to the negative in-plane pressure.} $\gamma := \frac{1}{2}\tr(\bsig)$ is obtained with the LDA and the GGA parameters and the results are compared in Fig.~\ref{f:pure_dilatation_example}. \textcolor{cgn}{The current continuum model is compared with quantum results from the literature}. The difference of LDA and GGA results is small for infinitesimal strains. But, the response is different for finite strain\textcolor{cgn2}{s}, where LDA shows a stiffer response. It should be noted that the material is unstable after the maximum value of the surface tension has been reached. The shear modulus is zero at this point and become negative afterwards. Displacement control is needed to go beyond that point.

\subsection{Uniaxial stretch}
\begin{figure}
    \begin{subfigure}[t]{0.49\textwidth}
        \centering
     \includegraphics[height=55mm]{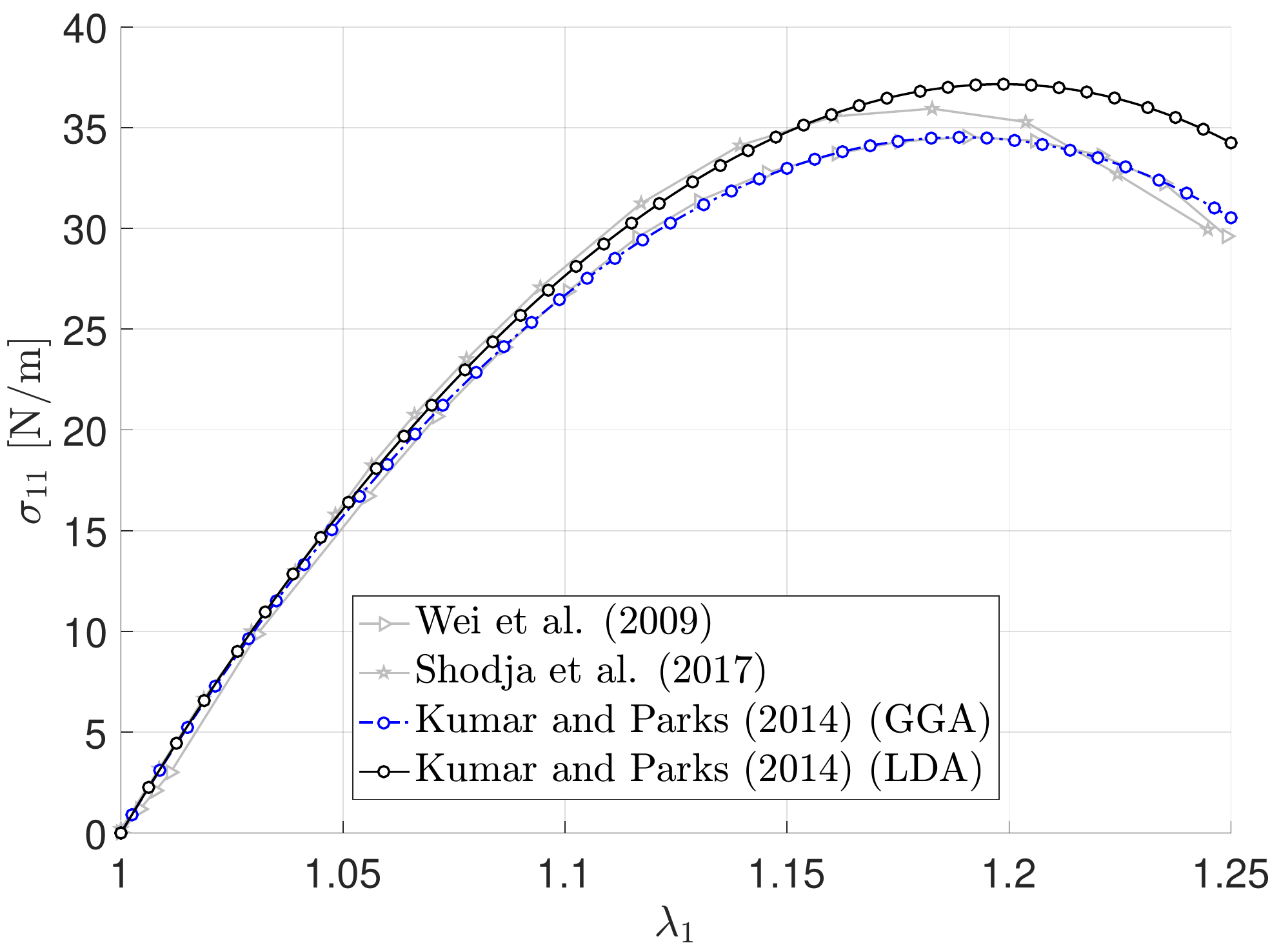}
        \subcaption{}
        \label{f:sigma11_uniaxial_arm_chair}
    \end{subfigure}
    \begin{subfigure}[t]{0.49\textwidth}
        \centering
 \includegraphics[height=55mm]{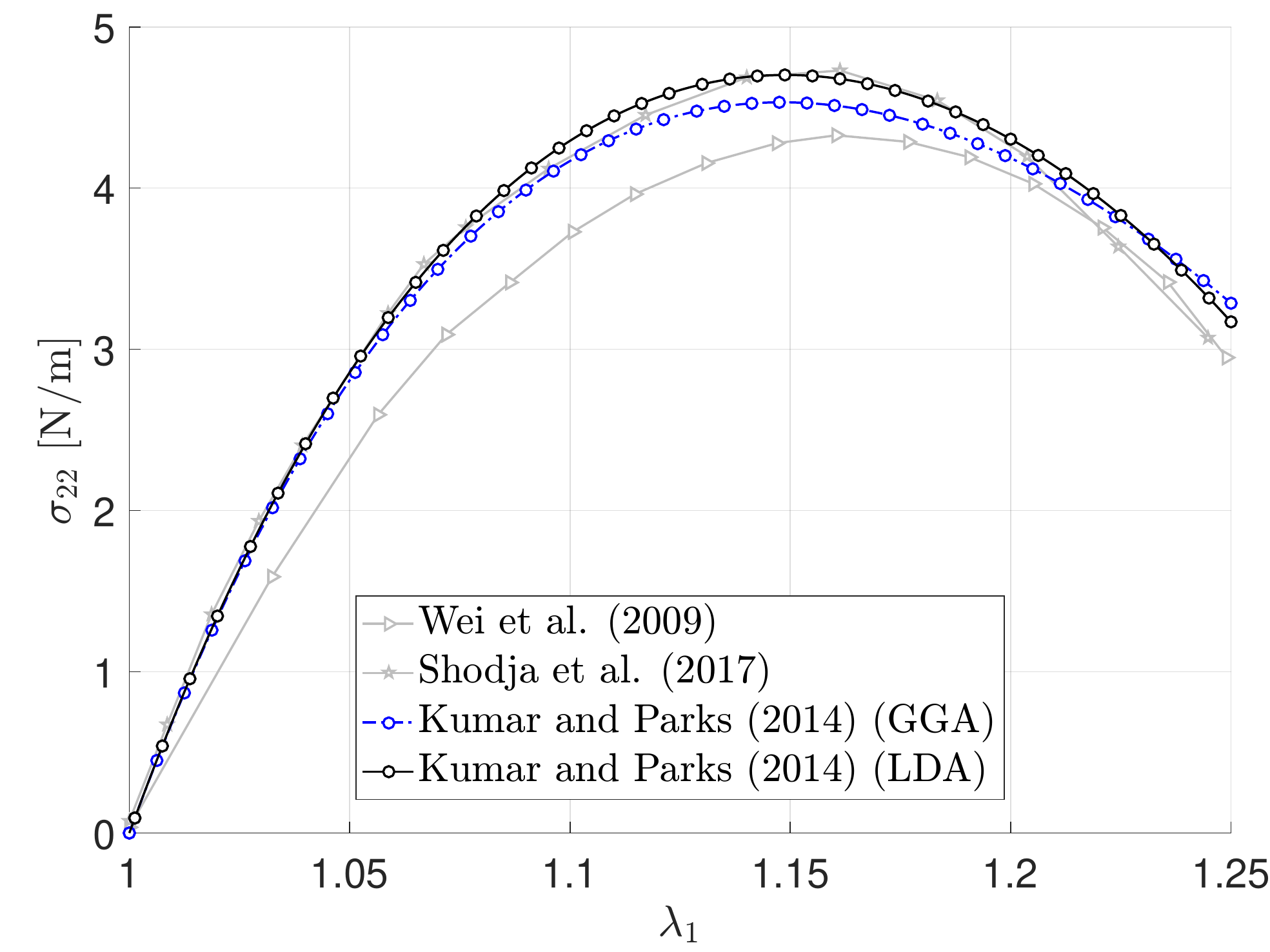}
        \subcaption{}
        \label{f:sigma22_uniaxial_arm_chair}
    \end{subfigure}
    \caption{Uniaxial stretch in the armchair direction: (\subref{f:sigma11_uniaxial_zig_zag}) Stress in the stretched direction, $\sigma_{11}$; (\subref{f:sigma22_uniaxial_zig_zag}) stress in the perpendicular direction, $\sigma_{22}$. \textcolor{cgn}{The current continuum model is based on the quantum mechanical results of \citet{Kumar2014_01}. Other quantum mechanical results are available in \citet{Wei2009_01} and \citet{Shodja2017_01}.}}
    \label{f:uniaxial_arm_chair}
\end{figure}

\begin{figure}
    \begin{subfigure}[t]{0.49\textwidth}
        \centering
    \includegraphics[height=55mm]{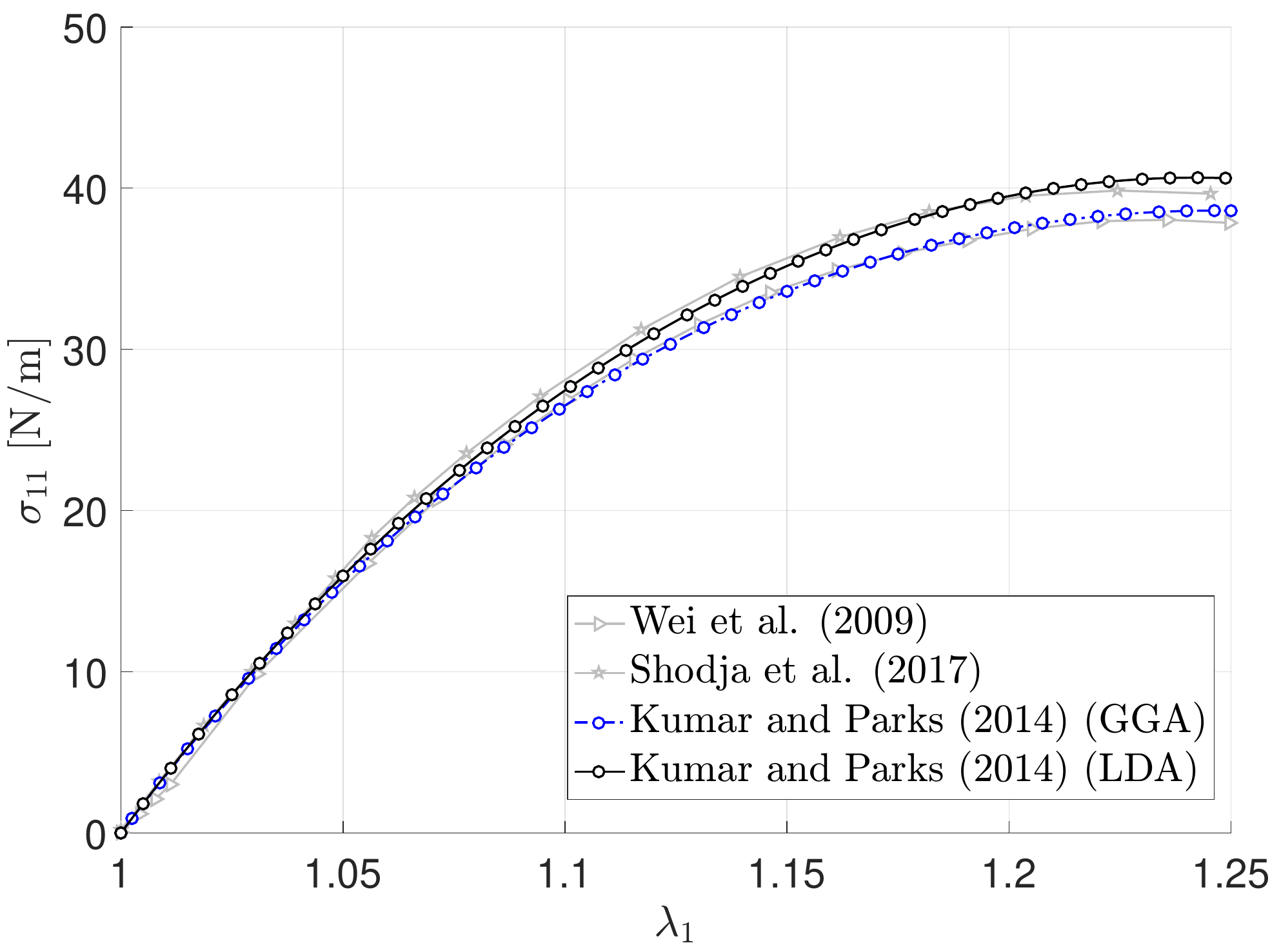}
        \subcaption{}
        \label{f:sigma11_uniaxial_zig_zag}
    \end{subfigure}
    \begin{subfigure}[t]{0.49\textwidth}
        \centering
    \includegraphics[height=55mm]{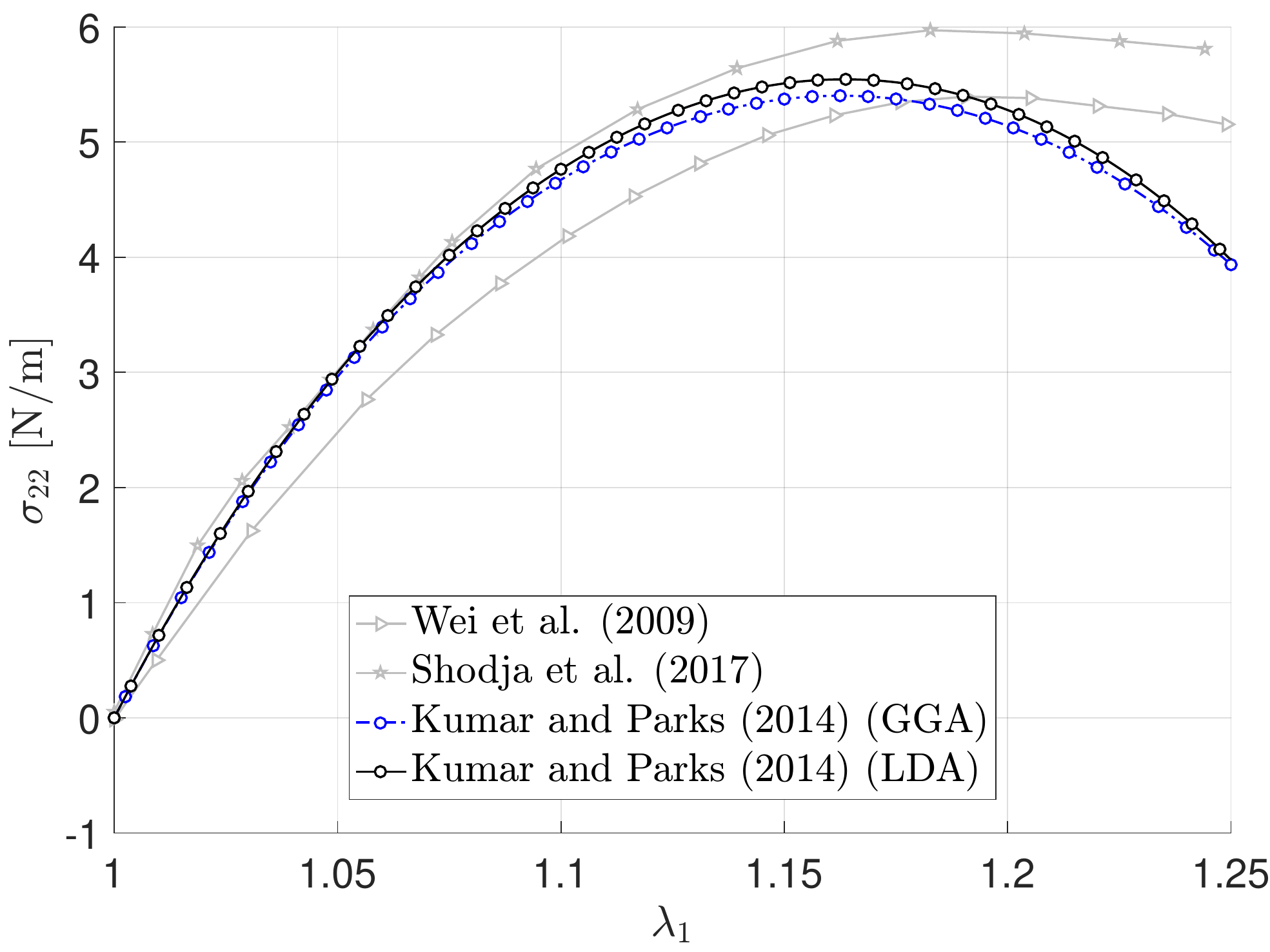}
        \subcaption{}
        \label{f:sigma22_uniaxial_zig_zag}
    \end{subfigure}
    \caption{Uniaxial stretch in the zigzag direction:  (\subref{f:sigma11_uniaxial_zig_zag}) Stress in the stretched direction, $\sigma_{11}$; (\subref{f:sigma22_uniaxial_zig_zag}) stress in the perpendicular direction, $\sigma_{22}$. \textcolor{cgn}{The current continuum model is based on the quantum mechanical results of \citet{Kumar2014_01}. Other quantum mechanical results are available in \citet{Wei2009_01} and \citet{Shodja2017_01}.}}
    \label{f:uniaxial_zig_zag}
\end{figure}

\begin{figure}
    \begin{subfigure}[t]{0.32\textwidth}
        \centering
    \includegraphics[height=38mm]{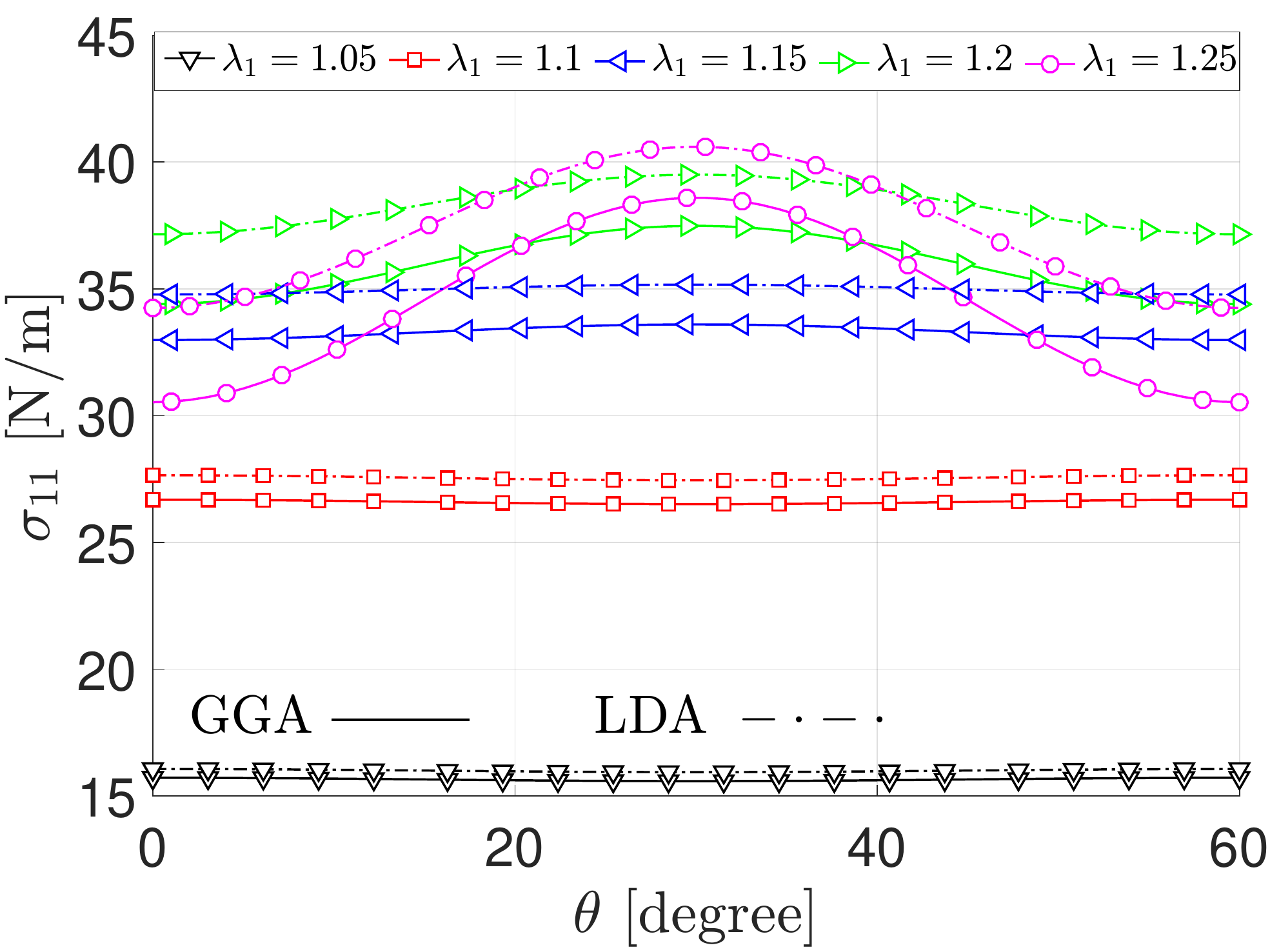}
        \subcaption{}
        \label{f:Uniaxial_Stress_cart_sig11}
    \end{subfigure}
    \begin{subfigure}[t]{0.32\textwidth}
        \centering
    \includegraphics[height=38mm]{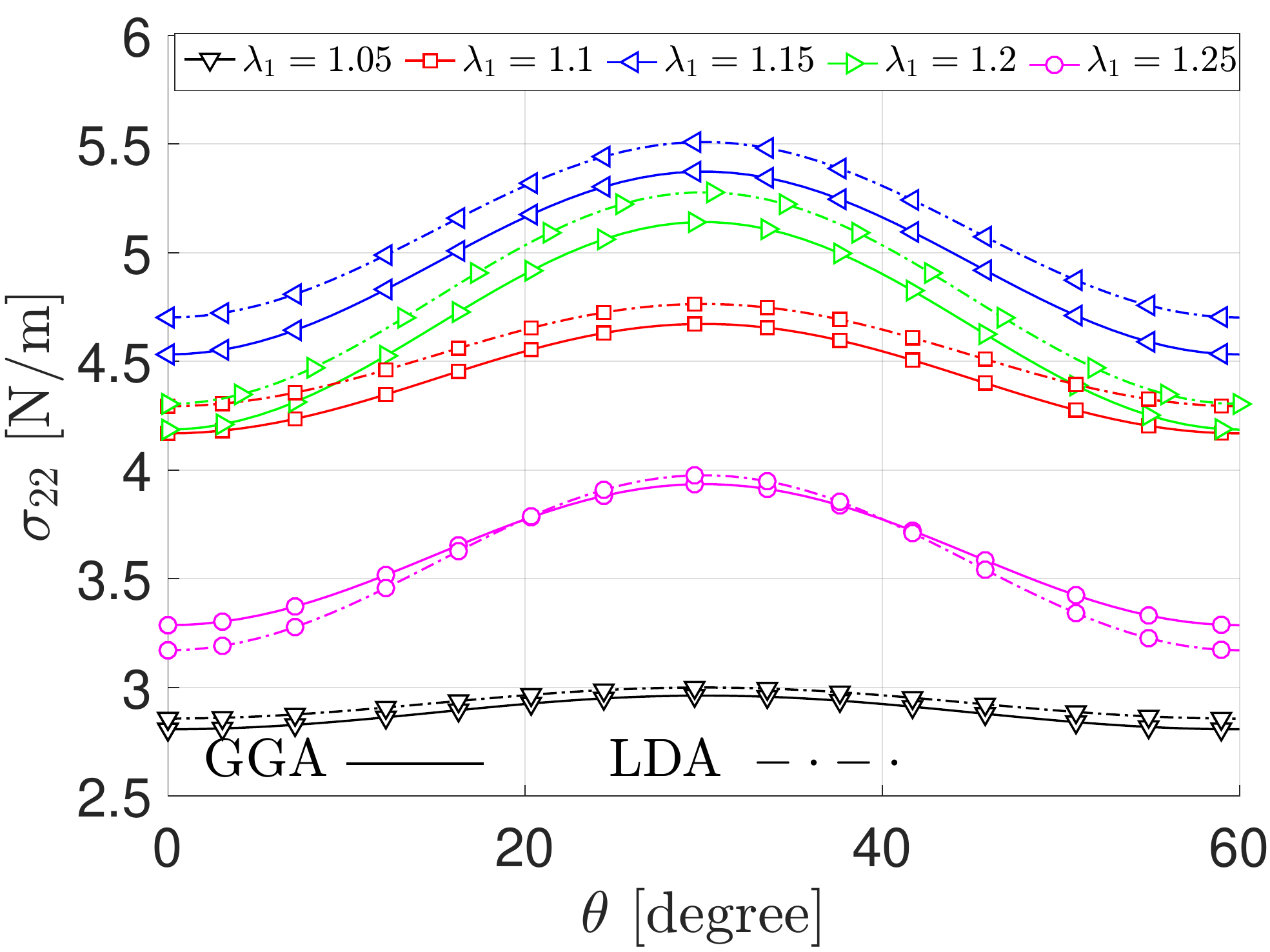}
        \subcaption{}
        \label{f:Uniaxial_Stress_cart_sig22}
    \end{subfigure}
    \begin{subfigure}[t]{0.32\textwidth}
        \centering
    \includegraphics[height=38mm]{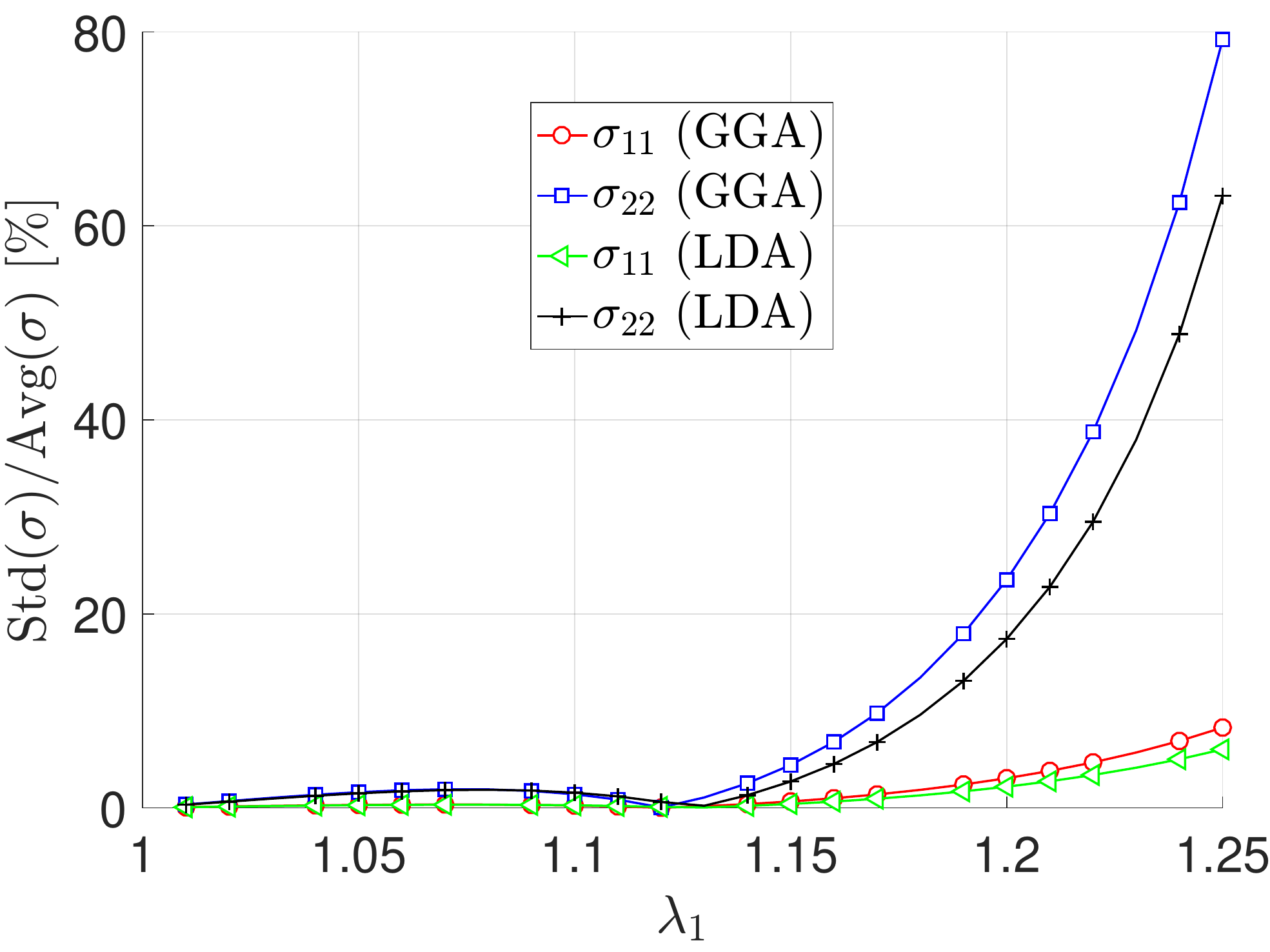}
        \subcaption{}
        \label{f:Uniaxial_Stress_std}
    \end{subfigure}
    \caption{Uniaxial stretch: Dependence of the stress components on stretch $\lambda_1$ and angle $\theta$,  (\subref{f:Uniaxial_Stress_cart_sig11}) $\sigma_{11}$; (\subref{f:Uniaxial_Stress_cart_sig22}) $\sigma_{22}$. (\subref{f:Uniaxial_Stress_std}) Dependence of the standard deviation of the stress on stretch $\lambda_1$.}
    \label{f:Uniaxial_Stress_cart_sig}
\end{figure}

Uniaxial stretching is considered along different directions, i.e. different chiralities. The model is stretched in one direction and restrained in the perpendicular direction (Fig.~\ref{f:Boundary_conditions}b). In order to assess the FE solution, the Cartesian stress components $\sigma_{11}$ and $\sigma_{22}$ are examined. They are the normal stress components in the direction of the stretch and perpendicular to it. They are obtained analytically in Appendix \ref{s:analytical_solution}. The results of stretching the model in the armchair and the zigzag directions are presented in Figs.~\ref{f:uniaxial_arm_chair} and \ref{f:uniaxial_zig_zag}. \textcolor{cgn}{The current continuum model is compared with quantum results from the literature}. Like for dilatational loading, a material instability occurs after the maximum stress has been reached and the elasticity tensor loses its ellipticity at this point.\\
\textcolor{cgm2}{Next}, the model is stretched in different directions, relative to the armchair direction, and the stress variation is examined for a set of stretch ratios in Fig.~\ref{f:Uniaxial_Stress_cart_sig}. As can be expected, the material response is periodic by $\pi/3$. Finally, the standard deviation and average of the stresses in one period is calculated for the different stretch ratios. The average and standard deviation can be calculated as
\eqb{lll}
\ds \text{Avg}(\sigma_{ij}) \is \ds \langle\sigma_{ij}\rangle~=~\frac{3}{\pi} \int_{0}^{\pi/3}{\sigma_{ij}~\textcolor{cgn}{\dif \theta~,}}
\eqe
\eqb{lll}
\ds \text{Std}(\sigma_{ij}) \is \ds \int_{0}^{\pi/3}{\sqrt{\frac{3}{\pi}(\sigma_{ij}-\langle\sigma_{ij}\rangle)^2}~\dif \theta}~.
\eqe

The standard deviation is normalized by the average stress and shown versus $\lambda_1$~ in Fig.~\ref{f:Uniaxial_Stress_std}. The graph can be interpreted as the \textcolor{cgm2}{error caused by assuming isotropy}. \textcolor{cgm2}{This error} can reach \textcolor{cgm2}{up} to 80\% of the average response. So, an anisotropic model is necessary to capture the material behavior for finite deformations.

\subsection{Pure shear}
\begin{figure}
 \centering
    \begin{subfigure}[t]{0.49\textwidth}
        \centering
    \includegraphics[height=55mm]{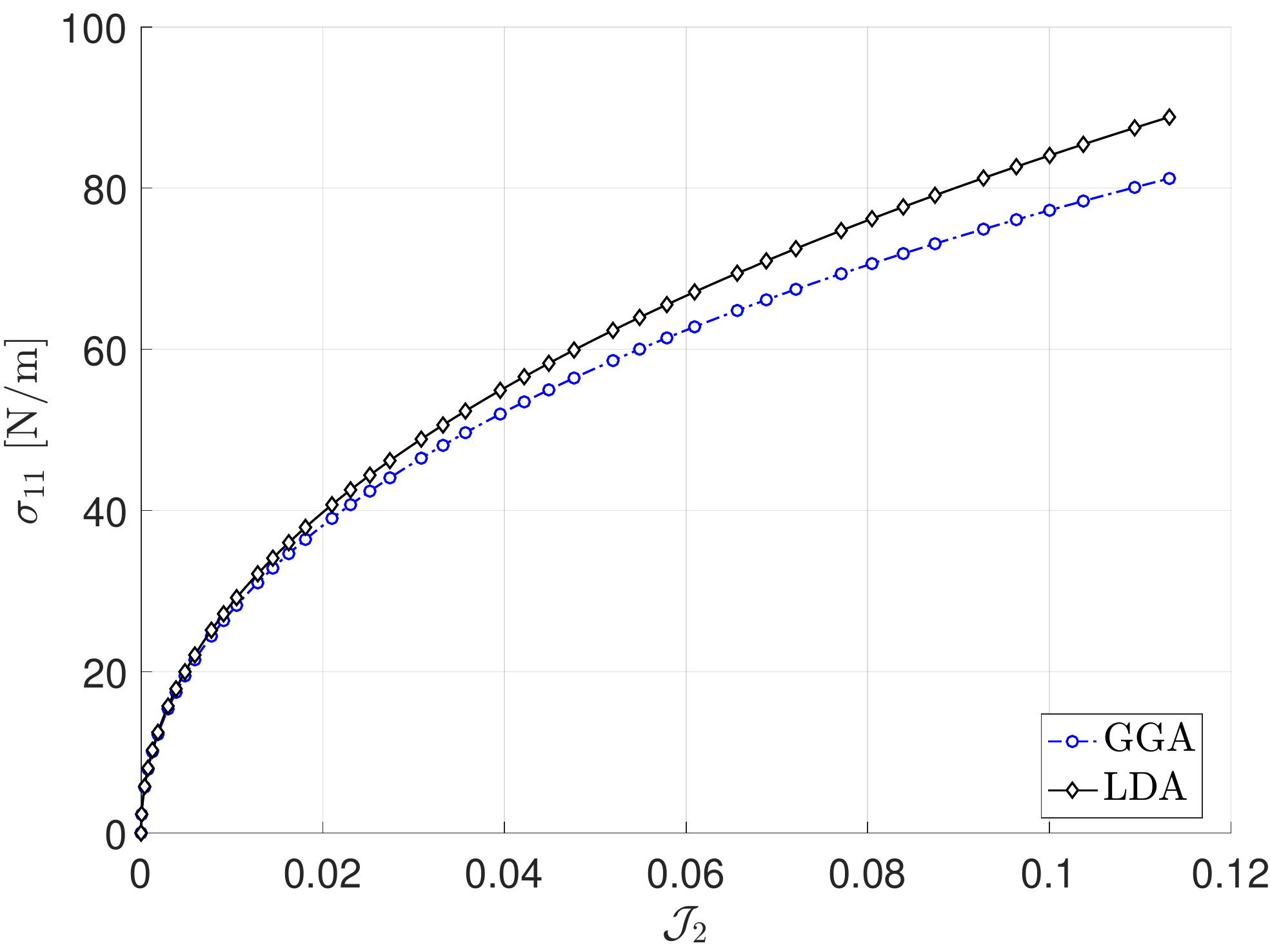}
        \subcaption{}
        \label{f:Analytical_solution_pure_shear_Sig11_armchair_fem}
    \end{subfigure}
    \begin{subfigure}[t]{0.49\textwidth}
        \centering
    \includegraphics[height=55mm]{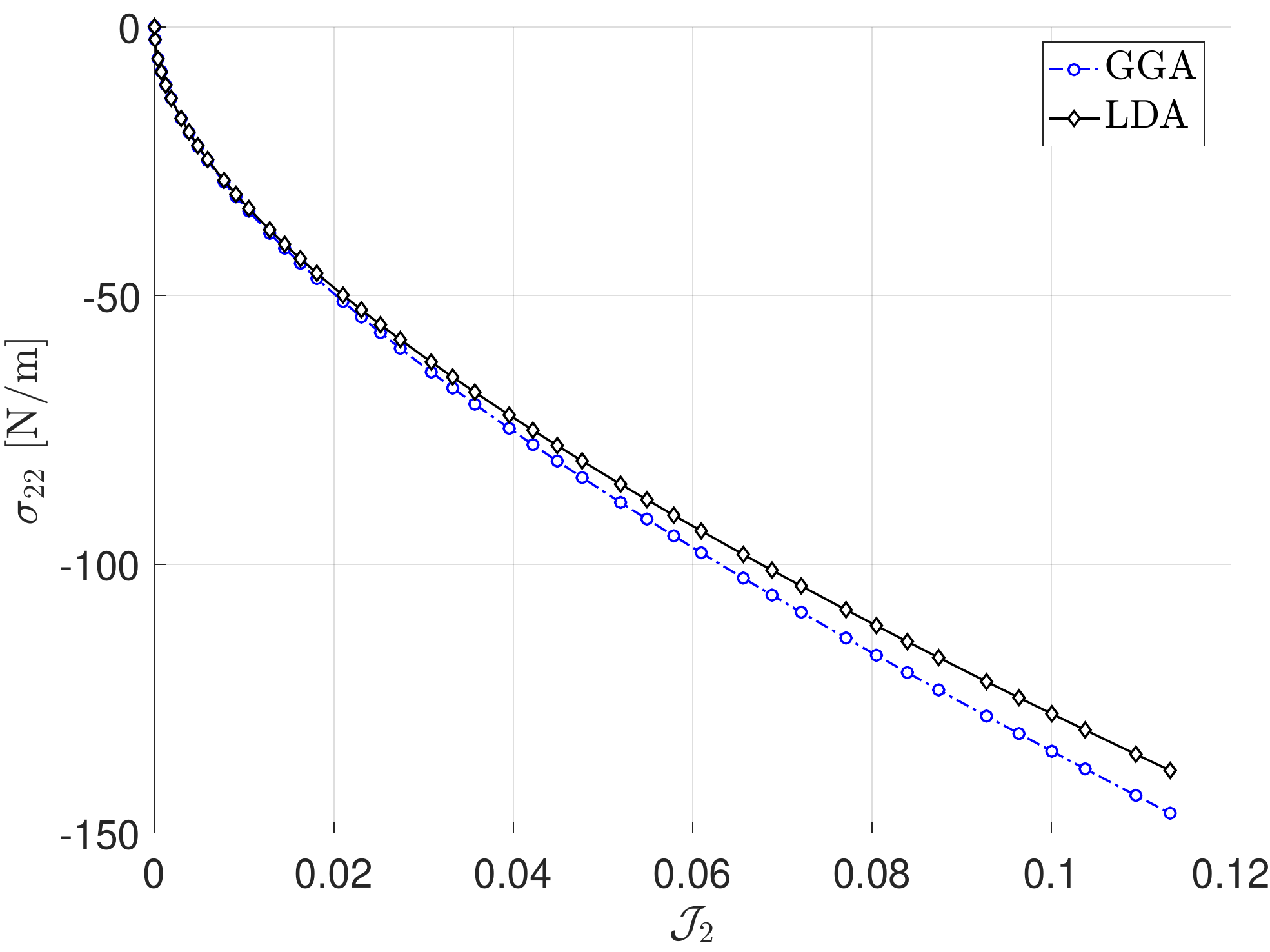}
        \subcaption{}
        \label{f:Analytical_solution_pure_shear_Sig22_armchair_fem}
    \end{subfigure}
    \caption{\textcolor{cgn2}{Pure shear with pull in the armchair and compression in the zigzag direction ($\theta=0$ in Fig.~3):} (\subref{f:Analytical_solution_pure_shear_Sig11_armchair_fem}) Stress in the pull direction, $\sigma_{11}$;
    (\subref{f:Analytical_solution_pure_shear_Sig22_armchair_fem}) stress in the compression direction, $\sigma_{22}$.}
    \label{f:solution_pure_shear_armchair_fem}
\end{figure}

\begin{figure}
 \centering
    \begin{subfigure}[t]{0.49\textwidth}
        \centering
    \includegraphics[height=55mm]{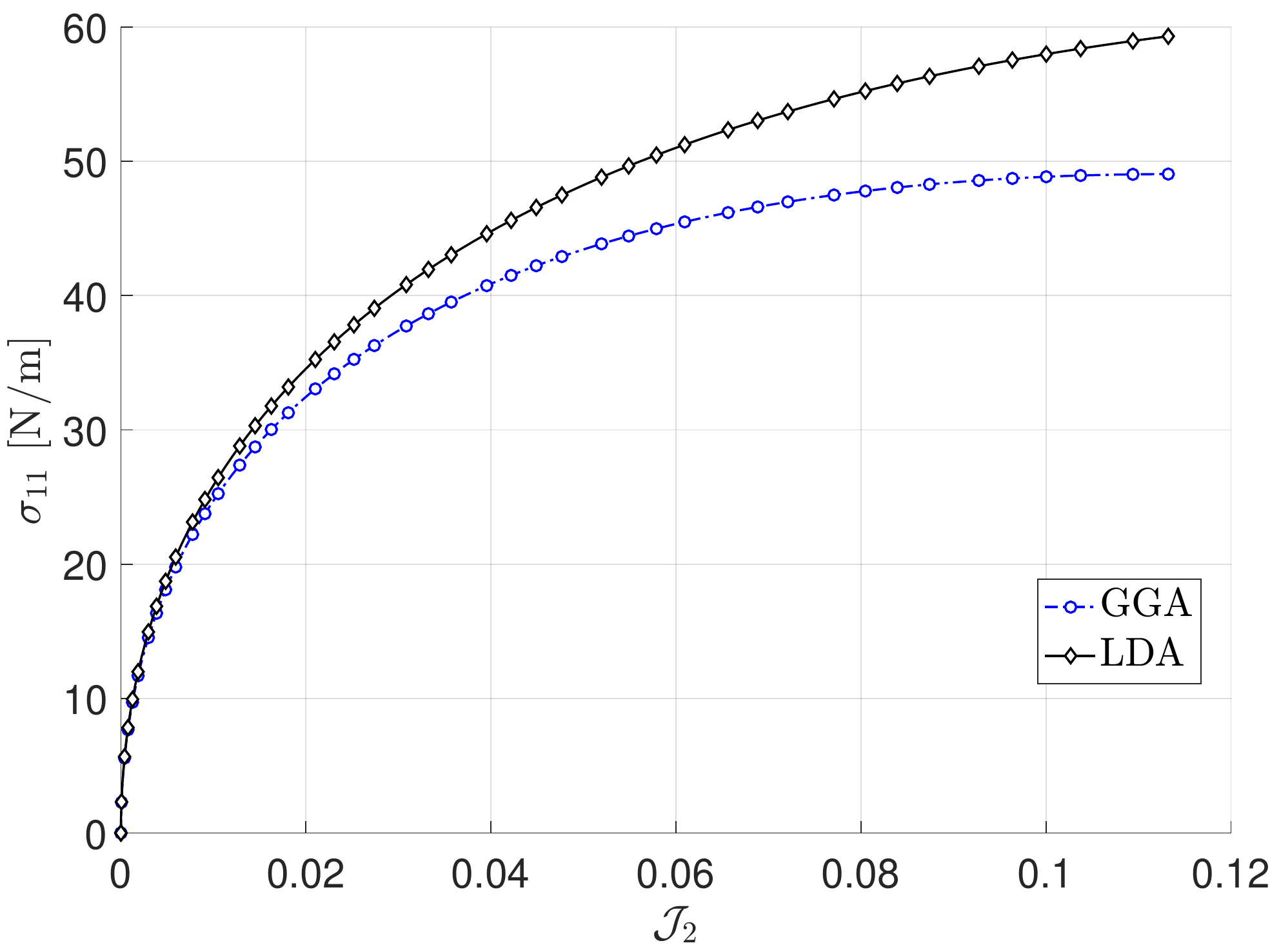}
        \subcaption{}
        \label{f:Analytical_solution_pure_shear_Sig11_zigzag_fem}
    \end{subfigure}
    \begin{subfigure}[t]{0.49\textwidth}
        \centering
    \includegraphics[height=55mm]{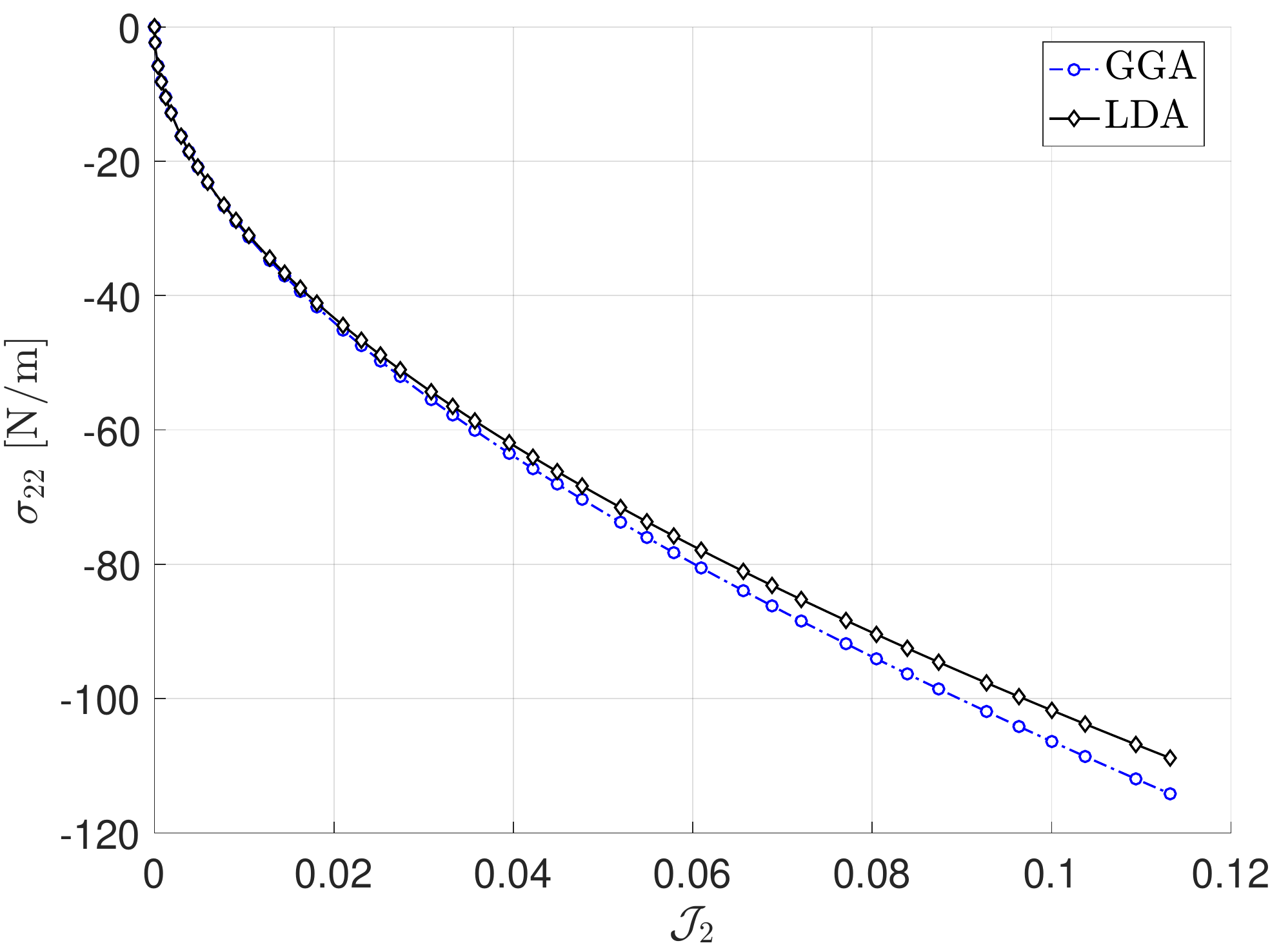}
        \subcaption{}
        \label{f:Analytical_solution_pure_shear_Sig22_zigzag_fem}
    \end{subfigure}
    \caption{\textcolor{cgn2}{Pure shear with pull in the zigzag and compression in the armchair direction ($\theta=\pi/6$ in Fig.~3):} (\subref{f:Analytical_solution_pure_shear_Sig11_zigzag_fem}) Stress in the pull direction, $\sigma_{11}$;
    (\subref{f:Analytical_solution_pure_shear_Sig22_zigzag_fem}) stress in the compression direction, $\sigma_{22}$.}
    \label{f:solution_pure_shear_zigzag_fem}
\end{figure}

\begin{figure}
    \begin{subfigure}{0.32\textwidth}
        \centering
    \includegraphics[height=38mm]{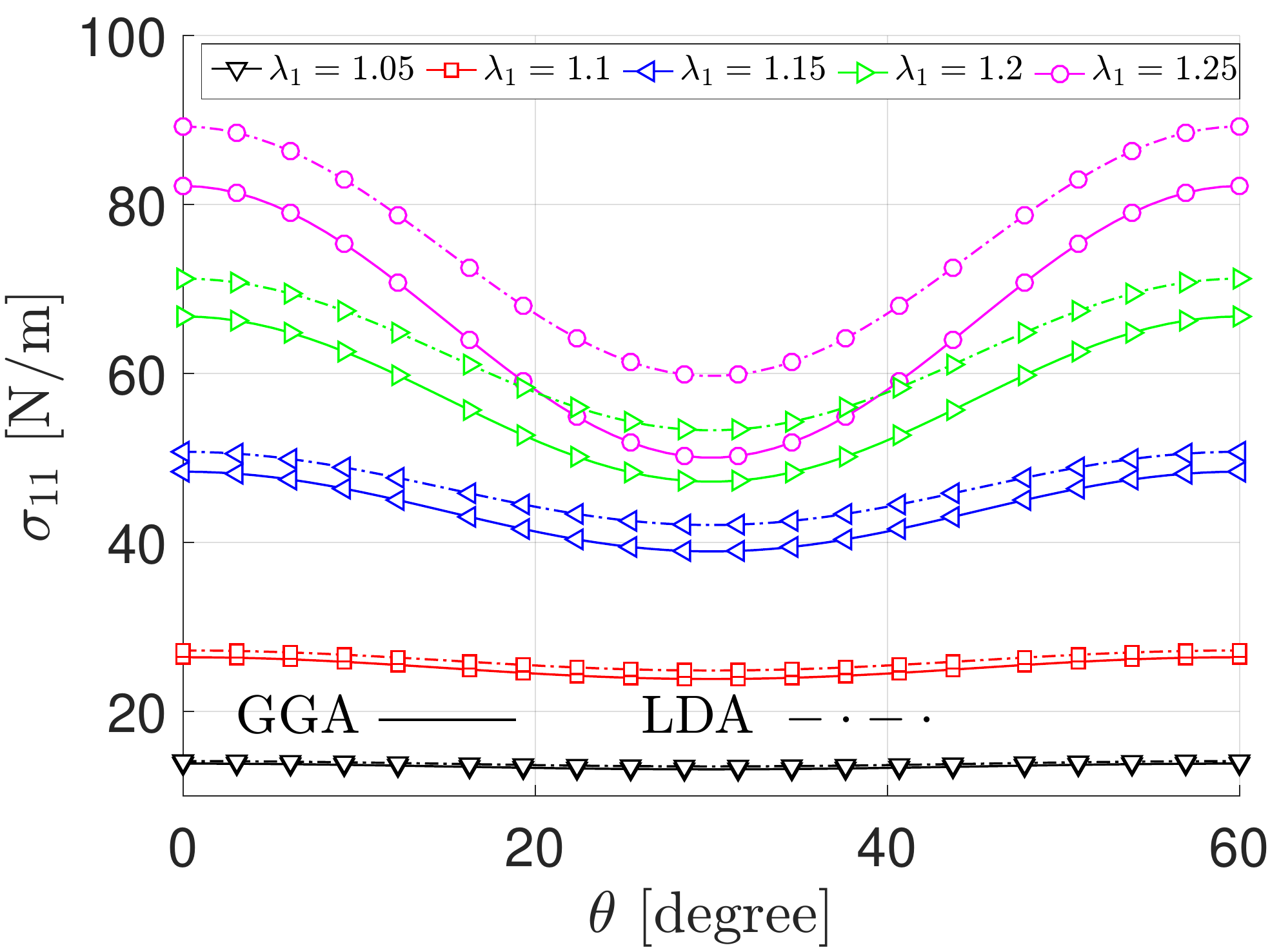}
        \subcaption{}
        \label{f:pure_shear_Stress_cart_sig11}
    \end{subfigure}
    \begin{subfigure}{0.32\textwidth}
        \centering
    \includegraphics[height=38mm]{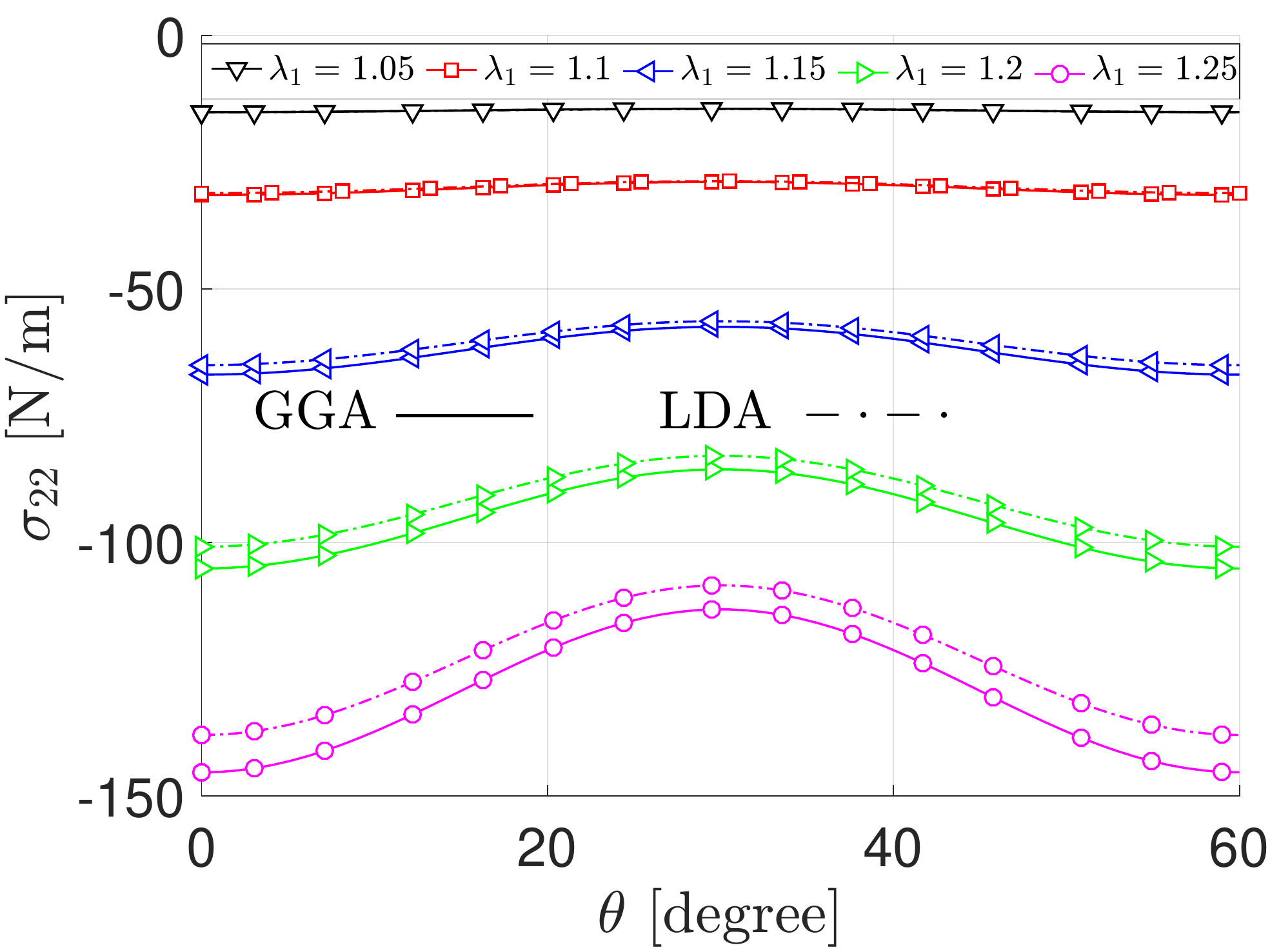}
        \subcaption{}
        \label{f:pure_shear_Stress_cart_sig22}
    \end{subfigure}
    \begin{subfigure}{0.32\textwidth}
        \centering
    \includegraphics[height=38mm]{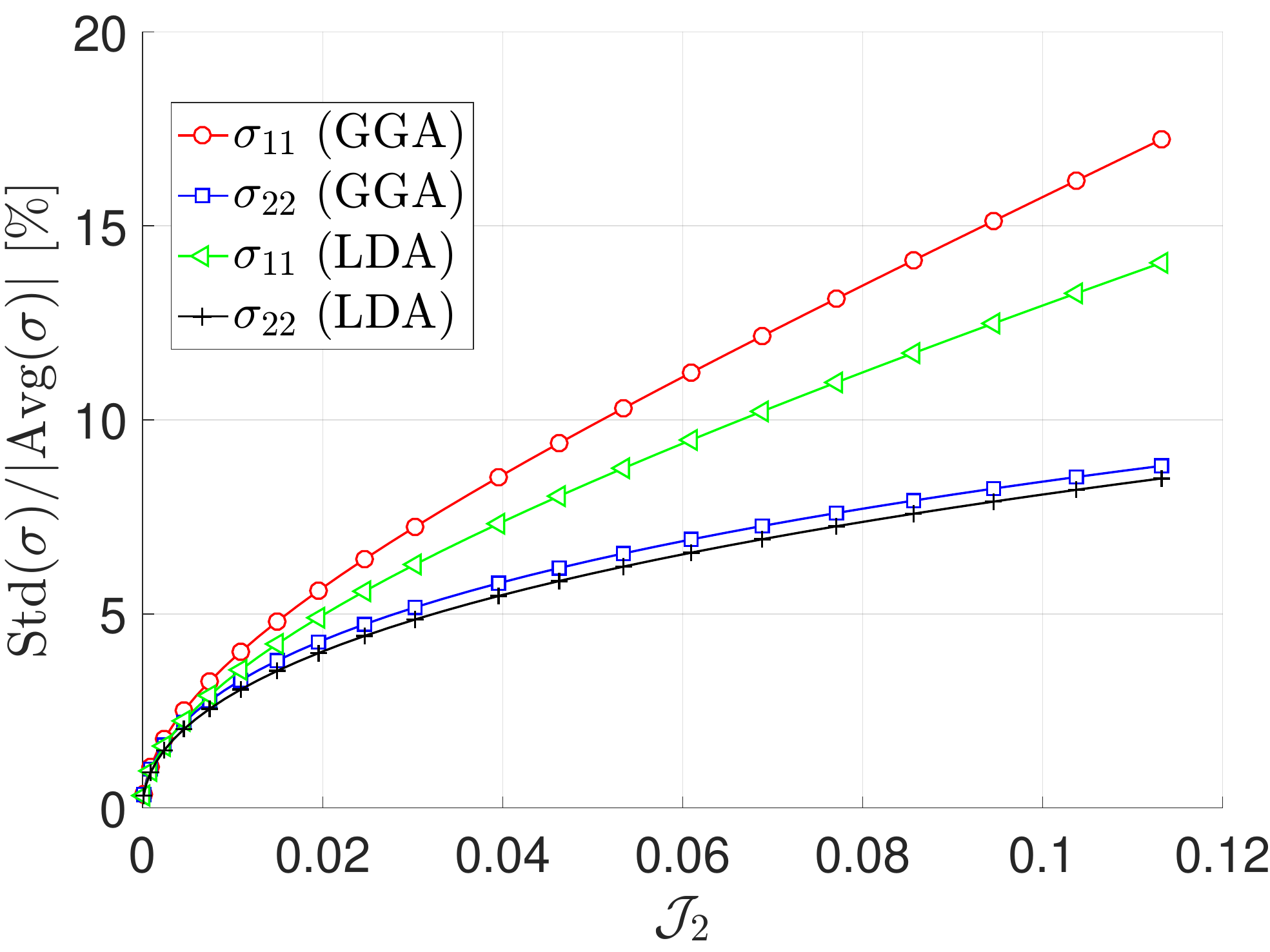}
        \subcaption{}
        \label{f:pure_shear_Stress_std}
    \end{subfigure}
    \vspace*{-0.3cm}
    \caption{Pure shear: Dependence of the stress components on stretch $\lambda_1$ and angle $\theta$,  (\subref{f:pure_shear_Stress_cart_sig11}) $\sigma_{11}$; (\subref{f:pure_shear_Stress_cart_sig22}) $\sigma_{22}$. (\subref{f:pure_shear_Stress_std}) Dependence of the standard deviation of the stress on $\sJ_2$.}
    \label{f:pure_shear_Stress_cart_sig}
\end{figure}

Pure shear is considered in different directions, i.e.~different chiralities. Therefore, the specimen is pulled in one direction and compressed in the perpendicular direction \textcolor{cgn}{by \textcolor{cgn2}{$\lambda_1 = \lambda_0>1$} and $\lambda_2 = 1/\lambda_0$ such that $J = \lambda_1\,\lambda_2 = 1$} (Fig.~\ref{f:Boundary_conditions}c).
The Cartesian stress components $\sigma_{11}$ and $\sigma_{22}$ are introduced and analytically derived in Appendix~ \ref{s:analytical_solution}. The stress variation versus $\sJ_2$, in the pull and compression direction, for the armchair and zigzag direction, are shown in Figs.~\ref{f:solution_pure_shear_armchair_fem} and \ref{f:solution_pure_shear_zigzag_fem}. \\
\textcolor{cgm2}{Next}, the loading is applied in different directions, relative to the armchair direction, and stress variation is demonstrated for a set of stretch ratio\textcolor{cgn2}{s} and GGA and LDA parameters in Fig.~\ref{f:pure_shear_Stress_cart_sig}. The graphes show the similar periodic behavior as noted in the previous section. Finally, the standard deviation and average of the stresses in one period are calculated for the different stretch ratios. The standard deviation is normalized by the average stress and it is shown versus $\sJ_2$ in Fig.~\ref{f:pure_shear_Stress_std}. Like for uniaxial stretch, the graph can be interpreted as the \textcolor{cgm2}{error caused by assuming isotropy}. As shown, this \textcolor{cgm2}{error} can reach up to 17 percent. The anisotropic behavior is increasing fast even for low values of $\sJ_2$.

\subsection{Pure bending}
\begin{figure}
\begin{subfigure}{0.495\textwidth}
\centering
     \includegraphics[height=50mm]{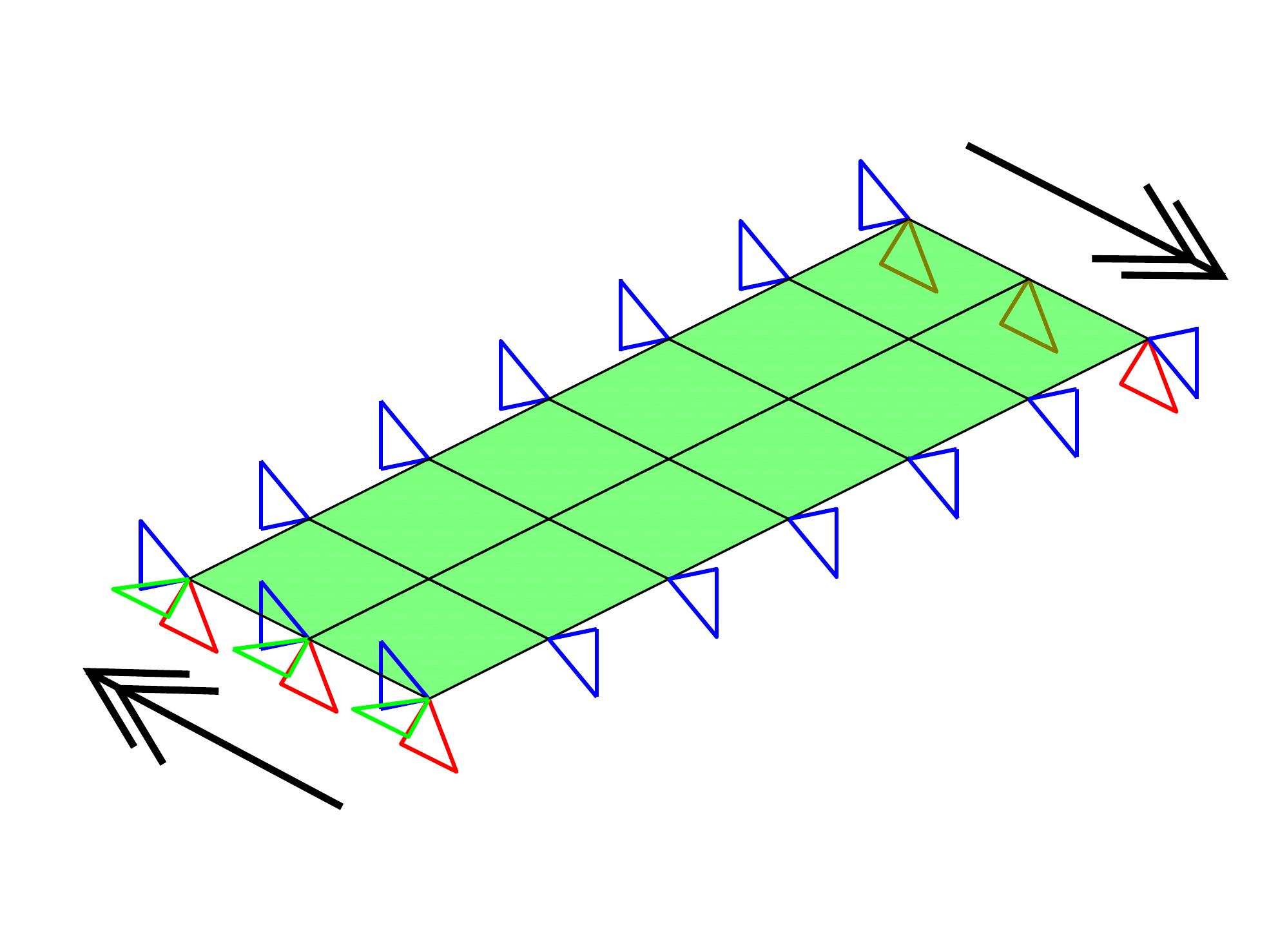}
        \subcaption{}
        \label{f:Pure_bending_example_bc}
        \end{subfigure}
     \begin{subfigure}{0.495\textwidth}
        \centering
 \includegraphics[height=50mm]{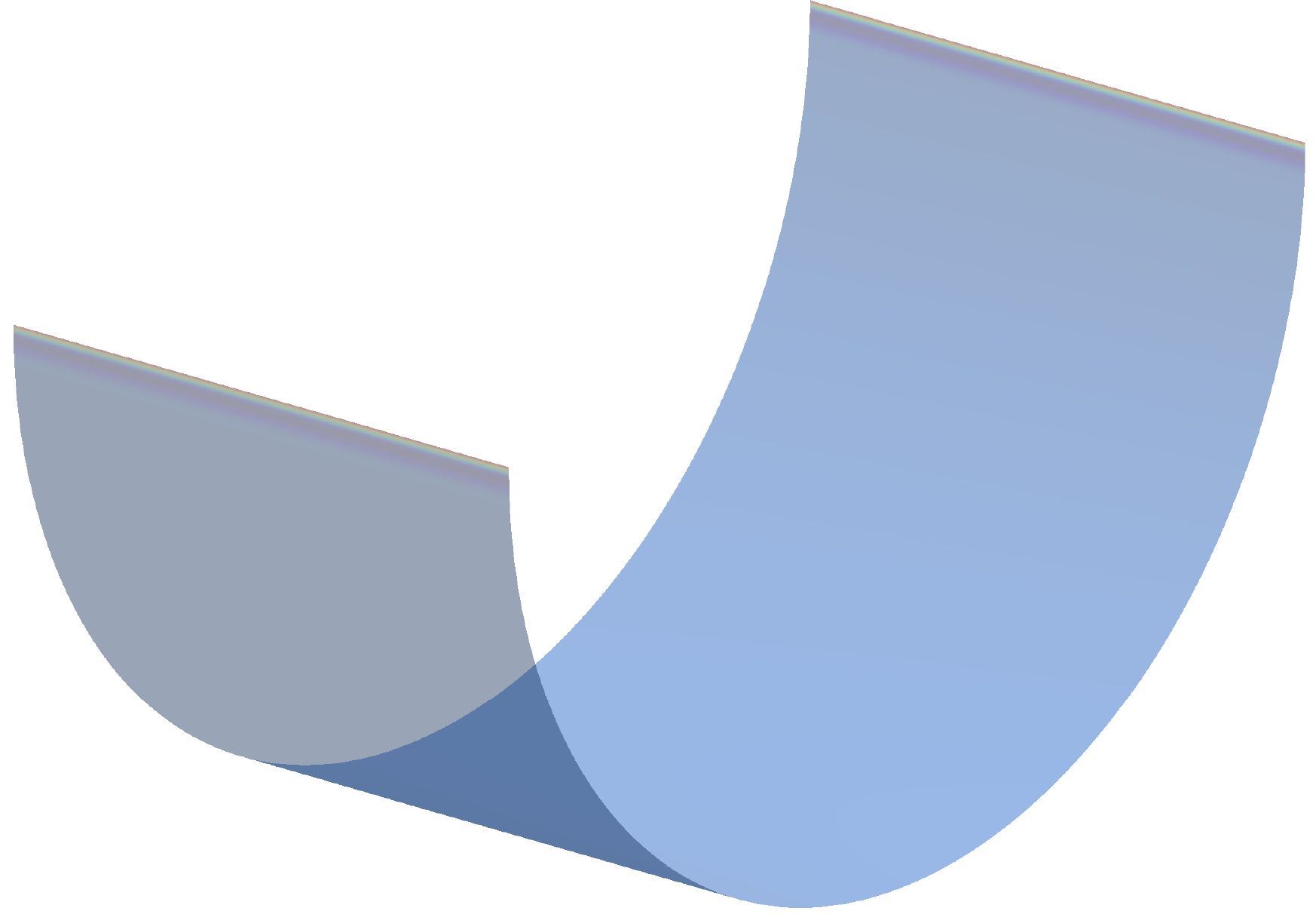}
        \subcaption{}
        \label{f:pure_bending_cutted}
    \end{subfigure}
    \vspace*{-0.3cm}
    \caption{Pure bending: (\subref{f:Pure_bending_example_bc}) boundary conditions (the green, blue and red triangles indicate restriction of displacement dofs in the longitudinal, lateral and perpendicular directions, respectively; the arrows indicate prescribed rotations); (\subref{f:pure_bending_cutted}) deformed geometry.}
\end{figure}

\begin{figure}
    \begin{subfigure}{0.49\textwidth}
        \centering
     \includegraphics[height=50mm]{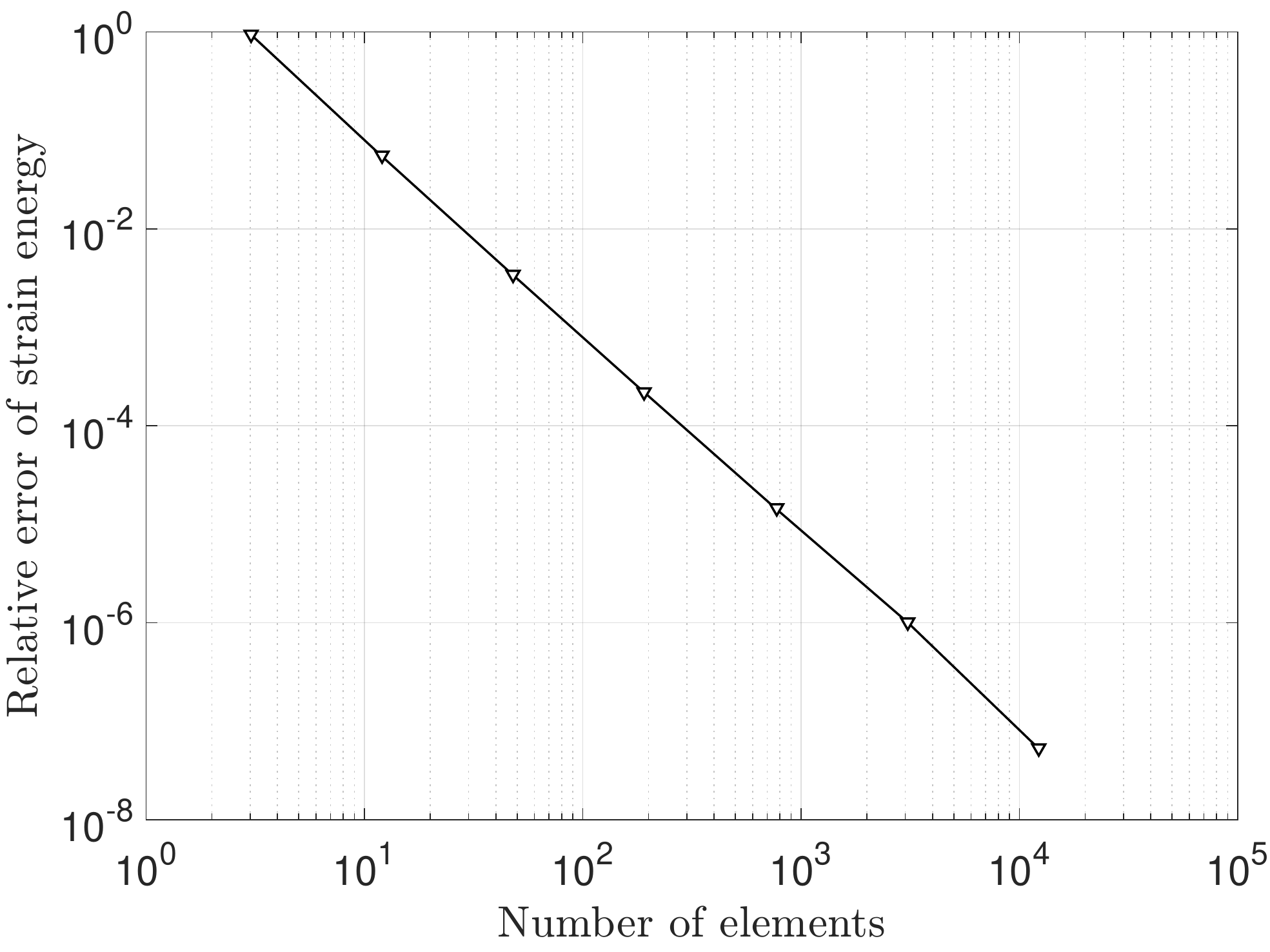}
        \subcaption{}
        \label{f:Pure_Bending_Energy_error}
    \end{subfigure}
    \begin{subfigure}{0.49\textwidth}
        \centering
     \includegraphics[height=50mm]{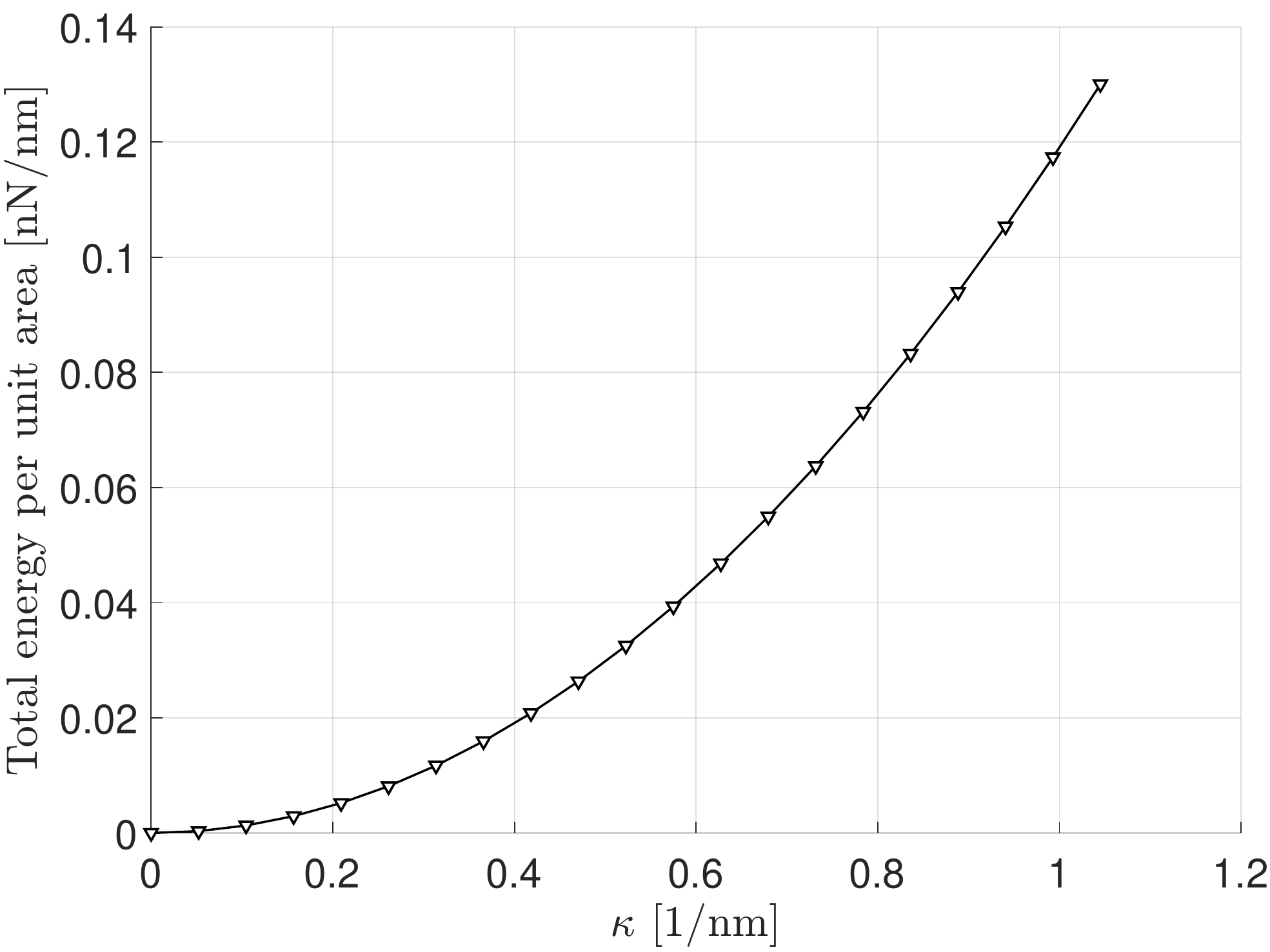}
        \subcaption{}
        \label{f:Pure_Bending_Energy}
    \end{subfigure}
    \vspace*{-0.3cm}
    \caption{Pure bending: (\subref{f:Pure_Bending_Energy_error}) Error of strain energy relative to the finest mesh ($384\times128$ quadratic NURBS elements); (\subref{f:Pure_Bending_Energy}) total energy per unit reference area.}
    \label{f:Pure_Bending}
\end{figure}

In this section, the bending formulation is validated by considering pure bending. The boundary conditions are shown in Fig.~\ref{f:Pure_bending_example_bc}. The specimen's width is restrained and it is stress free in the longitudinal direction. The details of the analytical solution for pure bending can be found in~\cite{Sauer2015_02}. The resultant force along the longitudinal direction, $N_1^1$ (see Eq.~(\ref{e:mix_N})), is zero for pure bending. $N_1^1=0$ is solved by the standard secant method to obtain the analytical solution. The final geometry is plotted in Fig.~\ref{f:pure_bending_cutted}. The convergence of the total energy is obtained with mesh refinement (Fig.~\ref{f:Pure_Bending_Energy_error}). The bending strain energy per unit reference area is shown in Fig.~\ref{f:Pure_Bending_Energy}. The relation between bending moment and curvature is linear.
\vspace*{-0.3cm}
\section{Numerical examples}\label{s:Numerical_example}
In the previous section, the proposed model is verified by standard tests. In the current section, several numerical benchmark examples are solved and compared with results from the literature. Considered are \textcolor{cgn}{five} examples: indentation and peeling of graphene, and torsion\textcolor{cgn}{, bending and uniaxial stretch} of carbon nanotubes (CNTs).
\begin{comment}
Finally, it should be noted that all simulations are conducted with finite elements, unless otherwise mentioned.
\end{comment}
\subsection{Indentation of graphene}
In order to compare the developed model with experimental results, an indentation problem is considered in this section. The experimental results are taken from a setup with \textcolor{cgn2}{multiple} cavities. Relative position of these cavities can \textcolor{cgn2}{affect} experimental measurements. In current numerical simulations, only one cavity is modeled and frictional effects are neglected. In the following subsections, first the specimen details, boundary conditions, and loading are discussed. Finally, a parameter study is conducted for the adhesion strength and indentor radius. \cite{Kumar2015_01} \textcolor{cgn2}{investigated} the same problem using the Morse potential to model substrate interactions and ABAQUS \citep{abaqus2016} in an explicit dynamic manner. In the current work, the Lennard-Jones potential and a quasi-static formulation is used.
\subsubsection{Specimen details, boundary conditions and substrate adhesion}
\begin{figure}
    \begin{subfigure}[t]{0.25\textwidth}
        \centering
     \includegraphics[height=65mm]{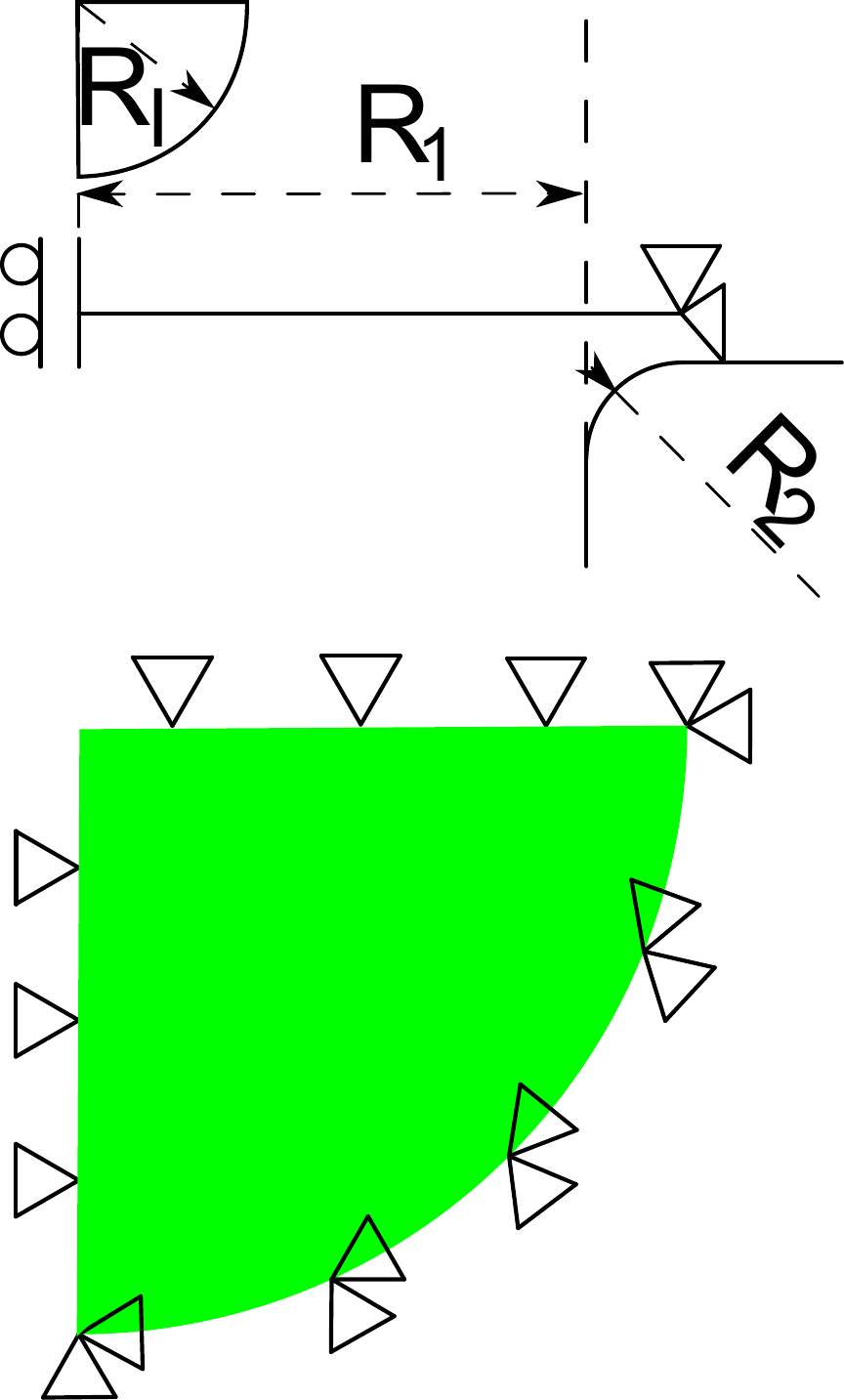}
        \subcaption{}
        \label{f:sketch_bc_indentation}
    \end{subfigure}
    \begin{subfigure}[t]{0.25\textwidth}
        \centering
     \includegraphics[height=65mm]{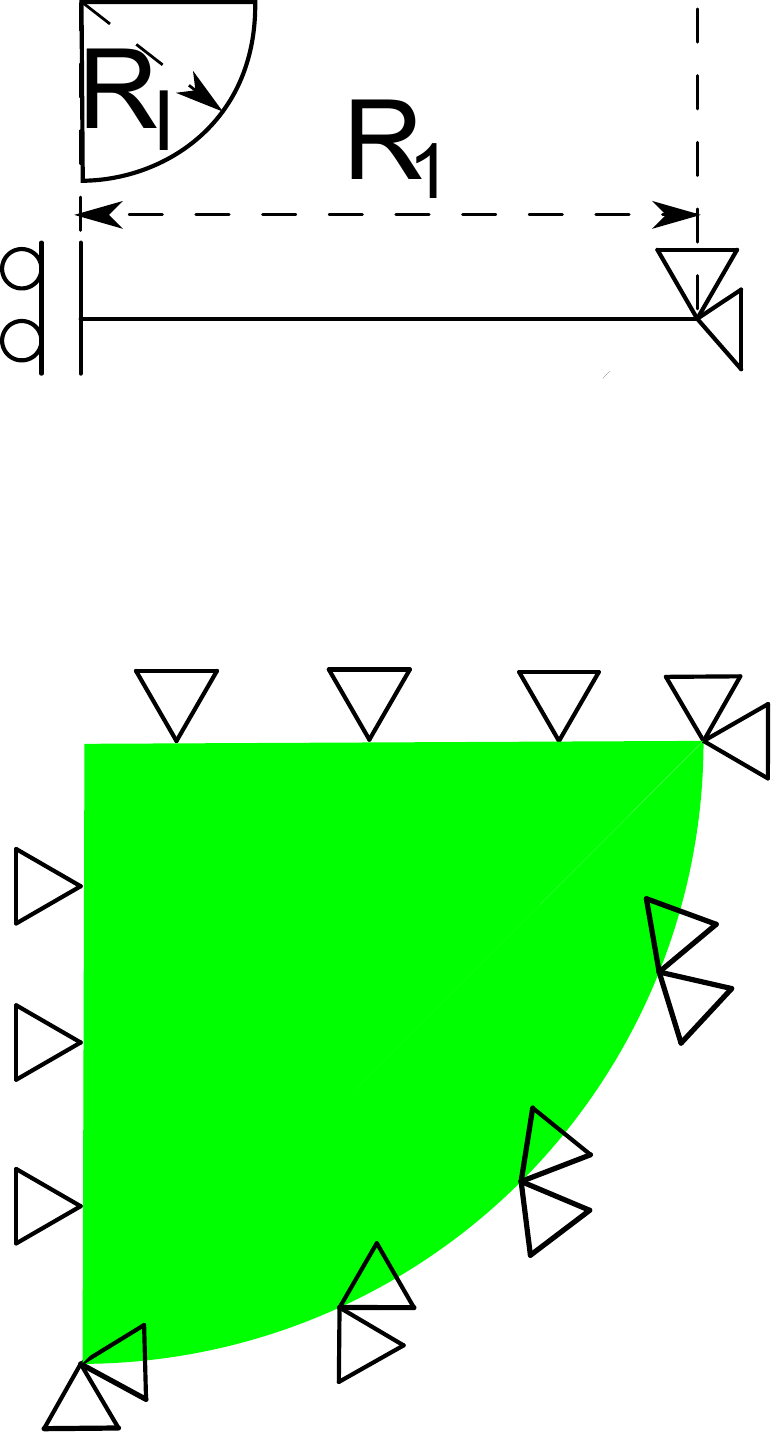}
        \subcaption{}
        \label{f:sketch_bc_indentation_woa}
    \end{subfigure}
    \begin{subfigure}[t]{0.49\textwidth}
 \includegraphics[height=60mm]{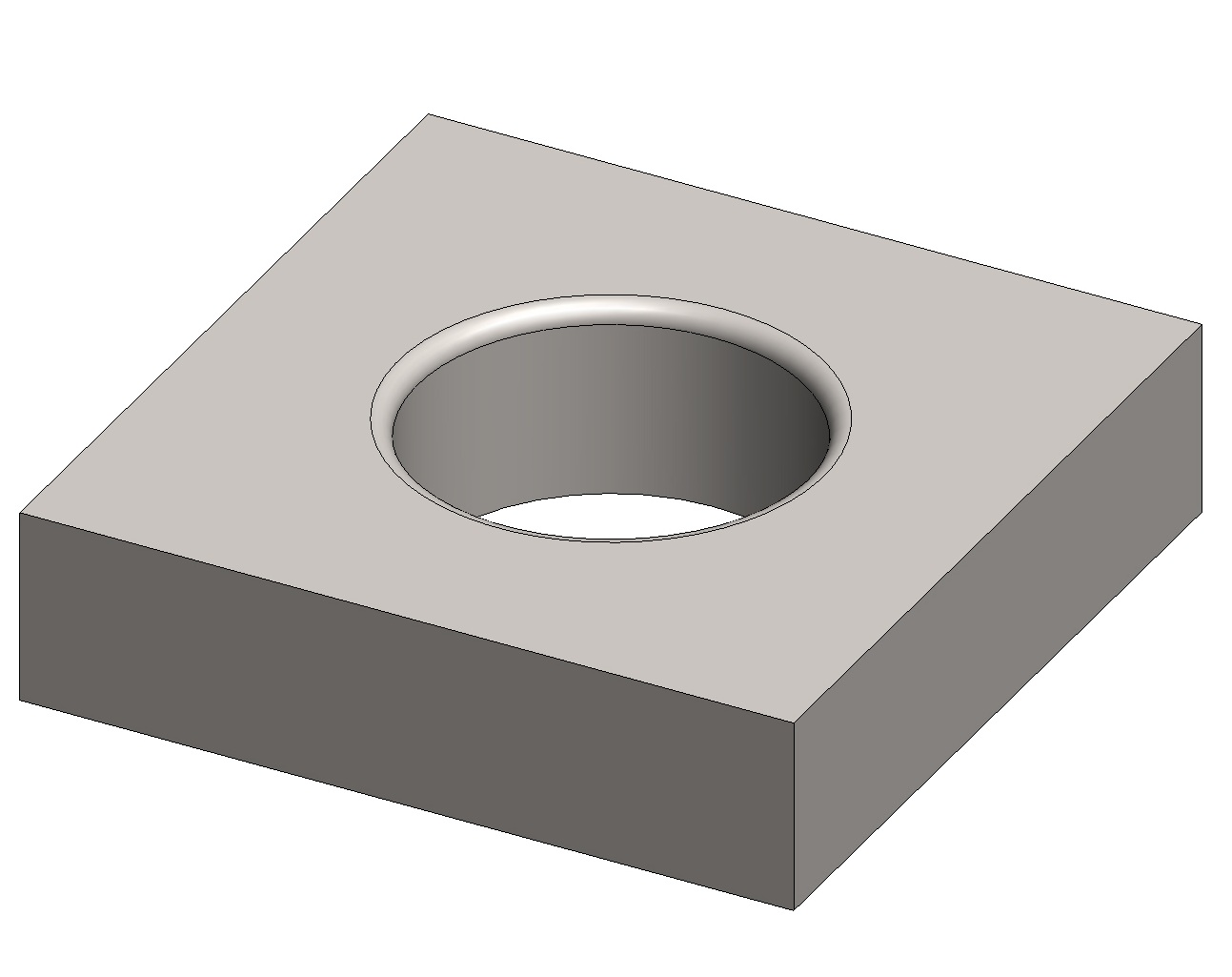}
        \subcaption{}
        \label{f:substrated_fillet}
    \end{subfigure}
    \caption{Indentation setup: Outer boundary conditions and symmetry boundary conditions for (\subref{f:sketch_bc_indentation}) adhesive substrate and (\subref{f:sketch_bc_indentation_woa}) substrate without adhesion; (\subref{f:substrated_fillet}) substrate geometry. $R_I$ is the indentor radius, $R_1$ the cavity radius and $R_2$ the fillet radius of the cavity.}
    \label{f:intendation_setup}
\end{figure}

\begin{figure}
    \centering
    \begin{subfigure}[t]{0.49\textwidth}
        \includegraphics[height=50mm]{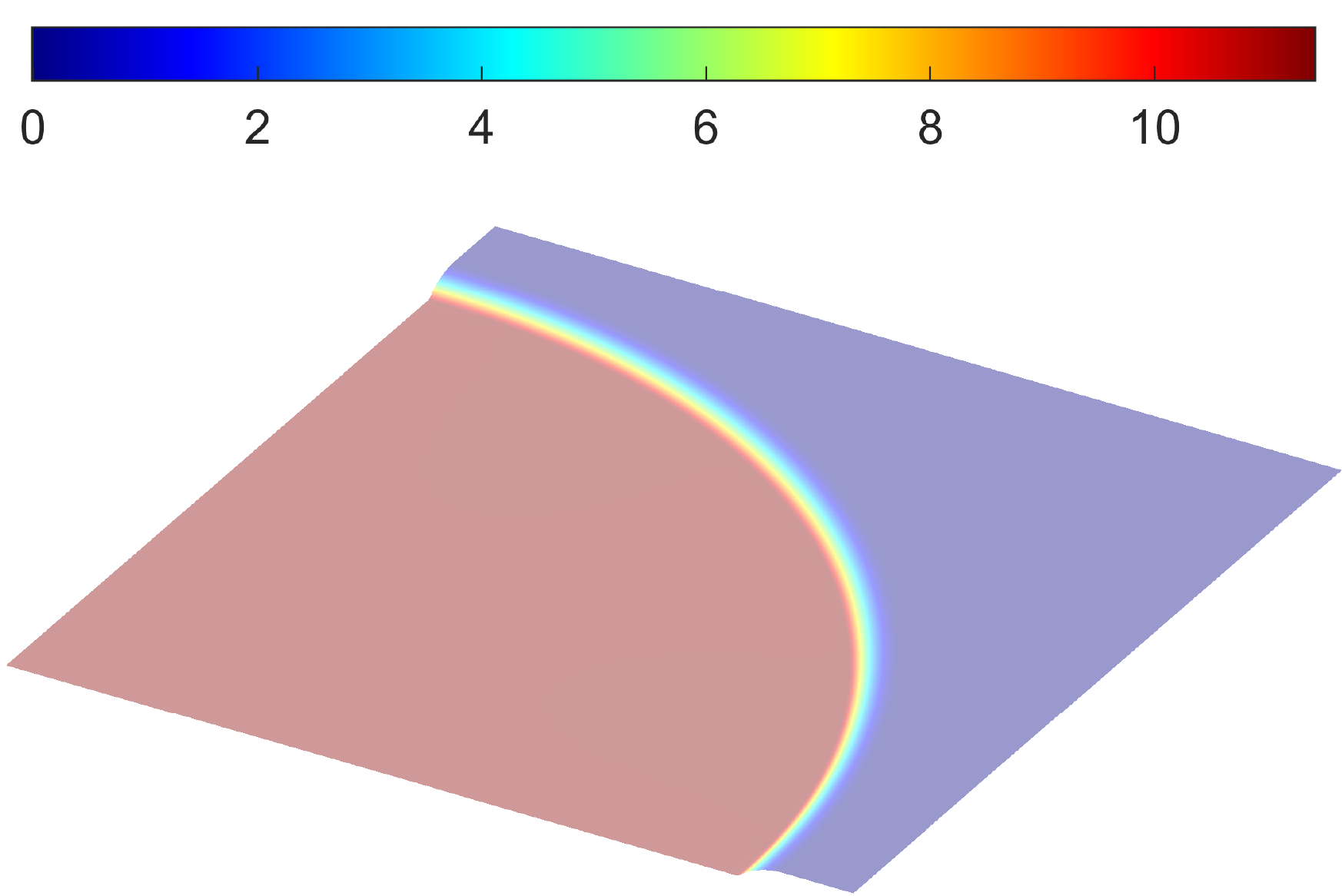}
        \subcaption{}
        \label{f:Graphene_Relaxation_disp}
    \end{subfigure}
    \begin{subfigure}[t]{0.49\textwidth}
        \includegraphics[height=50mm]{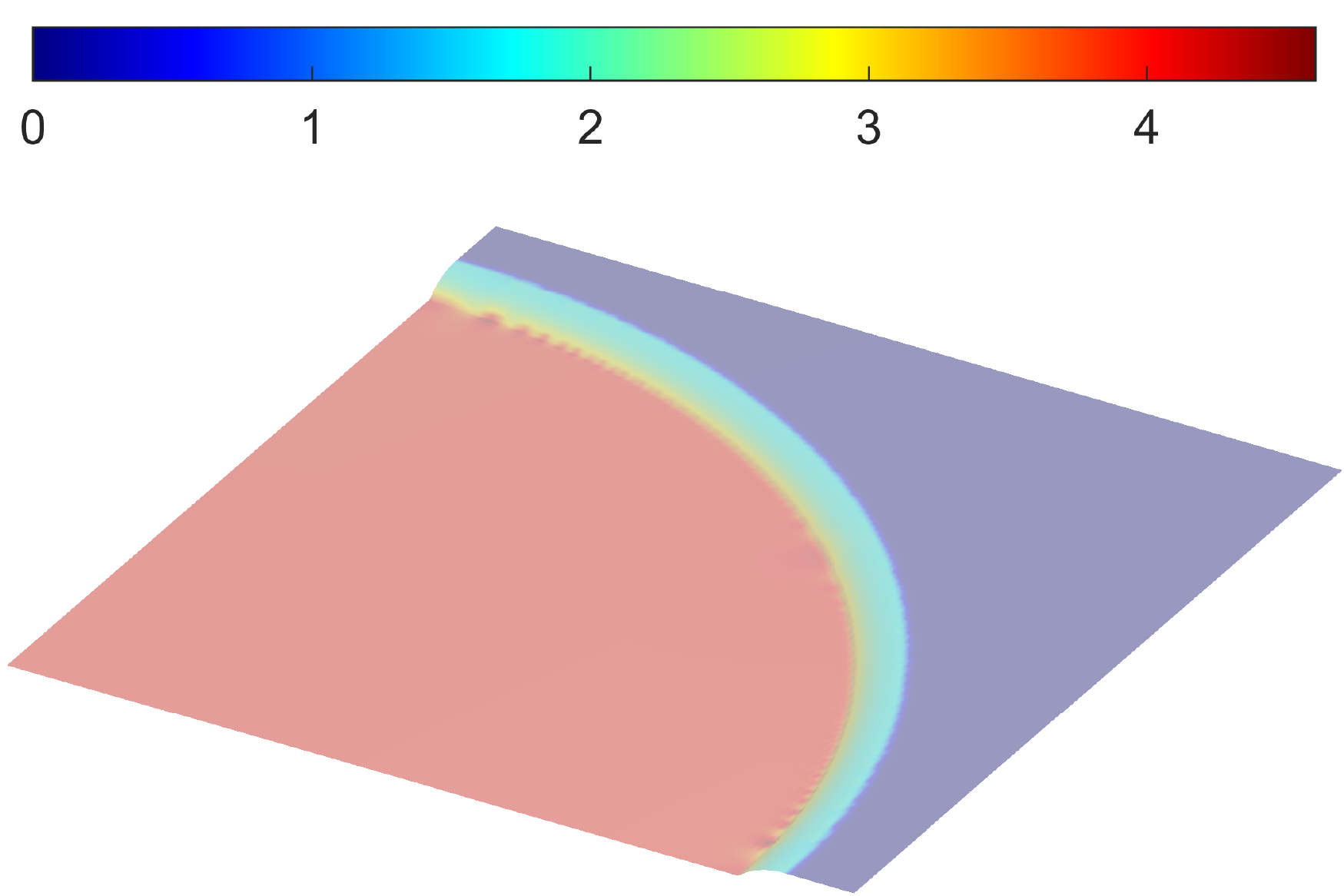}
        \subcaption{}
        \label{f:Graphene_Relaxation_stress}
    \end{subfigure}
    \caption{Substrate adhesion of graphene: (\subref{f:Graphene_Relaxation_disp}) Displacement in the perpendicular direction to the graphene sheet after relaxation (in $\text{nm}$); (\subref{f:Graphene_Relaxation_stress}) $\text{tr}(\bsig)$ [N/m]. The adhesion strength $\Gamma$ and $\text{R}_1$ (see Fig.~\ref{f:sketch_bc_indentation}) are set to 0.45 N/m and 500 nm, respectively. The maximum of tr($\bsig$) and the displacement are 4.95 N/m and 11.75 nm, respectively. In the simulation, all displacement dofs with a radial position greater than $R_1$ are fixed.}
    \label{f:Graphene_Relaxation}
\end{figure}

The problem setup consists of a sheet of graphene on a $\textnormal{SiO}_2$ substrate with a micro-cavity.
The boundary conditions and geometry of the substrate are depicted in \textcolor{cgn2}{Fig.~\ref{f:intendation_setup}.} For efficiency, only one quarter of the circular sheet is discretized. Thus, additional symmetry BCs have to be applied along the symmetry planes. In addition, the outer sheet boundary is fixed in all directions. The edge of the cavity is smoothed with a fillet radius of 50 nm. The adhesion plays an important role in the initial stress, stiffness and overall response of the sheet. A relaxation step is considered before the indentation loading. The adhesion between sheet and substrate is modeled via van der Waals (vdW) interaction (Appendix \ref{s:coarse_grain_contact_model}). In the relaxation step, the adhesion parameter is increased from zero to its final value. At each step, a standard Newton Raphson \textcolor{cgm2}{iteration} is utilized to determine equilibrium. The displacement and stress distribution of the relaxed geometry are shown in Fig.~\ref{f:Graphene_Relaxation}. This stress corresponds to a pre-stress within the graphene sheet that makes the structure stiffer.

\subsubsection{Indentation: Load step}
\begin{figure}
 \centering
    \begin{subfigure}[t]{1\textwidth}
        \centering
 \includegraphics[height=35mm]{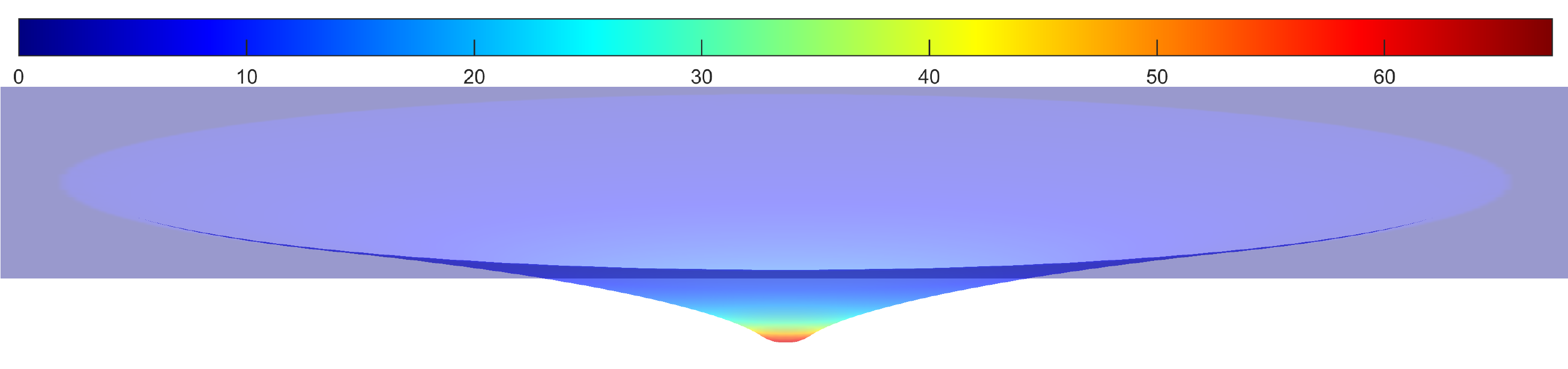}
        \subcaption{}
        \label{f:Indentation_R3_35}
    \end{subfigure}\\
        \begin{subfigure}[t]{1\textwidth}
        \centering
     \includegraphics[height=50mm]{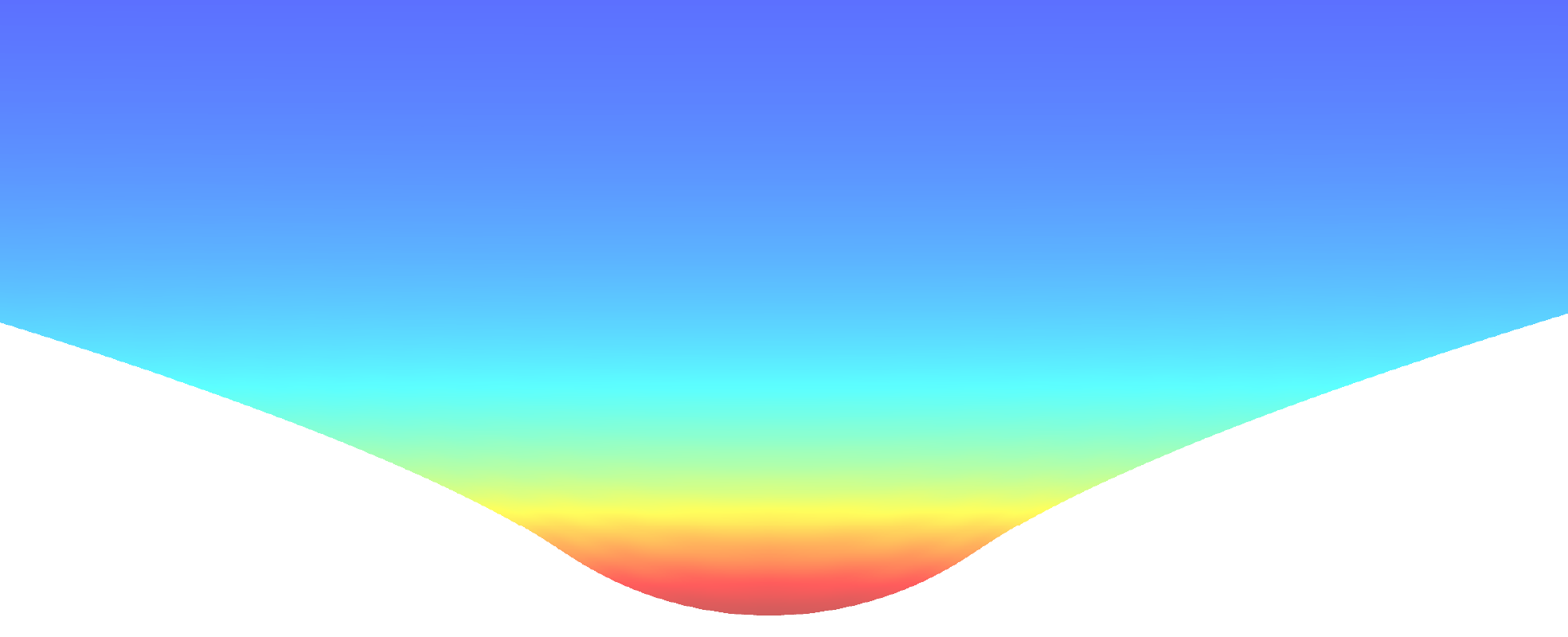}
        \subcaption{}
        \label{f:Indentation_R3_35_3D_cutted}
    \end{subfigure}
\caption{Graphene indentation:~(\subref{f:Indentation_R3_35}) Indented profile (colored by tr$(\bsig)~[\text{N/m}]$); (\subref{f:Indentation_R3_35_3D_cutted}) zoom of (\subref{f:Indentation_R3_35}). Adhesion parameter $\Gamma$, cavity radius $R_1$ and indentor radius $R_I$ are 0.45 N/m, 500 nm and 35 nm, respectively.}
\label{f:Indented_geometry}
\end{figure}

\begin{figure}
 \centering
    \begin{subfigure}[t]{0.49\textwidth}
        \centering
     \includegraphics[height=55mm]{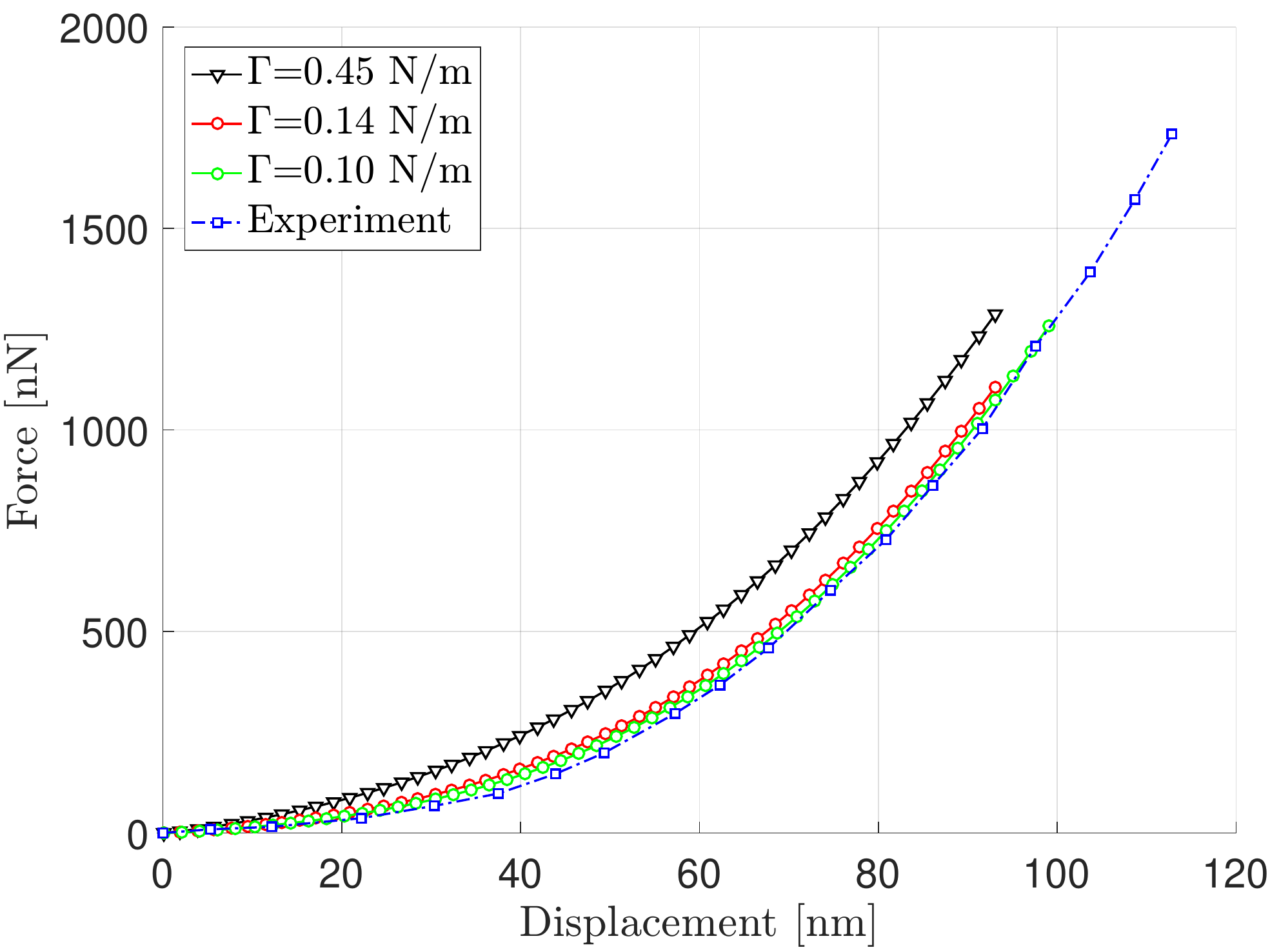}
        \subcaption{}
        \label{f:compare_ahdesion_effect_sphere_shell_R550}
    \end{subfigure}
    \begin{subfigure}[t]{0.49\textwidth}
        \centering
 \includegraphics[height=55mm]{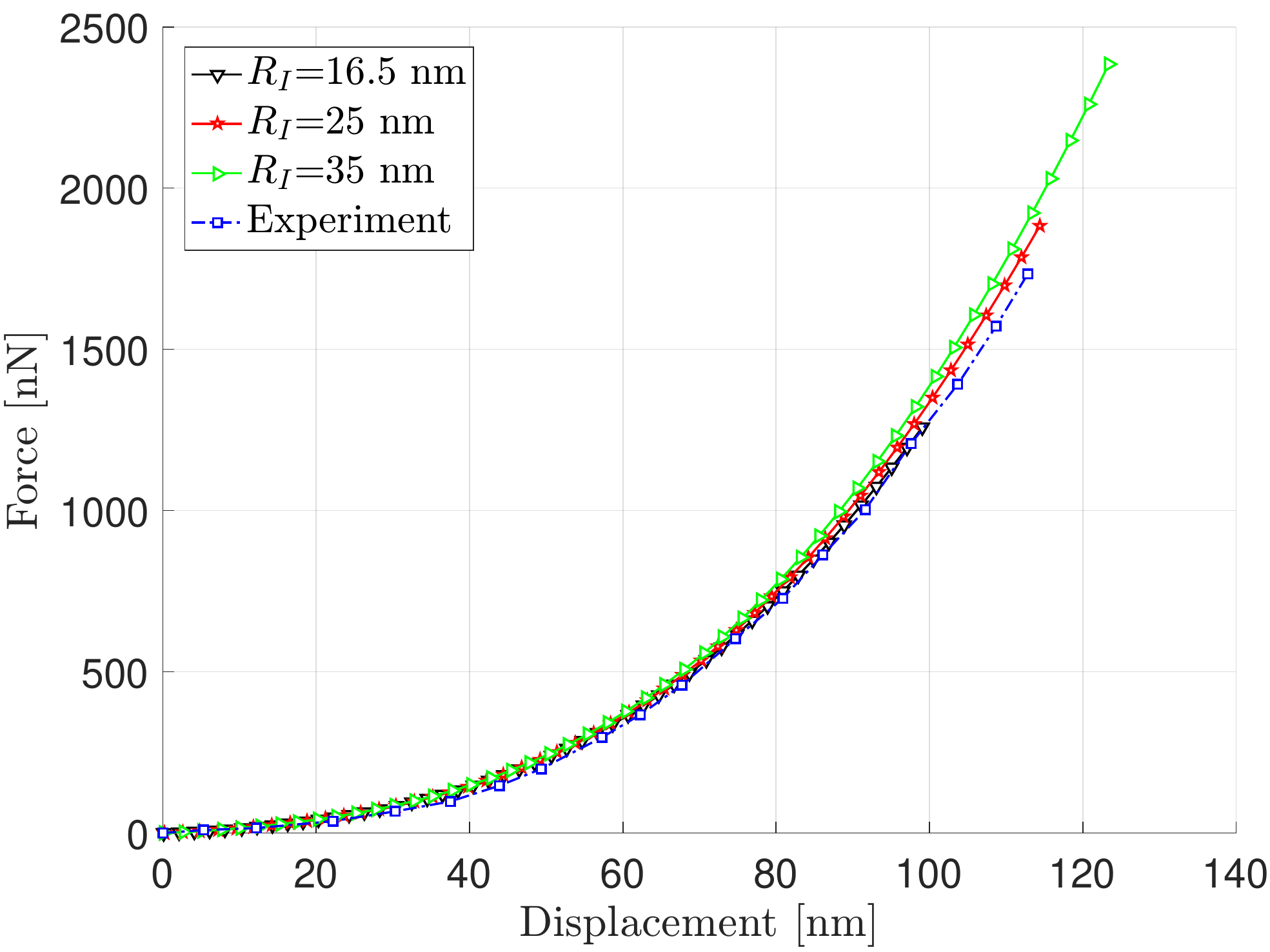}
        \subcaption{}
        \label{f:Indentation_force_disp_with_diff_indentor_sphere_shell_A01}
    \end{subfigure}
\caption{Graphene indentation, force-displacement curve of the indentor:~(\subref{f:compare_ahdesion_effect_sphere_shell_R550}) Comparison between different adhesion parameters (for fixed $R_1= 500$~nm and $R_I=16.5$~nm); (\subref{f:Indentation_force_disp_with_diff_indentor_sphere_shell_A01}) comparison between different indentor radii (for fixed $\Gamma= 0.1$ N/m and $R_1= 500$~nm). The experimental results are taken from \citet{Kumar2015_01}.}
\label{f:compare_model_with_and_without_ahdesion_sphere}
\end{figure}

\begin{comment}
As a first step to simulate the indentation problem, a convergence study is considered.
\end{comment}
10000 quadratic isogeometric finite elements\footnote{\textcolor{cgn}{The continuum model contains 122,412 nodes while the corresponding atomistic system has about 12 million atoms, i.e. about 100 times more.}} are used over a rectangular domain with the same number of elements in x and y directions.
Before the indentation phase, a relaxation step is conducted. A wide range for adhesion strengths ($\Gamma= 0.1\sim0.45~\text{N/m}$) can be found in the literature discussed in Sec.~\ref{s:Introduction}. To study the influence of adhesion, the indentation simulation is conducted for a set of different $\Gamma$. The indented geometry is presented in Fig.~\ref{f:Indented_geometry}. The indentor contact force is measured and compared with experimental results in Fig.~\ref{f:compare_ahdesion_effect_sphere_shell_R550}. A higher adhesion strength results in a stiffer structure and a higher indentor reaction force.\\
Next, to study the effect of the indentor size, a set of simulations are conducted for a series of indentors (Fig.~\ref{f:Indentation_force_disp_with_diff_indentor_sphere_shell_A01}). They have the same force-displacement curve in the beginning of indentation. But, they have a different response for large indentation. In addition, a larger indentor has a larger reaction force. \\
Finally, the problem is solved with zero bending stiffness (pure membrane model) and compared with the shell model for the substrate with and without adhesion. It is shown that the bending stiffness does not play an important role for the force-displacement curve (Fig.~\ref{f:compare_model_with_and_without_ahdesion_sphere}). In addition, in the cases without adhesion, it is assumed that the edge of the substrate cavity is not filleted and $R_1$ is selected to be consistent with the adhesive substrate setup (see Figs.~\ref{f:sketch_bc_indentation} and \ref{f:sketch_bc_indentation_woa}).

 \begin{figure}
 \centering
    \begin{subfigure}[t]{1\textwidth}
        \centering
     \includegraphics[height=55mm]{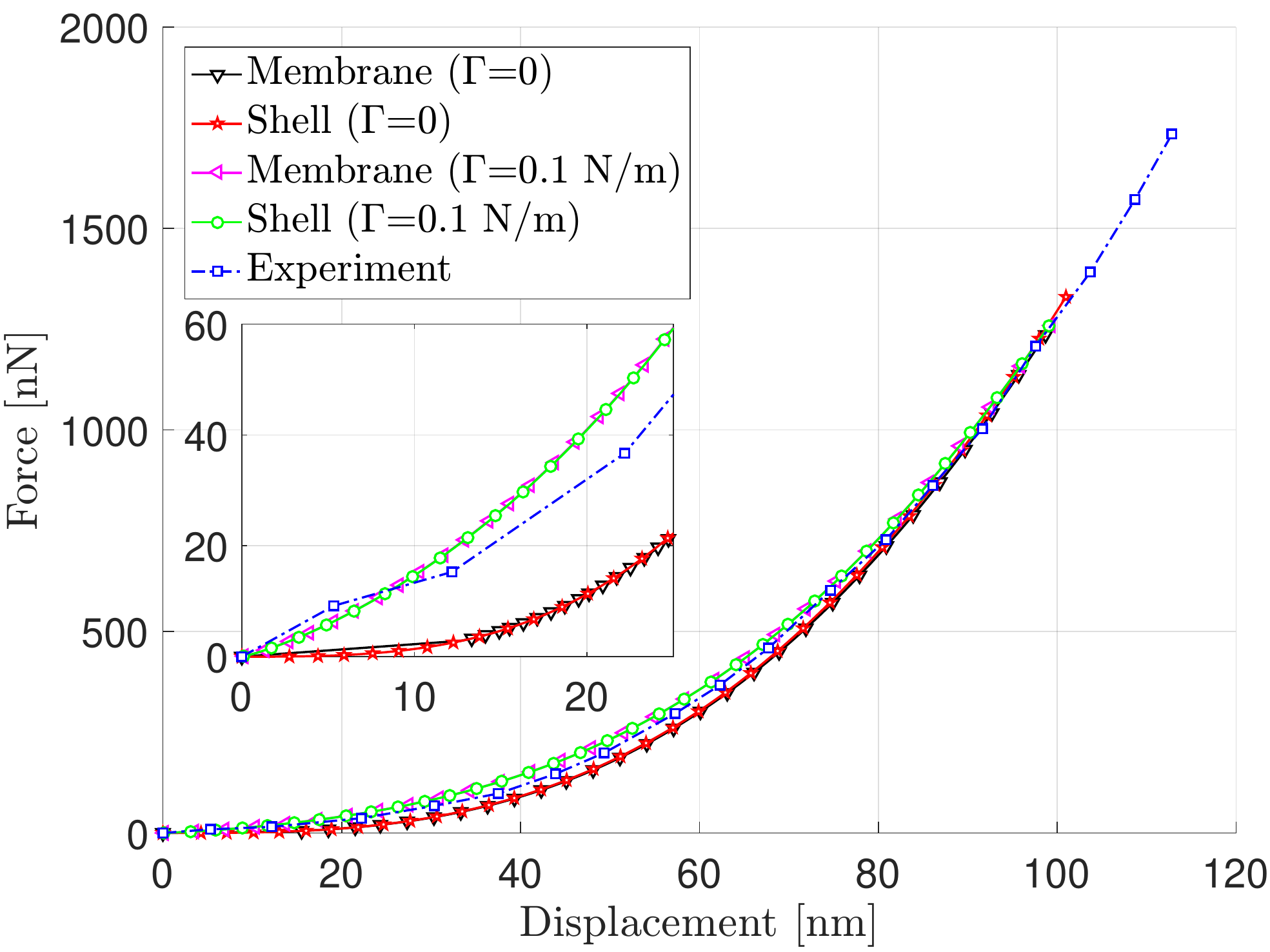}
        %\label{f:compare_model_with_and_without_ahdesion_sphere}
    \end{subfigure}
\caption{Graphene indentation: Comparison between membrane and shell models considering various values for $\Gamma$ (for fixed $R_1=500$~nm and $R_I=16.5$~nm). The experimental results are taken from \citet{Kumar2015_01}.}
\label{f:compare_model_with_and_without_ahdesion_sphere}
\end{figure}

\subsection{Peeling of graphene}
\begin{figure}[h]
    \begin{subfigure}[t]{0.5\textwidth}
        \centering
     \includegraphics[height=55mm]{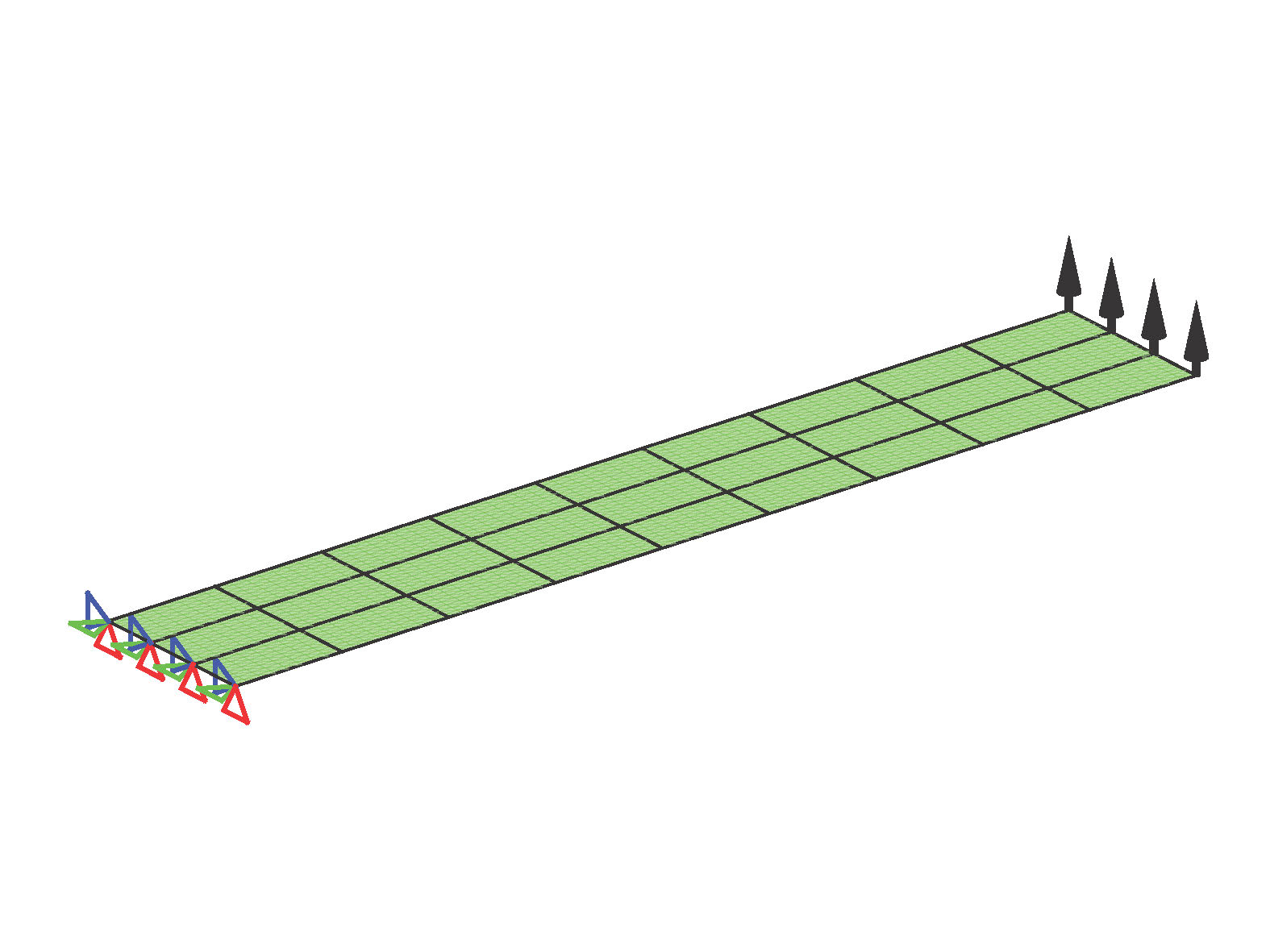}
        \subcaption{}
        \label{f:graphene_Loading_peeling}
    \end{subfigure}
    \begin{subfigure}[t]{0.5\textwidth}
        \centering
 \includegraphics[height=55mm]{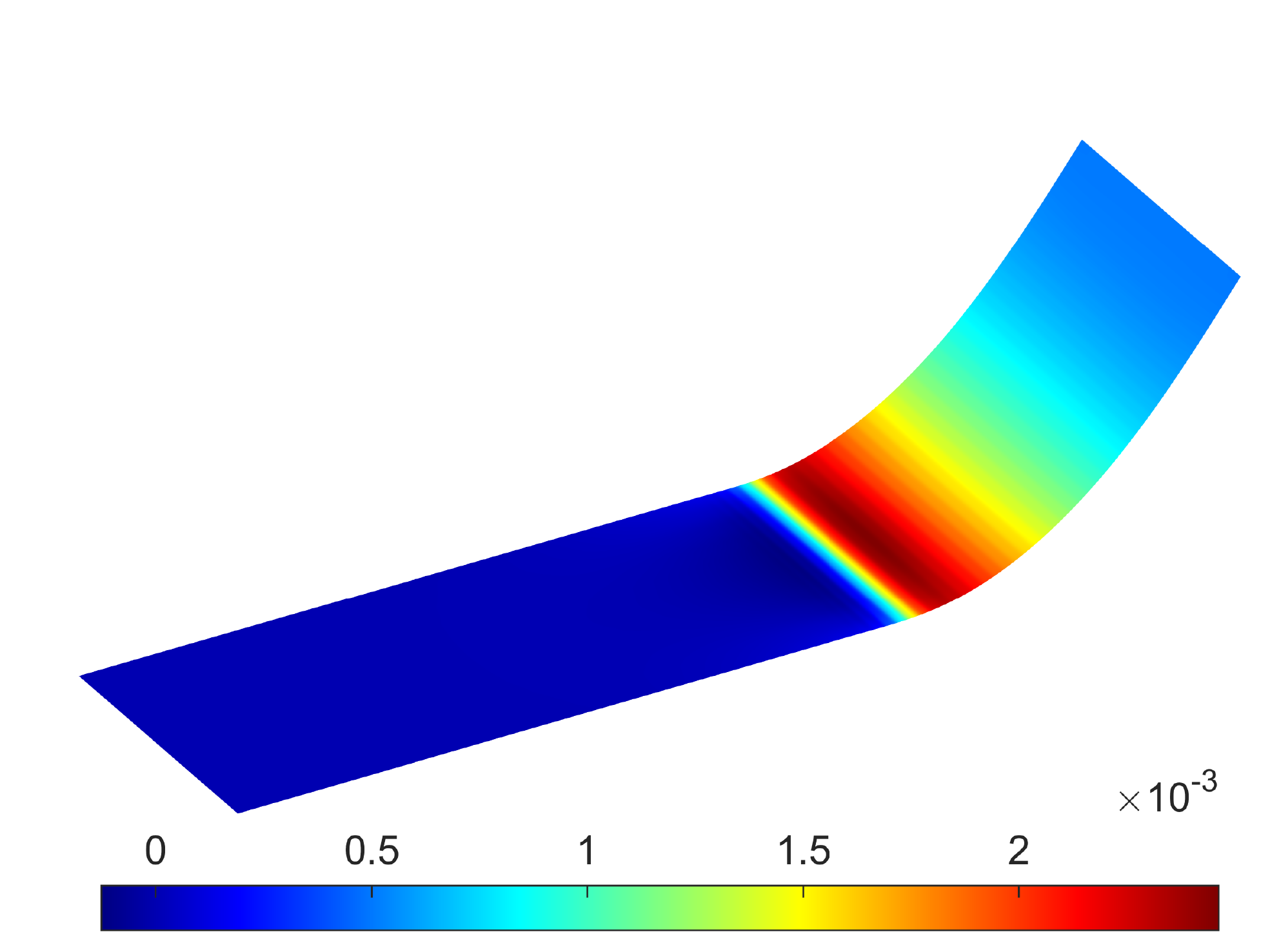}
        \subcaption{}
        \label{f:Peeled_Contour}
    \end{subfigure}
\caption{Graphene peeling:~(\subref{f:graphene_Loading_peeling}) The peeling setup (BC and loading). The size of the strip is $50 \mathrm{nm} \times 10 \mathrm{nm}$. (\subref{f:Peeled_Contour}) The deformed geometry of the peeled strip colored by tr$(\bsig)~[\text{N/m}]$.}
\end{figure}

\begin{figure}[h]
     \begin{subfigure}[t]{0.5\textwidth}
        \centering
 \includegraphics[height=55mm]{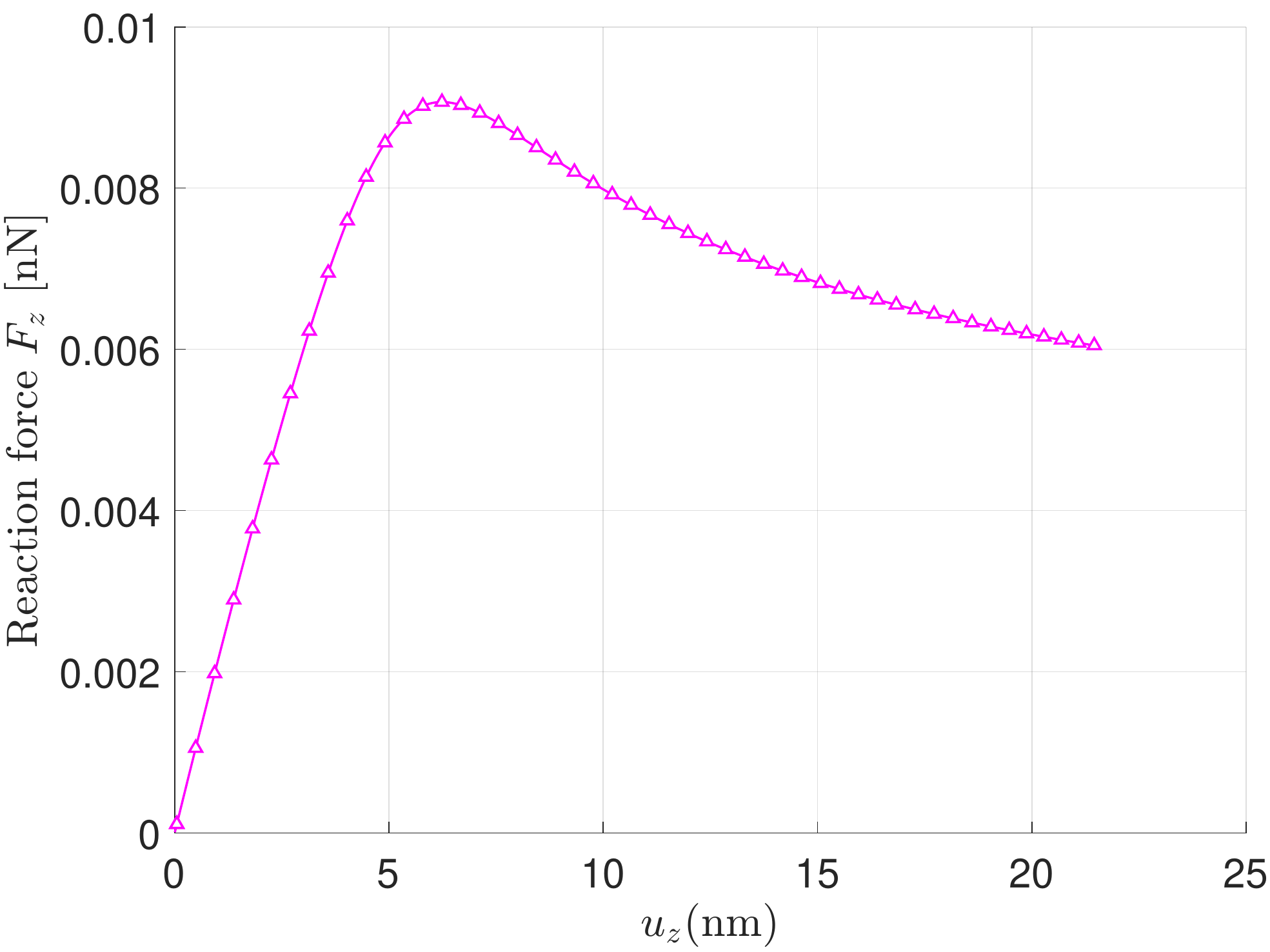}
        \subcaption{}
        \label{f:peeling_Lx50_Ly10_Reaction_force}
    \end{subfigure}
    \begin{subfigure}[t]{0.5\textwidth}
        \centering
     \includegraphics[height=55mm]{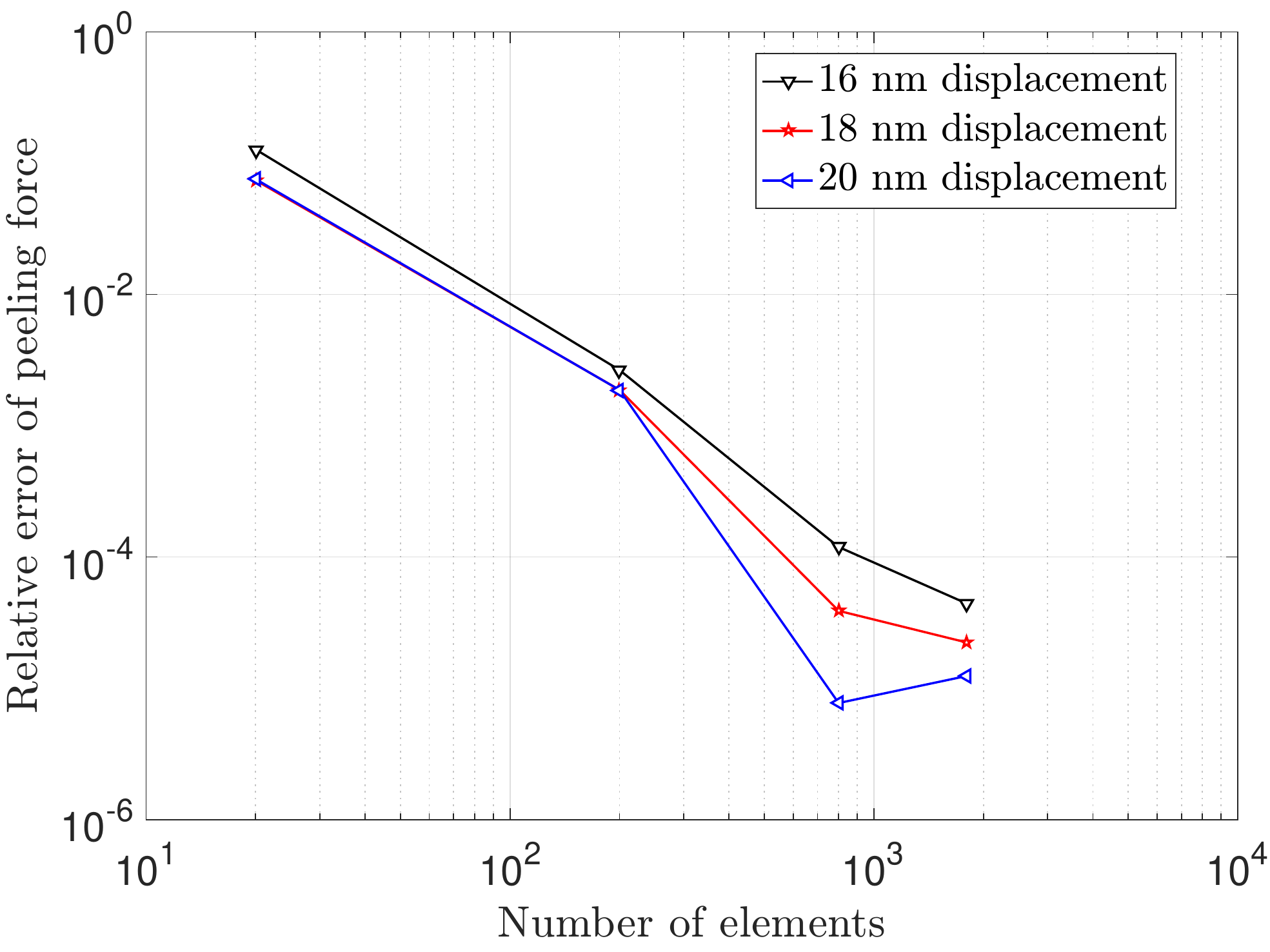}
        \subcaption{}
        \label{f:Peeling_force_shell_conv}
    \end{subfigure}
\caption{Graphene peeling: (\subref{f:peeling_Lx50_Ly10_Reaction_force}) Peeling force vs.~peeling displacement. (\subref{f:Peeling_force_shell_conv}) Error of the peeling force for various peeling displacements relative to the finest mesh ($90\times60$ quadratic NURBS elements). The LDA and FGBP parameter sets are used together with $\Gamma=0.14$~N/m.}
\end{figure}

Peeling between graphene and an adhesive substrate is considered next. A rectangular graphene strip is considered \textcolor{cgn2}{that} is initially positioned in equilibrium distance on a $\textnormal{SiO}_2$ substrate. The strip is restrained \textcolor{cgn2}{on} one side and the other side is displaced in the perpendicular direction to the substrate ($u_z$) (Fig.~\ref{f:graphene_Loading_peeling}). The first 20 percent of the length of the strip are not absorbed to the substrate (the adhesion parameter is assumed to be zero in this area). Quadratic NURBS elements with the same element size in both directions are used in the simulation. The deformed geometry is plotted in Fig.~\ref{f:Peeled_Contour}. The convergence of the peeling force with mesh refinement is shown Fig.~\ref{f:Peeling_force_shell_conv}. Fig.~\ref{f:peeling_Lx50_Ly10_Reaction_force} shows the peeling force against \textcolor{cgn2}{the} peeling displacement. The peeling force exhibits an increasing and a decreasing part. They capture the elastic deformation and adhesive debonding of the strip, respectively.

\subsection{Deformation of carbon nanotubes (CNTs)}
In this section, the proposed shell model is applied to CNTs. First, the relaxation process is discussed. Then, the CNT behavior under torsional and bending loading is investigated. The results for torsion are compared and validated with atomistic results from the literature. In both loading scenarios, the buckling of shell walls is captured and the buckling load is \textcolor{cgm2}{accurately} determined from the ratio of the membrane energy to the total energy. This energy ratio is determined from the strain energies introduced in Sec.~\ref{s:material_model}.
\subsubsection{Relaxation of CNTs}
\begin{figure}
        \centering
 \includegraphics[height=58mm]{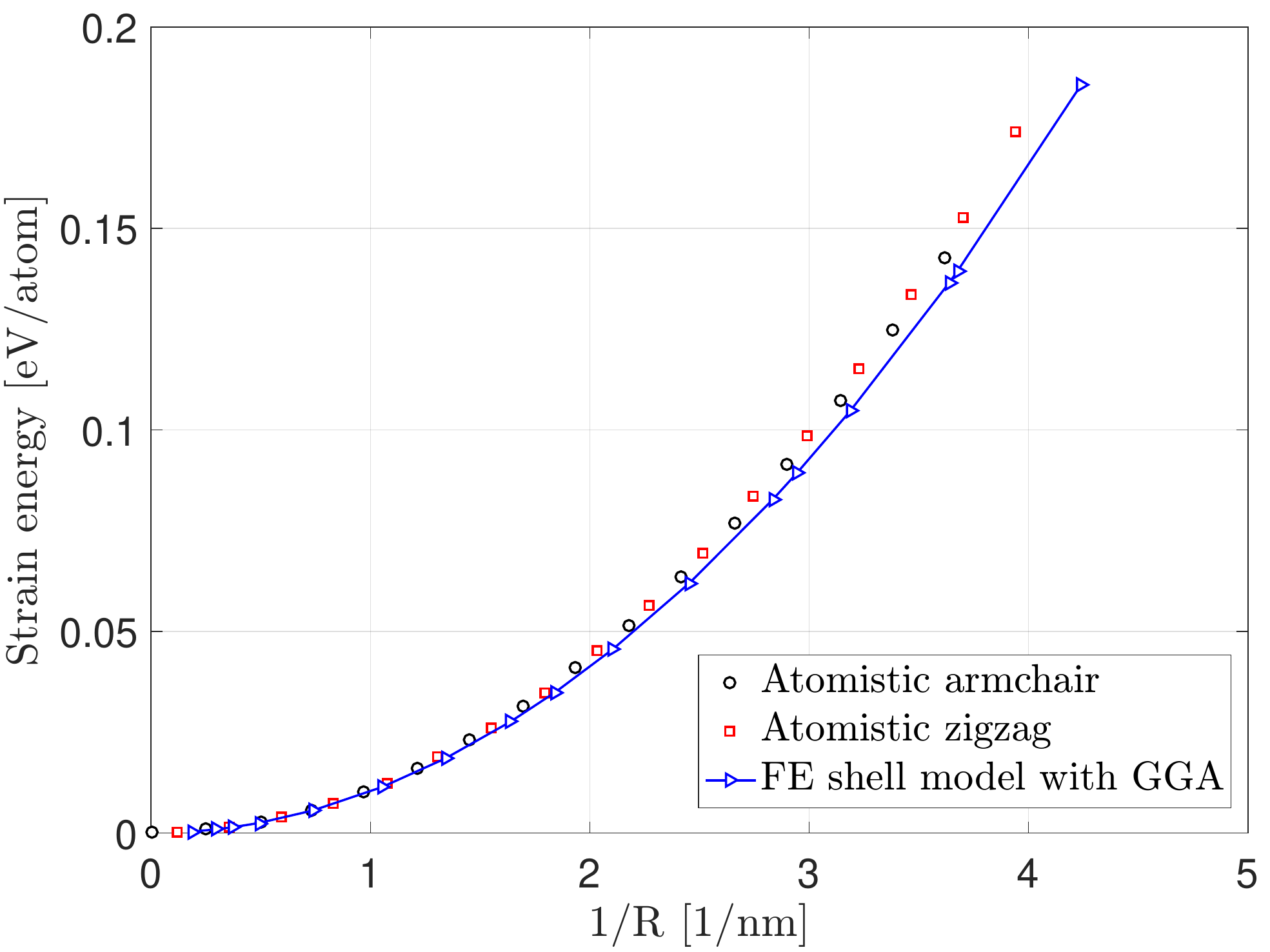}
\caption{CNT relaxation: the strain energy per atom obtained after relaxation. $R$ denotes the radius of the CNT according to Eq.~(\ref{e:CNT_radius}). For the finite element simulation different chiralities are used. The atomistic results are taken from \citet{Arroyo2004_01}.}
\label{f:Relaxation_Energy}
\end{figure}

\begin{figure}
        \centering
 \includegraphics[height=58mm]{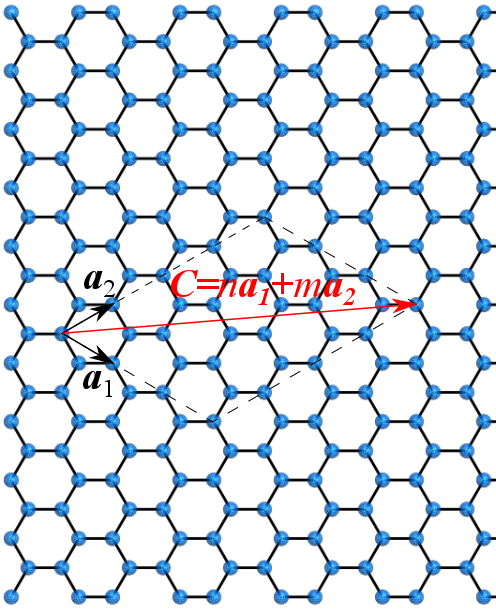}
\caption{CNT chirality: CNT($n$,0) and CNT($n$,$n$) indicate zigzag and armchair CNTs. $\ba_{\alpha}$ are the primary lattice vectors.}
\label{f:Graphene_chirality}
\end{figure}

The CNT model needs to be relaxed in order to obtain a stress-free initial configuration \citep{Favata2016}. The internal energy is minimized in this relaxation process. During relaxation the CNT deforms radially while the length remains almost constant. The minimized strain energy per atom for CNTs with different chirality is compared with results from the literature in Fig.~\ref{f:Relaxation_Energy}. The radius for CNT($n$,$m$) can be calculated as
\eqb{lll}
R \is \ds \frac{\sqrt{3}a_{\text{cc}}}{2\pi}\sqrt{n^2+n\,m+m^2}~,
\label{e:CNT_radius}
\eqe
where the chirality parameters $n$ and $m$ indicate the number of the unit cells along the primary lattice vectors (Fig.~\ref{f:Graphene_chirality}). $a_{\text{cc}}= 0.142$ nm is the equilibrium length of the carbon-carbon bond.

\subsubsection{Torsion of CNT}
\begin{figure}[h]
    \begin{subfigure}[t]{0.39\textwidth}
        \centering
     \includegraphics[width=70mm,trim=2cm 0cm 1cm 0cm,clip]{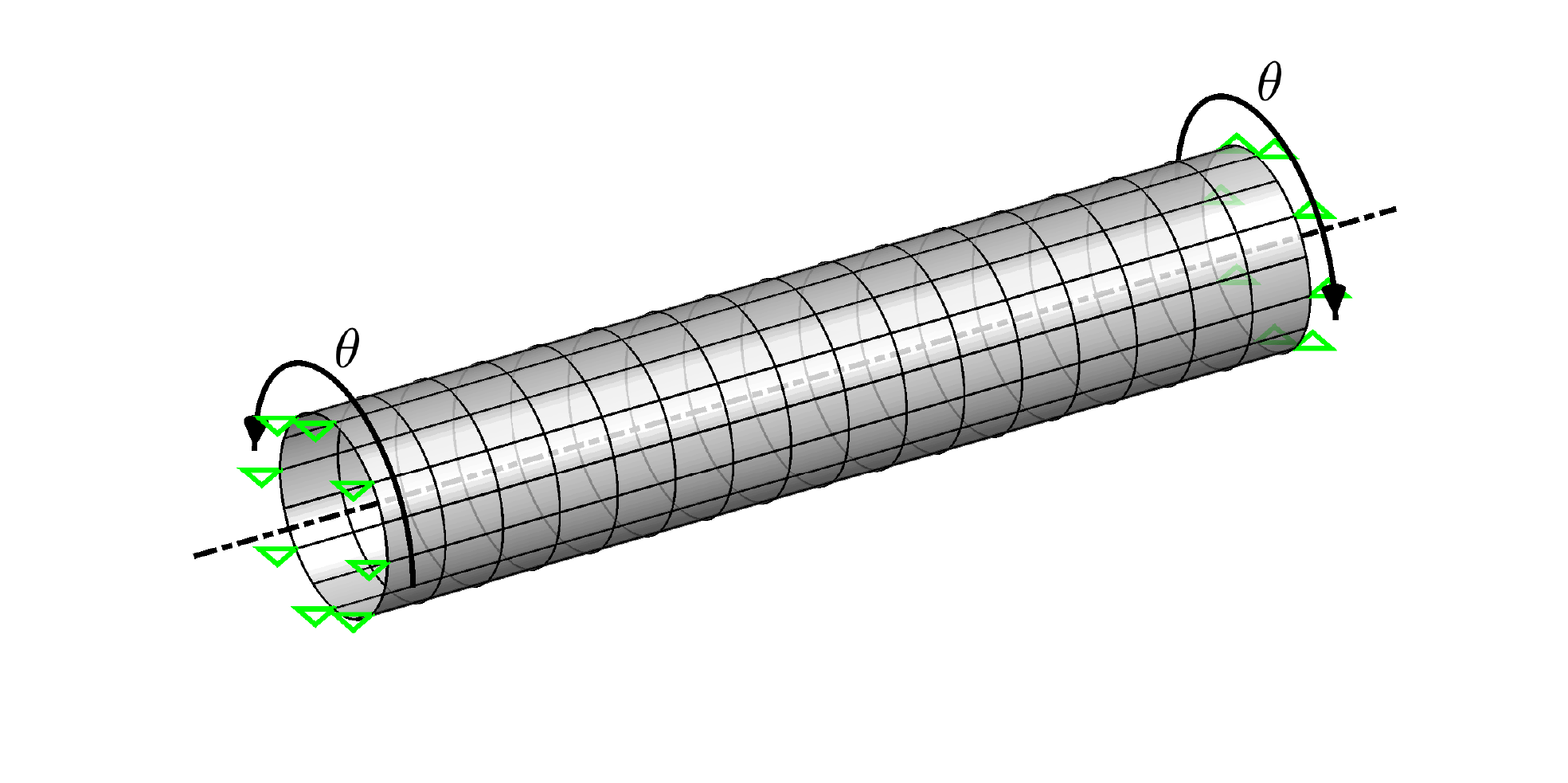}
        \vspace{-13mm}
        \subcaption{}
        \label{f:CNT_Loading_torsion}
    \end{subfigure}
     \begin{subfigure}[t]{0.39\textwidth}
        \centering
     \includegraphics[width=70mm,trim=2cm 0cm 1cm 0cm,clip]{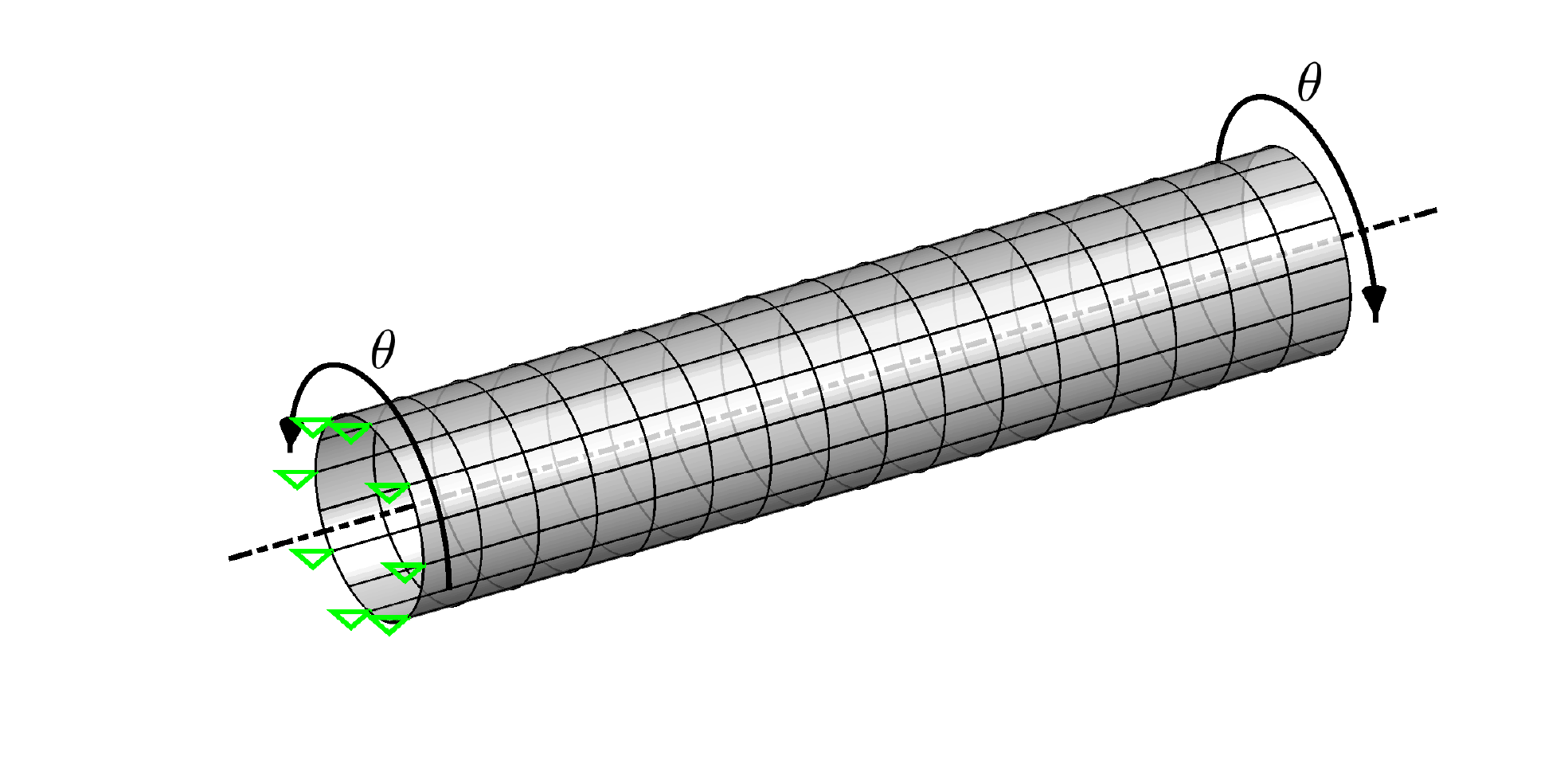}
        \vspace{-13mm}
        \subcaption{}
        \label{f:CNT_Loading_torsion_axial_free}
    \end{subfigure}
    \begin{subfigure}[t]{0.2\textwidth}
        \centering
     \includegraphics[width=40mm]{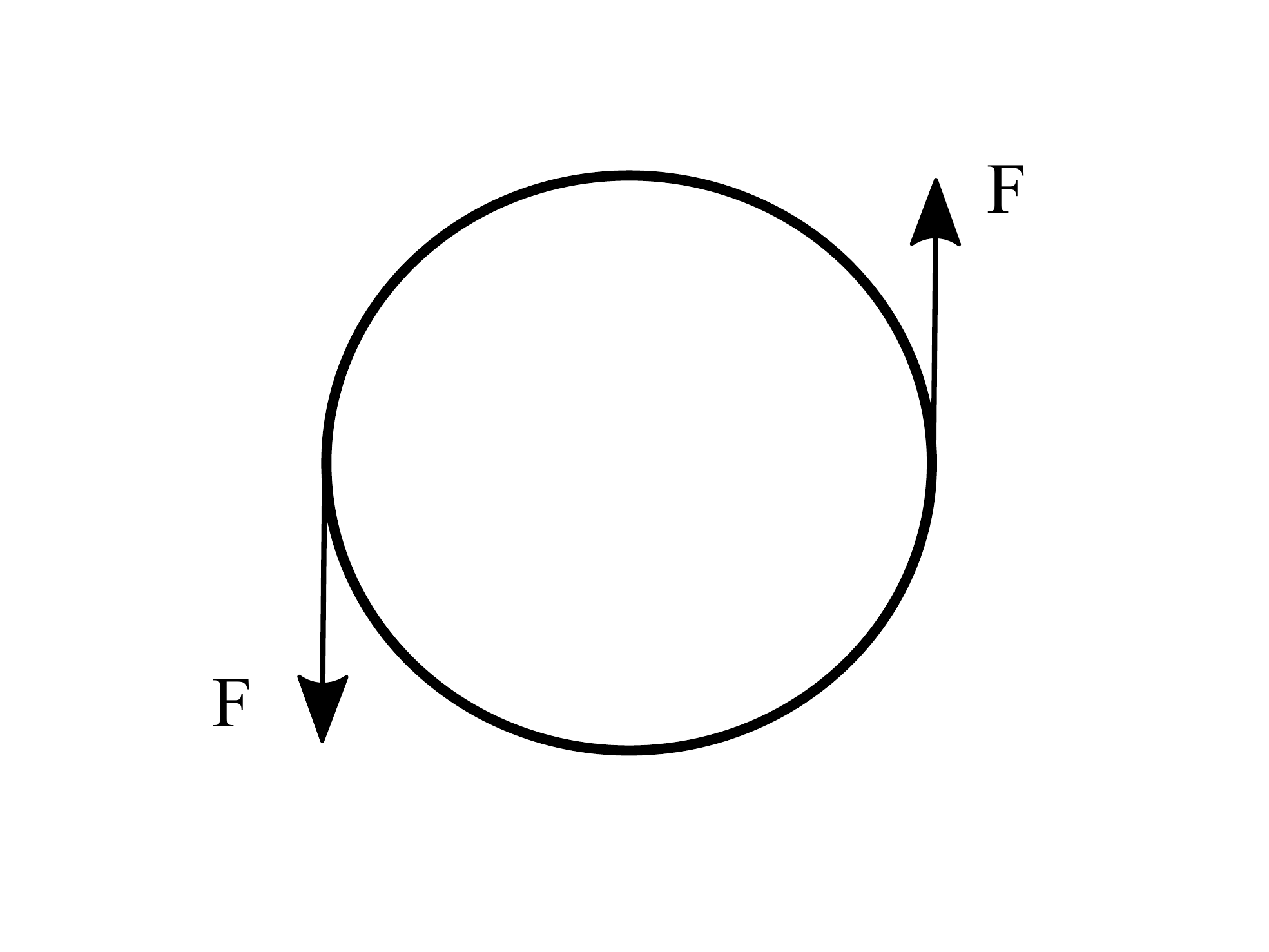}
        \vspace{-13mm}
        \subcaption{}
        \label{f:Imperfection_torque}
    \end{subfigure}
    \vspace{-10mm}
\caption{CNT twisting: BC for (\subref{f:CNT_Loading_torsion}) axially fixed case; (\subref{f:CNT_Loading_torsion_axial_free}) axially free case; (\subref{f:Imperfection_torque}) imperfection torque.}
\end{figure}

\begin{figure}[h]
         \centering
    \begin{subfigure}[t]{0.5\textwidth}
     \includegraphics[height=55mm]{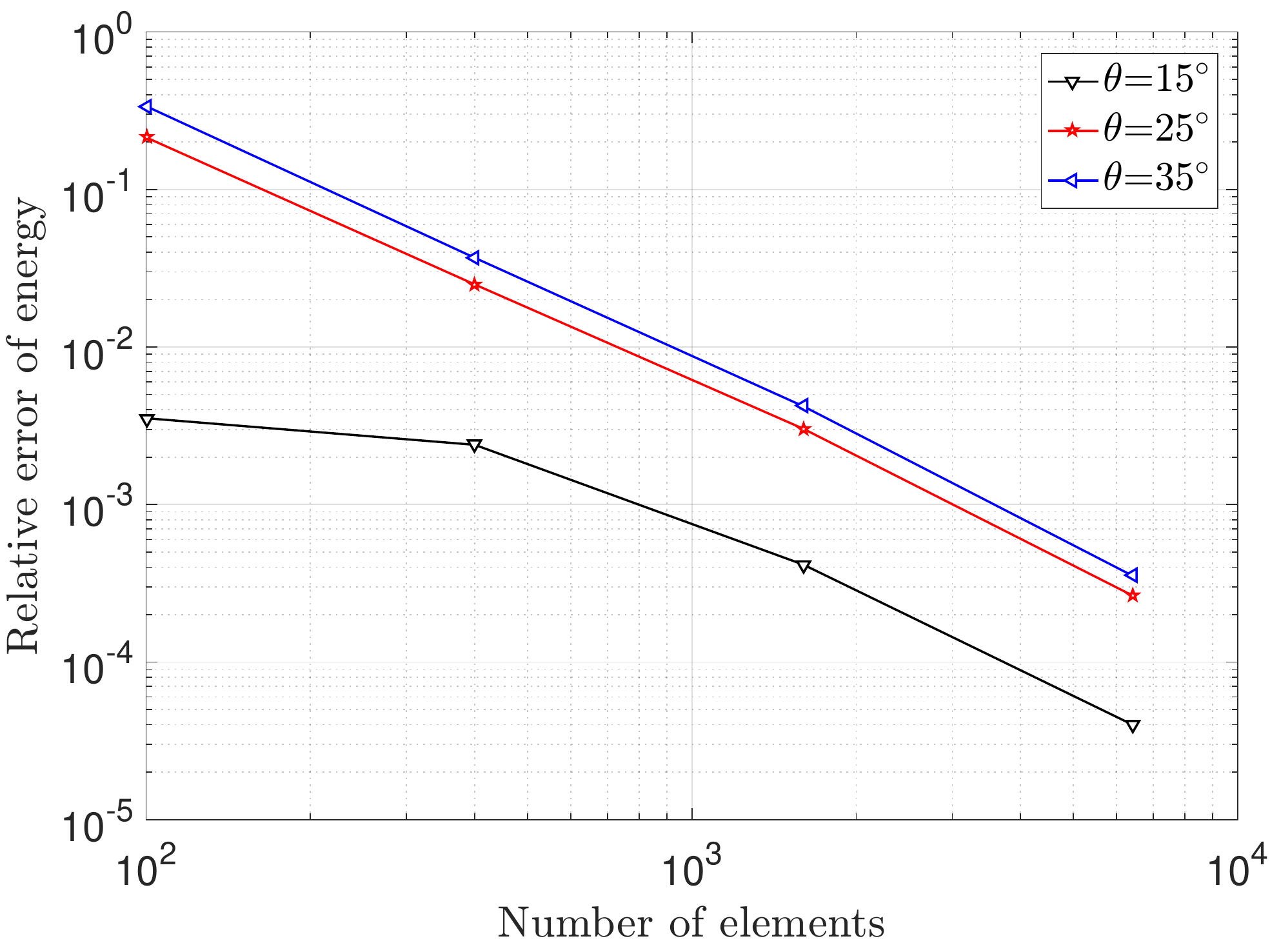}
    \end{subfigure}
\caption{CNT twisting: Error of strain energy relative to the finest mesh ($100\times100$ quadratic NURBS elements). CNT(12,6) with the length 6.74 nm is used for the axially fixed case (Fig.~\ref{f:CNT_Loading_torsion}). }
\label{f:Torsion_n12_m6_L10_6_74_strainEnergy_conv}
\end{figure}

\begin{figure}[h]
    \begin{subfigure}[t]{0.5\textwidth}
        \centering
     \includegraphics[height=55mm]{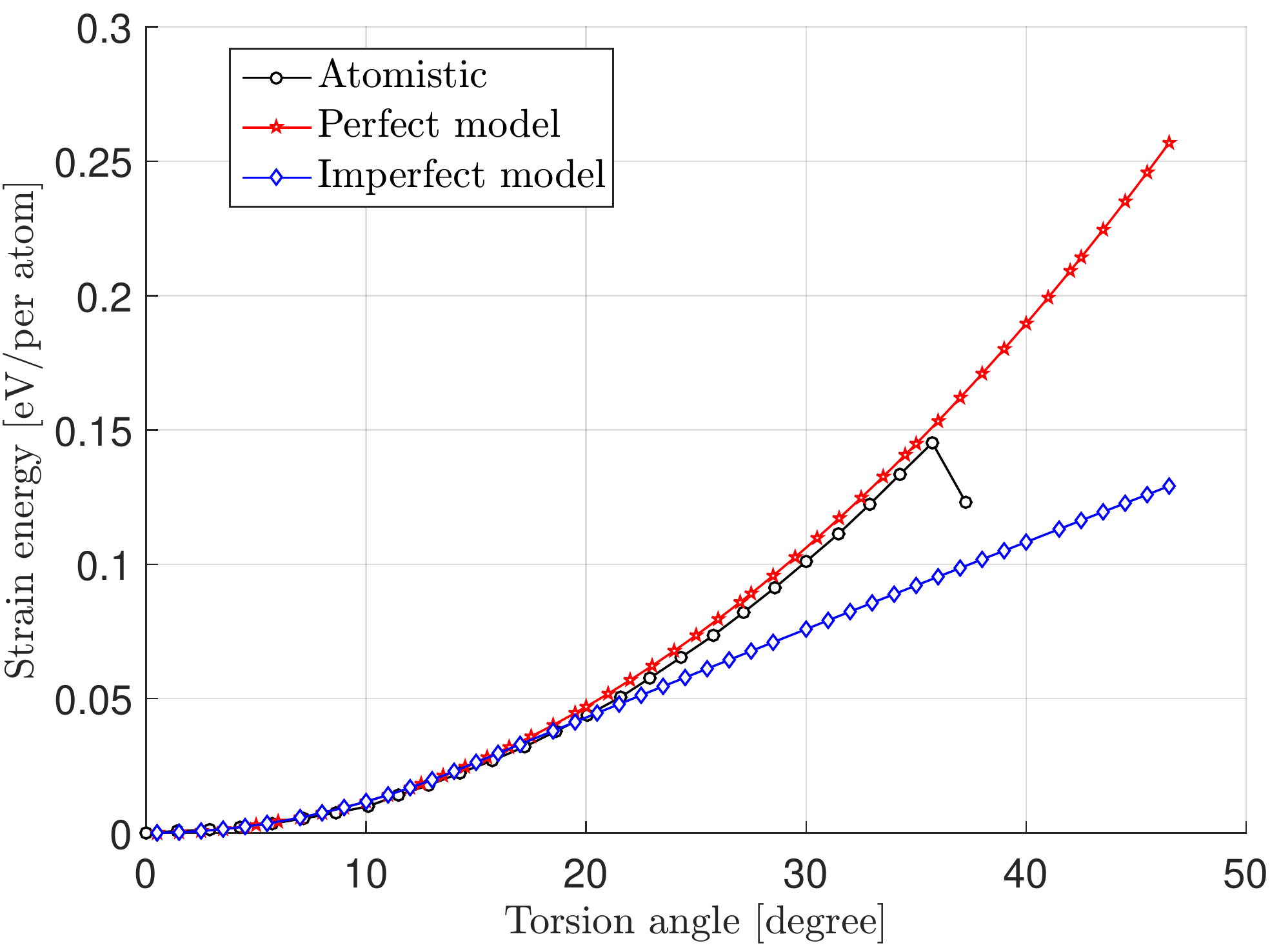}
        \subcaption{}
        \label{f:torsion_Energy_torsion_n12_m6_L6_74}
    \end{subfigure}
    \begin{subfigure}[t]{0.5\textwidth}
        \centering
     \includegraphics[height=55mm]{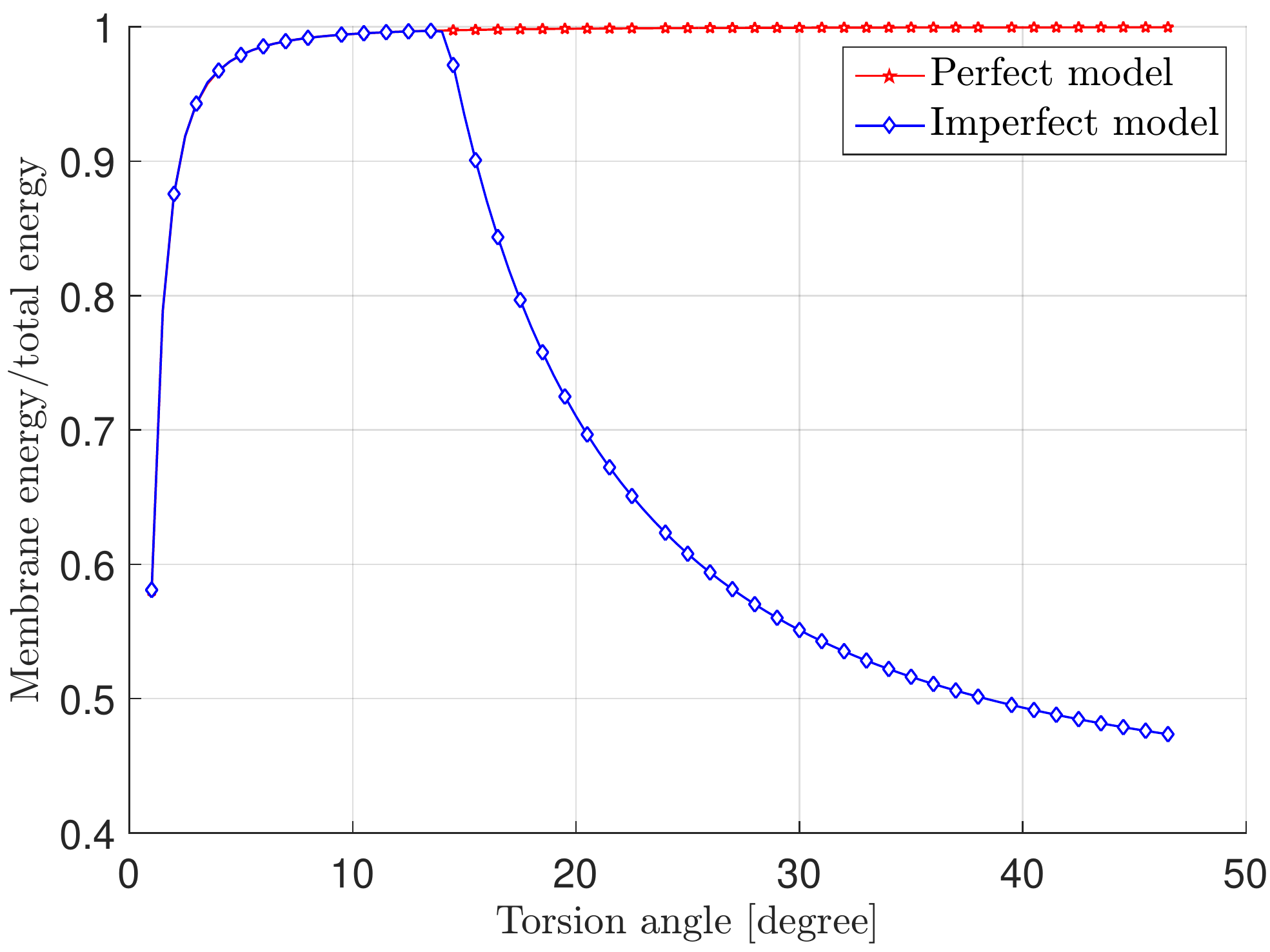}
        \subcaption{}
        \label{f:torsion_Energy_ratio_torsion_n12_m6_L6_74}
    \end{subfigure}
\caption{CNT twisting: (\subref{f:torsion_Energy_torsion_n12_m6_L6_74}) Strain energy per atom; (\subref{f:torsion_Energy_ratio_torsion_n12_m6_L6_74}) ratio of the membrane energy to the total energy (measured relative to relaxed geometry). CNT(12,6) with the length 6.74 nm is used for the axially fixed case (Fig.~\ref{f:CNT_Loading_torsion}). The perfect and imperfect models are used for the simulation. The atomistic results are taken from \citet{Sun_Torsional_strain_energy_evolution_of_carbon_nanotubes}.}
\end{figure}

\begin{figure}[h]
        \centering
   \includegraphics[height=55mm]{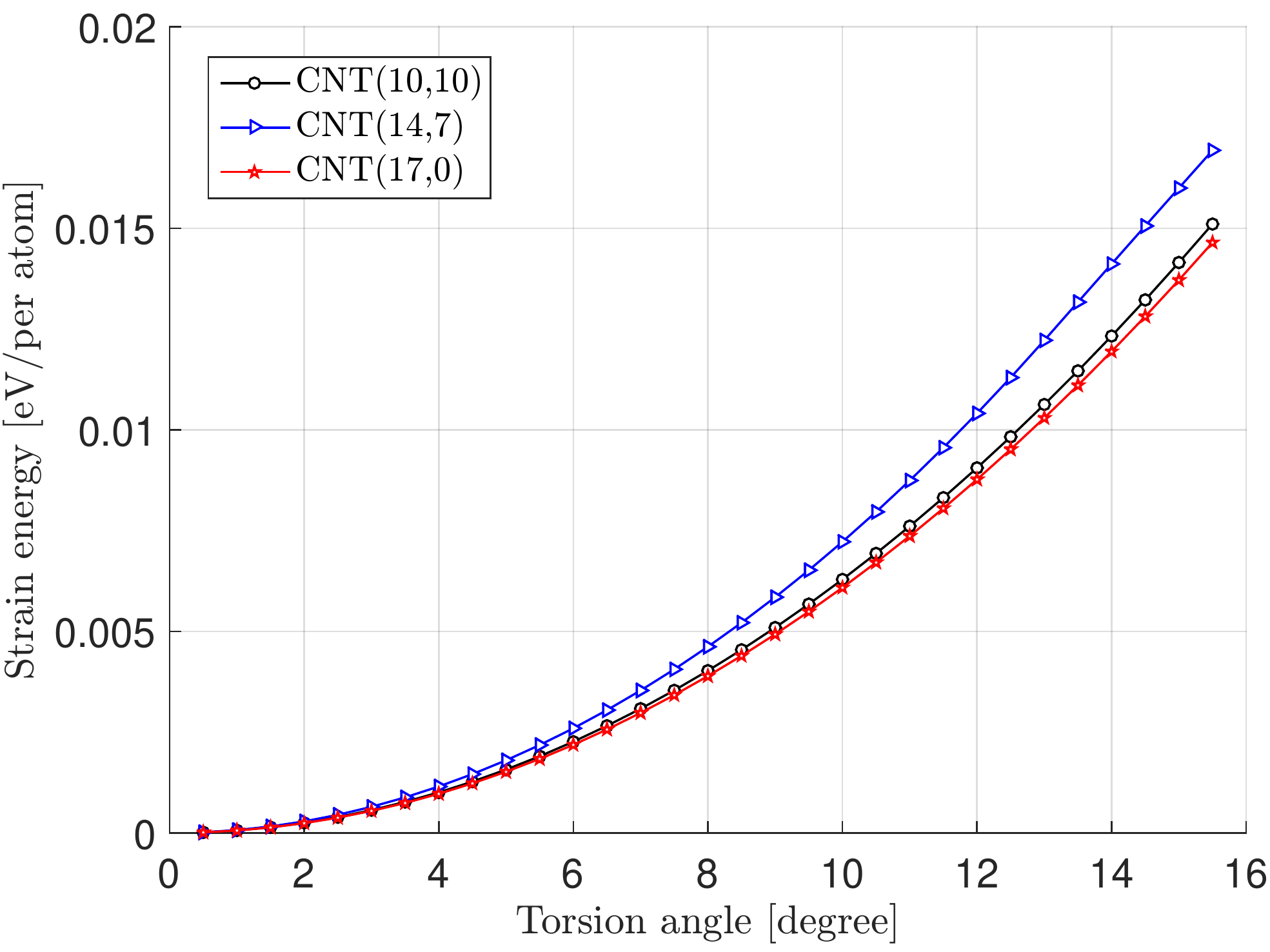}
\caption{CNT twisting: Strain energy per atom for different chiralities of CNTs with the length of 10 nm. CNTs are selected such that their radii are close to each other. The BC is set to the axially free case (Fig.~\ref{f:CNT_Loading_torsion_axial_free}) and imperfection is applied to the model.}
\label{f:Torsion_energy_different_chirality}
\end{figure}

\begin{figure}
\begin{center} \unitlength1cm
\begin{picture}(18,14)
\put(1.0,7.5){\includegraphics[height=40mm,trim=4cm 0 4cm 0,clip]{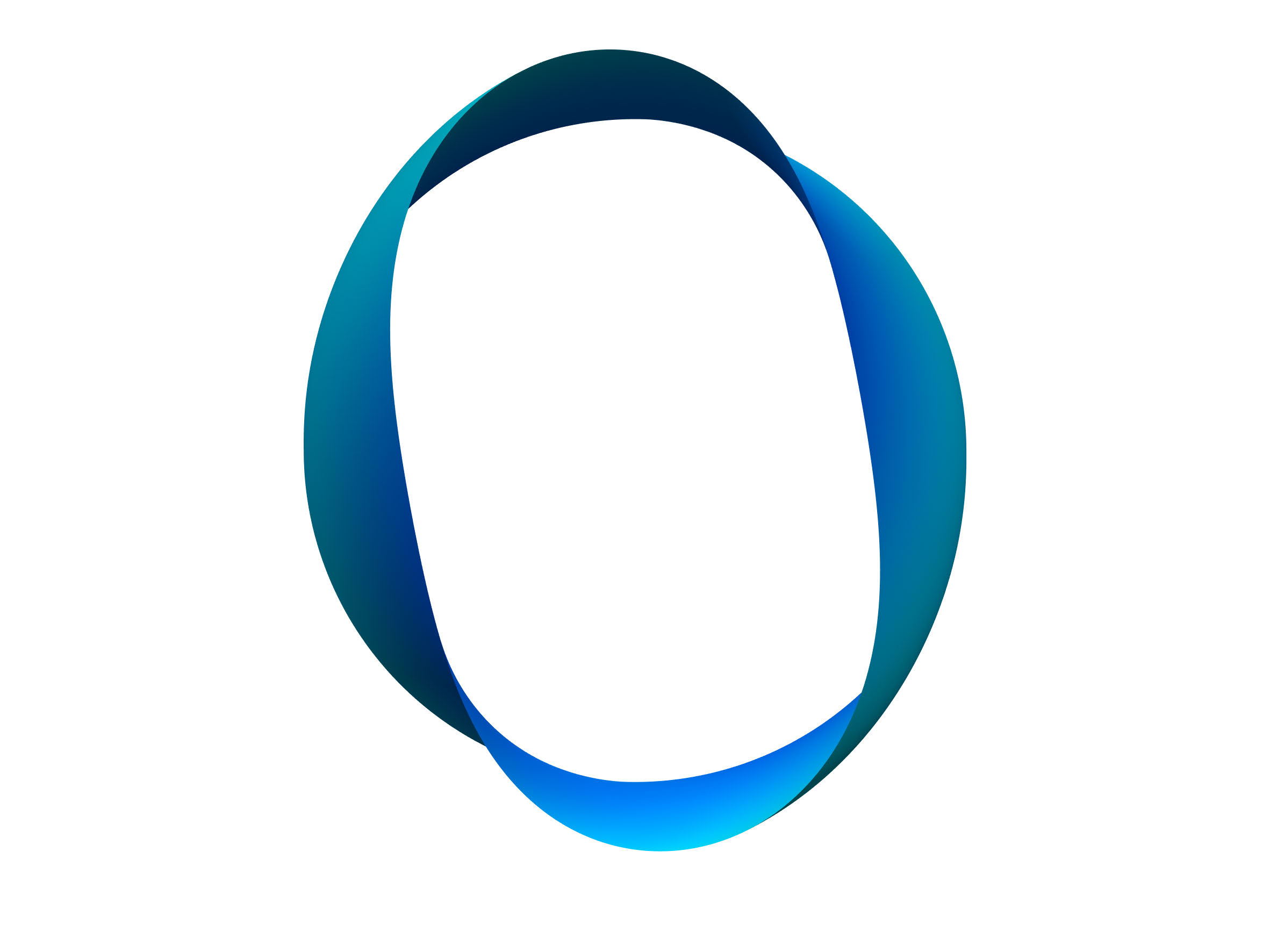}}
\put(5,7.5){\includegraphics[height=40mm,trim=4cm 0 4cm 0,clip]{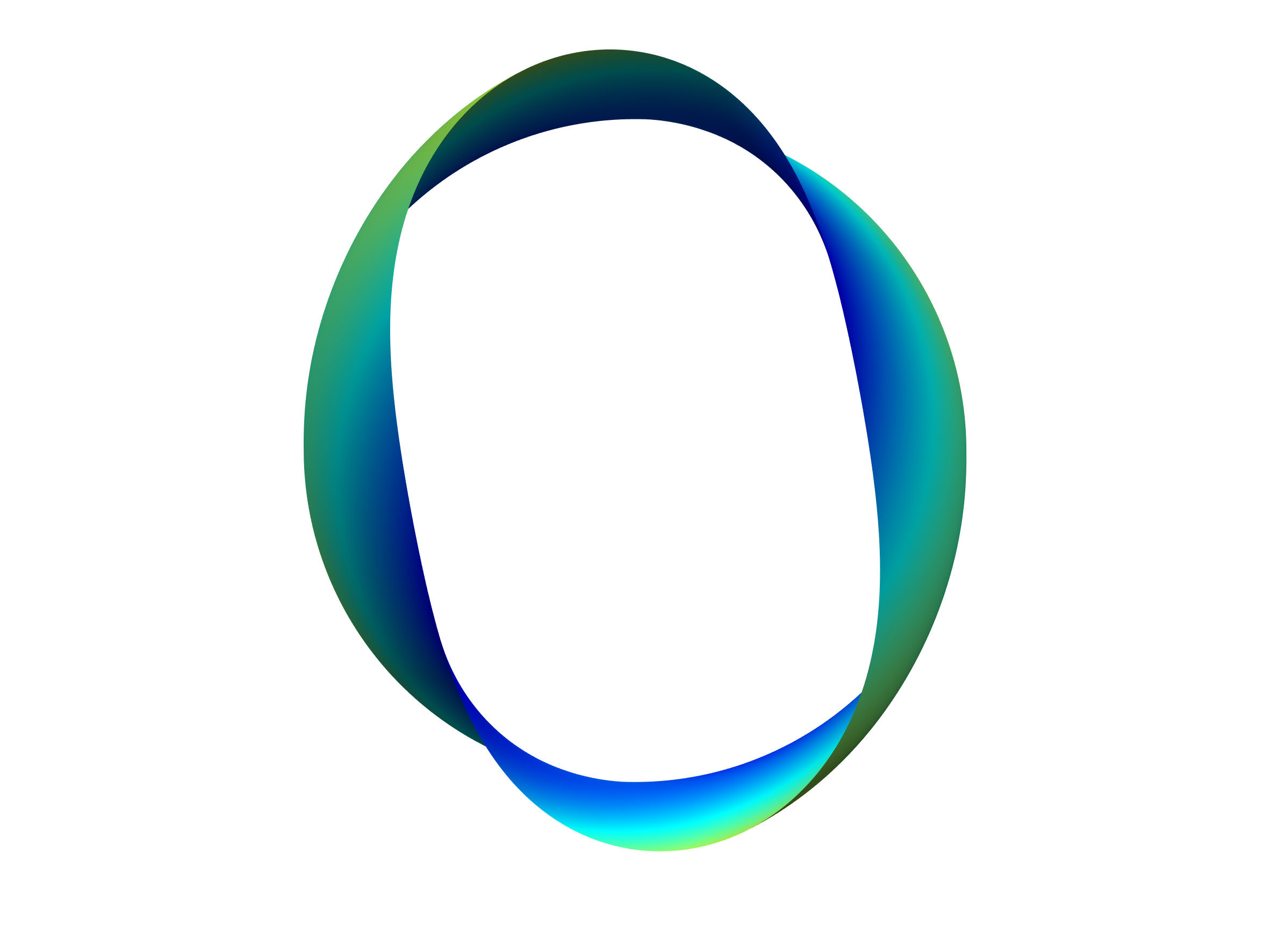}}
\put(9,7.5){\includegraphics[height=40mm,trim=4cm 0 4cm 0,clip]{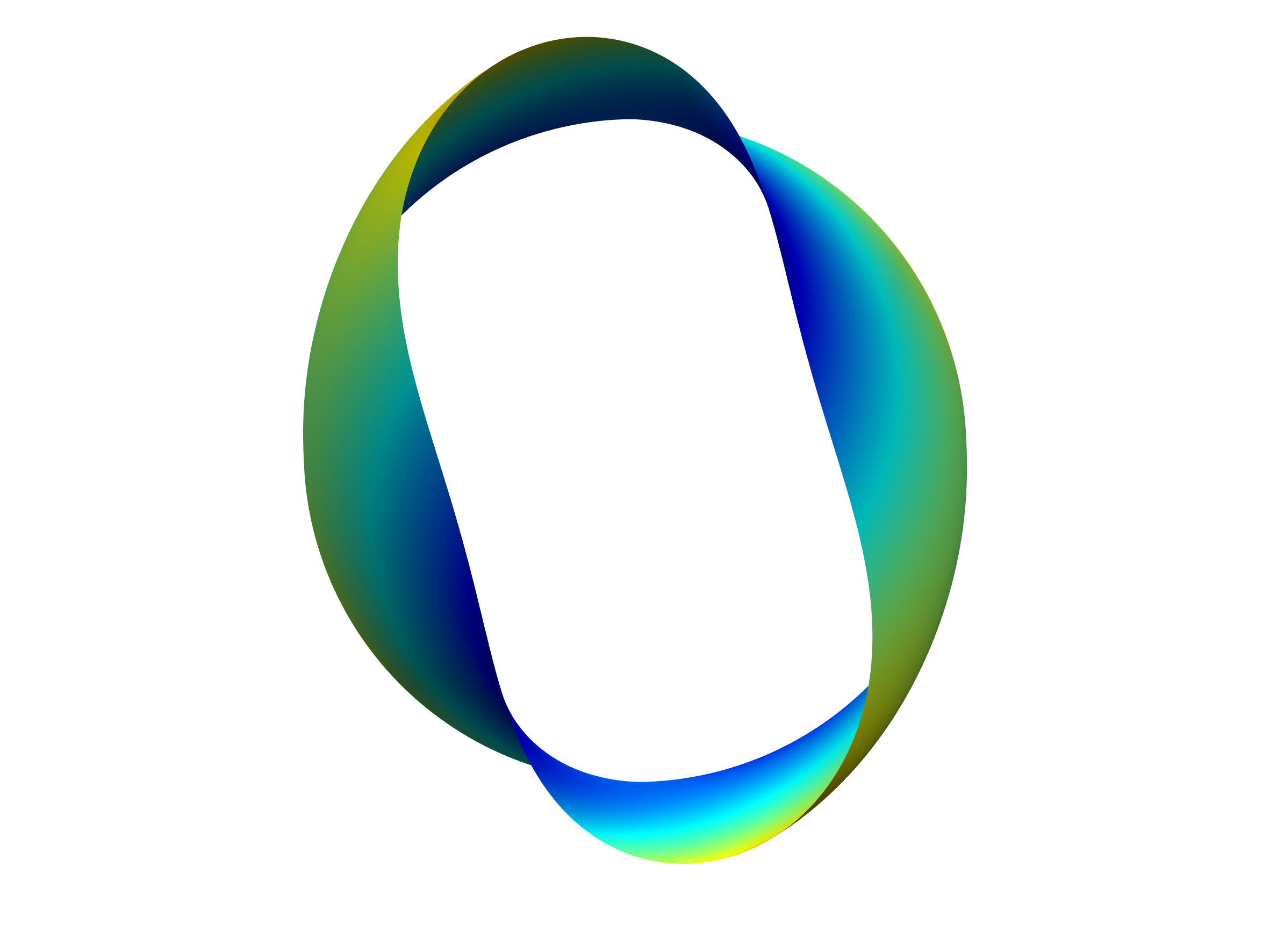}}
\put(13,7.5){\includegraphics[height=40mm,trim=4cm 0 4cm 0,clip]{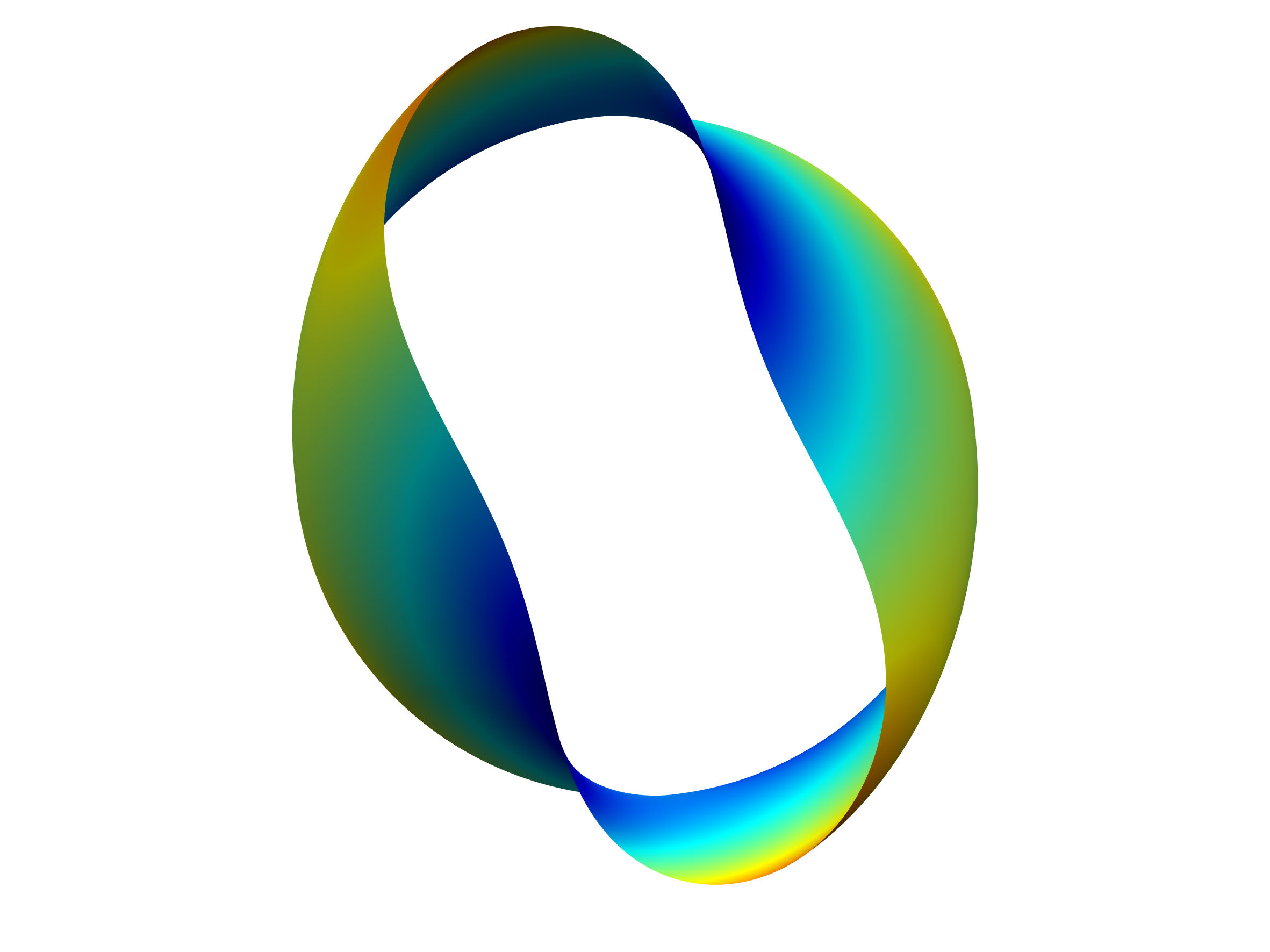}}

\put(1.0,3.5){\includegraphics[height=40mm,trim=4cm 0 4cm 0,clip]{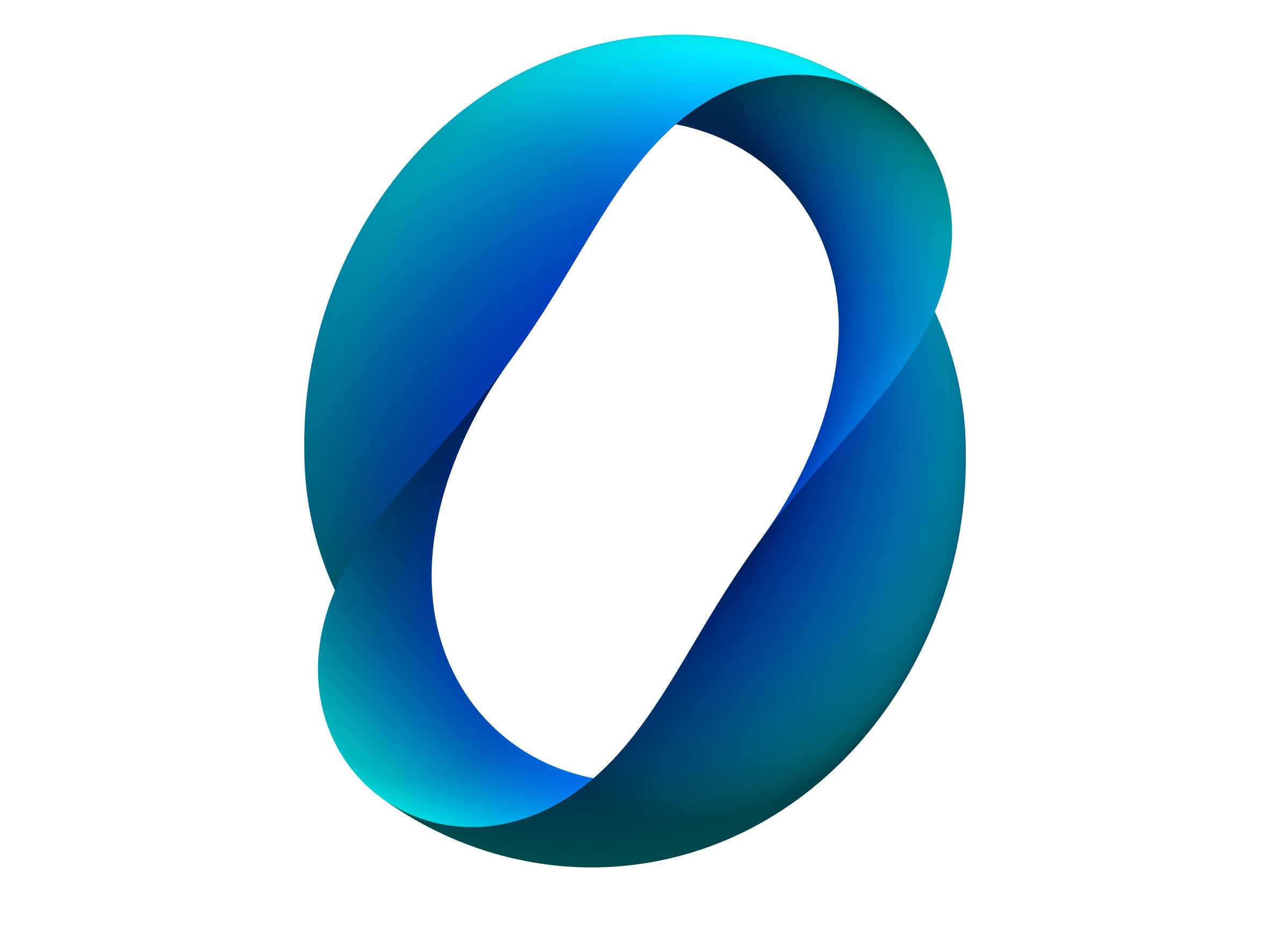}}
\put(5.0,3.5){\includegraphics[height=40mm,trim=4cm 0 4cm 0,clip]{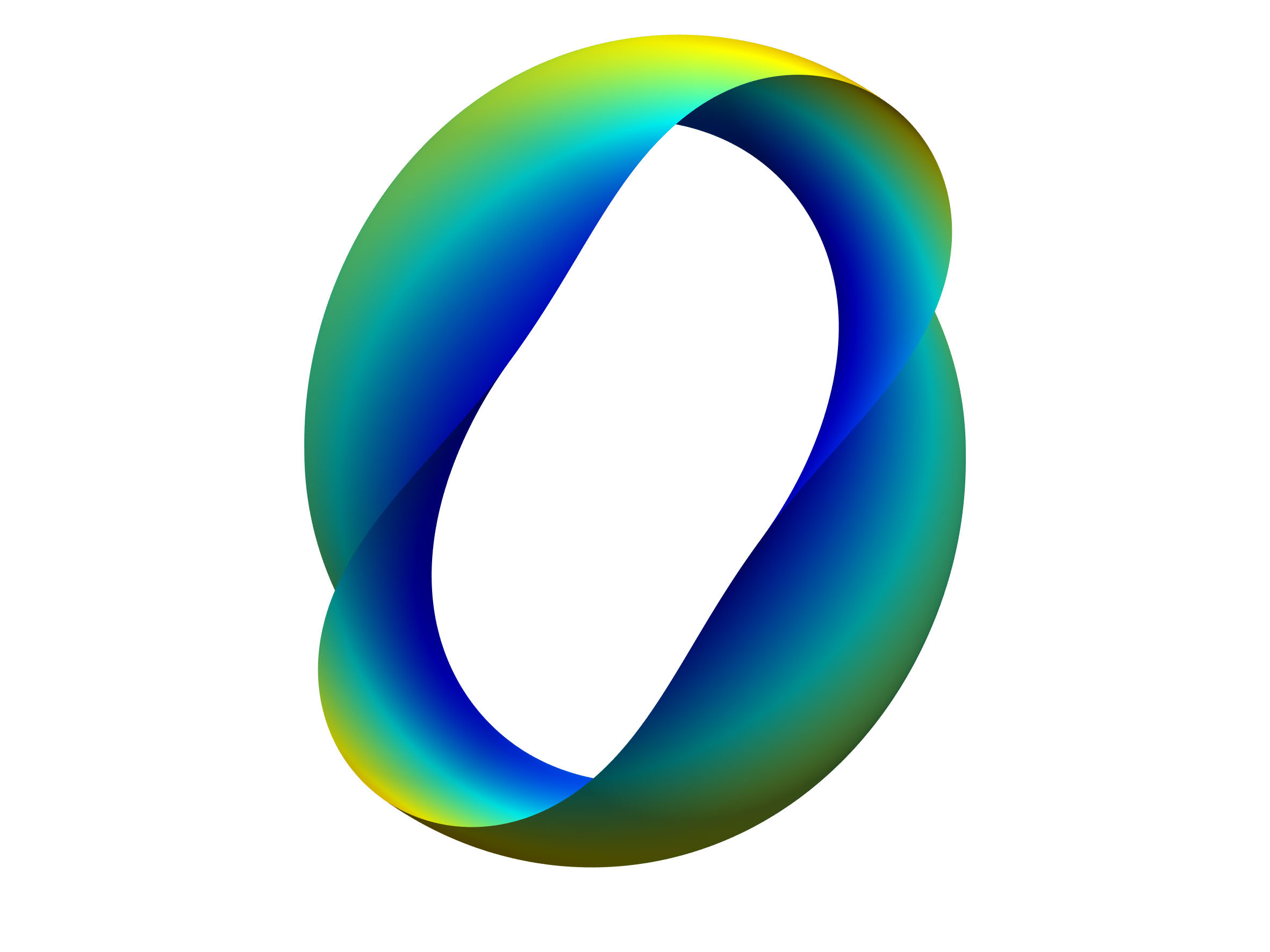}}
\put(9.0,3.5){\includegraphics[height=40mm,trim=4cm 0 4cm 0,clip]{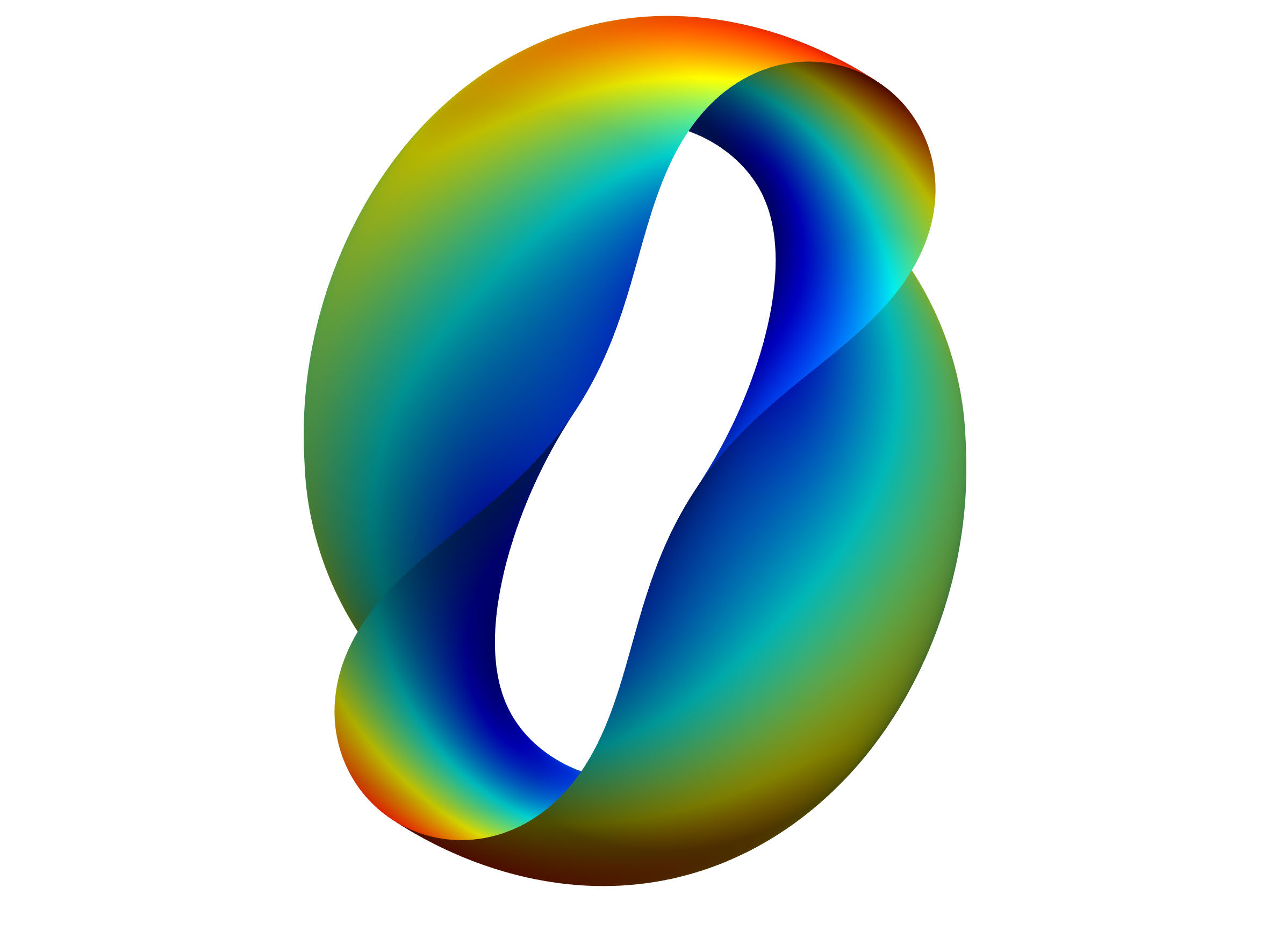}}
\put(13.0,3.5){\includegraphics[height=40mm,trim=4cm 0 4cm 0,clip]{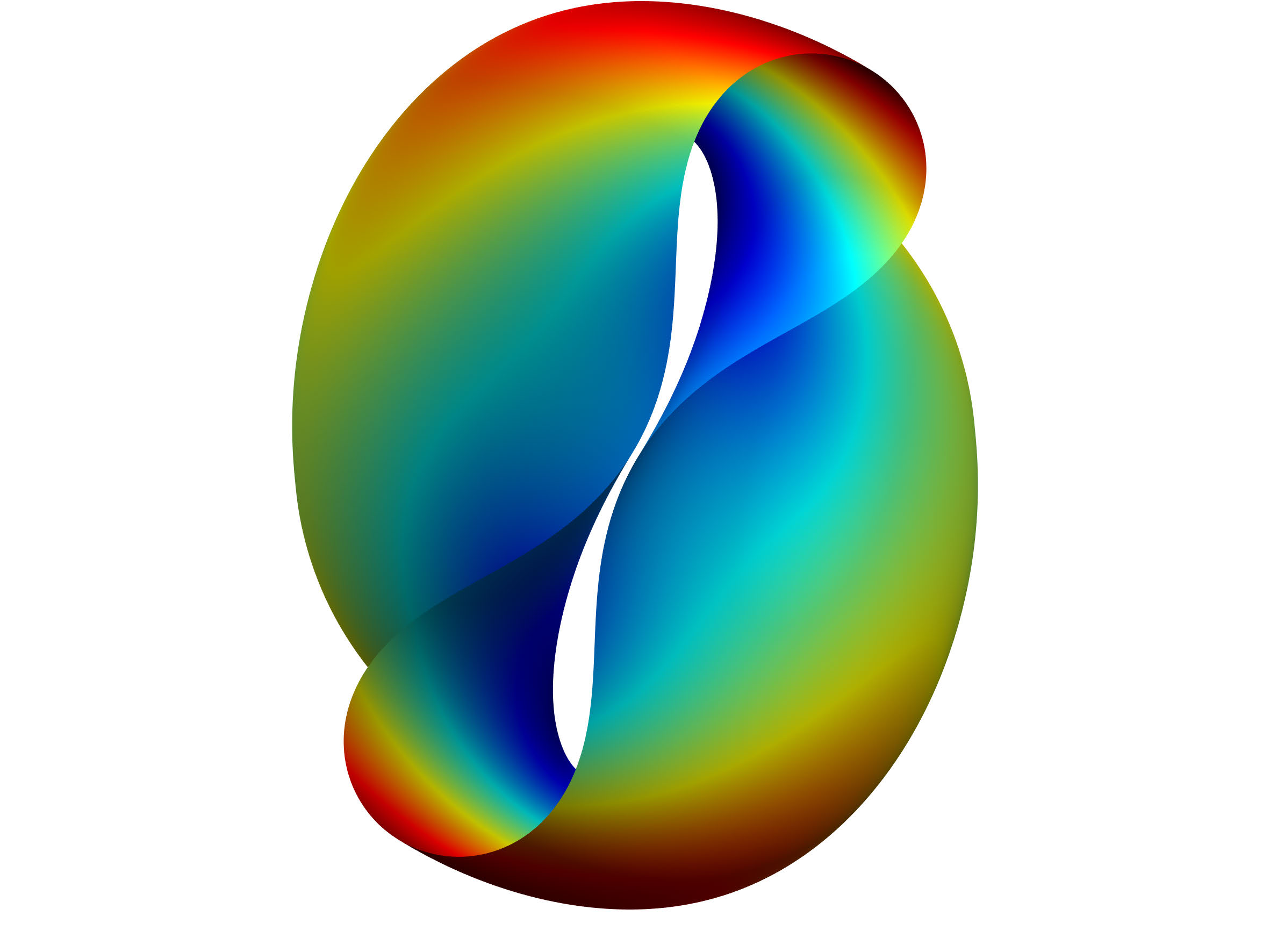}}

\put(0.5,0){\includegraphics[height=35mm,trim=2cm 0 2cm 0,clip]{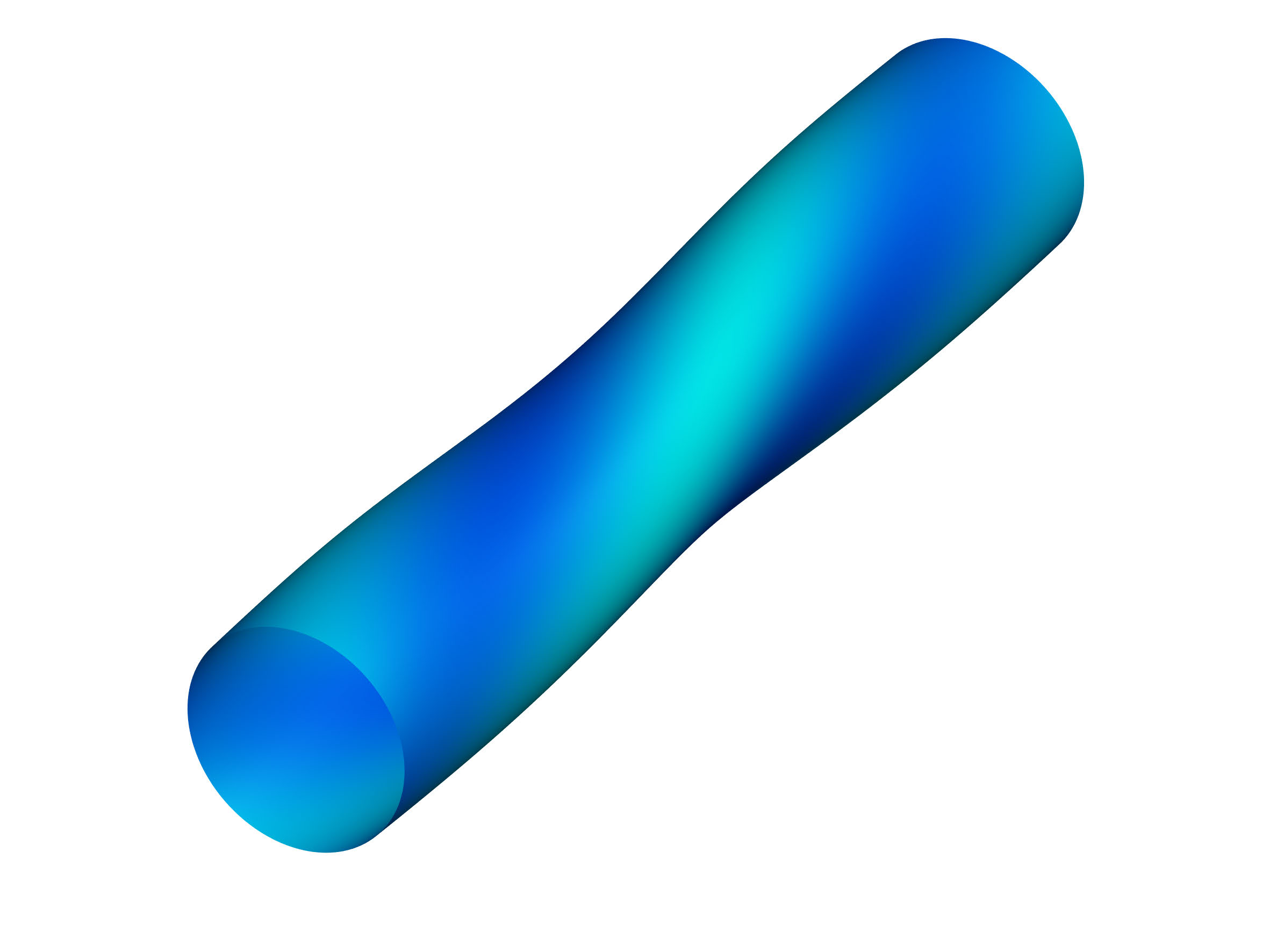}}
\put(4.5,0){\includegraphics[height=35mm,trim=2cm 0 2cm 0,clip]{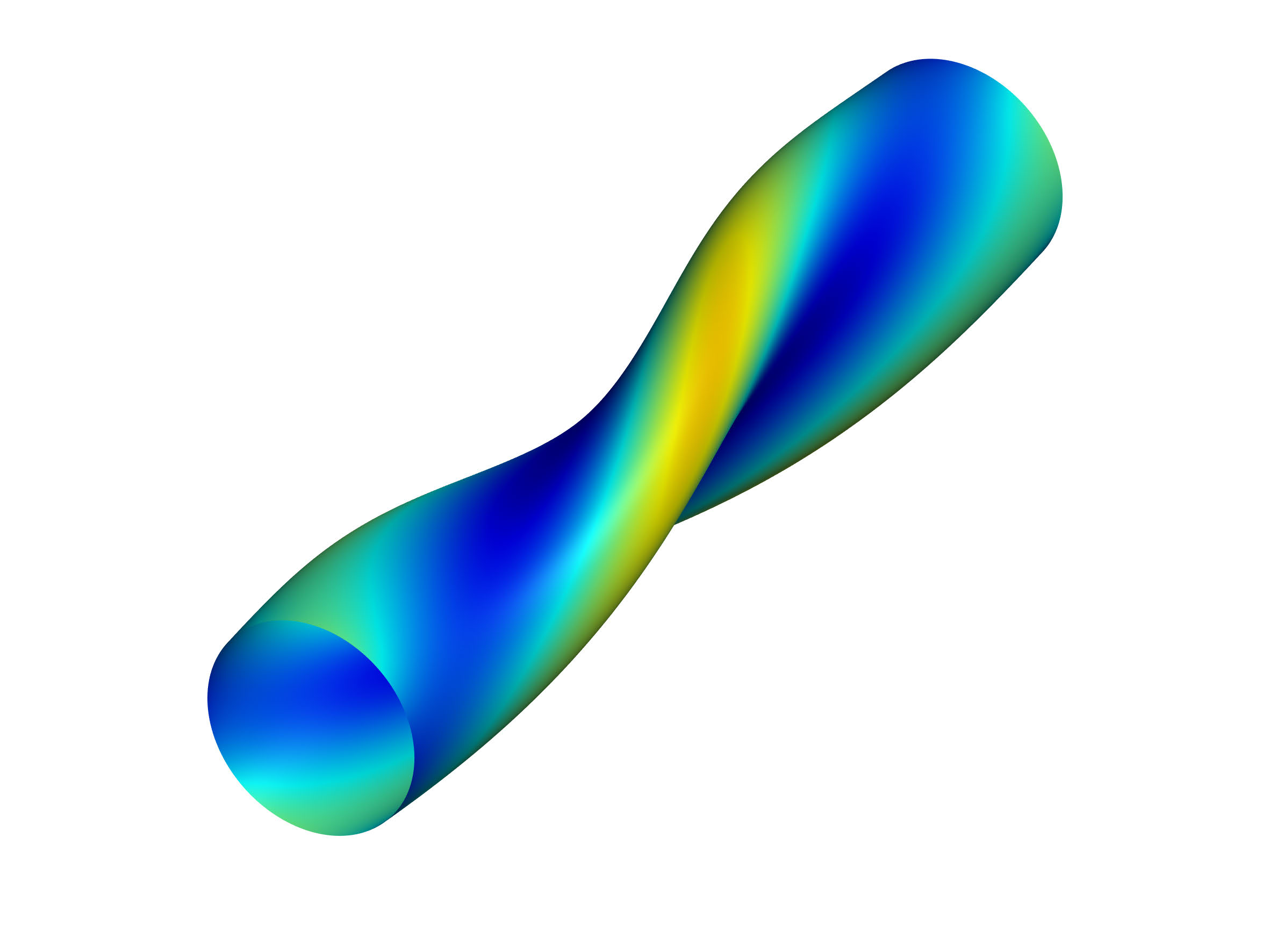}}
\put(8.5,0){\includegraphics[height=35mm,trim=2cm 0 2cm 0,clip]{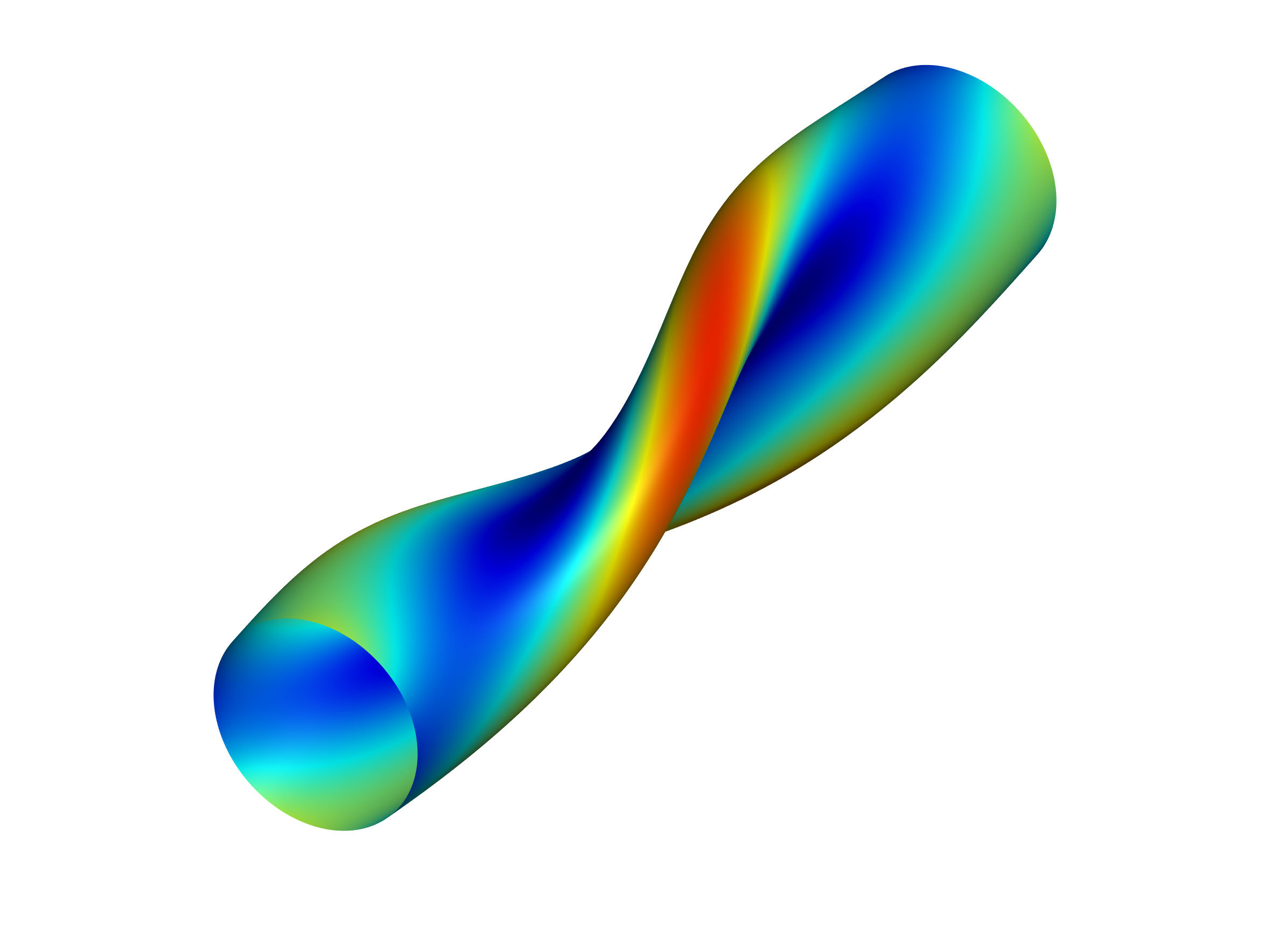}}
\put(12.5,0){\includegraphics[height=35mm,trim=2cm 0 2cm 0,clip]{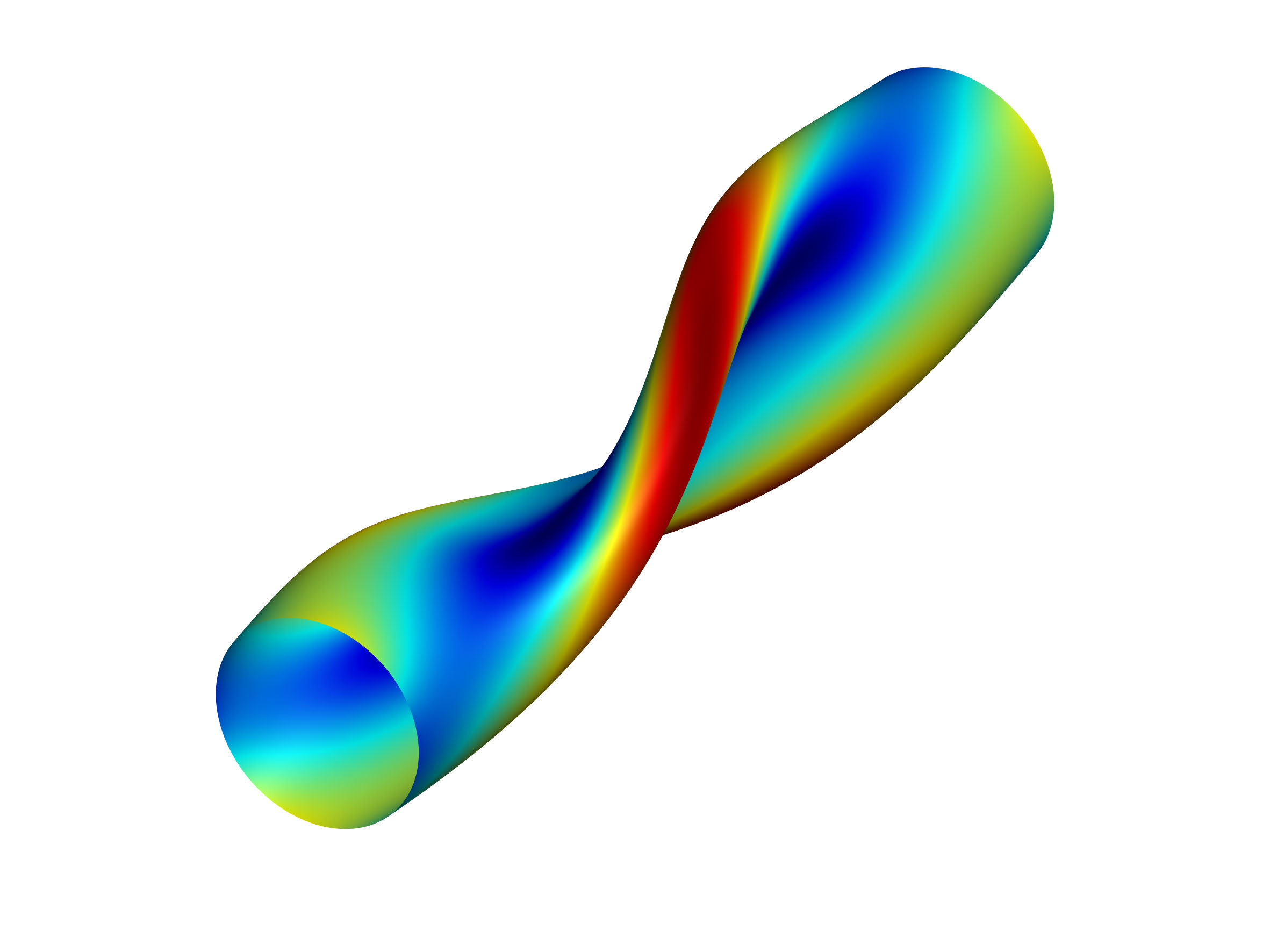}}

\put(2,11.5){(a)}
\put(6,11.5){(b)}
\put(10,11.5){(c)}
\put(14,11.5){(d)}
%\put(9.5,.1){(d)}
\put(2,12){\includegraphics[width=120mm]{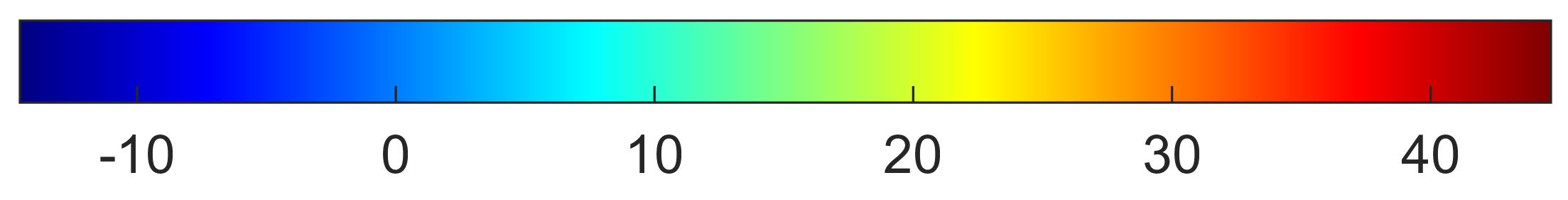}}
\put(0.0,7.3){\rotatebox{90}{{\small 50 percent of CNT length}}}
\put(0.0,3.0){\rotatebox{90}{{\small 75 percent of CNT length}}}
\end{picture}
\caption{CNT twisting: Comparison of cross sections and tr($\bsig$) [$\text{N/m}$] at the torsion angles
(a) $\theta=15^{\circ}$,
(b) $\theta=25^{\circ}$,
(c) $\theta=35^{\circ}$ and
(d) $\theta=46.5^{\circ}$. CNT(12,6) with the length 6.74 nm is used. The BCs are applied according to the axially fixed case (Fig.~\ref{f:CNT_Loading_torsion_axial_free}).}
\label{f:torsion_energy_contours_sliced_prependicular_to_axis}
\end{center}
\end{figure}

In this section, the behavior of an initially relaxed CNT under torsional loading is investigated. The loading can be conducted either by restraining the axial length of CNT (Fig.~\ref{f:CNT_Loading_torsion}) or letting the CNT deform in the axial direction (Fig.~\ref{f:CNT_Loading_torsion_axial_free}). A torsion angle is applied to both ends of CNT, and the simulation is conducted with and without imperfection. The imperfection is applied as a torque\footnote{by applying the force couple (F=1~nN)} in the middle of the CNT (Fig.~\ref{f:Imperfection_torque}). The convergence is studied for the total energy (Fig.~\ref{f:Torsion_n12_m6_L10_6_74_strainEnergy_conv}). The cross section of the model without imperfection remains cylindrical during torsion, i.e.~it does not buckle, and the results converge faster in comparison to the model with imperfection.
The strain energy per atom of CNT is plotted for the perfect and imperfect structures. The results from the presented model are compared to results from an atomistic simulation with a perfect crystal in Fig.~\ref{f:torsion_Energy_torsion_n12_m6_L6_74}.
In the loading process, the CNT with the imperfection buckles, but the exact point of the instability is not easily found from the variation of the total energy. To estimate the instability point more accurately, the ratio of the membrane energy to the total energy is used. The sharp decreasing of energy ratio determines the buckling angle around ($\theta_{\mathrm{b}}=14^{\circ}$) (Fig.~\ref{f:torsion_Energy_ratio_torsion_n12_m6_L6_74}). The strain energy is measured relative to the relaxed geometry. Next, a set of simulations is conducted for the different charities and the results are presented in Fig.~\ref{f:Torsion_energy_different_chirality}. The deformed and sliced geometries are compared for different torsion angles (Fig.~\ref{f:torsion_energy_contours_sliced_prependicular_to_axis}).

\subsubsection{Bending of CNT}
\begin{figure}
    \begin{subfigure}[t]{0.5\textwidth}
        \centering
     \includegraphics[height=55mm,trim=1.5cm 1cm 0cm 1cm,clip]{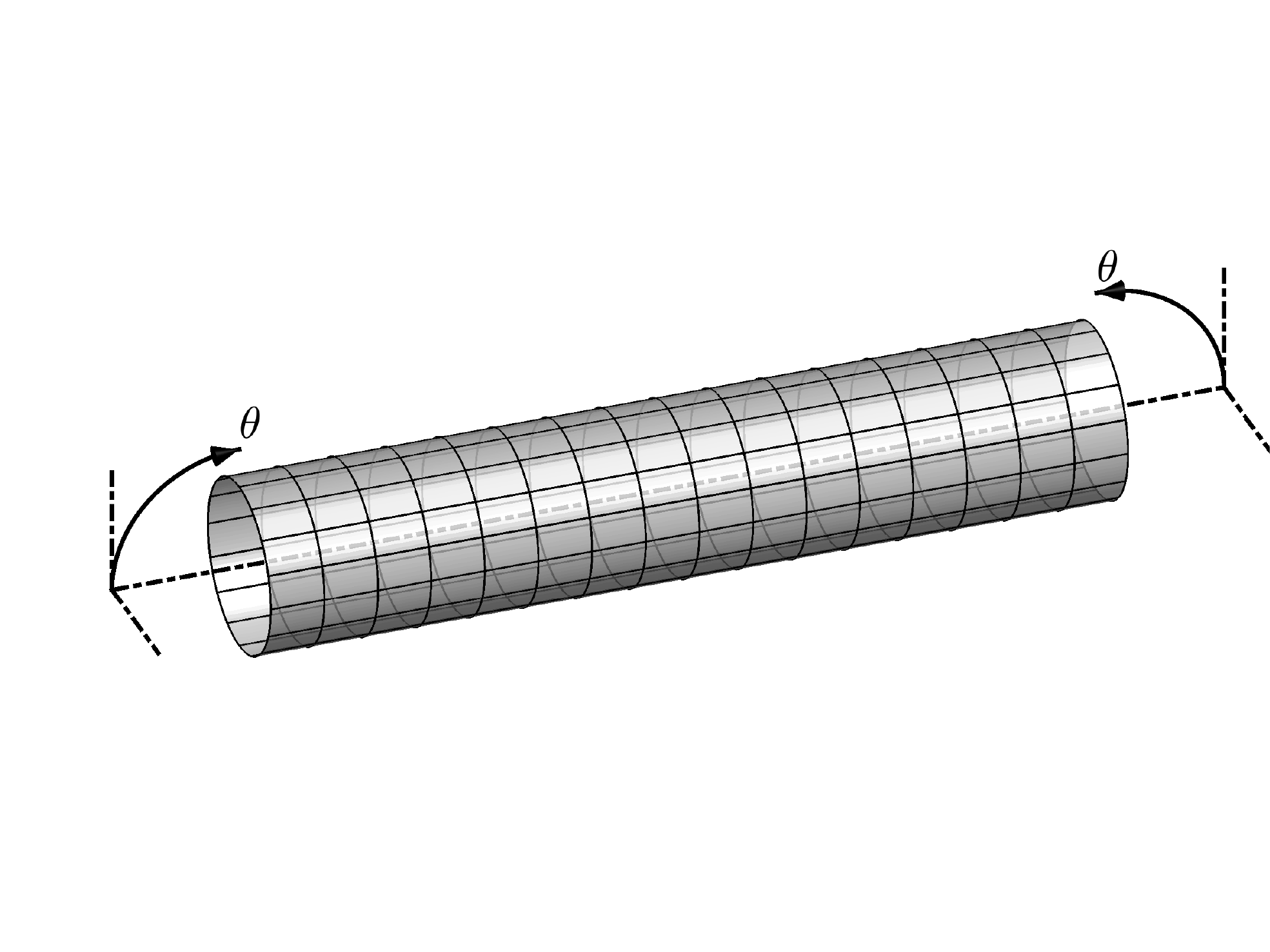}
    \subcaption{}
    \label{f:CNT_Loading_bending}
    \end{subfigure}
        \begin{subfigure}[t]{0.5\textwidth}
        \centering
     \includegraphics[height=55mm]{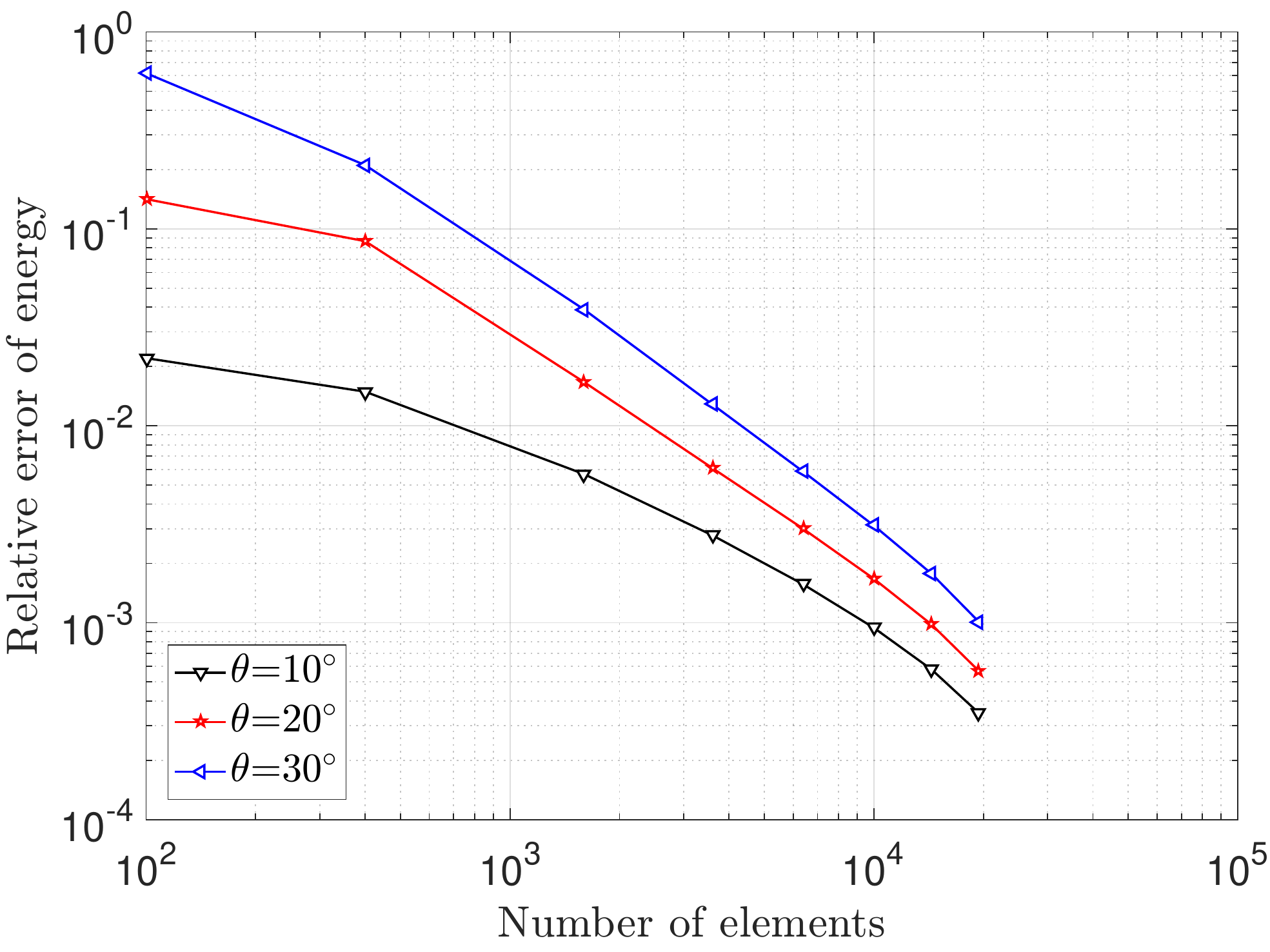}
        \subcaption{}
        \label{f:Bending_central_n10_m10_L10_strainEnergy_conv}
    \end{subfigure}
    \caption{CNT bending: (\subref{f:CNT_Loading_bending}) BC; (\subref{f:Bending_central_n10_m10_L10_strainEnergy_conv})  Error of strain energy relative to the finest mesh ($200\times200$ quadratic NURBS elements). CNT(10,10) with the length 10 nm and imperfection is used.}
\end{figure}

\begin{figure}[]
    \begin{subfigure}[t]{0.5\textwidth}
        \centering
     \includegraphics[height=55mm]{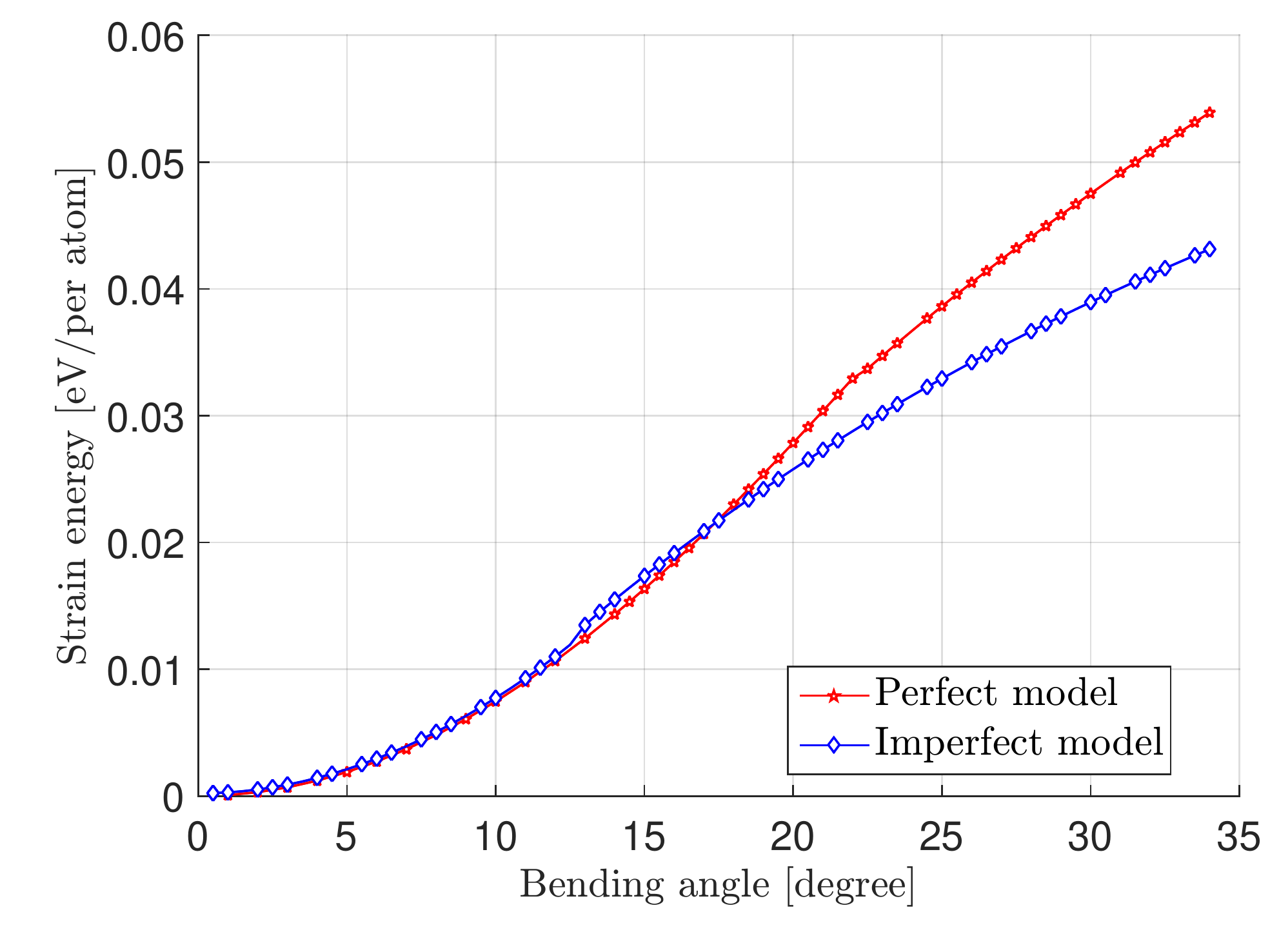}
        \subcaption{}
        \label{f:bending_Energy_central_n10_m10_L10_QI80_80}
    \end{subfigure}
        \begin{subfigure}[t]{0.5\textwidth}
        \centering
     \includegraphics[height=55mm]{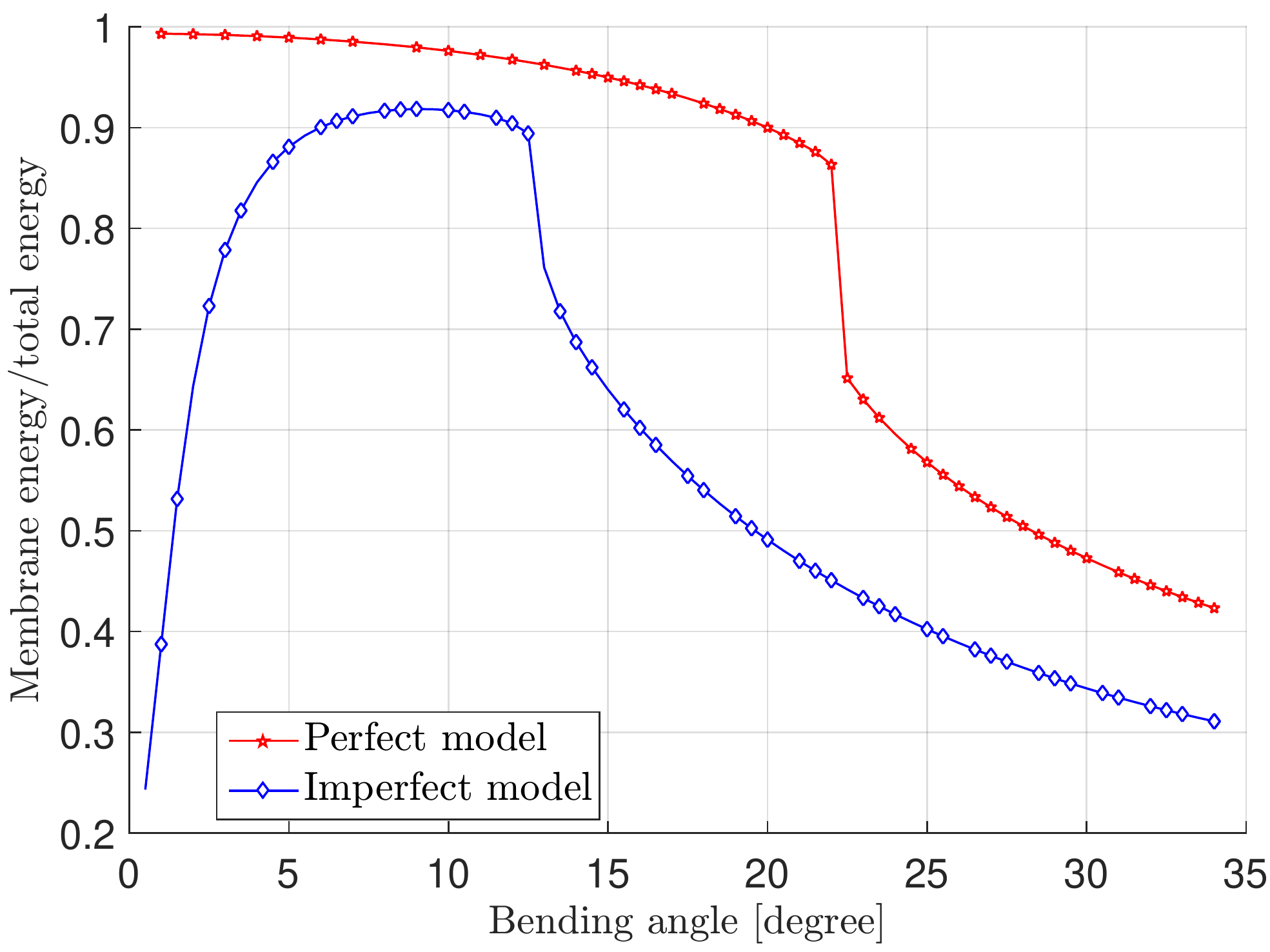}
        \subcaption{}
        \label{f:bending_Energy_ratio_central_n10_m10_L10_QI80_80}
    \end{subfigure}
\caption{CNT bending: (\subref{f:bending_Energy_central_n10_m10_L10_QI80_80}) Comparison of strain energy per atom in perfect and imperfect cases; (\subref{f:bending_Energy_ratio_central_n10_m10_L10_QI80_80}) comparison of the ratio of the membrane energy to the total in perfect and imperfect cases.  CNT(10,10) with the length 10 nm is used. }
\end{figure}

\begin{figure}[]
\begin{center} \unitlength1cm
\begin{picture}(18,11)
\put(0.0,7){\includegraphics[width=80mm,trim=0cm 8cm 0cm 8cm,clip]{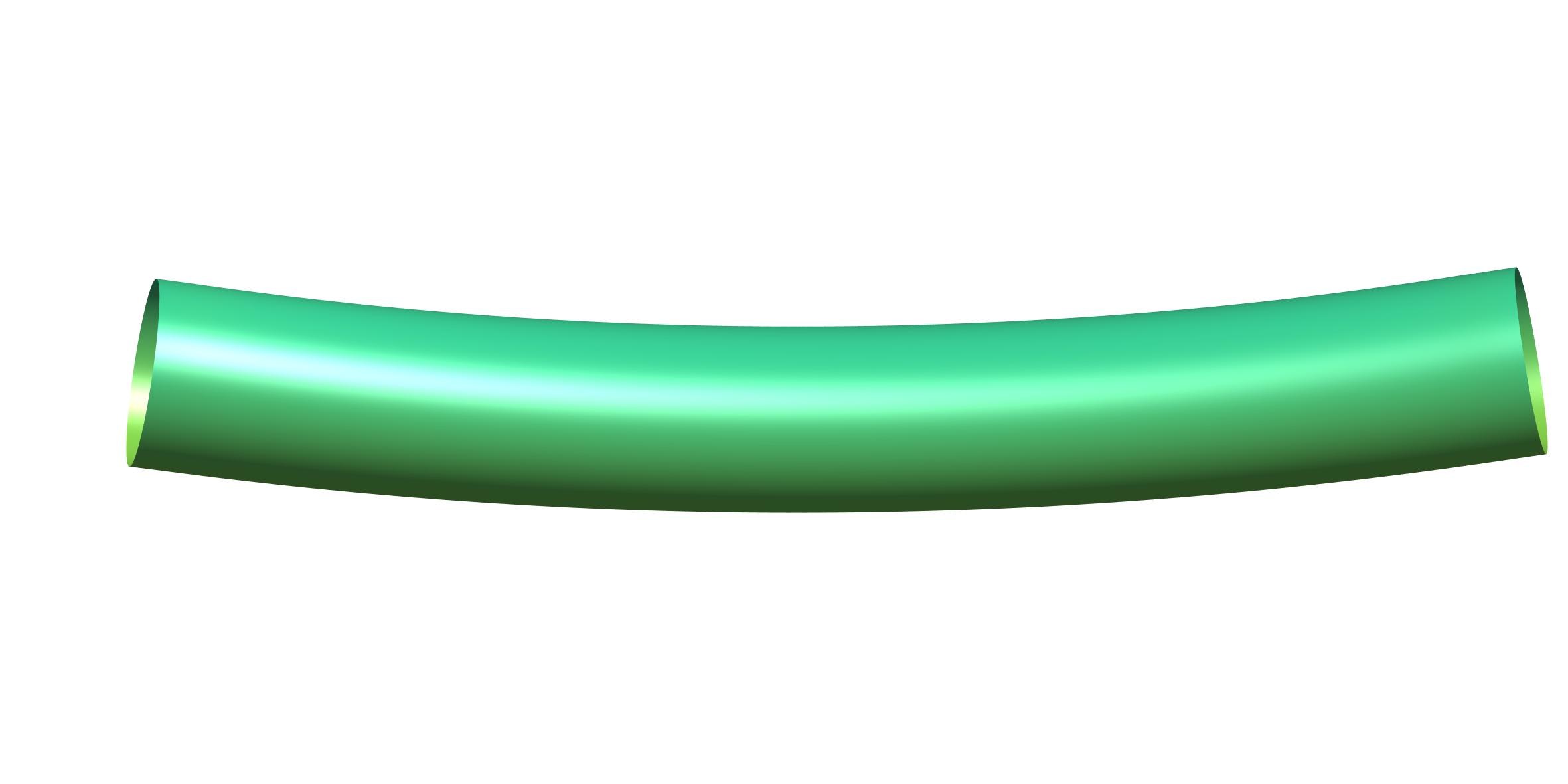}}
\put(8,7){\includegraphics[width=80mm,trim=0cm 8cm 0cm 8cm,clip]{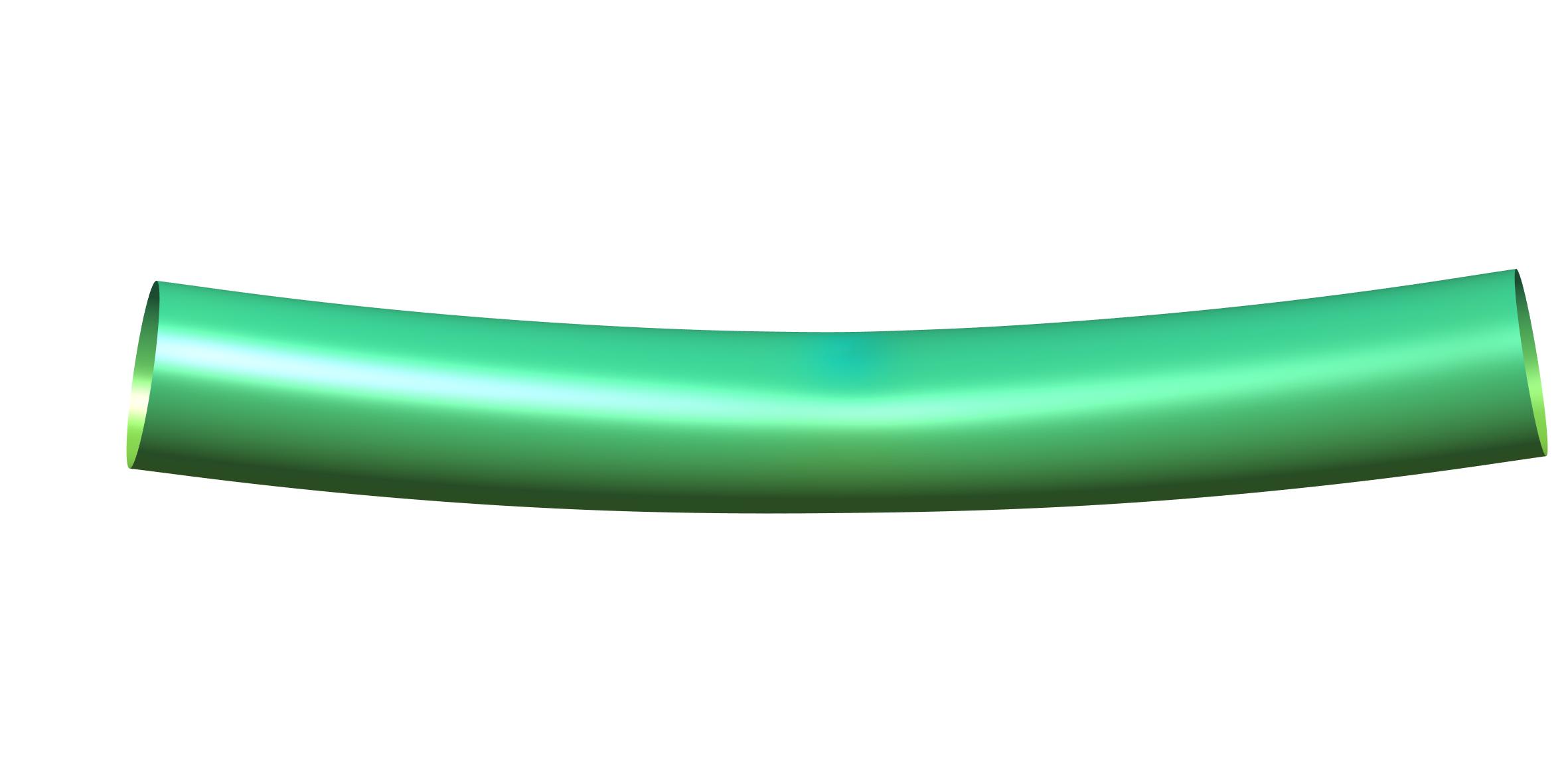}}
%\put(9.5,8.4){(b)}
\put(0.0,5){\includegraphics[width=80mm,trim=0cm 8cm 0cm 8cm,clip]{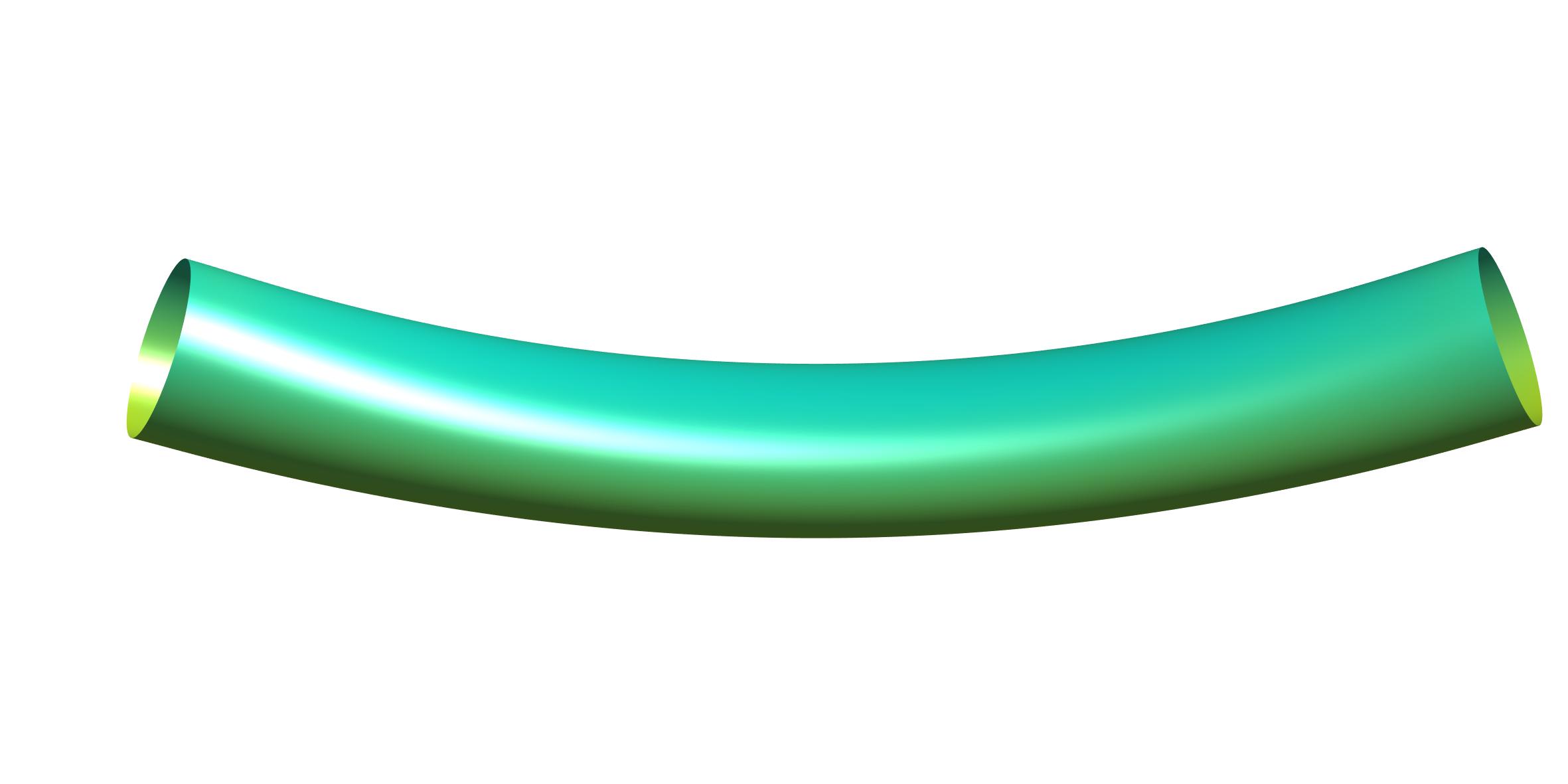}}
\put(8,5){\includegraphics[width=80mm,trim=0cm 8cm 0cm 8cm,clip]{Bending_central_n10_m10_L10_deg20_perf}}
%\put(9.5,3.1){(d)}
\put(0.0,2.5){\includegraphics[width=80mm,trim=0cm 8cm 0cm 8cm,clip]{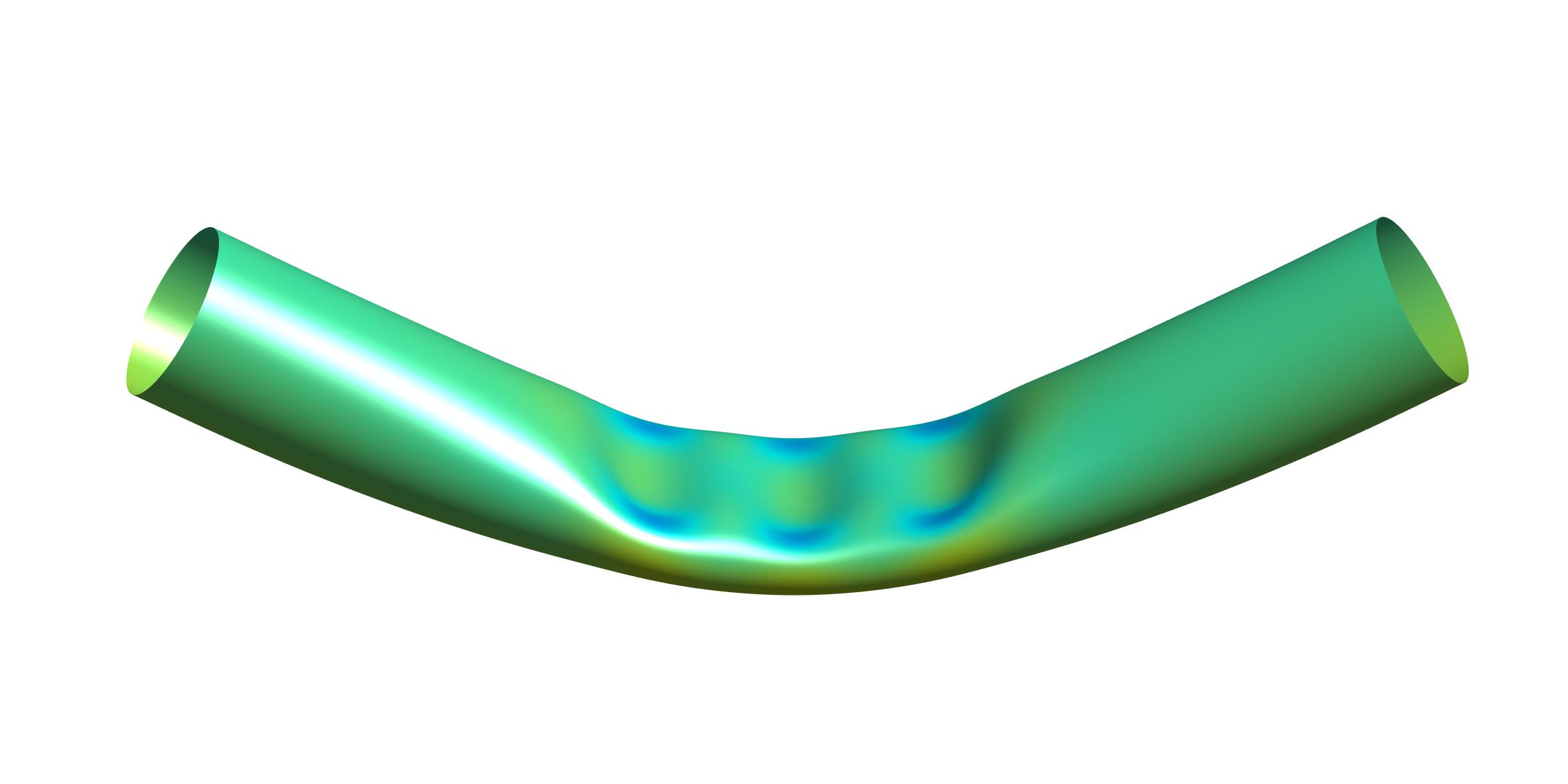}}
\put(8,2.5){\includegraphics[width=80mm,trim=0cm 8cm 0cm 8cm,clip]{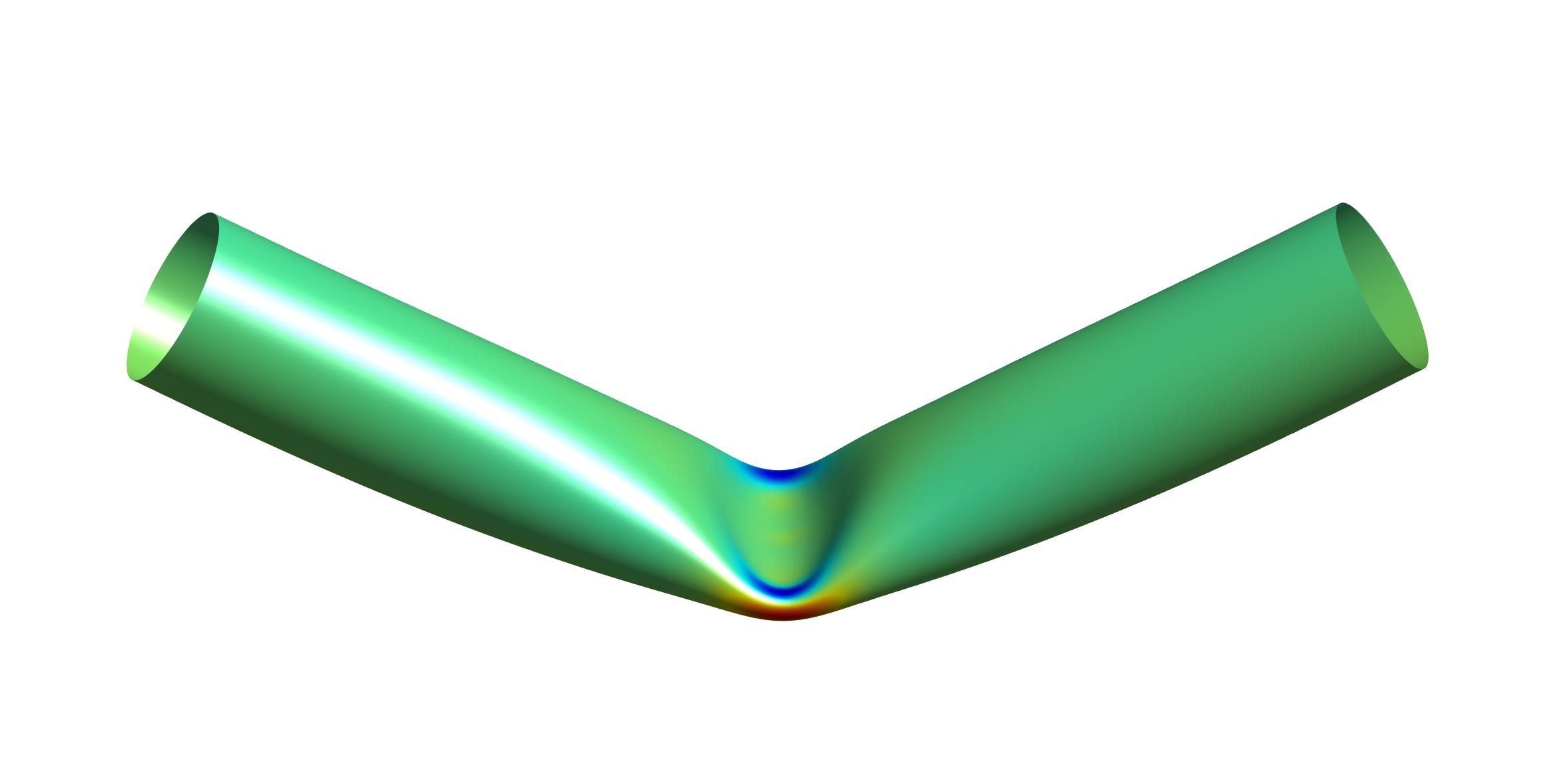}}
%\put(9.5,5.4){(b)}
\put(0.0,0.0){\includegraphics[width=80mm,trim=0cm 7cm 0cm 8cm,clip]{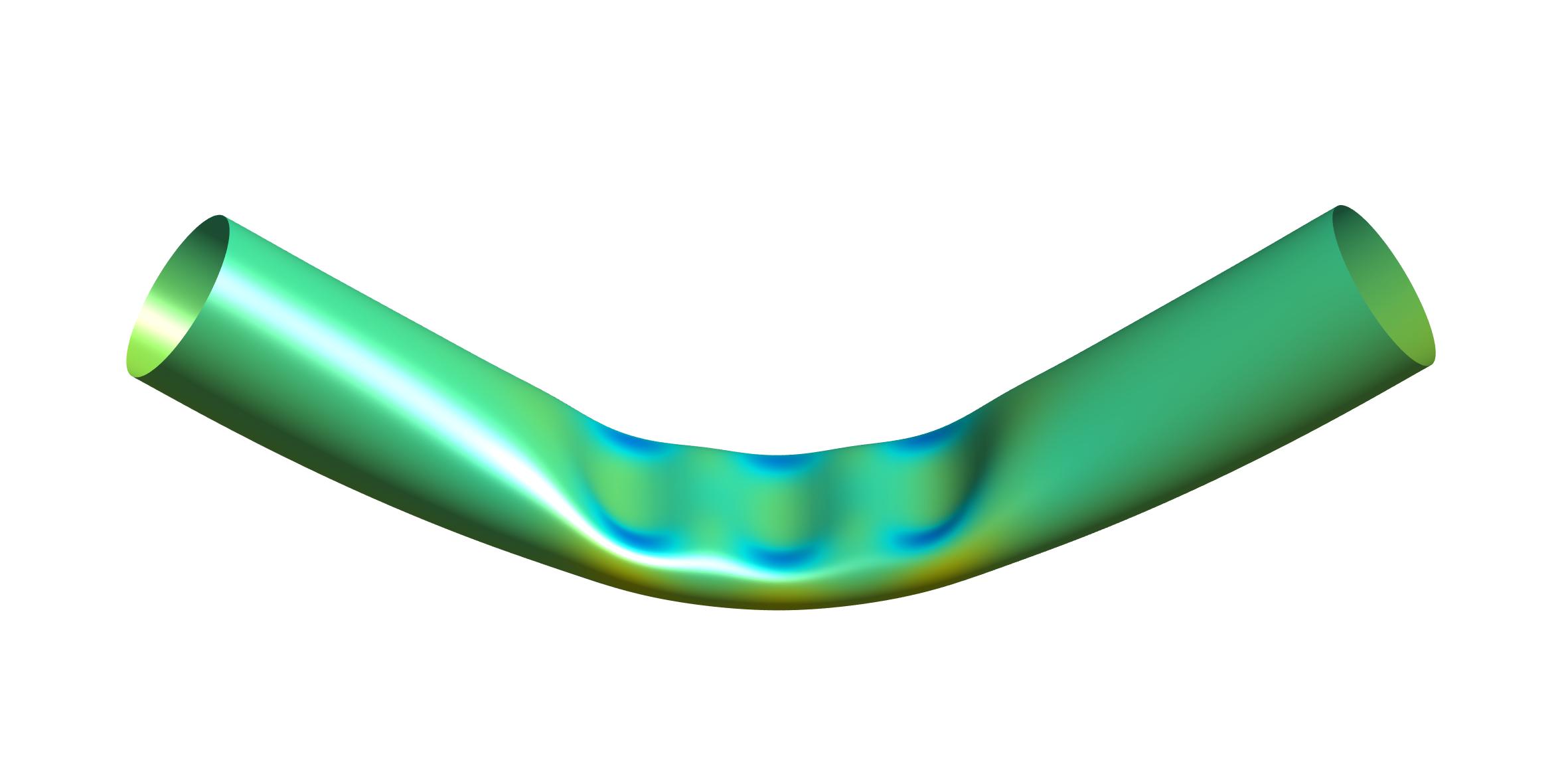}}
\put(8,0.0){\includegraphics[width=80mm,trim=0cm 7cm 0cm 8cm,clip]{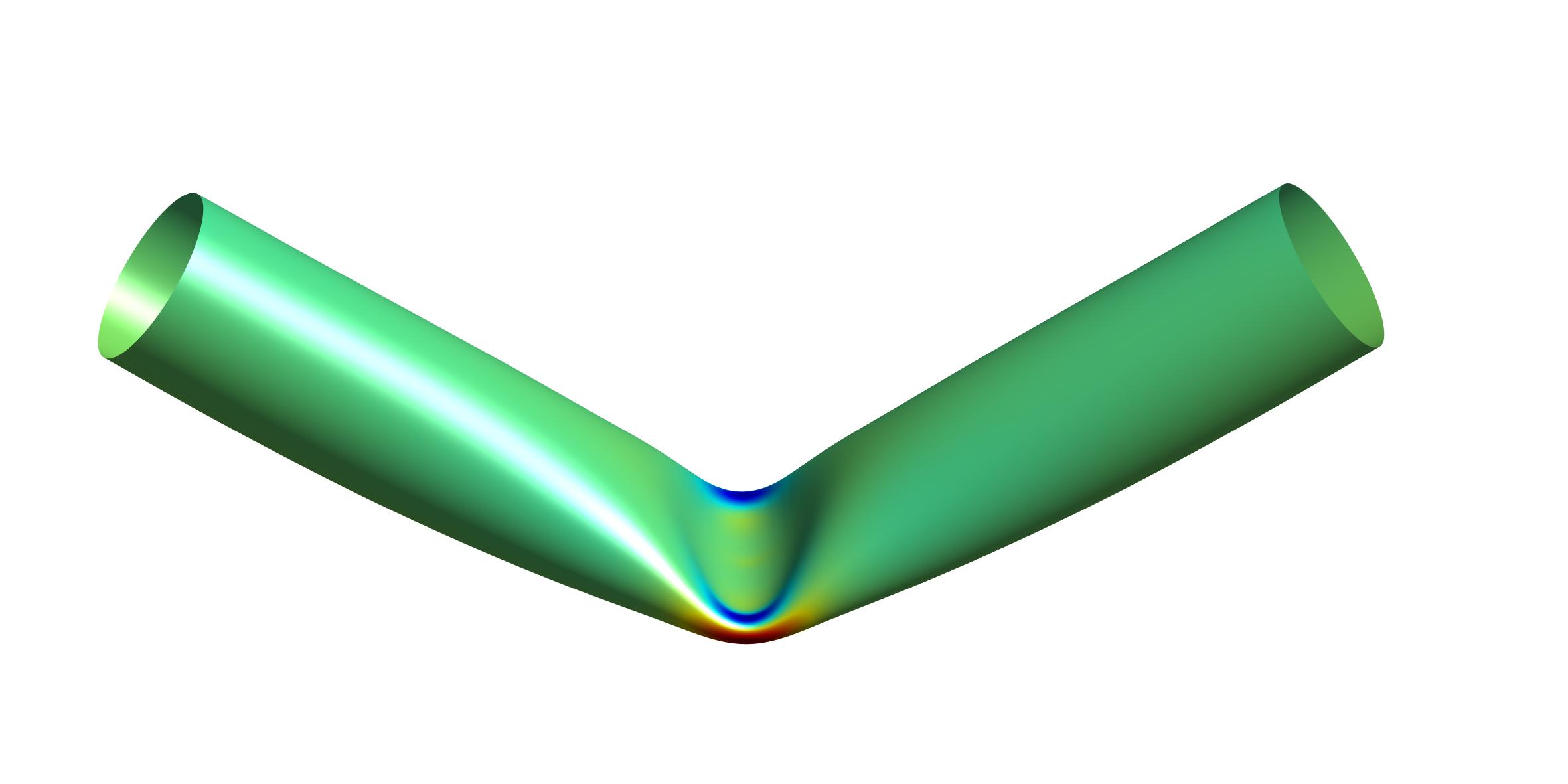}}
\put(0.0,8.0){(a)}
\put(0.0,6.5){(b)}
\put(0.0,4.0){(c)}
\put(0.0,1.5){(d)}
%\put(9.5,.1){(d)}
\put(2,9.5){\includegraphics[width=120mm]{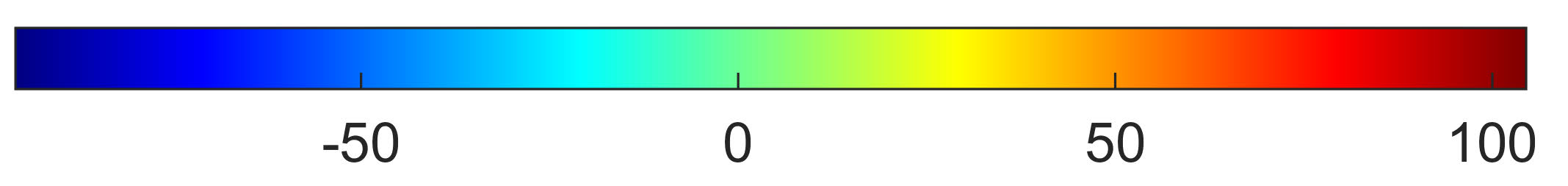}}
\put(3,9){Perfect model}
\put(10.5,9){Imperfect model}
\end{picture}
\caption{CNT bending: Comparison of tr($\bsig$)~$[\text{N/m}]$ of the perfect and imperfect models for the bending angles (a) $\theta=10^{\circ}$,
(b) $\theta=20^{\circ}$,
(c) $\theta=30^{\circ}$ and
(d) $\theta=34^{\circ}$. CNT(10,10) with the length 10 nm is used.
}\label{f:bending_energy_contours}
\end{center}
\end{figure}

\begin{figure}
     \begin{subfigure}{0.5\textwidth}
        \centering
     \includegraphics[width=80mm]{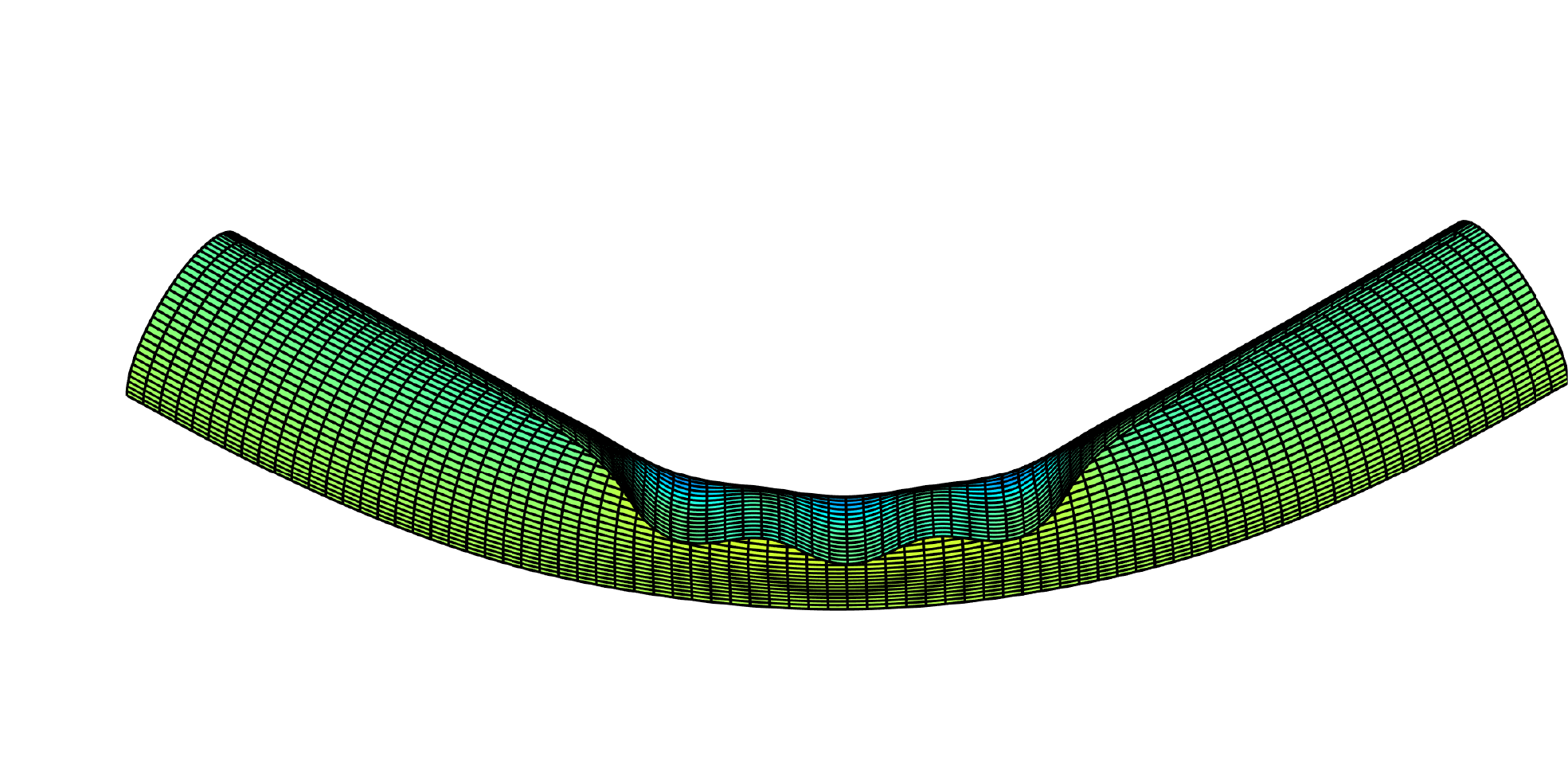}
      \vspace{-10mm}
        \subcaption{}
        \label{f:Bending_central_n10_m10_L10_deg34_perf_cuted}
    \end{subfigure}
    \begin{subfigure}{0.5\textwidth}
        \centering
     \includegraphics[width=80mm]{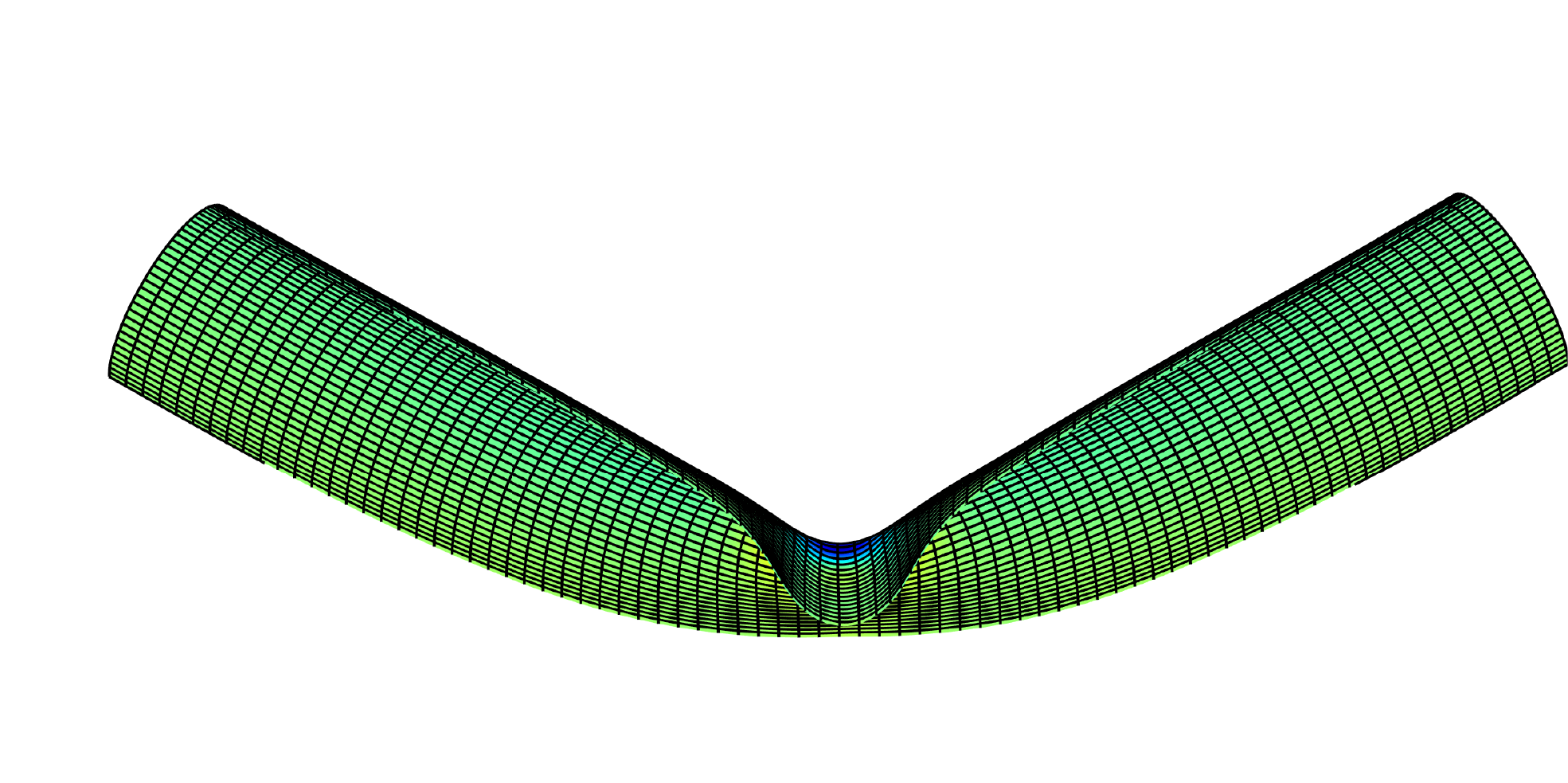}
      \vspace{-10mm}
        \subcaption{}
        \label{f:Bending_central_n10_m10_L10_deg34_cutted}
    \end{subfigure}
    \vspace{-3mm}
\caption{CNT bending: Cross sections of (\subref{f:Bending_central_n10_m10_L10_deg34_perf_cuted}) the perfect and (\subref{f:Bending_central_n10_m10_L10_deg34_cutted}) the imperfect models ($\theta=34^{\circ}$). CNT(10,10) with the length 10 nm is used.}\label{f:bending_contours_sliced}
\end{figure}

In this section, the behavior of a CNT under bending loading is investigated. CNT is bent by applying a bending angle equally at both ends (Fig.~\ref{f:CNT_Loading_bending}). During bending, the end faces of the CNT are assumed to be rigid and remain planar, and the CNT can \textcolor{cgn2}{deform} in axial direction in order to attain a state of pure bending loading. The strain energy convergence is studied with mesh refinement (Fig.~\ref{f:Bending_central_n10_m10_L10_strainEnergy_conv}). The strain energy per atom is presented in Fig.~\ref{f:bending_Energy_central_n10_m10_L10_QI80_80} for perfect and imperfect structures and the buckling point is determined from the sharp variation in the ratio of the membrane energy to the total energy (Fig.~\ref{f:bending_Energy_ratio_central_n10_m10_L10_QI80_80}). It is shown that for the imperfect model, buckling occurs earlier. tr($\bsig$) is compared at different bending angles for the perfect and imperfect models (Fig.~\ref{f:bending_energy_contours}). Finally, the models are cut open along the axial direction to compare the buckled geometry of the perfect and imperfect CNTs (Fig.~\ref{f:bending_contours_sliced}).


\begin{figure}
      \begin{subfigure}{1\textwidth}
        \centering
     \includegraphics[width=80mm]{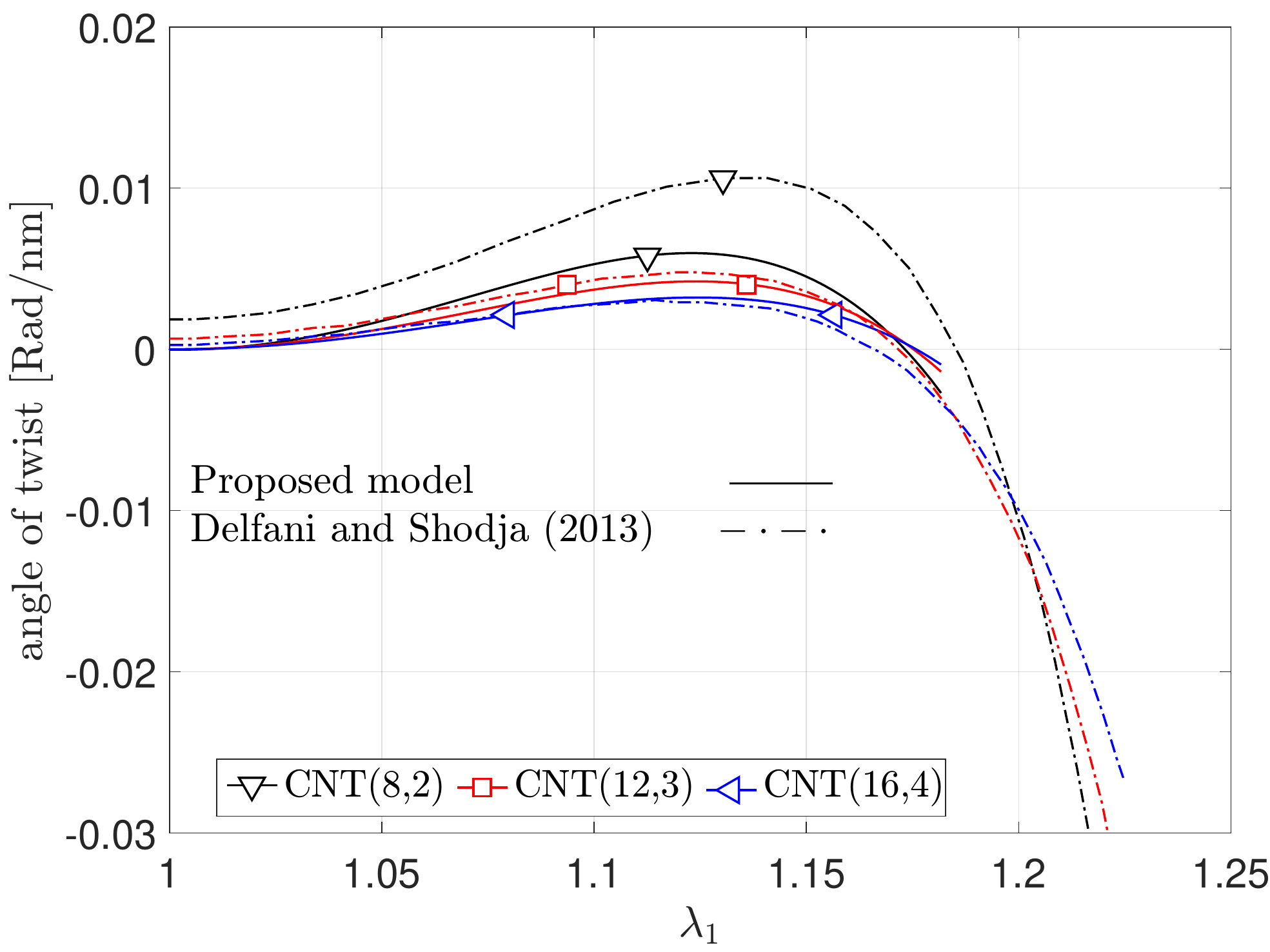}
      \vspace{-2mm}
        \subcaption{}
        \label{f:CNT_nat_twist_Ch12_82d}
    \end{subfigure}\\
        \begin{subfigure}{0.5\textwidth}
        \centering
     \includegraphics[width=80mm]{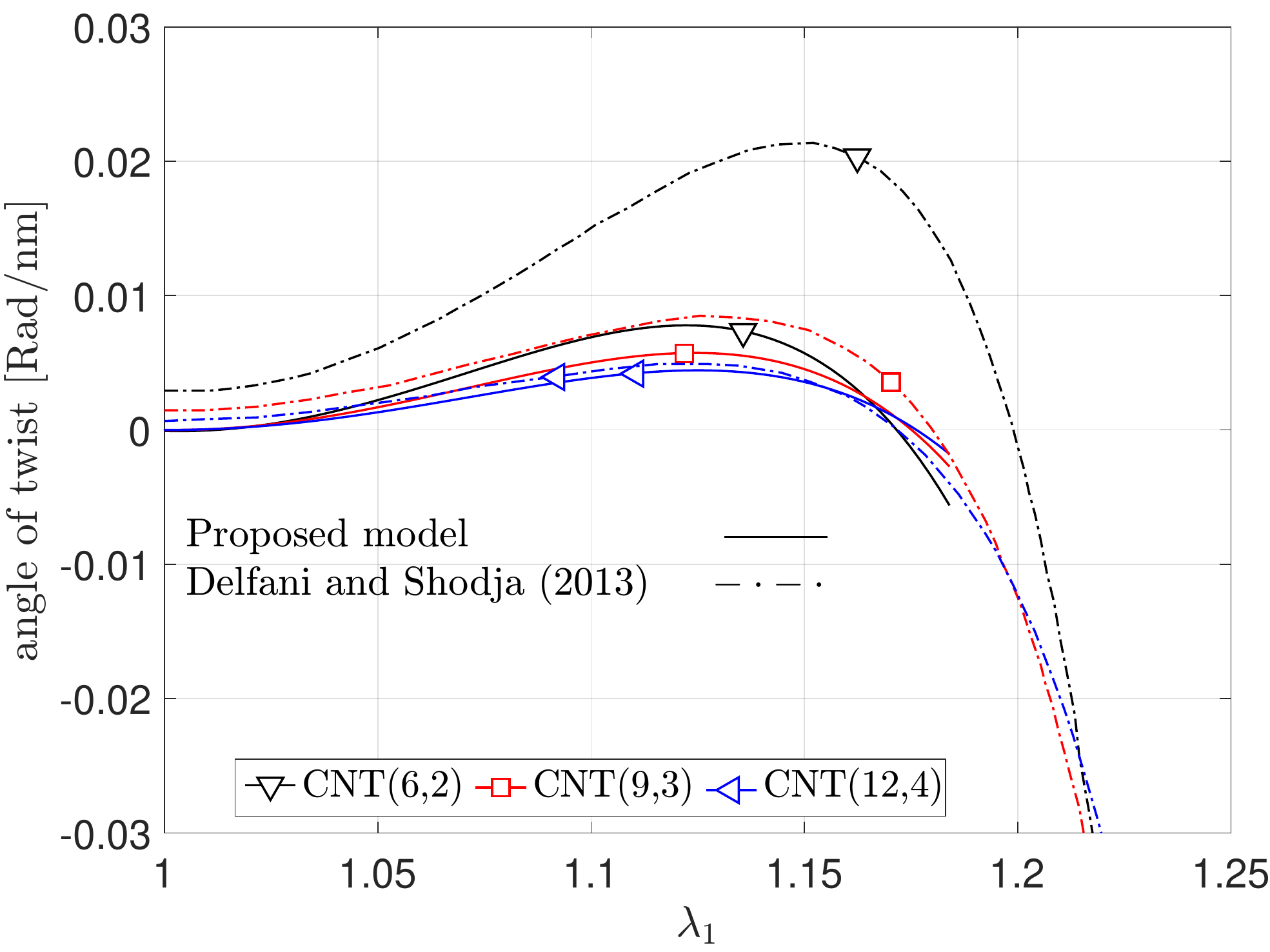}
      \vspace{-7mm}
        \subcaption{}
        \label{f:CNT_nat_twist_Ch13_90d}
    \end{subfigure}
     \begin{subfigure}{0.5\textwidth}
        \centering
     \includegraphics[width=80mm]{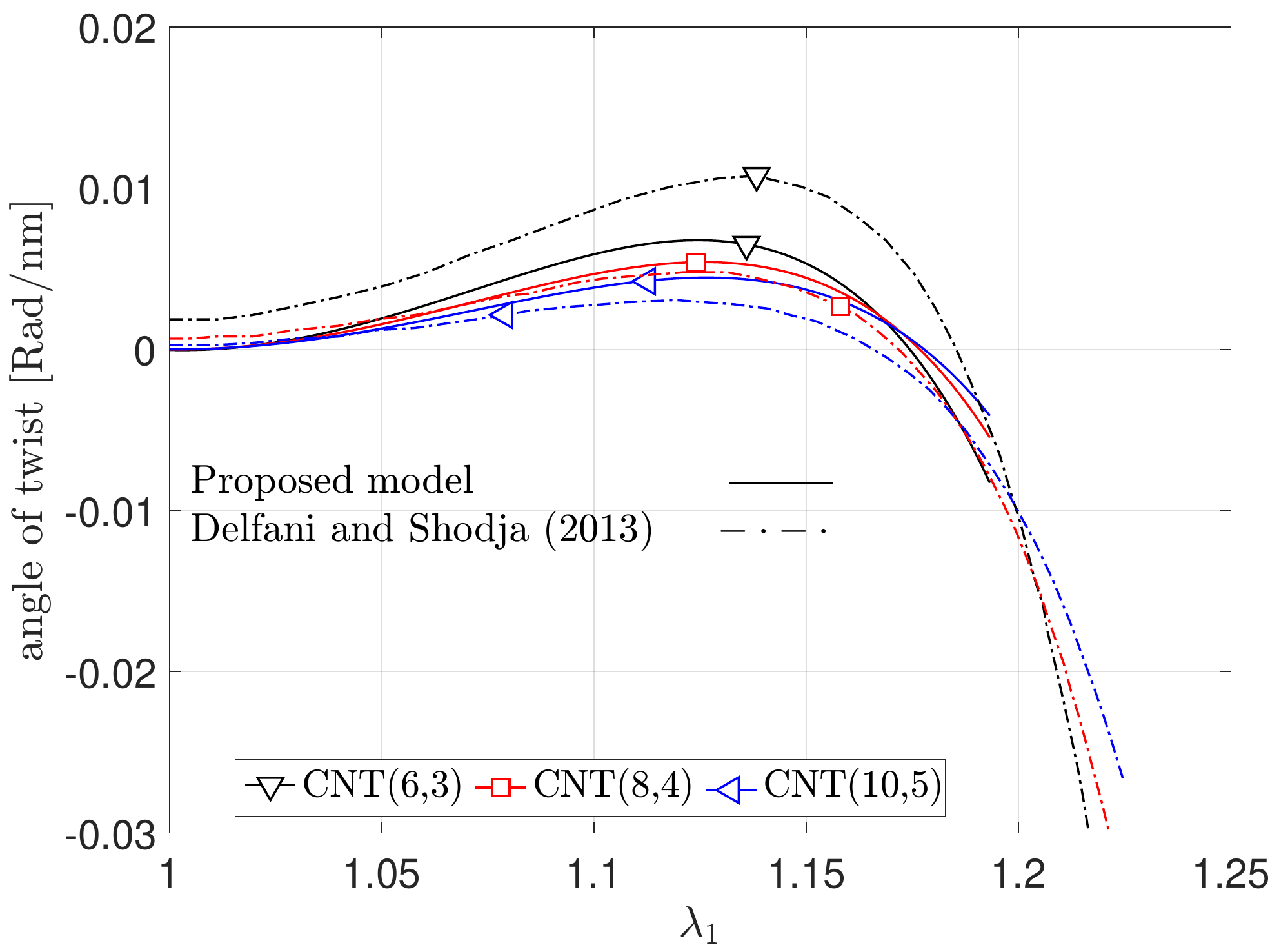}
      \vspace{-7mm}
        \subcaption{}
        \label{f:CNT_nat_twist_Ch19_11d}
    \end{subfigure}
    \vspace{-3mm}
\caption{\textcolor{cgn}{Stretch of chiral CNTs: Twist angle vs. stretch according to the proposed model and \citet{Delfani2013_02} for various chirality angles. (\subref{f:CNT_nat_twist_Ch12_82d}) $\varphi=12.82^{\circ}$; (\subref{f:CNT_nat_twist_Ch13_90d}) $\varphi=13.90^{\circ}$ and (\subref{f:CNT_nat_twist_Ch19_11d}) $\varphi=19.11^{\circ}$. The chirality angle is defined as $\tan(\varphi):=\sqrt{3}\,m/(2n+m)$.}}\label{f:CNT_nat_twist_Ch}
\end{figure}

\textcolor{cgn}{\subsection{Stretch of chiral CNTs}
The deformation of chiral CNTs under axial loading is investigated here. The deformation includes \textcolor{cgn2}{a} uniform extension and twisting along the axial direction of the CNT. The anisotropy of graphene is the source \textcolor{cgn2}{of this} twisting. Three chiralities are selected, and the variation of the twist angle against the stretch is presented in Fig.~\ref{f:CNT_nat_twist_Ch}. The results are compared with the results of \citet{Delfani2013_02}. There is very good agreement for large tube radii, but for small tube radii differences appear: \citet{Delfani2013_02} find an initial twist at zero stretch that they argue appears from relaxing the initial energy of the original tube (that is a rolled graphene sheet with non-zero bending energy). The proposed model, on the other hand, relaxes the initial bending energy by a change of radius and does not predict an initial twist.}
\begin{comment}
The proposed model correctly captures the behavior at zero stretch, where the twist should also vanish. The model of \citet{Delfani2013_02} on the other hand seems to overestimate the twist for small tube radii.}
\end{comment}

\section{Conclusion}\label{s:conclusion}
A new shell model is developed based on a new anisotropic membrane model and a new isogeometric shell formulation. The model is verified with standard biaxial, uniaxial and bending tests. It is used for the modeling of indentation and peeling of a graphene sheet, and the torsion, bending and stretching of a CNT. The new model is computationally \textcolor{cgn}{more efficient than atomistic methods when the problem size reaches micrometer scale.} In addition, the interatomic potentials presently used within the \textcolor{cgn}{exponential Cauchy-Born rule (ECB)} \textcolor{cgn}{tend to} underestimate the elastic modulus and bending stiffness obtained from quantum mechanics (QM). Also, molecular dynamics (MD) simulation have the same drawback as the ECB \textcolor{cgm2}{rule}. The current model uses \textcolor{cgn2}{existing} QM data to calibrate the material constants. The material \textcolor{cgn}{response of} the current model \textcolor{cgn2}{is} therefore closer to \textcolor{cgn2}{existing} ab-initio and experimental results.  \textcolor{cgn}{Thus the current model improves both accuracy and efficiency}. Finally, the current model is fully nonlinear and can handle large deformations, buckling and postbuckling. The buckling points in torsional and bending loading \textcolor{cgn2}{are} determined from the energy ratio, and the postbuckling behavior of CNTs is simulated.

\section*{Acknowledgment}{Financial support from the German Research Foundation (DFG) through grant GSC 111, is
gratefully acknowledged. The authors also thank Maximilian Harmel for checking the manuscript carefully.}
\FloatBarrier
%\clearpage
\appendix
\numberwithin{equation}{section}
\numberwithin{figure}{section}
\numberwithin{table}{section}
\section{Tensorial derivative and curvilinear description of logarithmic strain}\label{s:preliminary}
In this section, the derivative of eigenvalues and a formulation for the tensorial derivative of the logarithmic strain are given. Then, the logarithmic strain is described in curvilinear coordinates.
The derivative of eigenvalues can be written as \citep{itskov2015_01}
\eqb{l}
\ds \pa{\Lambda_i}{\bC} = \ds\pa{\Lambda_i}{\auab}\, \bA_\alpha\otimes\bA_\beta = P_i^{\alpha\beta}\,\bA_\alpha\otimes\bA_\beta~;~~~\text{with}~~~i=1,2~,
\eqe
where $P_i^{\alpha\beta}$ are
\eqb{l}
 P_i^{\alpha\beta} = A^{\alpha\gamma}\,P^i_{\gamma\delta}\,A^{\delta\beta}~.
\eqe
The derivative of the logarithmic strain can be written as \citep{itskov2015_01}

\eqb{l}
\ds\pa{\bE^{(0)}}{\bC} = \ds \sum_{i,j=1}^{2}{\ds f_{ij}\,\left[\mP_i\otimes\mP_j\right]^\mrs}~,
\label{e:Ec}
\eqe
where symmetrization operator $(\bullet)^\mrs$, transpose operator $(\bullet)^{\text{t}}$ and $f_{ij}$ are defined as
\eqb{lll}
\sA^\mrs :=\ds \frac{1}{2}\left(\sA+\sA^{\text{t}}\right)~,
\eqe
\eqb{lll}
  \left(\ba \otimes \bb \otimes \bc \otimes \bd\right)^{\text{t}} :=\left(\ba \otimes \bc \otimes \bb \otimes \bd\right)~,
\eqe
\eqb{l}
f_{ij} := \left\lbrace \begin{array}{l}
\ds\frac{1}{2\,\lambda_i^2}~;\quad $if $ i=j~, \\[4mm]
\ds\frac{\ln\lambda_i - \ln\lambda_j}{\lambda_i^2 - \lambda_j^2}~;\quad $if $ i\neq j~.
\end{array} \right.
\eqe
It is worth to note that there \textcolor{cgn2}{is} another transpose operator $(\bullet)^{\text{T}}$ defined as
\eqb{lll}
  (\ba \otimes \bb \otimes \bc \otimes \bd)^{\text{T}} :=(\bb \otimes \ba \otimes \bd \otimes \bc)~.
\eqe

Next, $\bE^{(0)}$ can be expressed in the curvilinear coordinate as
\eqb{l}
\bE^{(0)} = E^{(0)}_{\alpha\beta}\,\bA^\alpha\otimes\bA^\beta~,
\label{e:E0ab1}
\eqe
with
\eqb{l}
E^{(0)}_{\alpha\beta} :=\ds  \sum_{i=1}^{2}{\ln{\lambda_i}\,P^i_{\alpha\beta}}~.
\label{e:E0ab2}
\eqe
In addition, the co-variant and contra-variant components of the deviatoric logarithmic strain can be written as
\eqb{l}
\textcolor{cgm}{E^{(0)}_{\text{dev}\,\alpha\beta}} := \ln{\lambda}\,\left(P^1_{\alpha\beta}-P^2_{\alpha\beta}\right)~,
\eqe

\eqb{l}
\textcolor{cgm}{E^{(0)\,\alpha\beta}_{\text{dev}}} := \ln{\lambda}\,\left(P^{\alpha\beta}_1-P^{\alpha\beta}_2\right)~.
\eqe

\section{Derivation of membrane constitution for curvilinear coordinates}\label{s:Derivation_membrane_ constitutive_law_based_on_curvilinear_coordinate}
In this section, the Kirchhoff stress and corresponding elastic tensors are derived. They are related to the membrane energy. The repeated and distinct eigenvalue cases are investigated and corresponding constitutive laws are given in both cases. Some relations are repeated to make it easier to follow the derivation.
\subsection{Distinct eigenvalues}
The case of distinct eigenvalues is considered in this section. The Kirchhoff stress tensor can be obtained by the chain rule as
\eqb{lll}
\ds\tauab_{\text{m}} \is \ds \left( 2\pa{ W^{\mathrm{dil}}_{\text{m}}}{\sJ_1} + 2\,\pa{ W^{\mathrm{dev}}_{\text{m}}}{\sJ_1}\right) \pa{\sJ_1}{\auab} + 2\,\pa{ W^{\mathrm{dev}}_{\text{m}}}{\sJ_2}\pa{\sJ_2}{\auab} + 2\,\pa{ W^{\mathrm{dev}}_{\text{m}}}{\sJ_3}\pa{\sJ_3}{\auab}~,\\[4mm]
%\isd \ds\tauab_q + \tauab_{\sJ 1} + \tauab_{\sJ 2}  + \tauab_{\sJ 3}~,
\eqe
with
\eqb{lll}
\ds2\,\pa{ W^{\mathrm{dil}}_{\text{m}}}{\sJ_1} \is 2\,\varepsilon\,\hat{\alpha}^2\,\epsilon_\mra\,e^{-\hat{\alpha}\,\epsilon_\mra}~,\\[4mm]

\ds2\,\pa{ W^{\mathrm{dev}}_{\text{m}}}{\sJ_1} \is 4\,\sJ_2\,\mu'(\epsilon_\mra)  + 2\,\sJ_3\,\eta'(\epsilon_\mra) = -4\,(\mu_1\,\hat{\beta}\,\sJ_2\,e^{\hat{\beta}\,\epsilon_\mra} + \eta_1\,\epsilon_\mra\,\sJ_3)~,\\[4mm]

\ds2\,\pa{ W^{\mathrm{dev}}_{\text{m}}}{\sJ_2} \is 4\,\mu(\epsilon_\mra)~,\\[4mm]
\ds2\,\pa{ W^{\mathrm{dev}}_{\text{m}}}{\sJ_3} \is 2\,\eta(\epsilon_\mra)~,
\eqe
%$\pa{\sJ_1}{\auab}$ and $\pa{\sJ_2}{\auab}$  are
\eqb{lll}
\ds\pa{\sJ_1}{\auab} \is \ds\frac{1}{2}\,\aab~,\\[4mm]
\ds\pa{\sJ_2}{\auab} \is \ds\frac{1}{2}\, \left(\frac{1}{\lambda_1^2}\,P_1^{\alpha\beta} - \frac{1}{\lambda_2^2}\,P_2^{\alpha\beta}\right)\,\ln\lambda~,\\[4mm]
\eqe
\textcolor{cgn2}{where $\mu'$ and $\eta'$ are defined as}
\eqb{lll}
\textcolor{cgn2}{\mu'} \dis \ds \textcolor{cgn2}{\pa{\mu}{\epsilon_\mra}~,}\\[3mm]
\textcolor{cgn2}{\eta'} \dis \ds \textcolor{cgn2}{\pa{\eta}{\epsilon_\mra}~.}\\[3mm]
\eqe
Using Eq.~(\ref{e:defsJ}.3) and the chain rule, the derivative of $\sJ_3$  can be written as
\eqb{lll}
\ds\pa{\sJ_3}{\bC} := \ds\frac{1}{8}\pa{(\gamma_\theta^3)}{\bE^{(0)}}: \pa{\bE^{(0)}}{\bC}~,
\label{e:J3C}
\eqe
%where $\ds\pa{(\gamma_\theta^3)}{\bE^{(0)}}:=\bT^{(0)}$ is defined as
with
\eqb{lll}
\bT^{(0)} = \ds\pa{(\gamma_\theta^3)}{\bE^{(0)}} = \textcolor{cgm}{a_{\hat{M}}}\,\textcolor{cgm}{\hat{\bM}} + \textcolor{cgm}{a_{\hat{N}}}\,\textcolor{cgm}{\hat{\bN}}~.
\label{e:dgama}
\eqe
$\textcolor{cgm}{\hat{\bM}}$ and $\textcolor{cgm}{\hat{\bN}}$ can be written in curvilinear coordinates as
\eqb{lll}
\textcolor{cgm}{\hat{\bM}} = \textcolor{cgm}{\hat{M}_{\alpha\beta}}\, \bA^\alpha\otimes\bA^\beta;~\quad \textcolor{cgm}{\hat{M}_{\alpha\beta}}\dis \bA_\alpha\cdot\textcolor{cgm}{\hat{\bM}}\cdot\bA_\beta~,  \\[2mm]
\textcolor{cgm}{\hat{\bN}} = \textcolor{cgm}{\hat{N}_{\alpha\beta}}\, \bA^\alpha\otimes\bA^\beta;~\quad  \textcolor{cgm}{\hat{N}_{\alpha\beta}}\dis \bA_\alpha\cdot\textcolor{cgm}{\hat{\bN}}\cdot\bA_\beta~.
\label{e:Nab}
\eqe
Using Eq.~\eqref{e:E0ab2}, $\textcolor{cgm}{a_{\hat{M}}}$ and $\textcolor{cgm}{a_{\hat{N}}}$ can be written as
\eqb{lll}
\textcolor{cgm}{a_{\hat{M}}} \dis 3\,\left( \textcolor{cgm}{\hat{M}_{\alpha\beta}}\,\textcolor{cgm}{E_{\text{dev}}^{(0)\alpha\beta}}\right)^2 - 3\,\left(\textcolor{cgm}{\hat{N}_{\alpha\beta}}\,\textcolor{cgm}{E_{\text{dev}}^{(0)\alpha\beta}} \right)^2~,\\[2mm]
\textcolor{cgm}{a_{\hat{N}}} \dis -6\,\left(\textcolor{cgm}{\hat{M}_{\alpha\beta}}\,\textcolor{cgm}{E_{\text{dev}}^{(0)\alpha\beta}}\right) \left(\textcolor{cgm}{\hat{N}_{\gamma\delta}}\, \textcolor{cgm}{E_{\text{dev}}^{(0)\gamma\delta}}\right)~.
 \label{e:aMN}
\eqe
Using \cref{e:dgama,e:Nab,e:aMN}, the co-variant components of $\bT^{(0)}$ can be obtained as
\eqb{lll}
 T^{(0)}_{\alpha\beta} =  \textcolor{cgm}{a_{\hat{M}}}\,\textcolor{cgm}{\hat{M}_{\alpha\beta}} +  \textcolor{cgm}{a_{\hat{N}}}\,\textcolor{cgm}{\hat{N}_{\alpha\beta}}~.
\eqe

Substituting  Eqs.~\eqref{e:Ec} and \eqref{e:dgama} into Eq.~\eqref{e:J3C} and using $\bX:\left[\bY\otimes\bZ\right]^\mrs = \left[\bY^\mrT\bX\bZ^\mrT\right]^\mrs$, see \cite{itskov2015_01}, $\ds\pa{\sJ_3}{\bC}$ can be written as
\eqb{lll}
\ds\pa{\sJ_3}{\bC} = \ds\pa{\sJ_3}{\auab}\,\bA_\alpha\otimes\bA_\beta =   \ds\frac{1}{8}\,\sum_{i,j=1}^{2}{f_{ij}\,  \left[ P_i^{\alpha\gamma}\,T^{(0)}_{\gamma\delta}\,P_j^{\delta\beta}\,\bA_\alpha\otimes\bA_\beta \right]^\mrs} =: \ds\frac{1}{8}\,\mu^{\alpha\beta}\,\bA_\alpha\otimes\bA_\beta~,
\label{e:J3C2}
\eqe
where $\mu^{\alpha\beta}$ is defined as
\eqb{lll}
\mu^{\alpha\beta} :=  \ds \sum_{i,j=1}^{2}{\,f_{ij}\, P_i^{\alpha\gamma}\,T^{(0)}_{\gamma\delta}\,P_j^{\delta\beta}}~.
\eqe

Using the obtained relations, the Kirchhoff stress tensor can be written as
\eqb{lll}
\tauab_{\text{m}}  \is \left[ \varepsilon\,\hat{\alpha}^2\,\epsilon_\mra\,e^{-\hat{\alpha}\,\epsilon_\mra} - 2\,(\eta_1\,\epsilon_\mra\,\sJ_3 + \mu_1\,\hat{\beta}\,\sJ_2\,e^{\beta\,\epsilon_\mra})\right]\,\aab + 2\,\mu\,\ln\lambda\,\chi^{\alpha\,\beta} + \ds\frac{1}{4}\,\eta\,\mu^{\alpha\beta}~,
\eqe
where $\chi^{\alpha\,\beta}$ is defined as
\eqb{lll}
\ds \chi^{\alpha\beta} \dis \ds \left(\frac{1}{\lambda_1^2}\,P_1^{\alpha\beta}\, -\frac{1}{\lambda_2^2}\,P_2^{\alpha\beta}\right)~.
\eqe
In addition, the elastic tensor can be obtained as

\eqb{lll}
\cabgd_{\text{m}} \ds \is
\left[ \varepsilon\,\hat{\alpha}^2\,(1-\hat{\alpha}\epsilon_\mra)\,e^{-\hat{\alpha}\,\epsilon_\mra}\
-2(\eta_1\,\sJ_3 + \mu_1\,\hat{\beta}^2\,\sJ_2\,e^{\hat{\beta}\,\epsilon_\mra})\right]\,\aab\,\agd \\[4mm]
\mi \ds \left[\,\frac{1}{2}\,\eta_1\,\epsilon_\mra\,\mu^{\gamma\delta}
+2\,\mu_1\,\hat{\beta}\,e^{\hat{\beta}\,\epsilon_\mra}\ln\lambda\,\chi^{\gamma\,\delta}\right]\aab \\[4mm]
\plus \ds 2\big[ \varepsilon\,\hat{\alpha}^2\,\epsilon_\mra\,e^{-\hat{\alpha}\,\epsilon_\mra} - 2\,(\eta_1\,\epsilon_\mra\,\sJ_3 + \mu_1\,\hat{\beta}\,\sJ_2\,e^{\hat{\beta}\,\epsilon_\mra})\big]\,{a}^{\alpha\beta\gamma\delta}\\[4mm]
\mi 2\,\mu_1\,\hat{\beta}\,e^{\hat{\beta}\,\epsilon_\mra}\,\ln(\lambda)\,\chi^{\alpha\,\beta}\,\agd
+ \ds \mu\,\chi^{\alpha\,\beta}\,\chi^{\gamma\,\delta}\\[4mm]
\plus \ds 4\,\mu\,\ln\lambda \left(\frac{1}{\lambda_2^4}\,P_2^{\alpha\beta}\,P_2^{\gamma\delta}- \frac{1}{\lambda_1^4}\,P_1^{\alpha\beta}\,P_1^{\gamma\delta}\right)
+4\,\mu\,\ln\lambda \left(\frac{1}{\lambda_1^2}\,\pa{P_1^{\alpha\beta}}{\augd} -
\frac{1}{\lambda_2^2}\,\pa{P_2^{\alpha\beta}}{\augb}\right)\\[4mm]
\mi \ds\frac{1}{2}\,\eta_1\,\epsilon_\mra\,\mu^{\alpha\beta}\,\agd
+\ds\frac{1}{2}\,\eta\,\pa{f_{ij}}{\augd}\, \left[ P_i^{\alpha\xi}\,T^{(0)}_{\xi\upsilon}\,P_j^{\upsilon\beta}\right]^{\mathrm{s}}\\[4mm]
\plus  \ds \frac{1}{2}\,\eta\,f_{ij}\,  \left[ \pa{P_i^{\alpha\xi}}{\augd}\,T^{(0)}_{\xi\upsilon}\,{P_j^{\upsilon\beta}}\right]^{\text{s}}\\[4mm]
\plus  \ds \frac{1}{2}\,\eta\,f_{ij}\,  \left[ P_i^{\alpha\xi}\,T^{(0)}_{\xi\upsilon}\,\pa{P_j^{\upsilon\beta}}{{\augd}}\right]^{\text{s}}\\[4mm]
\plus \ds \frac{1}{2}\,\eta\,f_{ij}\,  \left[   P_i^{\alpha\xi}\,\pa{T^{(0)}_{\xi\upsilon}}{\augd}\,P_j^{\upsilon\beta}\right]^{\text{s}}~,
\eqe

%\pa{P_1^{\alpha\xi}}{\augd}\,T^{(0)}_{\xi\upsilon}\,\pa{P_j^{\upsilon\beta}}{{\augd}}\big
where $\pa{P_1^{\alpha\beta}}{\augd}$, $\pa{P_2^{\alpha\beta}}{\augd}$, $\pa{f_{ij}}{\augd}$,
${a}^{\alpha\beta\gamma\delta}$ and ${A}^{\alpha\beta\gamma\delta}$ can be written as

\eqb{lll}
\ds \pa{P_1^{\alpha\beta}}{\augd}  \is \ds
-\frac{\Aabgd+P_2^{\gamma\delta}\,\Aab\,+(P_1^{\gamma\delta}-P_2^{\gamma\delta})\,
P_1^{\alpha\beta}}{\Lambda_1 - \Lambda_2}~,
\eqe

\eqb{lll}
\ds \pa{P_2^{\alpha\beta}}{\augd}  \is \ds  -\frac{\Aabgd+P_1^{\gamma\delta}\,\Aab
+(P_2^{\gamma\delta}-P_1^{\gamma\delta})\,
P_2^{\alpha\beta}}
{\Lambda_2 - \Lambda_1}~,
\eqe

\eqb{lll}
\ds \pa{f_{ij}}{\augd} \is \ds \left\lbrace \begin{array}{l}
\ds-\frac{1}{\,2\lambda_i^4} P_i^{\gamma\delta}~;\quad $if $ i=j~, \\[4mm]
\ds\frac{\frac{\lambda_i^2 - \lambda_j^2}{\lambda_i^2}-2(\ln\lambda_i - \ln\lambda_j)}{2(\lambda_i^2 - \lambda_j^2)^2}P_i^{\gamma\delta}
-\ds\frac{\frac{\lambda_i^2 - \lambda_j^2}{\lambda_j^2}-2(\ln\lambda_i - \ln\lambda_j)}
{2(\lambda_i^2 - \lambda_j^2)^2}P_j^{\gamma\delta};\quad $if $ i\neq j~,
\end{array} \right.
\eqe

\eqb{lll}
\ds \aabgd \is \ds -\frac{1}{2}\left(\aag\abd+\aad\abg\right)~,
\eqe

\eqb{lll}
\ds \Aabgd \is \ds -\frac{1}{2}\left(\Aag\Abd+\Aad\Abg\right)~.
\eqe

In addition, $\pa{T^{(0)}_{\xi\upsilon}}{\augd}$ can be obtained as

\eqb{lll}
\ds \pa{T^{(0)}_{\xi\upsilon}}{\mC} \is \ds \pa{T^{(0)}_{\xi\upsilon}}{\bE^{(0)}}:\pa{\bE^{(0)}}{\mC} =\pa{T^{(0)}_{\xi\upsilon}}{\augd} \bA_\gamma \otimes \bA_\delta~,
\eqe

\eqb{lll}
\ds \pa{T^{(0)}_{\xi\upsilon}}{\augd} \is
\ds \sum_{k,l=1}^{2}{f_{kl}\,\left[P_k^{\gamma\rho}\,H_{\xi\upsilon\rho\omega}\,P_l^{\omega\delta}\right]^{\mathrm{s}}}~,
\eqe
where $H_{\xi\upsilon\rho\omega}$ can be related to the derivative of $\pa{T^{(0)}_{\xi\upsilon}}{\bE^{(0)}}$ as

\eqb{lll}
\ds \pa{T^{(0)}_{\xi\upsilon}}{\bE^{(0)}} \is
  H_{\xi\upsilon\rho\omega}\,\bA^\rho \otimes \bA^\omega~,
\eqe

\eqb{lll}
\ds  H_{\xi\upsilon\rho\omega} \is \ds
  6\,\left[\textcolor{cgm}{b_{\hat{M}\hat{M}}}\,\textcolor{cgm}{\hat{M}_{\xi\upsilon}} \,\textcolor{cgm}{\hat{M}_{\rho\omega}}
+ \textcolor{cgm}{b_{\hat{M}\hat{N}}}\,\textcolor{cgm}{\hat{M}_{\xi\upsilon}} \,\textcolor{cgm}{\hat{N}_{\rho\omega}}
+ \textcolor{cgm}{b_{\hat{N}\hat{M}}}\,\textcolor{cgm}{\hat{N}_{\xi\upsilon}} \,\textcolor{cgm}{\hat{M}_{\rho\omega}}
+ \textcolor{cgm}{b_{\hat{N}\hat{N}}}\,\textcolor{cgm}{\hat{N}_{\xi\upsilon}} \,\textcolor{cgm}{\hat{N}_{\rho\omega}}\right]~,
\eqe

where $\textcolor{cgm}{b_{\hat{M}\hat{M}}}$, $\textcolor{cgm}{b_{\hat{M}\hat{N}}}$, $\textcolor{cgm}{b_{\hat{N}\hat{M}}}$  and  $\textcolor{cgm}{b_{\hat{N}\hat{N}}}$ are defined as
\eqb{lll}
 \textcolor{cgm}{b_{\hat{M}\hat{M}}} \dis  \textcolor{cgm}{\hat{M}_{\xi\upsilon}}\,\textcolor{cgm}{E_{\text{dev}}^{(0)\xi\upsilon}}~,  \\[2mm]
  \textcolor{cgm}{b_{\hat{M}\hat{N}}} \dis - \textcolor{cgm}{\hat{N}_{\xi\upsilon}}\,\textcolor{cgm}{E_{\text{dev}}^{(0)\xi\upsilon}}~,\\[2mm]
  \textcolor{cgm}{b_{\hat{N}\hat{M}}} \dis - \textcolor{cgm}{\hat{N}_{\xi\upsilon}}\,\textcolor{cgm}{E_{\text{dev}}^{(0)\xi\upsilon}}~,\\[2mm]
  \textcolor{cgm}{b_{\hat{N}\hat{N}}} \dis - \textcolor{cgm}{\hat{M}_{\xi\upsilon}}\,\textcolor{cgm}{E_{\text{dev}}^{(0)\xi\upsilon}}~.
 \label{e:bMN}
\eqe
The elastic tensor can then be simplified as
\eqb{lll}
\cabgd_{\text{m}} \is \ds
\left[ \varepsilon\,\hat{\alpha}^2\,(1-\hat{\alpha}\epsilon_\mra)\,e^{-\hat{\alpha}\,\epsilon_\mra}\
-2(\eta_1\,\sJ_3 + \mu_1\,\hat{\beta}^2\,\sJ_2\,e^{\hat{\beta}\,\epsilon_\mra})\right]\,\aab\,\agd \\[4mm]
\mi \,\frac{1}{2}\,\eta_1\,\epsilon_\mra\,\left[\aab\,\mu^{\gamma\delta}+\mu^{\alpha\beta}\,\agd\right]
-2\,\mu_1\,\hat{\beta}\,e^{\hat{\beta}\,\epsilon_\mra}\,\ln(\lambda) \left[\chi^{\alpha\,\beta}\,\agd+\aab\,\chi^{\gamma\,\delta}\right]\\[4mm]
\plus 2\left[ \varepsilon\,\hat{\alpha}^2\,\epsilon_\mra\,e^{-\hat{\alpha}\,\epsilon_\mra} - 2\,(\eta_1\,\epsilon_\mra\,\sJ_3 + \mu_1\,\hat{\beta}\,\sJ_2\,e^{\hat{\beta}\,\epsilon_\mra})\right]{a}^{\alpha\beta\gamma\delta}
+ \ds \mu\, \chi^{\alpha\,\beta}\,\chi^{\gamma\,\delta}\\[4mm]
\plus \ds 4\,\mu\,\ln\lambda \left(\frac{1}{\lambda_2^4}\,P_2^{\alpha\beta}\,P_2^{\gamma\delta}- \frac{1}{\lambda_1^4}\,P_1^{\alpha\beta}\,P_1^{\gamma\delta}\right)
+4\,\mu\,\ln\lambda \left(\frac{1}{\lambda_1^2}\,\pa{P_1^{\alpha\beta}}{\augd} -
\frac{1}{\lambda_2^2}\,\pa{P_2^{\alpha\beta}}{\augb}\right)\\[4mm]
\plus \ds \frac{1}{2}\,\eta\,\biggl\{\,\pa{f_{ij}}{\augd}\,  \upsilon_{ij}^{\alpha\,\beta}
-  \frac{f_{ij}}{\Lambda_{i}-\Lambda_{\bar{i}}}\big[2\upsilon_{ij}^{\alpha\,\beta}\, (P_i^{\gamma\delta}-P_{\bar{i}}^{\gamma\delta})+ (\zeta_{j}^{\alpha\,\beta}+\zeta_{j}^{\beta\,\alpha})\,P_{\bar{i}}^{\gamma\,\delta}\\[4mm]
\mi\frac{1}{2}(\Aag\,\zeta_{j}^{\delta\,\beta}+ \Aad\,\zeta_{j}^{\gamma\,\beta}+ \zeta_{j}^{\gamma\,\alpha}\,\Abd+ \zeta_{j}^{\delta\,\alpha}\,\Abg)\big]+ \bar{H}^{\alpha\beta\gamma\delta}\biggr\}~,
\eqe
where $\bar{i}$ flips with $i$ ($\bar{i}$ is 2 when $i$ is 1 and vice versa) and  $\upsilon_{ij}^{\alpha\,\beta}$, $\zeta_{j}^{\alpha\,\beta}$ and $\bar{H}^{\alpha\beta\gamma\delta}$ are defined as
\eqb{lll}
\upsilon_{ij}^{\alpha\,\beta}
\dis P_i^{\alpha\xi}\,T_{\xi\upsilon}\,P_j^{\upsilon\,\beta}~,
\eqe
\eqb{lll}
\zeta_{j}^{\alpha\,\beta} \dis A^{\alpha\,\xi}\,T_{\xi\upsilon}\,P_j^{\upsilon\,\beta}~,
\eqe
\eqb{lll}
\bar{H}^{\alpha\beta\gamma\delta}
\dis 6[\textcolor{cgm}{b_{\hat{M}\hat{M}}}\,\kappa^{\alpha\beta}\,\kappa^{\gamma\delta} + \textcolor{cgm}{b_{\hat{M}\hat{N}}}\,\kappa^{\alpha\beta}\,\lambda^{\gamma\delta}+ \textcolor{cgm}{b_{\hat{N}\hat{M}}}\,\lambda^{\alpha\beta}\,\kappa^{\gamma\delta} +\textcolor{cgm}{b_{\hat{N}\hat{N}}}\,\lambda^{\alpha\beta}\,\lambda^{\gamma\delta}]~.
\eqe
$\kappa^{\alpha\beta}$ and $\lambda^{\alpha\,\beta}$ are defined as
\eqb{lll}
\ds \kappa^{\alpha\beta}
\dis \ds \sum_{i,j=1}^{2} {f_{ij}\,P_i^{\alpha\xi}\,\textcolor{cgm}{\hat{M}_{\xi\upsilon}}\,P_j^{\upsilon\,\beta}}~,
\eqe
\eqb{lll}
\ds \lambda^{\alpha\,\beta} \dis \ds \sum_{i,j=1}^{2} {f_{ij}\,P_i^{\alpha\xi}\,\textcolor{cgm}{\hat{N}_{\xi\upsilon}}\,P_j^{\upsilon\,\beta}}~.
\eqe

\subsection{Repeated eigenvalues}

In addition, the special case of repeated eigenvalues ($\Lambda_1 = \Lambda_2$) should be considered. In this case, the right Cauchy-Green tensor can be written as
\eqb{l}
\bC = \Lambda_1\, \bI~.
\eqe
$\pa{\Lambda_1}{\bC}$ can be written by the chain rule as
\eqb{lll}
\ds \pa{\Lambda_1}{\bC} \is \ds \pa{\Lambda_1}{\auab}\,\pa{\auab}{\bC}= \pa{\Lambda_1}{\auab}\, \bA_\alpha \otimes \bA_\beta~.
\eqe
In addition, $\pa{\Lambda_1}{\bC}$ can be written as
\eqb{lll}
\ds \pa{\Lambda_1}{\bC} \is \ds \bI~.
\eqe
So, the following relation for $\pa{\Lambda_1}{\auab}$ can be obtained \citep{SIMO1991_01,Miehe1993_01}
\eqb{lll}
\ds \pa{\Lambda_1}{\auab} \is \ds \Aab~.
\eqe
$\pa{\bE^{(0)}}{\bC}$ can be written as
\eqb{lll}
\ds \pa{\bE^{(0)}}{\bC} \is \ds \frac{1}{2\lambda_1^2}\left[\bI\otimes\bI\right]^s~.
\label{e:dev_E0_C}
\eqe
The obtained relation for $\pa{\sJ_1}{\auab}$ is the same for distinct and repeated eigenvalues. But, $\pa{\sJ_2}{\auab}$ and $\pa{\sJ_3}{\auab}$ should be derived for repeated eigenvalues. $\pa{\sJ_2}{\bC}$ can be expanded as
\eqb{lll}
\ds \pa{\sJ_2}{\bC} \is \ds \pa{\sJ_2}{\bE^{(0)}}:\pa{\bE^{(0)}}{\bC} = \ds \frac{1}{4}\,\pa{\gamma^2_i}{\bE^{(0)}}:\pa{\bE^{(0)}}{\bC}~,
\label{e:dev_J2_C}
\eqe
where $\pa{\gamma^2_i}{\bE^{(0)}}$ can be written as
\eqb{lll}
\ds \pa{\gamma^2_i}{\bE^{(0)}} \is \ds 4\textcolor{cgm}{\bE^{(0)}_{\text{dev}}}~,
\label{e:dev_gamm_E}
\eqe
Next, $\pa{\sJ_2}{\bC}$ and $\pa{\sJ_2}{\auab}$ are connected as
\eqb{lll}
\ds \pa{\sJ_2}{\bC} \is \ds \pa{\sJ_2}{\auab}  \bA_\alpha \otimes \bA_\beta~,
\label{e:dev_J2_C_expand}
\eqe
Substituting Eqs. (\ref{e:dev_gamm_E}) and (\ref{e:dev_E0_C}) into Eq. (\ref{e:dev_J2_C}) and using Eq. (\ref{e:dev_J2_C_expand}) results in
\eqb{lll}
\ds \pa{\sJ_2}{\auab} \is \ds \frac {1}{2\lambda^2_1}\,\textcolor{cgm}{E^{(0)\alpha\beta}_{\text{dev}}} = \eta^{\alpha\beta}~.
\label{e:dev_J2_aab}
\eqe
The same procedure should be repeated to obtain $\pa{\sJ_3}{\auab}$. $\pa{\sJ_3}{\bC}$ can be written as

\eqb{lll}
\ds \pa{\sJ_3}{\bC} \is \ds \pa{\sJ_3}{\bE^{(0)}}:\pa{\bE^{(0)}}{\bC}~.
\label{e:dev_J3_C}
\eqe
Substituting Eqs.~(\ref{e:dgama}) and (\ref{e:dev_E0_C}) into Eq.~(\ref{e:dev_J3_C}) results in
\eqb{lll}
\ds \pa{\sJ_3}{\bC} \is \ds \frac{1}{16\lambda_1^2}\left[\bI\,\bT^{(0)}\,\bI \right]^s =\frac{1}{16\lambda_1^2}\left[ T^{(0)\alpha\beta}\,\bA_\alpha \otimes \bA_\beta\right]^s~,
\eqe
where $T^{(0)\alpha\beta}$ is defined as
\eqb{lll}
\ds T^{(0)\alpha\beta} \dis \ds \Aag\,T^{(0)}_{\gamma\delta}\,\Abd~.
\eqe
Furthermore, $\pa{\sJ_3}{\bC}$ and $\pa{\sJ_3}{\auab}$ are connected as
\eqb{lll}
\ds \pa{\sJ_3}{\bC} \is \ds \pa{\sJ_3}{\auab}\,\bA_\alpha \otimes \bA_\beta~.
\eqe
So, $\pa{\sJ_3}{\auab}$ can be obtained as
\eqb{lll}
\ds \pa{\sJ_3}{\auab} \is \ds \frac{1}{16\lambda_1^2}\,T^{(0)\alpha\beta}= \frac{1}{8}\mu^{\alpha\beta}~.
\eqe
Using the obtained relations, the stress tensor can be written as
\eqb{lll}
\tauab_{\text{m}}  \is \left[ \varepsilon\,\hat{\alpha}^2\,\epsilon_\mra\,e^{-\hat{\alpha}\,\epsilon_\mra} - 2\,(\eta_1\,\epsilon_\mra\,\sJ_3 + \mu_1\,\hat{\beta}\,\sJ_2\,e^{\hat{\beta}\,\epsilon_\mra})\right]\,\aab + 4\,\mu\,\eta^{\alpha\beta} + \ds\frac{1}{4}\,\eta\,\mu^{\alpha\beta}~.
\eqe

It should be noted that all $\sJ_3$, $\sJ_2$, $\eta^{\alpha\beta}$ and $\mu^{\alpha\beta}$ are zero in the case of repeated eigenvalues. But, they should be kept to take the second derivative and obtain the elastic tensor. It is easy to show $\pa{\mu^{\alpha\beta}}{\augd}$ is zero, so only $\pa{\eta^{\alpha\beta}}{\augd}$ plays a role in the elastic tensor. $\pa{\eta^{\alpha\beta}}{\augd}$ can be written as

\eqb{lll}
\ds \pa{\eta^{\alpha\beta}}{\augd} \is \ds -\frac{1}{2\Lambda^2_1}\,\Agd\, \textcolor{cgm}{E^{(0)\,\alpha\beta}_{\text{dev}}}+\frac{1}{2\Lambda_1}\,\pa{\textcolor{cgm}{E^{(0)\,\alpha\beta}_{\text{dev}}}}{\augd}~.
\label{e:dev_mab_a_ab}
\eqe
The first term on the right hand side of Eq.~(\ref{e:dev_mab_a_ab}) is zero. The second term can be obtained from
\eqb{lll}
\ds \pa{\textcolor{cgm}{\bE^{(0)}_{\text{dev}}}}{\bC} \is \ds \pa{\textcolor{cgm}{\bE^{(0)}_{\text{dev}}}}{\bE^{(0)}}:\pa{\bE^{(0)}}{\bC}= \left(\left[\bI\otimes\bI\right]^s-\frac{\bI\odot\bI}{2}\right):\frac{1}{2\Lambda_1}\left[\bI\otimes\bI\right]^s~.
\label{e:E00_C_MID}
\eqe
Eq. (\ref{e:E00_C_MID}) can be further simplified by considering the following relations,
\eqb{lll}
  \bI \otimes \bI : \bX \is \bX : \bI \otimes \bI = \bX~, \\
  \bI \otimes \bI : \sA \is \sA : \bI \otimes \bI = \sA~, \\
  (\bI \otimes \bI)^{\text{t}} : \bX \is \bX : (\bI \otimes \bI)^{\text{t}} = \bX^{\text{T}}~,  \\
  \sA : (\bI \otimes \bI)^{\text{t}} \is \sA^{\text{t}}~, \\
  (\bI \otimes \bI)^{\text{t}} : \sA \is \sA^{\text{TtT}}~,\\
  (\bI \otimes \bI)^{\text{t}} : (\bI \otimes \bI)^t \is \bI \otimes \bI~,
\eqe
where $\bX$ and $\sA$ are second and fourth order tensors. The final simplified form of Eq. (\ref{e:E00_C_MID}) is
\eqb{lll}
\ds \pa{\textcolor{cgm}{\bE^{(0)}_{\text{dev}}}}{\bC} \is \ds \frac{1}{2\Lambda_1}\left(\left[\bI\otimes\bI\right]^s-\frac{\bI\odot\bI}{2}\right)~.
\label{e:E00_C}
\eqe
Next, $\pa{\textcolor{cgm}{\bE^{(0)}_{\text{dev}}}}{\bC}$ and $\pa{\textcolor{cgm}{E^{(0)\,\alpha\beta}_{\text{dev}}}}{\augd}$ are connected as
\eqb{lll}
\ds \pa{\textcolor{cgm}{E^{(0)\,\alpha\beta}_{\text{dev}}}}{\bC} \is \ds \pa{\textcolor{cgm}{E^{(0)\,\alpha\beta}_{\text{dev}}}}{\augd}\,(\bA_\gamma \otimes \bA_\delta)~.
\eqe
Finally, the basis vectors of $\pa{\textcolor{cgm}{E^{(0)\,\alpha\beta}_{\text{dev}}}}{\augd}$ should be rearranged as $\bA^{\alpha} \,\otimes\,\bA^\beta\,\odot\,\bA^\gamma\,\otimes\,\bA^\delta$ to be used in the FEM implementation. $\pa{\textcolor{cgm}{E^{(0)\,\alpha\beta}_{\text{dev}}}}{\augd}$, $\tauab_{\text{m}}$ and $\cabgd_{\text{m}}$ can be written as
\eqb{lll}
\ds \pa{\textcolor{cgm}{E^{(0)\,\alpha\beta}_{\text{dev}}}}{\augd}  \is \ds -\frac{1}{4\Lambda_1}\,\left[2\Aabgd+\Aab\,\Agd\right]~,
\eqe

\eqb{lll}
\tauab_{\text{m}}  \is \left[ \varepsilon\,\hat{\alpha}^2\,\epsilon_\mra\,e^{-\hat{\alpha}\,\epsilon_\mra}\right]\,\aab~,
\eqe

\eqb{lll}
\ds \cabgd_{\text{m}} \is \ds
\left[ \varepsilon\,\hat{\alpha}^2\,(1-\hat{\alpha}\epsilon_\mra)\,e^{-\hat{\alpha}\,\epsilon_\mra}
\right]\,\aab\,\agd+2\left[\varepsilon\,\hat{\alpha}^2\,\epsilon_\mra\,e^{-\hat{\alpha}\,\epsilon_\mra}\right] {a}^{\alpha\beta\gamma\delta}%\\[4mm]
- \ds \frac{\mu}{\Lambda^2_1}\left[2\Aabgd+\Aab\,\Agd\right]~.
\eqe
\section{Analytical solution for arbitrary stretch in Cartesian coordinates}\label{s:analytical_solution}
\begin{figure}
  \centering
  \includegraphics[width=50mm]{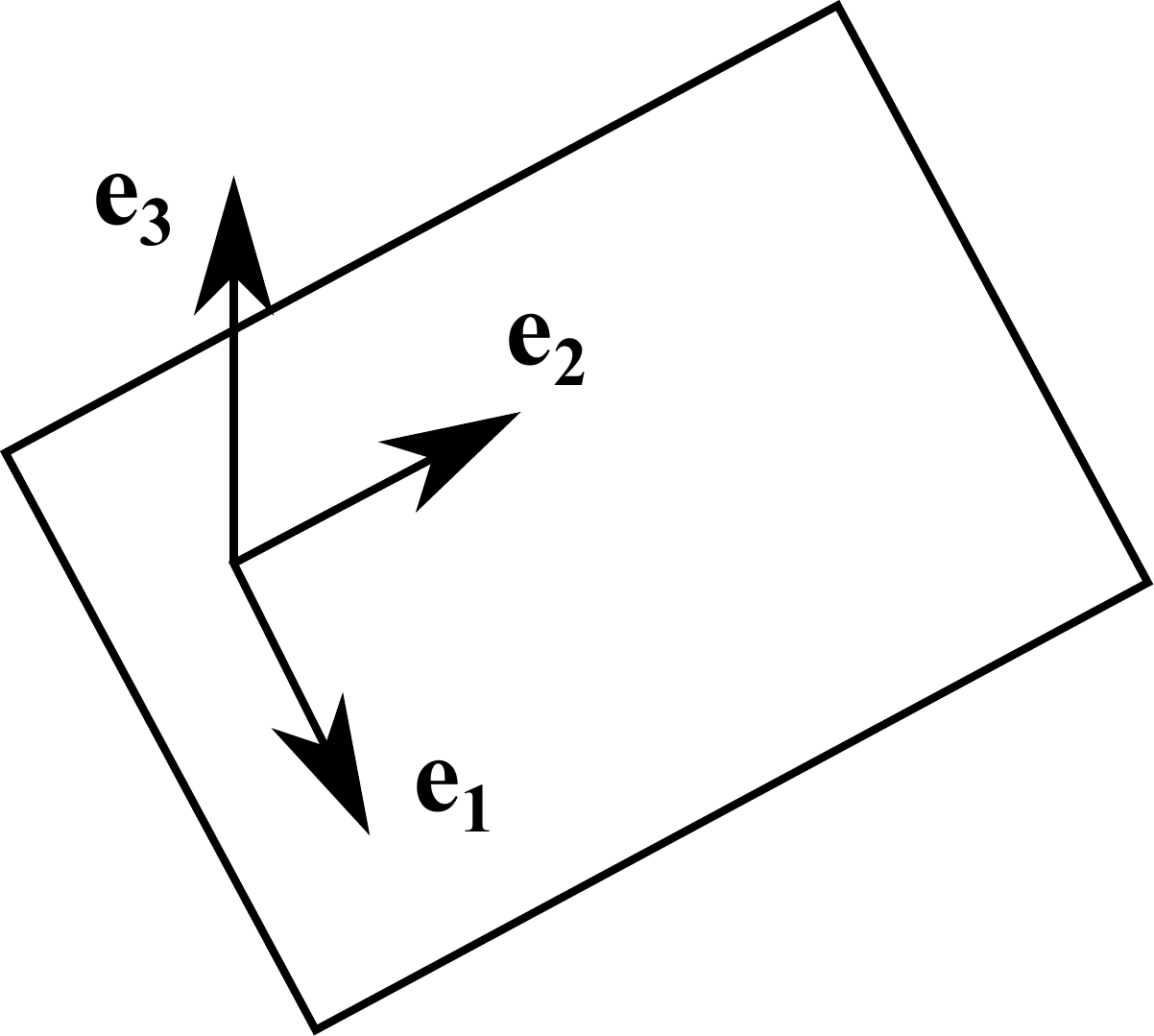}
  \caption{Local Cartesian coordinate system}\label{f:cart_coordinate}
\end{figure}

In this section, a analytical expression of the Cauchy stress tensor in a Cartesian coordinates system (Fig.~\ref{f:cart_coordinate}) is obtained for the case of arbitrary stretch along $\be_1$ and $\be_2$ directions. Then, a relation for pure shear is extracted from the solution of arbitrary stretch as a special case. This solution is used in the verification of the FE implementation in Sec.~\ref{s:Verification}. It should be noted that the general expression for the Cartesian stress components can be directly obtained from Eq.~(\ref{e:shell_stress}) as
\eqb{lll}
\sigma_{ij} \is \ds \be_{i}\cdot\bsig\cdot\be_{j}~.
\eqe
This expression fully defines the stress state of the shell. In the following, the particular expressions for $\sigma_{11}$, $\sigma_{22}$ and $\sigma_{12}$ are derived for the case of a homogeneous membrane stretch, i.e. the deformation gradient is given by
\eqb{lll}
\bF \is \lambda_1\,\be_1 \otimes \be_1 + \lambda_2\,\be_2 \otimes \be_2;~~~\lambda_1 > \lambda_2~,
\eqe
The right Cauchy-Green tensor, right stretch tensor and logarithmic volumetric and deviatoric strains can be written as
\eqb{lll}
\bC \is \Lambda_1\,\be_1 \otimes \be_1 +  \Lambda_2\,\be_2 \otimes \be_2~,
\eqe
\eqb{lll}
\bU \is  \lambda_1\,\be_1 \otimes \be_1 +  \lambda_2\,\be_2 \otimes \be_2~,
\eqe
\eqb{lll}
\ds \epsilon_\mra \is \ds \frac{1}{2}\ln(J)~,
\eqe
\eqb{lll}
\textcolor{cgm}{\bE^{(0)}_{\text{dev}}} \is \ln\lambda\,(\,\be_1 \otimes \be_1- \,\be_2 \otimes \be_2)~.
\eqe
The logarithmic stress can be written as
\eqb{lll}
\ds \bS^{(0)} \dis \ds \pa{W}{\bE^{(0)}} = \left[\varepsilon\,\hat{\alpha}^2\,\epsilon_\mra\,e^{-\hat{\alpha}\,\epsilon_\mra}+2\,\mu'(\epsilon_\mra)\,\sJ_2
+\,\eta'(\epsilon_\mra)\sJ_3\right] \bI + 2\mu\,(\epsilon_\mra)\,\textcolor{cgm}{\bE^{(0)}_{\text{dev}}}+\frac{\eta(\epsilon_\mra)}{8}\bS_{\textcolor{cgm}{\bE^{(0)}_{\text{dev}}}}~,
\label{e:T0_arbit}
\eqe
where $\bS_{\textcolor{cgm}{\bE^{(0)}_{\text{dev}}}}$ \textcolor{cgn2}{is} defined as
\eqb{lll}
\ds \bS_{\textcolor{cgm}{\bE^{(0)}_{\text{dev}}}} \dis \ds
3\left\{ \left[\left(\textcolor{cgm}{\hat{\bM}}:\textcolor{cgm}{\bE^{(0)}_{\text{dev}}}\right)^2-\left(\textcolor{cgm}{\hat{\bN}}:\textcolor{cgm}{\bE^{(0)}_{\text{dev}}}\right)^2\right]\bM -2\left[\left(\textcolor{cgm}{\hat{\bM}}:\textcolor{cgm}{\bE^{(0)}_{\text{dev}}}\right)\left(\textcolor{cgm}{\hat{\bN}}:\textcolor{cgm}{\bE^{(0)}_{\text{dev}}}\right)\right]\bN \right\}~.
\label{e:S_E0_arbit}
\eqe
$\textcolor{cgm}{\hat{\bM}}$ and $\textcolor{cgm}{\hat{\bN}}$ can be written as
\eqb{lll}
\textcolor{cgm}{\hat{\bM}} \is \hat{\bx} \otimes \hat{\bx} - \hat{\by} \otimes \hat{\by} = \cos(2\vphi)\,\be_1 \otimes \be_1 +\sin(2\vphi)\,(\be_1 \otimes \be_2 + \be_2 \otimes \be_1)-\cos(2\vphi)\,\be_2 \otimes \be_2~,
\eqe
\eqb{lll}
\ds \textcolor{cgm}{\hat{\bN}} \is \ds \hat{\bx} \otimes \hat{\by} + \hat{\by} \otimes \hat{\bx} = -\sin(2\vphi)\,\be_1 \otimes \be_1 +\cos(2\vphi)\,(\be_1 \otimes \be_2 + \be_2 \otimes \be_1)+\sin(2\vphi)\,\be_2 \otimes \be_2~,
\eqe
where $\hat{\bx}$ and $\hat{\by}$ are
\eqb{lll}
      \hat{\bx} \is  \cos(\vphi)\,\be_1 + \sin(\vphi)\,\be_2~, \\
      \hat{\by} \is  -\sin(\vphi)\,\be_1 + \cos(\vphi)\,\be_2~.
\eqe
$\vphi$ is the angle between $\be_1$ and $\hat{\bx}$.
Using the following relations
\eqb{lll}
      \textcolor{cgm}{\hat{\bM}}:\textcolor{cgm}{\bE^{(0)}_{\text{dev}}} \is \cos(2\vphi)\,\ln(\lambda^2)~, \\
      \textcolor{cgm}{\hat{\bN}}:\textcolor{cgm}{\bE^{(0)}_{\text{dev}}} \is -\sin(2\vphi)\,\ln(\lambda^2)~,
\eqe
Eq.~(\ref{e:S_E0_arbit}) can be simplified as
\eqb{lll}
\bS_{\textcolor{cgm}{\bE^{(0)}_{\text{dev}}}} \is
12\left( \cos(4\vphi)\,\sJ_2\,\textcolor{cgm}{\hat{\bM}}+\sin(4\vphi)\,\sJ_2\,\textcolor{cgm}{\hat{\bN}} \right)~.
\label{e:S_E0}
\eqe
In addition, the logarithmic stress can be simplified as
\eqb{lll}
\bS^{(0)} \is V_1\,\bI+V_2\,\textcolor{cgm}{\bE^{(0)}_{\text{dev}}}+V_3\,\bM+V_4\,\bN~,
\eqe
where $V_i$ are defined as
\eqb{lll}
\begin{array}{lll}
  V_1 \dis \varepsilon\,\hat{\alpha}^2\,\epsilon_\mra\,e^{-\hat{\alpha}\,\epsilon_\mra}+2\,\mu'(\epsilon_\mra)\,\sJ_2
+\,\eta'(\epsilon_\mra)\,\sJ_3~,\\[3mm]
  V_2 \dis 2\mu\,(\epsilon_\mra)~, \\[3mm]
  V_3 \dis \ds \frac{3}{2}\eta(\epsilon_\mra)\,\cos(4\vphi)\,\sJ_2~, \\[3mm]
  V_4 \dis \ds\frac{3}{2}\eta(\epsilon_\mra)\,\sin(4\vphi)\,\sJ_2~.
\end{array}
\eqe
$\bS^{(0)}$ can be converted to the second Piola-Kirchhoff stress tensor $\bS^{(2)}$ as
\eqb{lll}
\ds \bS^{(2)} \is \ds 2\pa{\bE^{(0)}}{\bC}:\bS^{(0)}~.
\label{e:S2_arbit}
\eqe
The eigenprojections for the assumed gradient deformation can be written as
\eqb{lll}
    \begin{array}{c}
      \bP_1 = \be_1 \otimes \be_1~, \\
      \bP_2 = \be_2 \otimes \be_2~.
    \end{array}
\eqe
Considering the following relations,
\eqb{lllllll}
      \bP_1 \otimes \bP_1 : \bI &=& \bP_1 &;&  \bP_2 \otimes \bP_2 : \bI &=& \bP_2~, \\
      \bP_1 \otimes \bP_2 : \bI &=& \mathbf{0} &;& \bP_2 \otimes \bP_1 : \bI &=& \mathbf{0}~,\\
      \bP_1 \otimes \bP_1 : \textcolor{cgm}{\bE^{(0)}_{\text{dev}}} &=& \ln(\lambda)\,\bP_1 &;&  \bP_2 \otimes \bP_2 : \textcolor{cgm}{\bE^{(0)}_{\text{dev}}} &=& -\ln(\lambda)\,\bP_2~, \\
      \bP_1 \otimes \bP_2 : \textcolor{cgm}{\bE^{(0)}_{\text{dev}}} &=& \mathbf{0} &;& \bP_2 \otimes \bP_1 : \textcolor{cgm}{\bE^{(0)}_{\text{dev}}} &=& \mathbf{0}~, \\
      \bP_1 \otimes \bP_1 : \textcolor{cgm}{\hat{\bM}} &=& \cos(2\vphi)\,\bP_1 &;&  \bP_2 \otimes \bP_2 : \textcolor{cgm}{\hat{\bM}} &=& -\cos(2\vphi)\,\bP_2~, \\
      \bP_1 \otimes \bP_2 : \textcolor{cgm}{\hat{\bM}} &=& \sin(2\vphi)\,(\be_1 \otimes \be_2) &;& \bP_2 \otimes \bP_1 : \textcolor{cgm}{\hat{\bM}} &=& \sin(2\vphi)\,(\be_2 \otimes \be_1)~,  \\
      \bP_1 \otimes \bP_1 : \textcolor{cgm}{\hat{\bN}} &=& -\sin(2\vphi)\,\bP_1 &;&  \bP_2 \otimes \bP_2 : \textcolor{cgm}{\hat{\bN}} &=& \sin(2\vphi)\,\bP_2~, \\
      \bP_1 \otimes \bP_2 : \textcolor{cgm}{\hat{\bN}} &=& \cos(2\vphi)\,(\be_1 \otimes \be_2) &;& \bP_2 \otimes \bP_1 : \textcolor{cgm}{\hat{\bN}} &=& \cos(2\vphi)\,(\be_2 \otimes \be_1)~,
\eqe
Eq. (\ref{e:S2_arbit}) can be simplified as
\eqb{lll}
\ds \bS^{(2)} \is \ds V_1\left[\frac{1}{\Lambda_1}\bP_1+\frac{1}{\Lambda_2}\bP_2\right] \\[4mm]
    \plus \ds V_2\left[\frac{\ln(\lambda)}{\Lambda_1}\bP_1-\frac{\ln(\lambda)}{\Lambda_2}\bP_2\right] \\[4mm]
    \plus \ds V_3\left[\frac{\cos(2\vphi)}{\Lambda_1}\bP_1-\frac{\cos(2\vphi)}{\Lambda_2}\bP_2 - \frac{\ln(\lambda^{4})}{\Lambda_2-\Lambda_1}\sin(2\vphi)\,(\be_1 \otimes \be_2+\be_2 \otimes \be_1)\right] \\[4mm]
    \plus \ds V_4\left[-\frac{\sin(2\vphi)}{\Lambda_1}\bP_1+\frac{\sin(2\vphi)}{\Lambda_2}\bP_2 - \frac{\ln(\lambda^{4})}{\Lambda_2-\Lambda_1}\cos(2\vphi)\,(\be_1 \otimes \be_2+\be_2 \otimes \be_1)\right]~.
\eqe
$\bS^{(2)}$ can be converted to $\bsig$ by using the following relation,
\eqb{lll}
   \ds \bsig \is \ds \frac{1}{J}\,\bF\,\bS^{(2)}\,\bF^{\text{T}}~.
\label{e:sig_arbit}
\eqe
Considering the following relations,
\eqb{lll}
      \bF\,\bP_1\,\bF^{\text{T}} \is \Lambda_1\,\bP_1~, \\
      \bF\,\bP_2\,\bF^{\text{T}} \is \Lambda_2\,\bP_2~,\\
      \bF\,(\be_1 \otimes \be_2)\,\bF^{\text{T}} \is \lambda_1\,\lambda_2\,\be_1\,\otimes\,\be_2 = J\,\be_1\,\otimes\,\be_2 ~,\\
      \bF\,(\be_2 \otimes \be_1)\,\bF^{\text{T}} \is \lambda_1\,\lambda_2\,\be_2\,\otimes\,\be_1 = J\,\be_2\,\otimes\,\be_1~,
\eqe
the Cartesian components of $\bsig$ then become
\eqb{lll}
    \sigma_{11} \is \ds \frac{1}{J}\left[V_1+V_2\,\ln(\lambda)+V_3\,\cos(2\vphi)-V_4\,\sin(2\vphi)\right]~,\\[3mm]
    \sigma_{22} \is \ds \frac{1}{J}\left[V_1-V_2\,\ln(\lambda)-V_3\,\cos(2\vphi)+V_4\,\sin(2\vphi)\right]~,\\[3mm]
    \sigma_{12} \is \ds -\frac{\ln(\lambda^4)}{\Lambda_2-\Lambda_1}\left[V_3\,\sin(2\vphi)+V_4\,\cos(2\vphi)\right]~.
    \label{e:arbit_stress}
\eqe
The analytical solution for pure shear can be easily obtained from Eq. (\ref{e:arbit_stress}) by assuming  $\lambda_2=\frac{1}{\lambda_1}$.
\section{Coarse grained contact model (CGCM)}\label{s:coarse_grain_contact_model}
This section focuses on the modeling of the substrate. The CGCM is introduced and based on that, the equivalent contact force and stiffness matrix are obtained for the substrate. Next, the substrate geometry is formulated based on analytical geometries. The closest point project (CPP) and principal curvatures are obtained, which are used in the contact formulation.
\subsection{Atomic interaction with a half space} \label{ss:Atomic_interaction_with_half_and_quarter_space}
The CGCM is used to avoid fully atomistic simulation of interactions between substrate and graphene. The type of interaction is van-der-Waals (vdW) interaction. It is assumed that the radius of curvature of any surface point of the substrate is much higher than the cut-off radius of the used potential. So, a half space approximation is a good estimation. In the following, an analytical solution for the interaction between graphene and the half space is given. The Lennard-Jones (L-J) potential is considered to model the atomistic interaction. The L-J potential between a pair of atoms $i$ and $j$ can be written as
\eqb{lll}
\ds \psi_{ij(\text{vdW})} \is \ds -\frac{C_1}{r^6}+\frac{C_2}{r^{12}}~,
\eqe
where $C_1$ and $C_2$ are constants and $r$ is the distance between two atoms. The half space potential can be written as \citep{israelachvili2011_01,sauer2007_01,Sauer2007_02, Aitken2010_01}
\eqb{lll}
\Psi_{ij\text{(vdW)h}} \is \ds -\Gamma\left[\frac{3}{2}\left(\frac{h_0}{r}\right)^3-\frac{1}{2}\left(\frac{h_0}{r}\right)^9\right]~,
\eqe
where $h_0$, $\Gamma$ and $r$ are the equilibrium distance, interfacial adhesion energy per unit area and normal distance of a point of the graphene sheet on the substrate. They are defined as
\eqb{lll}
h_0 \dis \ds \left(\frac{2C_2}{5C_1}\right)^{\frac{1}{6}}~,
\eqe
\eqb{lll}
\Gamma \dis \ds \frac{\pi\,\rho_{\mathrm{s}}\,\rho_{\mathrm{g}}\,C_1}{9h_0^3}~,
\eqe
where $\rho_{\mathrm{s}}$ and $\rho_{\mathrm{g}}$ are the number of atoms per unit volume and area of the substrate and graphene, respectively. Hence, $C_1$ and $C_2$ can be written based on $h_0$ and $\Gamma$ as
\eqb{lll}
C_1 \is \ds \frac{9h_0^3\,\Gamma}{\pi\,\rho_{\mathrm{s}}\,\rho_{\mathrm{g}}}~,
\eqe
\eqb{lll}
C_2 \is \ds \frac{45h_0^9\,\Gamma}{2\pi\,\rho_{\mathrm{s}}\,\rho_{\mathrm{g}}}~.
\eqe
The various $\Gamma$ and $h_0$ are provided in the literature (\cite{Zubaer2009_01}, \cite{Gao2014_01}, \cite{Aitken2010_01}). It is usual to estimate $h_0$ as the graphite interlayer spacing, which is 0.34 nm.

\subsection{Van der Waals contact force and its corresponding contact stiffness matrix}\label{s:contact_stiffness}
The required contact force and corresponding stiffness matrix for the FE formulation are determined in this section. The contact force can be obtained by taking the derivative from the half space potential. The contact stiffness matrix can be obtained by the linearization of the contact force. The details of the linearization of the contact force as well as the contact algorithm can be found in the literature \citep{Sauer2013_01, Sauer2015_01, Sauer2009_01}. Here the required formulation is shortly reviewed. The formulation is simplified by assuming one of the bodies to be rigid. Based on this assumption, the contact force per unit current area of graphene can be written as
\eqb{lll}
\ds \boldsymbol{f}_{\mathrm{g}}= \ds -\pa{\Psi_{ij\mathrm{(vdW)h}}}{\bx_{\mathrm{g}}}= -\pa{\Psi_{ij\mathrm{(vdW)h}}}{r}\pa{r}{\bx_{\mathrm{g}}}=f_{\mathrm{g}}\,\bn_{\mathrm{p}}~,
%\href{}{}\Psi_{ij(vdW)h} \is \ds -\Gamma_0[\frac{3}{2}(\frac{h_0}{z_0})^3-\frac{1}{2}(\frac{h_0}{z_0})^9]
\eqe
where $\bx_{\mathrm{g}}$ is a point on the graphene surface, and $\br$, $r$, $\bn_{\mathrm{p}}$ and $f_{\mathrm{g}}$ are defined as
\eqb{lll}
\ds \br \dis \ds \bx_{\mathrm{g}}-\bx_{\mathrm{p}}~,\\[3mm]
\ds r \dis \ds \|\br\|~,\\[3mm]
\ds \bn_{\mathrm{p}} \dis \ds \frac{\br}{r}~,\\[3mm]
\ds f_{\mathrm{g}} \dis \ds -\pa{\Psi_{ij\mathrm{(vdW)h}}}{r}~.
\eqe
$\bx_{\mathrm{p}}$ is the CPP of $\bx_{\mathrm{g}}$ onto the substrate.
The corresponding contact potential can be written as
\eqb{lll}
\ds {G}_{\mathrm{c}} \is \ds \int\limits_{\partial \mathcal{B}_{\mathrm{c}} }{\delta \bolds{\varphi}_{\mathrm{g}}\cdot\boldsymbol{f}_{\mathrm{g}}~\dif a}
= \int\limits_{\partial \mathcal{B}_{0\mathrm{c}} }{\delta \bolds{\varphi}_{\mathrm{g}}\cdot\boldsymbol{f}_{\mathrm{g}}^0~\dif A},
\eqe
where $\delta \bolds{\varphi}_{\mathrm{g}}$ is an admissible variation for the graphene displacement. $\partial \mathcal{B}_{\mathrm{c}}$ and $\partial \mathcal{B}_{0\mathrm{c}}$ are the contact boundaries between graphene and the substrate in the current and reference configurations, respectively. $\boldsymbol{f}_{\mathrm{g}}^0$ and $\boldsymbol{f}_{\mathrm{g}}$ can be connected by $J$ as
\eqb{lll}
\ds \boldsymbol{f}_{\mathrm{g}}^0=J\,\boldsymbol{f}_{\mathrm{g}}~.
\label{e:vdw_contact_bf}
\eqe
$G_{\mathrm{c}}^{e}$ can be obtained by changing \textcolor{cgn2}{the} integral domain to \textcolor{cgn2}{an} element. The nodal contact force can be obtained as
\eqb{lll}
\ds \boldsymbol{\mathrm{f}}_{\mathrm{c}}^{e} \is \ds -\int\limits_{\partial\mathcal{B}_{\mathrm{c}}^{e} }{\mN^{\text{T}}\,\boldsymbol{f}_{\mathrm{g}}~\dif a}=
-\int\limits_{\partial \mathcal{B}_{0\mathrm{c}}^{e} }{\mN^{\text{T}}\,\boldsymbol{f}_{\mathrm{g}}^0~\dif A}~.
\eqe
%where $\mN$ is shapes function array of graphene surface elements.
The substrate is assumed to be rigid and $\rho_{\mathrm{s}}$ is constant. Furthermore, $\rho_{\mathrm{g}}^0=J\rho_{\mathrm{g}}$ is the density in the reference configuration and can be considered constant in the linearization. Based on these assumptions, the contact stiffness matrix can be written as
\eqb{lll}
\ds \mathrm{\mk}_{\text{c}}^{e} = -\int\limits_{\partial\mathcal{B}_{0c}^{e}}{\mN^{\text{\text{T}}}\,\pa{\boldsymbol{f}_{\mathrm{g}}^0}{\bu_e}~\dif A}~.
\eqe
$\ds \pa{\boldsymbol{f}_{\mathrm{g}}^0}{\bu_e}$ can be written as
\eqb{lll}
\ds \pa{\boldsymbol{f}_{\mathrm{g}}^0}{\bu_e}=\pa{\boldsymbol{f}_{\mathrm{g}}^0}{\bx_{\mathrm{g}}}\,\pa{\bx_{\mathrm{g}}}{\bu_e} =\pa{\boldsymbol{f}_{\mathrm{g}}^0}{\bx_{\mathrm{g}}}\,\mN~,
\eqe
where $\pa{\boldsymbol{f}_{\mathrm{g}}^0}{\bx_{\mathrm{g}}}$ can be written as
%\eqb{lll}
%\ds \boldsymbol{f}_{\mathrm{g}}^0=f_{\mathrm{g}}^0\,\bn_{\mathrm{p}},
%\eqe
\eqb{lll}
\ds \pa{\boldsymbol{f}_{\mathrm{g}}^0}{\bx_{\mathrm{g}}} \is \ds \left(f_{\mathrm{g}}^0\right)^{\prime}\bn_{\mathrm{p}} \otimes \bn_{\mathrm{p}} + \frac{f_{\mathrm{g}}^0}{\bar{\kappa}_1^{-1}+r}\hat{\ba}_1^{\mathrm{p}} \otimes\hat{\ba}_1^{\mathrm{p}} +\frac{f_{\mathrm{g}}^0}{\bar{\kappa}_2^{-1}+r}\hat{\ba}_2^{\mathrm{p}} \otimes\hat{\ba}_2^{\mathrm{p}}~.
\eqe
$\hat{\ba}_{\alpha}^{\mathrm{p}}$ are the normalized tangent vectors along the principal curvatures of the surface of the substrate. $(f_{\mathrm{g}}^0)^{\prime}$ is defined as
\eqb{lll}
\ds \left(f_{\mathrm{g}}^0\right)^{\prime} \dis \ds \pa{f_{\mathrm{g}}^0}{r}~.
\eqe
Convex geometries are assumed to have a positive curvature in the computation of the contact stiffness. $\bar{\kappa}_{\alpha}$ are connected to the principal curvatures of the surface of the substrate $\kappa_{\alpha}$ by
\eqb{lll}
\bar{\kappa}_{\alpha} \is -\kappa_{\alpha}~.
\eqe
\subsection{Modeling of the substrate geometry}\label{s:Modeling_of_substrate_geometry}
In this section, an analytical description for the substrate geometry is presented. The substrate is divided into three area which are a flat, cylindrical and torus part. In each area, the normal unit vector to the surface, the principal curvatures and directions of the curvature tensor are given. These quantities are used in computation of the contact force and the contact stiffness matrix.
\subsubsection{Torus part}
The torus parametric description can be written as
\eqb{lll}
\ds \bolds{r}(\varphi,\omega) = \left[R+r\cos(\omega)\right]\,\cos(\varphi)\,\bolds{e}_1 +\left[R+r\cos(\omega)\right]\sin(\varphi)\,\bolds{e}_2 +R\sin(\omega)\,\bolds{e}_3~,
\eqe
where $R$ and $r$ are major and minor radii of the torus. $\varphi$ and $\omega$ are parametric coordinates. The tangent vectors then follow as
\eqb{lll}
\ds \bolds{a}_1 = -\left[R+r\cos(\omega)\right]\sin(\varphi)\,\bolds{e}_1+\left[R+r\cos(\omega)\right]\cos(\varphi)\,\bolds{e}_2~,\\[2mm]
\ds \bolds{a}_2 = -R\cos(\varphi)\sin(\omega)\,\bolds{e}_1-R\sin(\varphi)\sin(\omega)\,\bolds{e}_2+R\cos(\omega)\,\bolds{e}_3~.
\eqe
The co-variant components of curvature tensor become
\eqb{lll}
\ds b_{11} = -\left[R+r\cos(\omega)\right]\cos(\omega)~,\\
\ds b_{22} = -R~,\\
\ds b_{12} = b_{21} =0~.
\eqe
The principal curvatures of the torus can be written as
\eqb{lll}
\ds {\kappa}_1 = b_1^1 =\frac{b_{11}}{a_{11}}=-\frac{\cos(\omega)}{R+r\cos(\omega)}~,\\
\ds {\kappa}_2 = b_2^2 =\frac{b_{22}}{a_{22}}=-\frac{1}{r}~,
\eqe
and $\bar{\kappa}_{\alpha}$ can be written as
\eqb{lll}
\ds \bar{\kappa}_1 = \frac{\cos(\omega)}{R+r\cos(\omega)}~,\\
\ds \bar{\kappa}_2 = \frac{1}{r}~.
\eqe
It should be denoted that the principal directions of the curvature tensor and the tangent vectors are identical in the torus. The CPP of a point $(x_0, y_0, z_0)$ onto the surface of the torus can be written as \citep{Li2016_01}
\eqb{lll}
\bolds{x}_{\mathrm{p}}^{\text{T}} = [x_0\,w_0, y_0\,w_0, z_0\,w_1]~,
\eqe
where $w_0$ and $w_1$ are defined as
\eqb{lll}
\ds w_1 \dis \ds \frac{r}{\sqrt{(\sqrt{x_0^2+y_0^2}-R)^2+z_0^2}}~,\\
\ds w_0 \dis \ds \frac{R}{\sqrt{x_0^2+y_0^2}}+\left(1-\frac{R}{\sqrt{x_0^2+y_0^2}}\right)w_1~.
\eqe
The normal unit vector to the surface of the torus can be written as
\eqb{lll}
\ds \bolds{n}_{\mathrm{p}}^{\text{T}} = [\cos(\varphi)\cos(\omega),~\sin(\varphi)\,\cos(\omega),~\sin(\omega)]~.
\eqe
A point on the axis and major circle of the torus has infinite number of projections. They are not considered in the CPP computation.
\subsubsection{Cylinder and flat parts}
The normal unit vector to the internal area of a cylinder is
\eqb{lll}
\ds \bolds{n}_{\mathrm{p}}^{\text{T}} = -[\cos(\theta),\sin(\theta), 0]~,
\eqe
where $\theta$ is defined as
\eqb{lll}
\ds \theta \dis \ds \arctan\left(\frac{y_0}{x_0}\right)~.
\eqe
The principal curvatures $\kappa_{\alpha}$ of the internal surface of the cylinder are zero and $1/R$. The principal curvatures of a flat area are zero and the computation of its normal unit vector is trivial.

\bibliographystyle{apalike}
\bibliography{bibliography}
%\bibliography{bibliography}
%\bibliography{ShellBib,bibliography}

\end{document}